\documentclass[a4paper,11pt]{article}
\pdfoutput=1 
\usepackage{jheppub}
\usepackage{amsmath}
\usepackage{amsfonts}
\usepackage{amssymb}
\usepackage{slashed}
\usepackage{bm}
\usepackage{graphicx}%
\usepackage[T1]{fontenc} 
\usepackage{mathrsfs}
\usepackage{yfonts}
\usepackage{pifont}

\DeclareMathAlphabet{\mathpzc}{OT1}{pzc}{m}{it}

\newcommand{\ie}{i\epsilon}

\newcommand{\dhd}{{\textstyle d}
\lower.03ex\hbox{\kern-0.38em$^{\scriptstyle-}$}\kern-0.05em{}}
\newcommand{\dbar}{{\textstyle \delta}
\lower.03ex\hbox{\kern-0.38em$^{\scriptstyle-}$}\kern-0.05em{}}
\newcommand{\half}{{1\over 2}}
\newcommand{\bu}{{\bullet}}

\newcommand{\bare}{{\bar e}}
\newcommand{\barf}{{\bar f}}
\newcommand{\barg}{{\bar g}}
\newcommand{\barh}{{\bar h}}
\newcommand{\barj}{{\bar j}}

\newcommand{\bsi}{{\bar \psi}}

\newcommand{\bhi}{{\bar \chi}}

\newcommand{\barA}{{\bar A}}
\newcommand{\barB}{{\bar B}}

\newcommand{\Bsi}{{\bar \Psi}}
\newcommand{\Bxi}{{\bar \Xi}}

\newcommand{\cald}{{\cal D}}  
\newcommand{\cale}{{\cal E}}
\newcommand{\calf}{{\cal F}}

\newcommand{\calj}{{\cal J}} 
 
\newcommand{\kal}{{\cal L}} 
  
\newcommand{\calo}{{\cal O}}

\newcommand{\calW}{{\cal W}}

\newcommand{\hatp}{{\hat p}}

\newcommand{\hsi}{{\hat \psi}}
\newcommand{\hbsi}{\hat {\bar\psi}}

\newcommand{\tilj}{{\tilde j}} 

\newcommand{\tilq}{{\tilde q}}

\newcommand{\tilA}{{\tilde A}}
\newcommand{\tilB}{{\tilde B}}

\newcommand{\tilF}{{\tilde F}}

\newcommand{\tigma}{\tilde {\sigma}}

\newcommand{\brA}{\breve {A}} 
\newcommand{\breB}{\breve {B}}
\newcommand{\breF}{\breve {F}}

\newcommand{\cheV}{\check {V}} 
\newcommand{\chew}{\check {W}} 
\newcommand{\cheW}{\check {W}}

\newcommand{\chekalw}{\check {\cal W}}

\newcommand{\chepizw}{\check {\pizw}} 
\newcommand{\chepizW}{\check {\pizw}}

\newcommand{\matW}{\mathbb{W}}

\newcommand{\pizf}{\mathpzc{F}}
\newcommand{\pizh}{\mathpzc{H}}
\newcommand{\pizw}{\mathpzc{W}}
\newcommand{\pizW}{\mathpzc{W}}

\newcommand{\notk}{{\not\! k}}
\newcommand{\notp}{{\not\! p}}

\newcommand{\notA}{{\not\! \!A}}
\newcommand{\notB}{{\not\! \!B}}
\newcommand{\notD}{{\not \!\!D}}

\newcommand{\rma}{{\rm a}}
\newcommand{\rmb}{{\rm b}}
\newcommand{\rmA}{{\rm A}}
\newcommand{\rmF}{{\rm F}}
\newcommand{\rmH}{{\rm H}}
\newcommand{\rmS}{{\rm S}}

\newcommand{\dd}{{\Delta\Delta}}
\pdfoutput=1
\abstract{
The  Drell-Yan process is studied in the framework of TMD factorization  in the Sudakov region $s\gg Q^2\gg q_\perp^2$  
corresponding to recent LHC experiments with $Q^2$ of order of mass of Z-boson and transverse momentum 
of DY pair $\sim$ few tens GeV. The DY hadronic tensors are expressed in terms of quark and quark-gluon TMDs 
with ${1\over Q^2}$ and ${1\over N_c^2}$ accuracy. 
It is demonstrated that in the leading order in $N_c$ the higher-twist quark-quark-gluon TMDs reduce to 
leading-twist TMDs due to QCD equation of motion. The resulting hadronic tensors  
 depend on two leading-twist TMDs: $f_1$ responsible for total DY cross section,
 and Boer-Mulders function $h_1^\perp$.  The corresponding qualitative and semi-quantitative predictions
 seem to agree with LHC data on five angular coefficients $A_0-A_4$ of DY pair production. 
 The remaining three coefficients
$A_5-A_7$  are determined by quark-quark-gluon TMDs multiplied by extra ${1\over N_c}$ so they 
appear to be relatively small in accordance with LHC results. 
  }
\keywords{}
\arxivnumber{}
\affiliation{ Physics Department, Old Dominion University, Norfolk, VA 23529, USA and Thomas Jefferson National Accelerator Facility, Newport News, VA 23606, USA}

\emailAdd{balitsky@jlab.org}
\begin{document}

\title{\boldmath Drell-Yan angular lepton distributions at small $x$ from TMD factorization.}
\author{Ian Balitsky }
\preprint{JLAB-THY-21-3407}
\maketitle

\flushbottom

\section{Introduction\label{aba:sec1}}

The Drell-Yan (DY) process  \cite{Drell:1970wh}  has been extensively studied in high energy physics field for precise tests of QCD, investigation of the structure of the proton, 
and searches for possible new physics.  Since the advent of high-energy colliders, the attention shifted to  processes with
large invariant mass of DY pair  produced  both by photon and $Z$-boson. 
The important part of these studies is the  transverse-momentum dependence of angular  distribution  of DY lepton  pairs  with  l
arge invariant mass produced in hadronic collisions. It was extensively studied in the framework of collinear 
factorization \cite{Collins:1977iv,Collins:1978yt,Mirkes:1994eb,Mirkes:1994dp,Berger:2007si,Berger:2007jw}
leading to good agreement with experiment \cite{Lambertsen:2016wgj}.
If one considers, however, the DY process at transverse momentum of lepton pair much smaller than their invariant mass, the collinear factorization 
should be replaced by TMD factorization  \cite{Collins:2011zzd, Collins:1981uw, Collins:1984kg, Ji:2004wu, GarciaEchevarria:2011rb}. 
In this paper the DY process will be studied in so-called Sudakov plus small-$x$ region $s\gg Q^2\gg q_\perp^2$ where 
$Q^2$ is the invariant mass of DY pair and $q_\perp$ is the transverse momentum of produced leptons. The typical case is the lepton pair 
production at LHC with  DY invariant mass in the vicinity of $Z$-boson and transverse momentum of DY pair of order of ten or few tens of GeV.

 This paper is the third in a series of papers devoted to description of DY process in terms of TMD rapidity factorization. 
 In the first paper \cite{Balitsky:2017gis}, A. Tarasov and the author calculated power corrections
 to the total cross section of $Z$-boson production using the method developed in earlier paper  \cite{Balitsky:2017flc}. 
 The second paper \cite{Balitsky:2020jzt} was devoted to calculation of angular coefficients 
 for DY process mediated by virtual photon. Unfortunately, it is hard to compare these results with experiment due to the fact 
 that the  LHC measurements of angular distributions
 are performed at the invariant mass $\sim$ 100 GeV where the contribution of $Z$-boson is dominant. 
 The present paper is devoted to generalization of the approach of papers \cite{Balitsky:2017gis} and \cite{Balitsky:2020jzt}  to 
 angular distributions of DY pair production by unpolarized protons at LHC kinematics. 

The differential cross section of DY process is determined by the sum of products of leptonic tensors and hadronic tensors. 
The leptonic tensors are given by simple first-order EW diagrams while hadronic tensors are determined by QCD correlation functions
\begin{eqnarray}
\hspace{-1mm}
W_{\mu\nu}(q)~=~{1\over (2\pi)^4}\!\int\! d^4x~e^{-iqx}\langle p_A,p_B|j_\mu(x)j_\nu(0) |p_A,p_B\rangle  
\label{W}.
\end{eqnarray}
where $p_A,p_B$ are hadron momenta, $q$ is the momentum of DY pair,  and $j_\mu$ is either electromagnetic  or $Z$-boson current.

 As was mentioned above, a golden standard of QCD analysis of such hadronic tensors in the region where transverse momenta are much smaller than the invariant mass of the DY pair
  is the TMD factorization. The leading-twist hadronic tensors can be represented as
\begin{eqnarray}
&&\hspace{-2mm}
W_i~=~\sum_{\rm flavors}e_f^2\!\int\! d^2k_\perp
\cald_{f/A}^{(i)}(x_A,k_\perp)\cald_{f/B}^{(i)}(x_B,q_\perp-k_\perp)C_i(q,k_\perp)
\nonumber\\
&&\hspace{-2mm}
+~{\rm power ~corrections}~+~{\rm Y-terms}
\label{TMDf}
\end{eqnarray}
where $\cald_{f/A}(x_A,k_\perp)$ is the TMD density of  a parton $f$  in hadron $A$
with fraction of momentum $x_A$ and transverse momentum $k_\perp$, $\cald_{f/B}(x_B,q_\perp-k_\perp)$ is a similar quantity for hadron $B$, and  
 coefficient functions $C_i(q,k)$ are determined by the cross section $\sigma(ff\rightarrow \mu^+\mu^-)$  
 of production of DY pair of invariant mass $q^2$ in the scattering of two partons. 
 The DY angular distributions  are conventionally parametrized by eight angular coefficients $A_i$  in Collins-Soper frame \cite{Collins:1977iv}
\begin{eqnarray}
&&\hspace{-1mm}
{d\sigma\over dQ^2dy d\Omega_l}~=~{3\over 16\pi}{d\sigma\over dQ^2dy}\Big[(1+c_\theta^2)
+{A_0\over 2}(1-3c_\theta^2)+A_1s_{2\theta}c_\phi+{A_2\over 2}s_\theta^2c_{2\phi}
\nonumber\\
&&\hspace{-1mm}
+~A_3s_\theta c_\phi+A_4c_\theta+A_5s_\theta^2s_{2\phi}+A_6 s_{2\theta}s_\phi+A_7s_\theta s_\phi\Big]
\label{Ais}
\end{eqnarray}
where $y$ is the rapidity of DY pair and $c_\phi\equiv\cos\phi$,  $s_\phi\equiv\sin\phi$ etc.
The aim of this paper is to express $A_i(Q^2,q_\perp^2)$ at Sudakov kinematics $s\gg Q^2\gg q_\perp^2$ in terms of TMDs and compare
(at least qualitatively) to ATLAS \cite{Aad:2016izn} and CMS \cite{Khachatryan:2015paa} measurements. 
 
Unfortunately, the TMD analysis of Drell-Yan angular distributions $A_i$
is hindered by the fact that not all hadronic tensors are determined by leading-twist quark-antiquark TMDs which have parton interpretation. 
Some tensor structures in the r.h.s. of Eq. (\ref{W}) are determined by power corrections to leading-twist TMDs expressed in terms of quark-antiquark-gluon distributions
which are virtually unknown. 
Fortunately, as demonstrated recently in Ref. \cite{Balitsky:2017gis},  at the leading order in $N_c$ 
these power corrections are still determined by 
leading-twist TMDs.  
Moreover, at small $x_A$ and $x_B$ the majority of hadronic  tensors 
depends only on two leading-twist TMDs: $f_1$ responsible for total DY cross section,
 and Boer-Mulders function $h_1^\perp$. The rest of of hadronic tensors determined by power corrections 
due to  quark-antiquark-gluon distributions are down by at least one ${1\over N_c}$ factor which seems to qualitatively agree with LHC measurements
of angular distributions. 

Note that in addition to power corrections due to QCD dynamics, there are  fiducial power corrections arising from fiducial cuts on experimental measurements.
 These cuts introduce  linear power corrections in $q_\perp/Q$, see the discussion in Ref. \cite{Ebert:2020dfc}. In this paper we do not consider  fiducial power corrections.
 
 The paper is organized as follows. In Sect. \ref{sec:DYdef} I set up the notations, present the formula for differential cross section of DY lepton pair, and
 list the relevant hadronic tensors. In Sect. \ref{sec:rapfak} I briefly outline the method of calculation of power corrections to hadronic tensors 
 developed in Refs. \cite{Balitsky:2017gis} and  \cite{Balitsky:2017flc}. Sect. \ref{sec:foton} contains the streamlined calculation of photon-mediated DY
 process which is used as a reference point to calculation of $Z$-mediated and interference terms in Sections \ref{sec:Z} and \ref{sec:inter}. The results
 for hadronic tensors and angular coefficients are presented in Sect. \ref{sec:results} whic also contains the comparison to LHC measurements. 
 Conclusions are summarized in Sect. \ref{sec:konklu} and the necessary technical details are listed in the Appendix.

\section{Drell-Yan cross section in the Sudakov region \label{sec:DYdef}}

At high energies, the production of a neutral $e^+e^-$ (or $\mu^+\mu^-$) pair in hadron-hadron collisions is mediated by virtual photon or by $Z$-boson, see Fig. 1
\begin{figure}[htb]
\begin{center}

\vspace{-0mm}
\includegraphics[width=150mm]{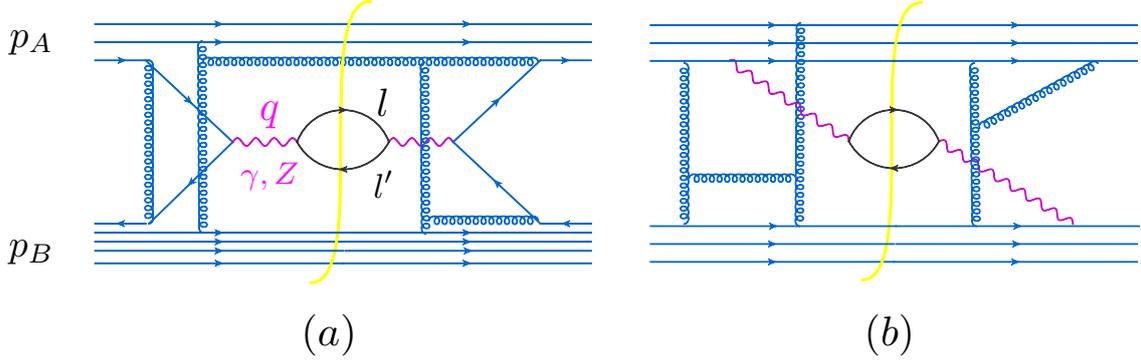}
\end{center}

\vspace{-66mm}
\caption{Typical annihilation-type (a) and exchange-type (b) diagrams for DY particle production. \label{fig:1}}
\end{figure}
\begin{eqnarray}
h_A(p_A) + h_B(p_B) \to \gamma,Z(q) + X \to l_1(l) + l_2(l') + X,
\end{eqnarray}
where $h_{A,B}$ denote the colliding hadrons with momenta $p_A$ and $P_B$ and $l_{1, 2}$ denote the outgoing lepton pair with total momentum $q = l + l'$. 
To avoid cluttering of $\mu$'s if our formulas, we will consider production of $e^+e^-$ pairs, the results  for $\mu^+\mu^-$ pairs are the same.

The relevant terms of the Lagrangian for quark fields $\psi^f$ are
\begin{eqnarray}
&&
\kal_\gamma~=~ e\!\int\! d^4x~J_\mu A^\mu(x),~~~~~~~~~J_\mu~=~\bare\gamma_\mu e-\sum_{\rm flavors}e_f\bsi^f \gamma_\mu \psi^f
\label{2.2}\\
&&
\kal_Z~=~ e\!\int\! d^4x~\calj_\mu Z^\mu(x),~~~~~~~~~\calj_\mu
~=~c_e\bare(a_e-\gamma_5)e-\sum_{\rm flavors}c_f\bsi^f \gamma_\mu(a_f-\gamma_5) \psi^f
\nonumber
\end{eqnarray}
where $e$ is the positron charge and
\begin{eqnarray}
&&
c_{u,c}~=~{1\over 4c_Ws_W},~~~ a_{u,c}~=~1-{8\over 3}s_W^2,~~~c_{d,s}~=~-{1\over 4c_Ws_W},~~~a_{d,s}~=~1-{4\over 3}s_W^2,
\nonumber\\
&&
c_e~=~{1\over 4c_Ws_W},~~~a_e~=~1-4s_W^2,~~~~~~~c_W\equiv\cos\theta_W,~~~s_W\equiv\sin\theta_W.
\end{eqnarray}

The differential cross section of  production of pair of leptons with momenta $l$ and $l'$ 
by scattering of two unpolarized protons is given by 
\begin{eqnarray}
&&\hspace{-1mm}
d\sigma~=~{d^4q\over 2s}\!\int\!{d^3ld^3l'\delta(q-l-l')\over (2\pi)^6E_lE'_l}e^4\!\int\! dx~e^{-iqx}
\langle p_A,p_B|[A^\lambda J_\lambda(x)+Z^\lambda\calj_\lambda(x)]
\nonumber\\
&&\hspace{-1mm}
\times~
\!\int\! dx' dy' \big[A^\mu\bare\gamma_\mu e(x')+c_eZ^\mu\bare\gamma_\mu(a_e-\gamma_5)e(x')\big]
\sum_X|X\rangle\langle  X|
\nonumber\\
&&\hspace{-1mm}
\times~\sum_{s,s'}|l,s;l',s'\rangle\langle l,s;l',s'|
\big[A^\nu\bare\gamma_\nu e(x')+c_eZ^\nu\bare\gamma_\nu(a_e-\gamma_5)e(x')\big]
\nonumber\\
&&\hspace{-1mm}
\times~[A^\rho J_\rho(0)+Z^\rho\calj_\rho(0)]|p_A,p_B\rangle 
\label{2.3}
\end{eqnarray}
where $\sum_X$ denotes summation over all intermediate hadron states.
Performing contractions  to get photon, Z-boson and lepton propagators, one obtains
\begin{eqnarray}
&&\hspace{-1mm}
d\sigma~=~{d^4q\over 2s}\!\int\!{d^3ld^3l'\delta(q-l-l')\over \pi^2E_lE'_l}{e^4\over N_c}
\Big[{1\over q^4}L^{\mu\nu}W_{\mu\nu}(q)
\nonumber\\
&&\hspace{-11mm}
+~{c_e^2\over |m_Z^2-q^2|^2+\Gamma_Z^2m_Z^2}\big[(a_e^2+1)L^{\mu\nu}
-2ia_e\epsilon^{\mu\nu\lambda\rho}l_\lambda {l'}_\rho\big]W^{\rm Z}_{\mu\nu}(q)
\nonumber\\
&&\hspace{-11mm}
+~2c_e{(q^2-m_Z^2)/q^2\over |m_Z^2-q^2|^2+\Gamma_Z^2m_Z^2} W^{\rm I1}_{\mu\nu}(q)
\big[a_eL^{\mu\nu}-i\epsilon^{\mu\nu\lambda\rho}l_\lambda {l'}_\rho\big]
\nonumber\\
&&\hspace{-11mm}
+~2c_e{i\Gamma_Zm_Z/q^2\over |m_Z^2-q^2|^2+\Gamma_Z^2m_Z^2} W^{\rm I2}_{\mu\nu}(q)
\big[a_eL^{\mu\nu}-i\epsilon^{\mu\nu\lambda\rho}l_\lambda {l'}_\rho\big]
\label{desigma}
\end{eqnarray}
where $L_{\mu\nu}=l_\mu l'_\nu+l'_\mu l_\nu-g_{\mu\nu}l\cdot l'$ and hadronic tensors $W^i_{\mu\nu}$ are 
defined as
\begin{eqnarray}
&&\hspace{-1mm}
W^i_{\mu\nu}(q)~=~{1\over (2\pi)^4}\!\int\!d^4x ~e^{-iqx} W^i_{\mu\nu}(x)~~~
\nonumber\\
&&\hspace{-1mm}
W_{\mu\nu}(x)~=~N_c\langle A,B|J_\mu(x)J_\nu(0)|A,B\rangle
\nonumber\\
&&\hspace{-1mm}
W^{\rm Z}_{\mu\nu}(x)~=~N_c\langle A,B|\calj_\mu(x)\calj_\nu(0)|A,B\rangle
\nonumber\\
&&\hspace{-1mm}
W^{\rm I1}_{\mu\nu}(x)~=~{N_c\over 2}\langle A,B|J_\mu(x)\calj_\nu(0)+\calj_\mu(x)J_\nu(0)|A,B\rangle
\nonumber\\
&&\hspace{-1mm}
W^{\rm I2}_{\mu\nu}(x)~=~{N_c\over 2}\langle A,B|\calj_\mu(x)J_\nu(0)-J_\mu(x)\calj_\nu(0)|A,B\rangle
\label{ws}
\end{eqnarray}
Hereafter $|p_A,p_B\rangle\equiv|A,B\rangle$ for brevity.
Note that  hadronic tensors are defined  with  an extra $N_c$ so 
that the leading-twist contribution will be $\sim N_c^0$.

To convolute with leptonic tensors, we need to find symmetric and antisymmetric parts separately so we define
\begin{equation}
\begin{array}{lll}
W^{\rm ZS}_{\mu\nu}~=~\half(W^{\rm Z}_{\mu\nu}+\mu\leftrightarrow\nu), &~~~~~~&
W^{\rm ZA}_{\mu\nu}~=~\half(W^{\rm Z}_{\mu\nu}-\mu\leftrightarrow\nu),
\\
 W^{\rm I1S}_{\mu\nu}~=~\half(W^{\rm I1}_{\mu\nu}+\mu\leftrightarrow\nu), &~~~~~~&
W^{\rm I1A}_{\mu\nu}~=~\half(W^{\rm I1}_{\mu\nu}-\mu\leftrightarrow\nu), 
\\
W^{\rm I2S}_{\mu\nu}~=~\half(W^{\rm I2}_{\mu\nu}+\mu\leftrightarrow\nu), &~~~~~~&
W^{\rm I2A}_{\mu\nu}~=~\half(W^{\rm I2}_{\mu\nu}-\mu\leftrightarrow\nu).
\end{array}
\label{warray}
\end{equation}
We will calculate these hadronic tensors with $O\big({1\over Q^2}\big)$ accuracy and express them in terms of TMDs like Eq. (\ref{TMDf}). 
It turns out that each $W^i$ is a sum of three parts
\begin{equation}
W^i_{\mu\nu}(q)~=~(W_1)^i_{\mu\nu}(q)+(W_2)^i_{\mu\nu}(q)+(W^{\rm ex})^i_{\mu\nu}(q)
\end{equation}
 The first  part is determined by leading-twist TMDs $f_1$ and $h_1^\perp$ as mentioned in the Introduction and satisfies the condition 
 \footnote{Strictly speaking, the Z-boson current is not conserved so one should not expect $q^\mu(W_1)^{\rm Z,I}_{\mu\nu}(q)=0$.
However, if we consider quarks to be massless, the  non-conservation is due to axial anomaly so the corresponding terms in  $q^\mu(W_1)^i_{\mu\nu}(q)$
will be proportional to $\langle p_A,p_B|\alpha_sF_{\mu\nu}\tilde{F}^{\mu\nu}(x)j_\nu(0)|p_A,p_B\rangle$. Such matrix elements  will be non-zero only at the two-loop level  $O(\alpha_s^2)$  which is beyond the accuracy of this paper.
}
\begin{equation}
q^\mu(W_1)^i_{\mu\nu}(q)=0
\end{equation}
The two remaining terms  $(W_2)^i_{\mu\nu}$ and $(W^{\rm ex})^i_{\mu\nu}$ are power corrections 
$\sim O\big({q_\perp^2 g^\perp_{\mu\nu}\over Q^2}, {q^\perp_\mu q^\perp_\nu\over Q^2}\big)$ which come from the diagrams of the Fig. \ref{fig:1}a,b type, respectively. 
They
are expressed in terms of quark-antiquark-gluon matrix elements
which cannot be reduced to leading-twist TMDs. The term $(W_2)^i_{\mu\nu}$ is $\sim {q_\perp^2\over Q^2}$ and $\sim N_c^0$ as 
while $(W^{\rm ex})^i_{\mu\nu}$ is $\sim {1\over N_c}$.  On the other hand,  since
 $(W^{\rm ex})^i_{\mu\nu}$ comes from exchange-type diagrams it may be numerically larger than $(W_2)^i_{\mu\nu}(q)$ coming from annihilation-type diagrams.

\section{TMD factorization from rapidity factorization \label{sec:rapfak}}

We use 
Sudakov variables $p=\alpha p_1+\beta p_2+p_\perp$, where $p_1$ and $p_2$ are light-like vectors close to $p_A$ and $p_B$ so that 
$p_A=p_1+{m^2\over s}p_2$ and $p_A=p_1+{m^2\over s}p_2$ with $m$ being the proton mass.
Also, we use the notations $x_\bu\equiv x_\mu p_1^\mu$ and $x_\star\equiv x_\mu p_2^\mu$ 
for the dimensionless light-cone coordinates ($x_\star=\sqrt{s\over 2}x_+$ and $x_\bu=\sqrt{s\over 2}x_-$). Our metric is $g^{\mu\nu}~=~(1,-1,-1,-1)$ 
which we will frequently rewrite as a sum of longitudinal part and transverse part: 
\begin{equation}
g^{\mu\nu}~=~g_\parallel^{\mu\nu}+g^{\mu\nu}_\perp~=~{2\over s}\big(p_1^\mu p_2^\nu+p_2^\mu p_1^\nu)+g_\perp^{\mu\nu}
\label{delta}
\end{equation}
Consequently,  $p\cdot q~=~(\alpha_p\beta_q+\alpha_q\beta_p){s\over 2}-(p,q)_\perp$ where $(p,q)_\perp\equiv -p_iq^i$. 
Throughout the paper, the sum over the Latin indices $i$, $j$, ... runs over two transverse components while the sum over Greek indices $\mu$, $\nu$, ... runs over four components as usual.

\begin{figure}[htb]
\begin{center}

\vspace{-4mm}
\includegraphics[width=155mm]{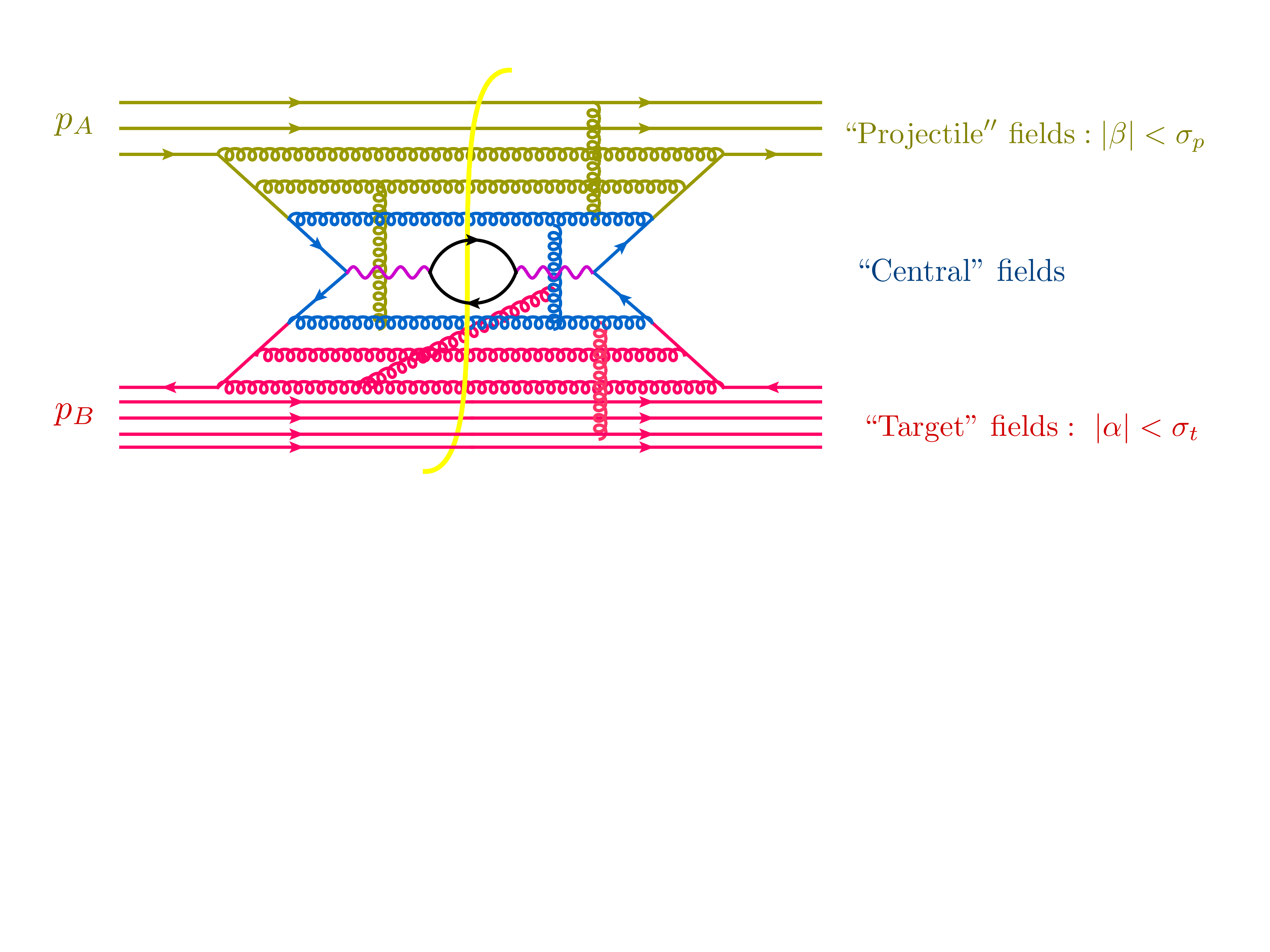}
\end{center}

\vspace{-64mm}
\caption{Rapidity factorization for DY particle production \label{fig:2}}
\end{figure}

Following Ref. \cite{Balitsky:2017flc}  we separate quark and gluon fields into three sectors (see figure \ref{fig:2}): 
``projectile'' fields $A_\mu, \psi_A$ 
with $|\beta|<\sigma_p$, 
``target'' fields $B_\mu, \psi_B$ with $|\alpha|<\sigma_t$ and ``central rapidity'' fields $C_\mu,\psi_C$ with $|\alpha|>\sigma_t$ and $|\beta|>\sigma_p$, 
see Fig. \ref{fig:2}. 
\footnote{Although the kinematics is best suited for LHC collider, I call $A$ hadron  ``projectile''  and $B$  hadron 
``target'' for convenience.}
Our goal is to integrate over central fields and get
the amplitude in the factorized form, i.e. as a product of functional integrals over $A$ fields representing projectile matrix elements (TMDs of the projectile) 
and functional integrals over $B$ fields representing target matrix elements (TMDs of the target).
In the spirit of background-field method, we ``freeze'' projectile and target fields and get a sum of diagrams in these external fields. 
Since  $|\beta|<\sigma_p$ in the projectile fields and $|\alpha|<\sigma_t$  in the target fields, at the  tree level 
one can set with power accuracy $\beta=0$ for the  projectile fields and $\alpha=0$ for the target fields - the corrections will
be $O\big({m^2\over\sigma_p s}\big)$ and  $O\big({m^2\over\sigma_t s}\big)$. 
In the coordinate space, projectile fields depend on $x_\bu$ and $x_\perp$ and target ones on $x_\star$ and $x_\perp$. 
Beyond the tree level, the integration over $C$ fields produces
 logarithms of the cutoffs $\sigma_p$ and $\sigma_t$ which match  the corresponding
logs in TMDs of the projectile and the target, see the discussion in Ref. \cite{Balitsky:2020jzt}

\begin{figure}[htb]
\begin{center}
\includegraphics[width=99mm]{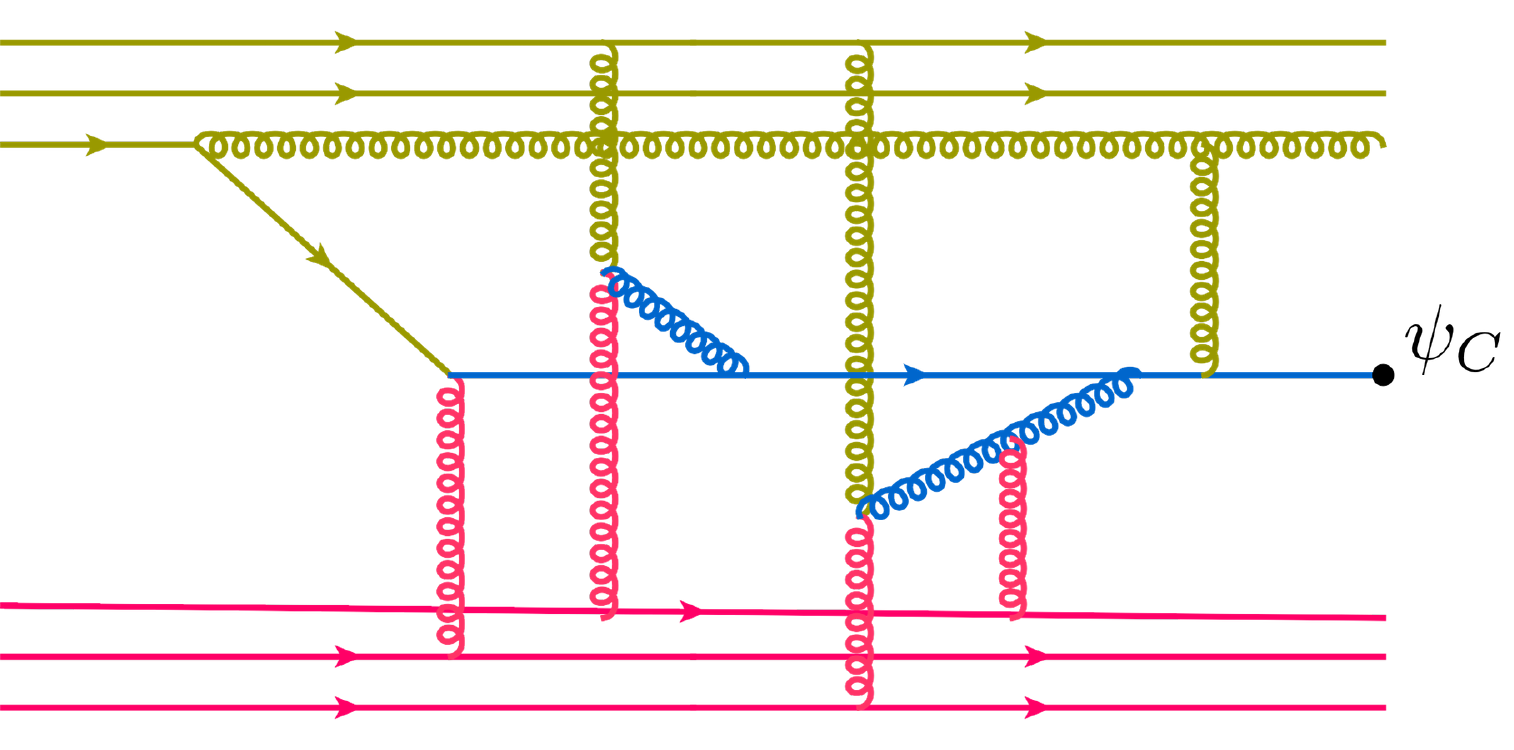}
\end{center}
\caption{Typical diagram for the classical field with projectile/target sources. The Green functions of central fields are given by retarded propagators.   \label{fig:3}}
\end{figure}

As discussed in Ref. \cite{Balitsky:2020jzt} ,  central fields at the tree level are given by a set of Feynman diagrams with {\it retarded} propagators 
in background field $A + B$ and $\psi_A+\psi_B$,  see figure \ref{fig:3}.
\footnote{We take into account only $u,d,s,c$ quarks and consider them massless. In principle, one can include ``massless'' $b$-quark for $q_\perp^2\gg m_b^2$.}
The set of such ``retarded'' diagrams represent the solution of QCD equations of motion with sources being projectile and target fields.
After summation of these diagrams the hadronic tensor (\ref{W}) can  be represented as
\begin{equation}
\hspace{-1mm}  
W_{\mu\nu}~=~\frac{1}{(2\pi)^4}\!\int \! d^4x  e^{-iqx}
\sum_{m,n}\! \int\! dz_m c_{m,n}(q,x)
\langle p_A|\hat\Phi_A(z_m)|p_A\rangle\!\int\! dz'_n\langle p_B| \hat\Phi_B(z'_n)|p_B\rangle.   
\label{W5}
\end{equation}
where $c_{m,n}$ are coefficients and $\Phi$ can be any of  the background fields promoted to operators after integration over
projectile and target fields.

In general, the summation of diagrams of Fig. \ref{fig:3} type is a formidable task which still awaits its solution. Fortunately, as demonstrated in Ref. \cite{Balitsky:2020jzt} ,
at our kinematics we have a small parameter ${q_\perp^2\over Q^2}\ll 1$ and it is possible to expand classical solution for central fields in powers of this parameter.

Now we expand the classical  quark and gluon fields  in powers of  ${p_\perp^2\over p_\parallel^2}\sim{m_\perp^2\over s}$.
It is convenient to choose a gauge where $A_\star=0$ for projectile fields and $B_\bu=0$ for target fields.
(The existence of such gauge was proved in appendix B of Ref. \cite{Balitsky:2017flc} by explicit construction.)
As demonstrated in Ref. \cite{Balitsky:2017gis},
expanding it in powers of $p^2_\perp/p_\parallel^2$ we obtain
\begin{eqnarray}
&&\hspace{-1mm}
\Psi(x)~=~\Psi_1(x)+\Psi_2(x)
+\dots,
\label{klfildz}
\end{eqnarray}
where
\begin{eqnarray}
&&\hspace{-5mm}
\Psi_1~=~\psi_A+\Xi_{1},~~~~
\Xi_{1}~=~-{\slashed{p}_2\over s}\gamma^iB_i{1\over \alpha+i\epsilon}\psi_A
~=~{i\over s}\sigma_{\star i}B^i{1\over \alpha+i\epsilon}\psi_A,
\nonumber\\
&&\hspace{-5mm}
\Bsi_1~=~\bar\psi_A+\Bxi_{1},~~~~
\Bxi_{1}~=~-\big(\bar\psi_A{1\over\alpha-i\epsilon}\big)\gamma^iB_i{\slashed{p}_2\over s}
~=~-{i\over s}\big(\bar\psi_A{1\over\alpha-i\epsilon}\big)B^i\sigma_{\bu i}
\nonumber\\
&&\hspace{-5mm}
\Psi_2~=~\psi_B+\Xi_{2},~~~~
\Xi_{2}~=~-{\slashed{p}_1\over s}\gamma^iA_i{1\over\beta+i\epsilon}\psi_B
~=~{i\over s}\sigma_{\bu i}A^i{1\over\beta+i\epsilon}\psi_B,
\nonumber\\
&&\hspace{-5mm}
\Bsi_2~=~\bar\psi_B+\Bxi_{2},~~~~
\Bxi_{2}~=~-\big(\bar\psi_B{1\over \beta-i\epsilon}\big)\gamma^iA_i{\slashed{p}_1\over s}
~=~-{i\over s}\big(\bar\psi_B{1\over \beta-i\epsilon}\big)A_i\sigma_{\bu i}
\label{fildz0}
\end{eqnarray}
and dots stand for terms subleading in ${q_\perp^2\over Q^2}$ and/or $\alpha_q,\beta_q$ parameters.  In this formula
\begin{eqnarray}
&&\hspace{-1mm}
{1\over \alpha+i\epsilon}\psi_A(x_\bu,x_\perp)~\equiv~-i\!\int_{-\infty}^{x_\bu}\! dx'_\bu~\psi_A(x'_\bu,x_\perp),
\nonumber\\
&&\hspace{-1mm}
\Big(\bsi_A{1\over \alpha-i\epsilon}\Big)(x_\bu,x_\perp)~\equiv~i\!\int_{-\infty}^{x_\bu}\! dx'_\bu~\bsi_A(x'_\bu,x_\perp)
\label{3.25}
\end{eqnarray}
and similarly for ${1\over\beta\pm i\epsilon}$. 
For brevity,  in what follows we denote $\big(\bar\psi_A{1\over\alpha}\big)(x)\equiv \big(\bsi_A{1\over \alpha-i\epsilon}\big)(x)$ and
$\big(\bar\psi_B{1\over\beta}\big)(x)\equiv \big(\bsi_B{1\over \beta-i\epsilon}\big)(x)$. 
The corresponding expansion of classical gluon fields is presented in  Ref. \cite{Balitsky:2017flc}, but we do not need it here. 
\footnote{Since we are dealing with tree approximation and quark equations of motion, it is convenient to include coupling constant $g$
in the definition of gluon fields.}

Let us estimate the relative size of corrections $\Xi$ in Eq. (\ref{fildz0}) at small $x$. As we will see, ${1\over\alpha}$ and ${1\over\beta}$ transform
to ${1\over\alpha_q}$ and ${1\over\beta_q}$ in our TMDs so
\begin{eqnarray}
&&\hspace{-5mm}
\Xi_{1}~\sim~\psi_A{m_\perp\over \alpha_q\sqrt{s}}~\sim~\psi_A{q_\perp\over Q},~~~~\Xi_{2}~\sim~\psi_B{m_\perp\over \beta_q\sqrt{s}}~\sim~\psi_B{q_\perp\over Q}
\label{psi0}
\end{eqnarray}
if $\alpha_q\sim\beta_q\sim {Q\over\sqrt{s}}$ (recall that we assume that the DY pair is emitted in the central region of rapidity).
For example, the correction $\sim [\bsi_A\gamma_\mu\Xi_{2}][\bsi_B\gamma_\nu\Xi_{1}]$ will be of order of ${q_\perp^2\over Q^2}$ in comparison to leading-twist contribution $[\bsi_A\gamma_\mu\psi_B][\bsi_B\gamma_\nu\psi_A]$.
\footnote{The reader may wonder why there are no corrections $\sim {q_\perp^2\over Q^2}$ coming from next terms in the expansion (\ref{klfildz})
like $[\bsi_A(x)\gamma_\mu\psi_B(x)][\bsi_B(0)\gamma_\nu{\gamma^i\over s}\slashed{p}_2{1\over \beta}{1\over \alpha}\gamma^j\partial_iB_j\Psi_1(0)]$. The reason is that 
${1\over \beta}$ between $\bsi_B(0)$ and $B_j(0)$  does \underline{not} transform to ${1\over\beta_q}$ and remains $\sim O(1)$, see the discussion 
in the Appendix 8.3.4 of Ref. \cite{Balitsky:2017gis}.
}
As demonstrated in Ref. \cite{Balitsky:2020jzt} , the relevant terms contributions to hadronic tensors (\ref{ws}) with this accuracy are
\begin{eqnarray}
&&\hspace{-1mm}
g_\perp^{\mu\nu}\Big\{1,~{q_\perp^2\over \alpha_q\beta_qs}\Big\},~~ {q_\perp^\mu q_\perp^\nu\over q_\perp^2}
\Big\{1,~{q_\perp^2\over \alpha_q\beta_qs}\Big\},~~g_\parallel^{\mu\nu}\Big\{1,~{q_\perp^2\over \alpha_q\beta_qs}\Big\},~~
\nonumber\\
&&\hspace{-1mm}
{1\over \beta_qs}\big(p_1^\mu q_\perp^\nu\pm ~p_1^\nu q_\perp^\mu\big),~~
{1\over \alpha_qs}\big(p_2^\mu q_\perp^\nu\pm ~p_2^\nu q_\perp^\mu\big),~~
q_\perp^2{p_1^\mu p_1^\nu \over \beta_q^2s^2},~~
q_\perp^2{p_2^\mu p_2^\nu\over \alpha_q^2s^2}
\label{pc}
\end{eqnarray}
Let us also specify the terms which we do not calculate.  Roughly speaking, they correspond to terms in Eq. (\ref{pc}) multiplied by 
${q_\perp^2\over Q^2}$ or by either $\alpha_q$ or $\beta_q$. 

In addition, in this paper we will consider only leading-$N_c$ power corrections. As we will see below, leading-twist hadronic tensor is $\sim N_c^0$
and power corrections can be $\sim N_c^0$, $\sim{1\over N_c}$, or ${1\over N_c^2}$. The corrections  corrections  $\sim{1\over N_c^2}$ were found in Ref.  \cite{Balitsky:2017gis}
for the case of total cross section, i.e. for  Eq. (\ref{desigma}) integrated over $l$. 
In this paper such corrections $\sim{1\over N_c^2}$ will be neglected. 

Thus, the calculation of power corrections with our accuracy boils down to calculation of tensors (\ref{ws}) with $\psi\rightarrow \Psi_1+\Psi_2$.
In the next sections we will
consider five lines in Eq. (\ref{desigma}) for the differential cross section.

\section{Hadronic tensor for photon-mediated DY process\label{sec:foton}}
In this Section I briefly summarize the calculation of $W^f_{\mu\nu}$ performed in Ref. \cite{Balitsky:2020jzt}  paying attention only to terms
giving non-negligible contributions listed in Eq. (\ref{pc}) at the leading-$N_c$ level. 
The reason is that hadronic tensors listed in Eq. (\ref{ws}) differ from $W_{\mu\nu}$ by
replacement(s) $\gamma_{\mu(\nu)}\rightarrow \gamma_{\mu(\nu)}\gamma_5$ and/or $\mu,\nu$ antisymmetrization instead of symmetrization. Both operations
do not change power counting in ${q_\perp^2\over Q^2}$ and $\alpha_q,\beta_q$ parameters so the calculation of the rest of the terms in  (\ref{ws}) will be based 
on the calculation  of (non-negligible) contributions to in $W_{\mu\nu}$ outlined in this Section.

In this Section we take into account hadronic tensor due to electromagnetic currents of $u,d,s,c$ quarks and consider these quarks to be massless. 
It is convenient to define coordinate-space hadronic tensor multiplied by ${2\over s}$ (and denoted by extra ``check'' mark) as follows
\begin{eqnarray}
&&\hspace{-1mm}
\cheW_{\mu\nu}(x)~\equiv~
{2\over s}\langle p_A,p_B|J_\mu(x)J_\nu(0)|p_A,p_B\rangle
\label{defW}\\
&&\hspace{-1mm}
W_{\mu\nu}(q)~=~{s/2\over(2\pi)^4}\int\!d^4x ~e^{-iqx} \cheW_{\mu\nu}(x).
\nonumber
\end{eqnarray}
For future use, let us also define the hadronic tensor in mixed representation: in momentum longitudinal 
space but in transverse coordinate space
\begin{eqnarray}
&&\hspace{-1mm}
W_{\mu\nu}(q)~=~\int\!d^2x_\perp ~e^{i(q,x)_\perp} W_{\mu\nu}(\alpha_q,\beta_q, x_\perp),
\label{defWcoord}\\
&&\hspace{-1mm}
W_{\mu\nu}(\alpha_q,\beta_q,x_\perp)~\equiv~
{1\over(2\pi)^4}\!\int\! dx_\bu dx_\star ~e^{-i\alpha_qx_\bu-i\beta_q x_\star}
\cheW_{\mu\nu}(x_\bu,x_\star,x_\perp).
\nonumber
\end{eqnarray}
With the definition (\ref{defW}), power counting of contributions to $\cheW_{\mu\nu}(x_\bu,x_\star,x_\perp)$ 
will mirror that of $W_{\mu\nu}(q)$ terms without extra ${1\over s}$ factor.

After integration over central fields in the tree approximation we obtain
\begin{equation}
\hspace{-1mm}
\cheW_{\mu\nu}(x)~\equiv~
N_c{2\over s}\langle A,B|J_\mu(x_\bu,x_\star,x_\perp)J_\nu(0)|A,B\rangle
\label{4.3}
\end{equation}
where
\begin{eqnarray}
&&\hspace{-1mm}
-J^\mu~=~J^\mu_1+J^\mu_2+J^\mu_{12}+J^\mu_{21},
\nonumber\\
&&\hspace{-1mm}
J^\mu_1~=~\sum_f e_f\bar\Psi_1^f\gamma^\mu\Psi_1^f,~~~
J^\mu_{12}~=~\sum_f e_f\bar\Psi_1^f\gamma^\mu\Psi_2^f
\label{jeiz}
\end{eqnarray}
and similarly for $J^\mu_2$ and $J^\mu_{21}$.  Here 
$\langle A,B|\calo(\psi_A,A_\mu,\psi_B,B_\mu)|A,B\rangle$ denotes 
double functional integral over $A$ and $B$ fields  which gives matrix elements between projectile and target states of Eq. (\ref{W5}) type.

The leading-twist contribution to $W_{\mu\nu}(q)$  comes only from annihilation-type
product $J_{12}^\mu(x)J_{21}^\nu(0)+1\leftrightarrow 2$  
while power corrections may come also from $J_1^\mu(x)J_2^\nu(0)+1\leftrightarrow 2$.  
As demonstrated in Ref. \cite{Balitsky:2020jzt} , 
power corrections from 
 $J_1^\mu(x)J_2^\nu(0)$ terms are down by  one power of $N_c$ in comparison to leading-$N_c$ terms. On the other hand,
 they come from exchange-type diagrams like Fig. \ref{fig:1}b so they are determined by product of two quark distributions (one with additional gluon) 
 rather than from annihilation-type diagrams in Fig. \ref{fig:1}a proportional to product of quark and antiquark distributions.
 
 We will first calculate annihilation-type contributions coming from 
  $J_{12}^\mu(x)J_{21}^\nu(0)+1\leftrightarrow 2$ terms. Since leptonic tensor $L_{\mu\nu}$ is symmetric, 
  we consider
\begin{eqnarray}
&&\hspace{-5mm}
\cheW^{\rma}_{\mu\nu}(x)~=~{N_c\over s}\langle A,B|J_{12}^\mu(x)J_{21}^\nu(0)
+J_{21}^\mu(x)J_{12}^\nu(0)+\mu\leftrightarrow\nu|A,B\rangle 
~=~\sum_f e_f^2\cheW^{f}_{\mu\nu}(x),
\nonumber\\
&&\hspace{-5mm}
\cheW^{f}_{\mu\nu}(x)
~=~{N_c\over s}\langle A,B|[\Bsi_1^f(x)\gamma_\mu\Psi_2^f(x)][\Bsi_2^f(0)\gamma_\mu\Psi_1^f(0)]
+\mu\leftrightarrow\nu|A,B\rangle
+x\leftrightarrow 0
\label{wefdef}
\end{eqnarray}
After  Fierz transformations (\ref{fierz}) and (\ref{fierz5sym}) they can be sorted out as
\begin{eqnarray}
&&\hspace{-5mm}
\cheW^{f}_{\mu\nu}(x)~=~\cheW_{\mu\nu}^{\rmF f}(q)+\cheW_{\mu\nu}^{\rmH f}(q)
\nonumber\\
&&\hspace{-5mm}
 \cheW_{\mu\nu}^{\rm F}(x)~
=~
{N_c\over 2s}
\big(g_{\mu\nu}g^{\alpha\beta}-\delta_\mu^\alpha\delta_\nu^\beta-\delta_\nu^\alpha\delta_\mu^\beta\big)
\langle A,B|[\Bsi_1^m(x)\gamma_\alpha\Psi_1^n(0)][\Bsi_2^n(0)\gamma_\beta\Psi_2^m(x)]
\nonumber\\
&&\hspace{22mm}
+~\gamma_\alpha\otimes\gamma_\beta\leftrightarrow\gamma_\alpha\gamma_5\otimes\gamma_\beta\gamma_5|A,B\rangle
+x\leftrightarrow 0,
\label{fotonf}\\
\nonumber\\
&&\hspace{-5mm}
\cheW_{\mu\nu}^{\rm H}(x)~=~\cheW_{\mu\nu}^{\rm G}(x)+\cheW_{\mu\nu}^{\rm T}(x)
\label{hoton}\\
&&\hspace{-5mm}
\cheW_{\mu\nu}^{\rm G}(x)~=~-g_{\mu\nu}{N_c\over 2s}\langle A,B|[\Bsi_1^m(x)\Psi_1^n(0)][\Bsi_2^n(0)\Psi_2^m(x)]
\nonumber\\
&&\hspace{22mm}
-~[\Bsi_1^m(x)\gamma_5\Psi_1^n(0)][\Bsi_2^n(0)\gamma_5\Psi_2^m(x)]|A,B\rangle+x\leftrightarrow 0,
\label{goton}\\
&&\hspace{-5mm}
\cheW_{\mu\nu}^{\rm T}(x)~=~{N_c\over 2s}\big(\delta_\mu^\alpha\delta_\nu^\beta+\delta_\nu^\alpha\delta_\mu^\beta-\half g_{\mu\nu}g^{\alpha\beta}\big)
\nonumber\\
&&\hspace{22mm}
\times~\langle A,B|[\Bsi_1^m(x)\sigma_{\alpha\xi}\Psi_1^n(0)]\Bsi_2^n(0)\sigma_\beta^{~\xi}\Psi_2^m(x)]|A,B\rangle
 +x\leftrightarrow 0
 \label{toton}
\end{eqnarray}
for flavor $f$ which we are considering.  As discussed in Sect. \ref{sec:rapfak},  $x_\star$ in projectile matrix elements is set to be zero and similarly 
for $x_\bu=0$ in target matrix elements. To save space, we will often assume this instead of explicitly displaying.

In the remainder of this Section we will outline calculation of leading power corrections to  the above equations starting with $W^{\rm F}$ terms. 

\subsection{$W^\rmF$ contribution\label{sec:wf}}
As we discussed in Sect. \ref{sec:rapfak}, 
to calculate (\ref{fotonf}) one needs to plug in $\Psi_i$ in the form (\ref{fildz0}).  First, the leading-twist contribution is 

\begin{eqnarray}
&&\hspace{-1mm}
 \cheW_{\mu\nu}^{\rm F,lt}(x)~
=~
{N_c\over 2s}
\big(g_{\mu\nu}g^{\alpha\beta}-\delta_\mu^\alpha\delta_\nu^\beta-\delta_\nu^\alpha\delta_\mu^\beta\big)
\langle p_A,p_B|\big\{[\bsi_A^m(x)\gamma_\alpha\psi_A^n(0)][\bsi_B^n(0)\gamma_\beta\psi_B^m(x)]
\nonumber\\
&&\hspace{14mm}
+~\gamma_\alpha\otimes\gamma_\beta\leftrightarrow\gamma_\alpha\gamma_5\otimes\gamma_\beta\gamma_5 \big\}|p_A,p_B\rangle
+x\leftrightarrow 0
\nonumber\\
&&\hspace{-1mm}
=~
{1\over 2s}
\big(g_{\mu\nu}g^{\alpha\beta}-\delta_\mu^\alpha\delta_\nu^\beta-\delta_\nu^\alpha\delta_\mu^\beta\big)
\langle\bsi(x)\gamma_\alpha\psi(0)\rangle_A\langle\bsi(0)\gamma_\beta\psi(x)\rangle_B
\nonumber\\
&&\hspace{14mm}
+~(\psi(0)\otimes\psi(x)\leftrightarrow\gamma_5\psi(0)\otimes\gamma_5\psi(x) \big\}
+x\leftrightarrow 0
\label{chewlt}
\end{eqnarray}
Hereafter we use notations $\langle\calo\rangle_A\equiv\langle p_A|\calo|p_A\rangle$ and 
$\langle\calo\rangle_B\equiv\langle p_B|\calo|p_B\rangle$ for brevity. Using parametrizations (\ref{Amael}) and (\ref{baramael}) 
we can write down the corresponding contribution to $W^F_{\mu\nu}$ in the form
\begin{equation}
\hspace{-0mm}
W_{\mu\nu}^{\rm F,lt}(q)
~=~{1\over 16\pi^4}\!\int\! dx_\bu dx_\star d^2x_\perp~e^{-i\alpha_qx_\bu-i\beta_qx_\star+i(q,x)_\perp}\cheW^{\rm lt}_{\mu\nu}(x)
~=~-g_{\mu\nu}^\perp\!\int\! d^2k_\perp~F(q,k_\perp)
\label{WLTF}
\end{equation}
where
\begin{equation}
F^f(q,k_\perp)~=~f_1^f\big(\alpha_q,k_\perp\big)\barf_1^f\big(\beta_q,(q-k)_\perp\big)~+~f_1^f\leftrightarrow\barf_1^f
\label{F}
\end{equation}
Here the  term with $f\leftrightarrow\barf$ comes from $x\leftrightarrow 0$ contribution.

\subsubsection{Terms with one quark-quark-gluon TMD (one-gluon terms)}
Next, here will be terms with one or two gluon fields in Eq. (\ref{fotonf}) coming from replacement(s) (\ref{fildz0}). 
Terms with one gluon are
\begin{eqnarray}
&&\hspace{-1mm}
 \cheW_{\mu\nu}^{\rm 1F}(x)~
=~
{N_c\over 2s}
\big(g_{\mu\nu}g^{\alpha\beta}-\delta_\mu^\alpha\delta_\nu^\beta-\delta_\nu^\alpha\delta_\mu^\beta\big)
\langle p_A,p_B|\big\{[\psi_A^m(x)\gamma_\alpha\Xi_1^n(0)][\psi_B^n(0)\gamma_\beta\psi_B^m(x)]
\nonumber\\
&&\hspace{-1mm}
+~
[\Bxi_1^m(x)\gamma_\alpha\psi_A^n(0)][\psi_B^n(0)\gamma_\beta\psi_B^m(x)]
+[\psi_A^m(x)\gamma_\alpha\psi_A^n(0)][\psi_B^n(0)\gamma_\beta\Xi_2^m(x)]
\label{fodingluon}\\
&&\hspace{-1mm}
+~[\psi_A^m(x)\gamma_\alpha\psi_A^n(0)][\Bxi_2^n(0)\gamma_\beta\psi_B^m(x)]
~+~\gamma_\alpha\otimes\gamma_\beta\leftrightarrow\gamma_\alpha\gamma_5\otimes\gamma_\beta\gamma_5 \big\}|p_A,p_B\rangle
+x\leftrightarrow 0
\nonumber
\end{eqnarray}

Let us consider the first term in the r.h.s. of the above equation. As demonstrated in Ref. \cite{Balitsky:2020jzt} , the only non-negligible contribution 
comes from longitudinal  $\mu$ and transverse  $\nu$ (or {\it vice versa}), the term $\sim g_{\mu\nu}$ vanishes, and we obtain
\begin{eqnarray}
&&\hspace{-1mm}
\cheW^{(1)\rmF}_{1\mu\nu}(x)~=~{N_c\over 2s}\langle A,B|[\psi_A^m(x)\gamma_\mu\Xi_1^n(0)][\psi_B^n(0)\gamma_\nu\psi_B^m(x)]
\nonumber\\
&&\hspace{11mm}
+~\gamma_\mu\otimes\gamma_\nu\leftrightarrow\gamma_\mu\gamma_5\otimes\gamma_\nu\gamma_5 
|A,B\rangle+\mu\leftrightarrow\nu+x\leftrightarrow 0
\nonumber\\
&&\hspace{5mm}
=~{p_{2\mu}\over s^3}
\Big\{\Big[
\langle \bar\psi(x)\gamma_{\nu_\perp}\notp_2\gamma_i{1\over \alpha}\psi(0)\rangle_A
\langle \bar\psi B^i(0)\notp_1\psi(x)\rangle_B
+\langle\bar\psi(x)\notp_1\notp_2\gamma_i{1\over \alpha}\psi(0)\rangle_A
\nonumber\\
&&\hspace{11mm}
\times~\langle\bar\psi B^i(0)\gamma_{\nu_\perp}\psi(x)\rangle_B
~+~(\psi(0)\otimes\psi(x)\leftrightarrow\gamma_5\psi(0)\otimes\gamma_5\psi(x)\Big]
\nonumber\\
&&\hspace{5mm}
=~{p_{2\mu}\over s^3}\big[
\langle 
\bar\psi(x_\bu,x_\perp)\notp_2{1\over\alpha}\psi(0)\rangle_A
\langle\bar\psi\notB(0)\notp_1\gamma^\perp_\nu\psi(x_\star,x_\perp)\rangle_B
\nonumber\\
&&\hspace{11mm}
+~i\langle\bar\psi(x_\bu,x_\perp)\sigma_{\star\nu}{1\over\alpha}\psi(0)\rangle_A
\langle\bar\psi\notB(0)\hatp_1\psi(x_\star,x_\perp)\rangle_B
\big]~+~\mu\leftrightarrow\nu
~+~x\leftrightarrow 0
\label{onexi19}
\end{eqnarray}
Here we separated color-singlet contributions and used Eq. (\ref{gammas17}) to reduce number of $\gamma$-matrices.

Using formulas (\ref{maelqg1}), (\ref{maelqg2}), (\ref{tw3mael1}), and (\ref{8.48}) for quark-antiquark-gluon operators  and parametrizations 
from Sect. \ref{sec:paramlt} we get the contribution  to $W_{\mu\nu}$ in the form
\begin{eqnarray}
&&\hspace{-1mm}
W^{(1)\rmF}_{1\mu\nu}(q)~=~{N_c\over 16\pi^4}{1\over s}
\!\int\!dx_\bu dx_\star d^2x_\perp~e^{-i\alpha x_\bu-i\beta x_\star+i(q,x)_\perp}
\label{onexi20}\\
&&\hspace{33mm}
\times~
\langle A,B|\big[\bar\psi_A(x)\gamma_\mu\psi_B(x)\big]\big[\bar\psi_B(0)\gamma_\nu\Xi_{1}(0)\big]+x\leftrightarrow 0|A,B\rangle
~+~\mu\leftrightarrow\nu
\nonumber\\
&&\hspace{-1mm}
=~{1\over 64\pi^6}{p_{2\mu}\over s^3}\!\int d^2k_\perp
\!\int\!dx_\bu d^2x_\perp~e^{-i\alpha x_\bu+i(k,x)_\perp}\!\int\!dx_\star d^2x'_\perp e^{-i\beta x_\star+i(q-k,x')_\perp}
\nonumber\\
&&\hspace{15mm}
\times~\big[
\langle 
\bar\psi(x_\bu,x_\perp)\notp_2{1\over\alpha}\psi(0)\rangle_A
\langle\bar\psi\notB(0)\notp_1\gamma^\perp_\nu\psi(x_\star,x'_\perp)\rangle_B
+x\leftrightarrow 0\big]~+~\mu\leftrightarrow\nu
\nonumber\\
&&\hspace{-1mm}
=~{p_{2\mu}\over \alpha_qs}\!\int\! d^2k_\perp(q-k)_\nu F^f(q,k_\perp)
~+~\mu\leftrightarrow\nu
\nonumber
\end{eqnarray}
where $f_1\leftrightarrow\barf_1$ term in $F$ comes from $x\leftrightarrow 0$ contribution.

As demonstrated in  Ref. \cite{Balitsky:2020jzt} , contribution of the second term in r.h.s. of Eq. (\ref{fodingluon}) 
\begin{eqnarray}
&&\hspace{-1mm}
\cheW^{(1)\rmF}_{2\mu\nu}(x)~=~{N_c\over 2s}\langle A,B|[\Bxi_1^m(x)\gamma_\mu\psi_A^n(0)][\psi_B^n(0)\gamma_\nu\psi_B^m(x)]
+~...
\label{chewf12}
\end{eqnarray}
doubles the result (\ref{onexi20}) and the result for the 
third and the fourth terms is obtained from Eq. (\ref{onexi20}) by replacement 
${p_{2\mu}(q-k)^\perp_\nu\over \alpha_qs}\leftrightarrow {p_{1\mu}k^\perp_\nu\over \beta_qs}$ so
\begin{eqnarray}
&&\hspace{-1mm}
W^{(1)\rmF}_{\mu\nu}(q)~=~
2\!\int\! d^2k_\perp\Big({p_{1\mu}k^\perp_\nu\over\beta_qs}+{p_{2\mu}(q-k)^\perp_\nu\over\alpha_qs}
\Big)F^f(q,k_\perp)
~+~\mu\leftrightarrow\nu 
\label{w1fotvet}
\end{eqnarray}
This result agrees with the corresponding $1/Q$ terms in Ref. \cite{Mulders:1995dh}.

\subsubsection{Terms with two quark-quark-gluon TMDs (two-gluon terms) \label{sec:f2g}}
Let us now consider terms in $W^F_{\mu\nu}$ from Eq. (\ref{fotonf}) with two gluon operators. The first of such terms is
\begin{eqnarray}
&&\hspace{-1mm}
 \cheW_{1\mu\nu}^{\rm (2a)F}(x)~
=~
{N_c\over 2s}
\big(g_{\mu\nu}g^{\alpha\beta}-\delta_\mu^\alpha\delta_\nu^\beta-\delta_\nu^\alpha\delta_\mu^\beta\big)
\langle p_A,p_B|\big\{[\psi_A^m(x)\gamma_\alpha\Xi_1^n(0)][\psi_B^n(0)\gamma_\beta\Xi_2^m(x)]
\label{chew2odin}\\
&&\hspace{15mm}
+~[\Xi_1^m(x)\gamma_\alpha\psi_A^n(0)][\Xi_2^n(0)\gamma_\beta\psi_B^m(x)]+\gamma_\alpha\otimes\gamma_\beta\leftrightarrow\gamma_\alpha\gamma_5\otimes\gamma_\beta\gamma_5 \big\}|p_A,p_B\rangle
+x\leftrightarrow 0
\nonumber
\end{eqnarray}
It is convenient to start from the  contribution 
\begin{eqnarray}
&&\hspace{-1mm}
\cheV_{\mu\nu}^{\rm 1F}(x)~
=~
{N_c\over 2s}
\langle A,B|\big\{[\psi_A^m(x)\gamma_\mu\Xi_1^n(0)][\psi_B^n(0)\gamma_\nu\Xi_2^m(x)]
\nonumber\\
&&\hspace{16mm}
+~\gamma_\mu\otimes\gamma_\nu\leftrightarrow\gamma_\mu\gamma_5\otimes\gamma_\nu\gamma_5 \big\}|A,B\rangle+\mu\leftrightarrow\nu
+x\leftrightarrow 0
\nonumber\\
&&\hspace{5mm}
=~-{1\over 2s^3}\big\{\langle\bsi A_i(x)\gamma_\mu\notp_2\gamma^j{1\over \alpha}\psi(0)\rangle_A
\langle\bsi B_j(0)\gamma_\nu\notp_1\gamma^i{1\over\beta}\psi(x)\rangle_B
\nonumber\\
&&\hspace{16mm}
+~\psi(0)\otimes\psi(x)\leftrightarrow\gamma_5\psi(0)\otimes\gamma_5\psi(x)+\mu\leftrightarrow\nu\}
~+~x\leftrightarrow 0
\label{veodinf}
\end{eqnarray}
As demonstrated in Ref. \cite{Balitsky:2020jzt}, the non-negligible contribution comes only from transverse $\mu$ and $\nu$. In this case
 we can use formula (\ref{formula9}) and get
\begin{eqnarray}
&&\hspace{-1mm}
\cheV^{\rm 1F}_{\mu_\perp\nu_\perp}~
=~-{1\over 2s^3}\big\{\langle\bsi A_i(x)\gamma_{\mu_\perp}\notp_2\gamma^j{1\over \alpha}\psi(0)\rangle_A
\langle\bsi B_j(0)\gamma_{\nu_\perp}\notp_1\gamma^i{1\over\beta}\psi(x)\rangle_B
\nonumber\\
&&\hspace{10mm}
+~\psi(0)\otimes\psi(x)\leftrightarrow\gamma_5\psi(0)\otimes\gamma_5\psi(x)+\mu\leftrightarrow\nu\}
~+~x\leftrightarrow 0
\nonumber\\
&&\hspace{10mm}
=~-{g_{\mu\nu}^\perp\over s^3}\langle\bsi\notA(x)\notp_2\gamma_i{1\over \alpha}\psi(0)\rangle_A
\langle\bsi\notB(0)\notp_1\gamma^i{1\over\beta}\psi(x)\rangle_B~+~x\leftrightarrow 0
\label{kalvedvaperp}
\end{eqnarray}
This gives the contribution of the first matrix element in the r.h.s. of Eq. (\ref{chew2odin}) to $W_{\mu\nu}(q)$ in the form
\begin{eqnarray}
&&\hspace{-1mm}
\half\big(g_{\mu\nu}g^{\alpha\beta}-\delta_\mu^\alpha\delta_\nu^\beta-\delta_\nu^\alpha\delta_\mu^\beta\big)
{1\over 16\pi^4}\!\int\! dx_\bu dx_\star d^2x_\perp~e^{-i\alpha_qx_\bu-i\beta_qx_\star+i(q,x)_\perp}\cheV_{2\mu_\perp\nu_\perp}(x) 
\nonumber\\
&&\hspace{-1mm}
=~{g_{\mu\nu}^\parallel\over Q_\parallel^2}\!\int\! d^2k_\perp (k,q-k)_\perp F^f(q,k_\perp)
\label{kalv2glav}
\end{eqnarray}
where we again used formulas from  Appendices \ref{sec:paramlt} and \ref{sec:qqgparam}.  Next, as shown in  Ref. \cite{Balitsky:2020jzt} ,
the contribution of the second matrix element in the r.h.s. of Eq. (\ref{chew2odin}) is equal to that of the first one so we get
\begin{eqnarray}
&&\hspace{-1mm}
W_{\mu\nu}^{\rm (2a)F}(q)~
=~{2g_{\mu\nu}^\parallel\over Q_\parallel^2}\int\! d^2k_\perp (k,q-k)_\perp F^f(q,k_\perp)
\label{w2afotvet}
\end{eqnarray}

The second two-gluon contribution to $\cheW^F$ in Eq. (\ref{fotonf}) is
\begin{eqnarray}
&&\hspace{-1mm}
\cheW^{\rm (2b)F}_{\mu\nu}(x)~=~-{N_c\over 2s}(\delta_\mu^\alpha\delta_\nu^\beta+\delta_\nu^\alpha\delta_\mu^\beta-g_{\mu\nu}g^{\alpha\beta})
\langle A,B|\big\{[\bar\psi_A^m(x)\gamma_\alpha\psi_A^n(0)][\Bxi_{2}^n(0)\gamma_\beta\Xi_{2}^m(x)]
\nonumber\\
&&\hspace{-1mm}
+~[\Bxi_1^m(x)\gamma_\alpha\Xi_1^n(0)][\bsi_B^n(0)\gamma_\beta\psi_B^m(x)]+\gamma_\alpha\otimes\gamma_\beta\leftrightarrow\gamma_\alpha\gamma_5\otimes\gamma_\beta\gamma_5 \big\}|A,B\rangle
+x\leftrightarrow 0
\label{chew2dva}
\end{eqnarray}
Let us consider the first matrix element in the r.h.s. of the above equation. We get
\begin{eqnarray}
&&\hspace{-1mm}
\cheW^{\rm (2b)F}_{1\mu\nu}(x)~=~-{N_c\over 2s}(\delta_\mu^\alpha\delta_\nu^\beta+\delta_\nu^\alpha\delta_\mu^\beta-g_{\mu\nu}g^{\alpha\beta})
\nonumber\\
&&\hspace{20mm}
\times~
\langle A,B|\big\{[\bar\psi_A^m(x)\gamma_\alpha\psi_A^n(0)][\Bxi_{2}^n(0)\gamma_\beta\Xi_{2}^m(x)]|p_A,p_B\rangle
+x\leftrightarrow 0
\nonumber\\
&&\hspace{-1mm}
=~-{1\over s^3}(\delta_\mu^\alpha p_{1\nu}+\delta_\nu^\alpha p_{1\mu}-g_{\mu\nu}p_1^\alpha)
\Big(\langle \bar\psi(x)A_j(x)\gamma_\alpha A_i(0)\psi(0)\rangle_A
\nonumber\\
&&\hspace{20mm}
\times~\langle\big(\bar\psi{1\over \beta}\big)(0)\gamma^i\notp_1
\gamma^j{1\over\beta}\psi(x)\rangle_B
+\psi(0)\otimes\psi(x)\leftrightarrow \gamma_5\psi(0)\otimes\gamma_5\psi(x)\Big)~+~x\leftrightarrow 0
\nonumber\\
&&\hspace{-1mm}
=~-{4p_{1\mu} p_{1\nu}\over s^4}\Big(
\langle \bar\psi(x)A_j(x)\notp_2A_i(0)\psi(0)\rangle_A\langle\big(\bar\psi{1\over \beta}\big)(0)\gamma^i\notp_1
\gamma^j{1\over\beta}\psi(x)\rangle_B
\nonumber\\
&&\hspace{20mm}
+~\psi(0)\otimes\psi(x)\leftrightarrow \gamma_5\psi(0)\otimes\gamma_5\psi(x)\Big)
\label{v4}
\end{eqnarray}
up to the terms which are negligible as shown in Ref. \cite{Balitsky:2020jzt} .
Using Eq. (\ref{flagamma}) this can be rewritten as
\begin{eqnarray}
&&\hspace{-1mm}
\cheW^{\rm (2b)F}_{1\mu\nu}(x)~=~-{4p_{1\mu} p_{1\nu}\over s^4}\Big(
\langle \bar\psi(x)\notA(x)\notp_2\notA(0)\psi(0)\rangle_A\langle\big(\bar\psi{1\over \beta}\big)(0)\notp_1{1\over\beta}\psi(x)\rangle_B
\nonumber\\
&&\hspace{-1mm}
+~\psi(0)\otimes\psi(x)\leftrightarrow \gamma_5\psi(0)\otimes\gamma_5\psi(x)\Big)
~+~x\leftrightarrow 0
\label{v4longa}
\end{eqnarray}
The corresponding contribution to $W_{\mu\nu}(q)$ is obtained from QCD equation of motion (\ref{AA1}) and formula (\ref{maelqg2}) from 
Appendix \ref{sec:qqgparam}:
\begin{eqnarray}
&&\hspace{-1mm}
W^{\rm (2b)F}_{1\mu\nu}(q)~=~{4p_{1\mu} p_{1\nu}\over \beta_q^2s^2}\int\! d^2k_\perp
k_\perp^2F^f(q,k_\perp)
\label{w2botvet}
\end{eqnarray}
Next, as shown in  Ref. \cite{Balitsky:2020jzt} , the contribution of the second matrix element in the r.h.s. of the Eq. (\ref{chew2dva}) 
\begin{eqnarray}
&&\hspace{-1mm}
\cheW^{\rm (2b)F}_{2\mu\nu}(x)~=~-{N_c\over 2s}(\delta_\mu^\alpha\delta_\nu^\beta+\delta_\nu^\alpha\delta_\mu^\beta-g_{\mu\nu}g^{\alpha\beta})
\label{chew2dva}\\
&&\hspace{-1mm}
\langle A,B|[\Bxi_1^m(x)\gamma_\alpha\Xi_1^n(0)][\bsi_B^n(0)\gamma_\beta\psi_B^m(x)]+\gamma_\alpha\otimes\gamma_\beta\leftrightarrow\gamma_\alpha\gamma_5\otimes\gamma_\beta\gamma_5 \big\}|A,B\rangle
+x\leftrightarrow 0
\nonumber
\end{eqnarray}
differs from Eq. (\ref{w2botvet}) by  replacements 
$p_1\leftrightarrow p_2$, $\alpha_q\leftrightarrow\beta_q$ and exchange of  projectile matrix elements and the target ones so we finally get
\begin{eqnarray}
&&\hspace{-1mm}
W^{\rm (2b)F}_{\mu\nu}(q)~=~
\!\int\! d^2k_\perp\Big[{4p_{1\mu}p_{1\nu}\over \beta_q^2s^2}k_\perp^2
+{4p_{2\mu}p_{2\nu}\over \alpha_q^2s^2}
(q-k)_\perp^2\Big]F^f(q,k_\perp)
\label{w2bfotvet}
\end{eqnarray}

The third two-gluon contribution to $\cheW^F$ in Eq. (\ref{fotonf}) has the form
\begin{eqnarray}
&&\hspace{-1mm}
 \cheW_{\mu\nu}^{\rm (2c)F}(x)~
=~
{N_c\over 2s}
\big(g_{\mu\nu}g^{\alpha\beta}-\delta_\mu^\alpha\delta_\nu^\beta-\delta_\nu^\alpha\delta_\mu^\beta\big)
\langle p_A,p_B|\big\{[\Bxi_1^m(x)\gamma_\alpha\psi_A^n(0)][\psi_B^n(0)\gamma_\beta\Xi_2^m(x)]
\label{chew2tri}\\
&&\hspace{15mm}
+~[\psi_A^m(x)\gamma_\alpha\Xi_1^n(0)][\Xi_2^n(0)\gamma_\beta\psi_B^m(x)]+\gamma_\alpha\otimes\gamma_\beta\leftrightarrow\gamma_\alpha\gamma_5\otimes\gamma_\beta\gamma_5 \big\}|p_A,p_B\rangle
+x\leftrightarrow 0
\nonumber
\end{eqnarray}
It is easy to see that after separating color-singlet matrix elements  this term is $O\big({1\over N_c^2}\big)$  in comparison to 
(\ref{veodinf}) so we neglect it.

Thus, the result for $W_{\mu\nu}^{\rm F}(q)$ is the sum of Eqs. (\ref{WLTF}), (\ref{w1fotvet}), (\ref{w2afotvet}), and (\ref{w2bfotvet}).
After some algebra, it can be rewritten as
\begin{eqnarray}
&&\hspace{-1mm}
W^{\rm F}_{\mu\nu}(q)~=~\sum_f e_f^2W^{\rm fF}_{\mu\nu}(q),~~~~
W^{\rm fF}_{\mu\nu}(q)~=~\!\int\!d^2k_\perp F^f(q,k_\perp)\calW^F_{\mu\nu}(q,k_\perp),
\label{resultf}
\end{eqnarray}
where
\begin{eqnarray}
&&\hspace{-1mm}
\calW^{\rmF}_{\mu\nu}(q,k_\perp)~=~
-g_{\mu\nu}^\perp +{1\over Q_\parallel^2}(q^\parallel_\mu q^\perp_\nu+q^\parallel_\nu q^\perp_\mu)
+{q_\perp^2\over Q_\parallel^4}q^\parallel_\mu q^\parallel_\nu+{\tilq_\mu\tilq_\nu\over Q_\parallel^2}[q_\perp^2-4(k,q-k)_\perp]
\nonumber\\
&&\hspace{20mm}
-~\Big[{\tilq_\mu\over  Q_\parallel^2}\Big(g^\perp_{\nu i}-{q^\parallel_\nu q_i\over Q_\parallel^2}\Big)(q-2k)_\perp^i
+\mu\leftrightarrow\nu\Big]
\label{WF}
\end{eqnarray}
It is easy to see that $q^\mu\calW^{\rmF}_{\mu\nu}(q,k_\perp)~=~0$.

\subsection{$W^{\rm G}$ term of Eq. (\ref{goton})}
In this section we will repeat the above calculations for the $W^{\rm G}_{\mu\nu}(q)$.
Let us start from
\begin{eqnarray}
&&\hspace{-1mm}
\cheW^{\rm G}_{\mu\nu}(x)~=~
-{N_cg_{\mu\nu}\over 2s}\langle A,B|[\Bsi_1^m(x)\Psi_1^n(0)][\Bsi_2^n(0)\Psi_2^m(x)]
\nonumber\\
&&\hspace{33mm}
-~[\Bsi_1^m(x)\gamma_5\Psi_1^n(0)][\Bsi_2^n(0)\gamma_5\Psi_2^m(x)]|A,B\rangle
~+~x\leftrightarrow 0
\end{eqnarray}
First, as seen  from parametrizations (\ref{Amael}) and (\ref{baramael}), the leading-twist contribution can be neglected.
Second, as demonstrated in Ref. \cite{Balitsky:2020jzt} , the contribution of one-gluon gluon operators also vanishes with our accuracy.
Let us now consider two-gluon terms and start with
\begin{eqnarray}
&&\hspace{-1mm}
\cheW^{\rm G}_{\mu\nu}(x)
~=~-{N_cg_{\mu\nu}\over 2s}\langle A,B|[\bsi_A^n(x)\Xi_{1}^m(0)][\bsi_B^n(0)\Xi_{2}^m(x)]
\nonumber\\
&&\hspace{33mm}
-~[\bsi_A^m(x)\gamma_5\Xi_{1}^n(0)][\bsi_B^n(0)\gamma_5\Xi_{2}^m(x)]|A,B\rangle~+~x\leftrightarrow 0
\nonumber\\
&&\hspace{13mm}
=~{g_{\mu\nu}\over 2s^3}\Big[\langle \bsi A^i(x)\sigma_{\star j}{1\over \alpha}\psi(0)\rangle_A
\langle\bsi B^j(0)\sigma_{\bu i}{1\over\beta}\psi(x)\rangle_B
\nonumber\\
&&\hspace{33mm}
-~\psi(0)\otimes\psi(x)\leftrightarrow\gamma_5\psi(0)\otimes\gamma_5\psi(x)\Big]
~+~x\leftrightarrow 0
\nonumber\\
&&\hspace{13mm}
=~-{g_{\mu\nu}\over 2s^3}\langle \bsi{\notA}(x)\notp_2{1\over \alpha}\psi(0)\rangle_A
\langle\bsi{\notB}(0)\notp_1{1\over\beta}\psi(x)\rangle_B\Big[1+O\big({q_\perp^2\over s}\big)\Big]
\label{twoxia1}
\end{eqnarray}
where we used formula (\ref{gammas14a})  and the fact that  
\footnote{A rigorous argument goes like that: the matrix element (\ref{formula2})  can be rewritten as
$\epsilon_{\nu_\perp j}\epsilon_{kl}\langle \bar\psi(0)[A_k(0)\sigma_{\bu l}\psi(x)\rangle
~=~\epsilon_{j\nu_\perp}\langle \bar\psi(0)\notA(0)\notp_1\gamma_5\psi(x)\rangle$.
As demonstrated in Sect. \ref{sec:qqgparam}, $\notA$ in this formula can be replaced by $\notk_\perp$ so the contribution is
proportional to matrix element $k^i\langle \bar\psi(0)i\sigma_{\bu i}\gamma_5\psi(x)\rangle=k^i\epsilon_{ij}
\langle \bar\psi(0)\sigma_{\bu j}\psi(x)\rangle$ which vanishes as seen from the parametrization (\ref{hmael}).
\label{zanulil}
} 
\begin{equation}
\langle\bar\psi(x)\big[A_k\sigma_{\star j}-A_j(x)\sigma_{\star k}\big]\psi(0)\rangle_A~=~0
\label{formula2}
\end{equation}
It is easy to see that with our accuracy the above equation is the only two-gluon contribution to 
$\cheW^{\rm G}_{\mu\nu}(x)$ since  $\Bxi_1\Xi_{1}=\Bxi_2\Xi_{2}=0$ and the matrix element
$\langle[\Bxi_1^m(x)\gamma_\alpha\psi_A^n(0)][\psi_B^n(0)\gamma_\beta\Xi_2^m(x)]\rangle$ is 
$\sim O\big({1\over N_c^2}\big)$ in comparison to Eq. (\ref{formula2}) similarly to Eq. (\ref{chew2tri}).

Now, using equations (\ref{maelqg1}), (\ref{9.56}), (\ref{9.58}) and parametrizations (\ref{hmael}) we obtain the contribution to 
photon-mediated hadronic tensor in the form
\begin{eqnarray}
&&\hspace{-1mm}
W^{\rm G}_{\mu\nu}(q)~=~{g_{\mu\nu}\over 16\pi^4}\!\int\! dx_\bu dx_\star d^2x_\perp~e^{-i\alpha_qx_\bu-i\beta_qx_\star+i(q,x)_\perp}
\cheW^{\rm G}_{\mu\nu}(x) 
\nonumber\\
&&\hspace{14mm}
=~-{g_{\mu\nu}\over 2Q_\parallel^2}\int\! d^2k_\perp {1\over m^2} k_\perp^2(q-k)_\perp^2H(q,k_\perp)
\label{gotvet}
\end{eqnarray}
where 
\begin{equation}
H^f(q,k_\perp)~=~h_1^{\perp f}\big(\alpha_q,k_\perp\big)\barh_1^{\perp f}\big(\beta_q,(q-k)_\perp\big)
~+~h_1^{\perp f}\leftrightarrow\barh_1^{\perp f}
\label{H}
\end{equation}
 for the flavor that we are considering.

\subsection{$W^{\rm T}$ contribution of Eq. (\ref{toton}) \label{sec:wt}}
In this Section we calculate the  $W^{\rm T}_{\mu\nu}$ term of Eq. (\ref{toton}). 
The leading-twist contribution 
\begin{eqnarray}
&&\hspace{-5mm}
\cheW_{\mu\nu}^{\rm T,lt}(x)~=~\cheW_{\mu\nu}^{\rm H,lt}(x)
\label{fotonhlt}\\
&&\hspace{-1mm}
=~{N_c\over 2s}\big(\delta_\mu^\alpha\delta_\nu^\beta+\delta_\nu^\alpha\delta_\mu^\beta-\half g_{\mu\nu}g^{\alpha\beta}\big)
\langle A,B|[\psi_A^m(x)\sigma_{\alpha\xi}\psi_A^n(0)]\psi_B^n(0)\sigma_\beta^{~\xi}\psi_B^m(x)]|A,B\rangle
 +x\leftrightarrow 0
 \nonumber
\end{eqnarray}
is easily obtained from parametrizations (\ref{hmael}) \cite{Tangerman:1994eh}:
\begin{eqnarray}
&&\hspace{-1mm}
W_{\mu\nu}^{\rm H,lt}(\alpha_q,\beta_q,q_\perp)
~=~{1\over 16\pi^4}\!\int\! dx_\bu dx_\star d^2x_\perp~e^{-i\alpha_qx_\bu-i\beta_qx_\star+i(q,x)_\perp}\cheW^{\rm lt}_{\mu\nu}(x)
\nonumber\\
&&\hspace{-1mm}
~=~-\sum_f e_f^2\!\int\! d^2k_\perp\big[k^\perp_\mu(q-k)^\perp_\nu+k^\perp_\nu(q-k)^\perp_\mu+g_{\mu\nu}^\perp(k,q-k)_\perp\big]
H^f(q,k_\perp)
\label{WLTH}
\end{eqnarray}

For the calculation of one- and two-gluon terms it is convenient to consider
\begin{equation}
\hspace{-0mm}
\cheV_{\mu\nu}^{\rm H}(x)~=~{N_c\over 2s}
\langle A,B|[\Bsi_1^m(x)\sigma_{\mu\xi}\Psi_1^n(0)][\Bsi_2^n(0)\sigma_\nu^{~\xi}\Psi_2^m(x)|A,B\rangle
+\mu\leftrightarrow\nu+x\leftrightarrow 0
\label{vh1}
\end{equation}
and subtract trace to get $\cheW_{\mu\nu}^{\rm T}(x)$ afterwards.

\subsubsection{One-gluon terms in  $\cheV_{\mu\nu}^{\rm H}(x)$ \label{sec:onegluonhoton}}

Let us start from one-gluon term coming from $\Xi_2$. Sorting out 
color-singlet matrix elements, we get
\begin{equation}
\hspace{-0mm}
\cheV_{1\mu\nu}^{\rm (1)H}(x)~=~
-{1\over 2s^2}
\langle\bar\psi(x)\sigma_{\mu\xi}\notp_2\gamma^i{1\over \alpha}\psi(0)\rangle_A
\langle\bar\psi B^i(0)\sigma_\nu^{~\xi}\psi(x)\rangle_B~+~\mu\leftrightarrow\nu~+~x\leftrightarrow 0
\label{onexi9}
\end{equation}
As demonstrated in Ref. \cite{Balitsky:2020jzt} , the only non-negligible contributions are those with one of the indices 
in Eq. (\ref{onexi9}) longitudinal and one transverse. For example,
let $\mu$ be longitudinal and $\nu$ transverse, the opposite case
will differ by replacement $\mu\leftrightarrow\nu$. 
Using the decomposition of $g^{\mu\nu}$ in longitudinal and transverse part (\ref{delta}) 
we get
\begin{eqnarray}
&&\hspace{-1mm}
\Big({2p_1^\mu p_2^{\mu'}\over s}+\mu\leftrightarrow\mu'\Big)\cheV_{1\mu\nu_\perp}^{\rm (1)H}(x)~=~
-\Big({p_{2\mu} p_1^{\mu'}\over s^3}+\mu\leftrightarrow\mu'\Big)
\nonumber\\
&&\hspace{-1mm}
\times~
\big[\langle\bar\psi(x)\sigma_{\mu'\xi}\notp_2\gamma^i{1\over \alpha}\psi(0)\rangle_A
\langle\bar\psi B^i(0)\sigma_{\nu_\perp}^{~\xi}\psi(x)\rangle_B~+~\mu'\leftrightarrow\nu\big]~+~x\leftrightarrow 0
\label{onexi12}
\end{eqnarray}
As demonstrated in Ref. \cite{Balitsky:2020jzt} , the term  proportional to $p_{1\mu}$ is small so we get
\begin{eqnarray}
&&\hspace{-3mm}
\cheV_{1\mu_\parallel\nu_\perp}^{\rm (1)H}(x)~=~
\nonumber\\
&&\hspace{-3mm}
=~{p_{2\mu}\over s^3}\Big\{
i\langle\bar\psi(x)\sigma_{\nu_\perp j}\sigma_{\star i}{1\over \alpha}\psi(0) \rangle_A
\langle \bar\psi B^i(0)\sigma_\bu^{~j}\psi(x)\rangle_B
-~\langle\bar\psi(x)\sigma_{\star i}{1\over \alpha}\psi(0) \rangle_A
\langle \bar\psi B^i(0)\sigma_{\bu \nu_\perp}\psi(x)\rangle_B
\nonumber\\
&&\hspace{15mm}
+~i\langle\bar\psi(x)\sigma_{\bu j}\sigma_{\star i}{1\over \alpha}\psi(0) \rangle_A
\langle \bar\psi B^i(0)\sigma_{\nu_\perp}^{~j}\psi(x)\rangle_B
\nonumber\\
&&\hspace{33mm}
+~{2i\over s}\langle\bar\psi(x)\sigma_{\bu \nu_\perp}\sigma_{\star i}{1\over \alpha}\psi(0) \rangle_A
\langle \bar\psi B^i(0)\sigma_{\star\bu}\psi(x)\rangle_B\Big\}~+~x\leftrightarrow 0
\label{4.31}
\end{eqnarray}
Using Eq. (\ref{sigmasigmas}) one can prove that the contribution of last two terms in the r.h.s. is $\sim{q_\perp^2\over s}$ 
and therefore 
\begin{eqnarray}
&&\hspace{-1mm}
\cheV_{1\mu_\parallel\nu_\perp}^{\rm (1)H}(x)
~=~{p_{2\mu}\over s^3}\Big[i\langle\bar\psi(x)\sigma_{\nu_\perp j}\sigma_{\star i}{1\over \alpha}\psi(0) \rangle_A\langle \bar\psi B^i(0)\sigma_\bu^{~j}\psi(x)\rangle_B
\label{onexi16}\\
&&\hspace{44mm}
-~\langle\bar\psi(x)\sigma_{\star i}{1\over \alpha}\psi(0) \rangle_A\langle \bar\psi B^i(0)\sigma_{\bu \nu_\perp}\psi(x)\rangle_B\Big]
+\leftrightarrow 0
\nonumber\\
&&\hspace{-1mm}
=~{p_{2\mu}\over s^3}\langle\bar\psi(x)\sigma_\star^{~ j}{1\over \alpha}\psi(0) \rangle_A
\langle \bar\psi(0)[B_\nu(0)\sigma_{\bu j}-\nu\leftrightarrow j]\psi(x)\rangle_B~-~{p_{2\mu}\over s^3}\langle\bar\psi(x)\sigma_{\star \nu_\perp}{1\over \alpha}\psi(0) \rangle_A
\nonumber\\
&&\hspace{-1mm}
\times~
\langle \bar\psi B^j(0)\sigma_{\bu j}\psi(x)\rangle_B+x\leftrightarrow 0
~=~{ip_{2\mu}\over s^3}\langle\bar\psi(x)\sigma_{\star \nu_\perp}{1\over \alpha}\psi(0) \rangle_A
\langle \bar\psi(0) \notB(0)\notp_1\psi(x)\rangle_B+x\leftrightarrow 0
\nonumber
\end{eqnarray}
where we used formulas (\ref{formula2}) and  (\ref{sigmasigmas}).

Using formulas (\ref{maelqg1}), (\ref{maelqg2}), (\ref{9.56}),  and (\ref{9.58}) 
for quark-antiquark-gluon operators  and parametrizations 
from Sect. \ref{sec:paramlt} we get the contribution  to $W_{\mu\nu}$ in the form
\begin{eqnarray}
&&\hspace{-1mm}
V^{\rm (1)H}_{1\mu\nu}(q)~=~{N_c\over 16\pi^4}{1\over s}
\!\int\!dx_\bu dx_\star d^2x_\perp~e^{-i\alpha x_\bu-i\beta x_\star+i(q,x)_\perp}\cheV^{(1)H}_{1\mu\nu}(x)
\label{v1h}\\
&&\hspace{-1mm}
=~-{p_{2\mu}\over \alpha_qs}\!\int\! d^2k_\perp ~k_\nu {(q-k)_\perp^2\over m^2}H(q,k_\perp)
~+~\mu\leftrightarrow\nu
\nonumber
\end{eqnarray}
where terms with replacement  $h_{1f}^\perp\leftrightarrow\barh_{1f}^\perp$ come from $x\leftrightarrow 0$ contribution 
as usually.

Next, as  proved in Ref. \cite{Balitsky:2020jzt} ,  the term with with $\Bxi_1(x)$ doubles the contribution (\ref{v1h}) and
the terms with $\Xi_2(0)$ and $\Bxi_2(x)$ are obtained by the projectile $\leftrightarrow$ target replacement, namely $p_1\leftrightarrow p_2$, 
$\alpha_q\leftrightarrow\beta_q$ and $k_\perp\leftrightarrow (q-k)_\perp$. Thus, the contribution of one-gluon terms to $W^{\rm T}_{\mu\nu}(q)$
has the form
\begin{equation}
\hspace{-0mm}
W^{\rm (1)T}_{\mu\nu}(q)~
=~-2\!\int\! d^2k_\perp ~ \Big[{p_{1\mu}(q-k)_\nu\over \beta_qs}{k_\perp^2\over m^2}
+{p_{2\mu}k_\nu\over \alpha_qs}{(q-k)_\perp^2\over m^2}\Big]H(q,k_\perp)
~+~\mu\leftrightarrow\nu
\label{w1t}
\end{equation}

\subsubsection{Two-gluon terms in  $\cheV_{\mu\nu}^{\rm H}(x)$ \label{sec:twogluonshoton}}

Let us start with the term 
\begin{equation}
\hspace{-0mm}
\cheV_{1\mu\nu}^{\rm (2a)H}(x)~=~{N_c\over 2s}
\langle A,B|[\psi_A^m(x)\sigma_{\mu\xi}\Xi_1^n(0)][\psi_B^n(0)\sigma_\nu^{~\xi}\Xi_2^m(x)|A,B\rangle
+\mu\leftrightarrow\nu+x\leftrightarrow 0
\label{vh21}
\end{equation}
Separating color-singlet contributions, we get
\begin{equation}
\hspace{0mm}
\cheV_{1\mu\nu}^{\rm (2a)H}(x)~=~{1\over 2s^3}\langle\bsi A_i(x)\sigma_{\mu\alpha}\notp_2\gamma^j{1\over \alpha}\psi(0)\rangle_A
\langle\bsi B_j(0)\sigma_\nu^{~\alpha}\notp_1\gamma^i{1\over\beta}\psi(x)\rangle_B+\mu\leftrightarrow\nu
+x\leftrightarrow 0
\label{4.44}
\end{equation}
As demonstrated in Ref. \cite{Balitsky:2020jzt} , the contributions from longitudinal $\mu$ and transverse $\nu$ (or {\it vice versa}) are small. 
For transverse $\mu$ and $\nu$ we obtain
\begin{eqnarray}
&&\hspace{-1mm}
\cheV_{1\mu_\perp\nu_\perp}^{\rm (2a)H}(x)
~=~-{1\over 2s^3}\langle\bsi A^i(x)\sigma_{\mu_\perp k}\sigma_{\star j}{1\over \alpha}\psi)0)\rangle_A
\langle\bsi B^j(0)\sigma_{\nu_\perp}^{~k}\sigma_{\bu i}{1\over\beta}\psi(x)\rangle_B
\nonumber\\
&&\hspace{-1mm}
-~{1\over s^4}\langle\bsi A^i(x)\sigma_{\bu\mu_\perp}\sigma_{\star j}{1\over \alpha}\psi(0)\rangle_A
\langle\bsi B^j(0)\sigma_{\star\nu_\perp}\sigma_{\bu i}{1\over\beta}\psi(x)\rangle_B
+\mu\leftrightarrow\nu
+x\leftrightarrow 0
\label{4.45}
\end{eqnarray}
Using Eq.  (\ref{sigmasigmas}) and  (\ref{formula2}), it is possible to demonstrate the second term in the r.h.s. is small and  the first term 
can be rewritten as
\begin{eqnarray}
&&\hspace{-1mm}
\cheV_{1\mu_\perp\nu_\perp}^{\rm (2a)H}(x)
~=~-{g_{\mu\nu}^\perp\over 2s^3}\langle\bsi A^i(x)\sigma_{\star i}{1\over \alpha}\psi(0)\rangle_A
\langle\bsi B^j(0)\sigma_{\bu j}{1\over\beta}\psi(x)\rangle_B
\nonumber\\
&&\hspace{9mm}
+~{1\over s^3}\big\{\langle\bsi\big(A_k\sigma_{\star \mu_\perp}-\half g_{\mu k}\sigma_{\star j}A^j\big)(x){1\over \alpha}\psi(0)\rangle_A
\nonumber\\
&&\hspace{18mm}
\times~\langle\bsi\big(B^k\sigma_{\bu \nu_\perp}-\half \delta_\nu^k\sigma_{\bu j}B^j\big)(0){1\over\beta}\psi(x)\rangle_B
+\mu\leftrightarrow\nu\big\}
+x\leftrightarrow 0
\label{kalv3perpee}
\end{eqnarray}
The corresponding contribution to $\cheV_{\mu_\perp\nu_\perp}^{\rm H}(q)$ can be obtained 
from QCD equations of motion (\ref{9.56}), (\ref{9.58}) and parametrization (\ref{maelsa}). We get 
\begin{eqnarray}
&&\hspace{-11mm}
\cheV_{1\mu_\perp\nu_\perp}^{\rm (2a)H}(q)~=~{1\over Q_\parallel^2N_c}\!\int d^2k_\perp 
\Big[g^\perp_{\mu\nu}{k_\perp^2(q-k)_\perp^2\over 2m^2}
H^f(q,k_\perp)
\label{kalvklad3perp}\\
&&\hspace{8mm}
+~ \big[(k^\perp_\mu(q-k)^\perp_\nu+\mu\leftrightarrow\nu)(k,q-k)_\perp
-k_\perp^2(q-k)^\perp_\mu(q-k)^\perp_\nu
\nonumber\\
&&\hspace{22mm}
-~(q-k_\perp)^2k^\perp_\mu k^\perp_\nu
-{g_{\mu\nu}^\perp\over 2}k_\perp^2(q-k_\perp)^2\big]
{1\over m^2}H_A^f(q,k_\perp)\Big]
\nonumber
\end{eqnarray}
where we introduced the notation
\begin{eqnarray}
&&\hspace{-11mm}
H_A^f(q,k_\perp)~\equiv~h_A^f (\alpha_q,k_\perp)\barh_A^f(\beta_q,(q-k)_\perp)+h_A^f\leftrightarrow\barh_A^f
\label{HA}
\end{eqnarray}

Next, consider the case when both $\mu$ and $\nu$ are longitudinal. The non-vanishing terms are
\begin{eqnarray}
&&\hspace{-1mm}
\cheV_{1\mu_\parallel\nu_\parallel}^{\rm (2a)H}(x)~=~
{4p_{1\mu}p_{2\nu}\over s^2}{1\over 2s^3}\langle\bsi A_i(x)\sigma_{\star\alpha}\notp_2\gamma^j{1\over \alpha}\psi(0)\rangle_A
\langle\bsi B_j(0)\sigma_\bu^{~\alpha}\notp_1\gamma^i{1\over\beta}\psi(x)\rangle_B
\nonumber\\
&&\hspace{-1mm}
+~{4p_{1\nu}p_{2\mu}\over s^2}{1\over 2s^3}\langle\bsi A_i(x)\sigma_{\bu\alpha}\notp_2\gamma^j{1\over \alpha}\psi(0)\rangle_A
\langle\bsi B_j(0)\sigma_\star^{~\alpha}\notp_1\gamma^i{1\over\beta}\psi(x)\rangle_B
\label{kalv3long}\\
&&\hspace{-1mm}
-~{4p_{2\mu}p_{1\nu}\over s^2}{1\over 2s^3}\langle\bsi A^i(x)\sigma_{\bu k}\sigma_{\star j}{1\over \alpha}\psi(0)\rangle_A
\langle\bsi B^j(0)\sigma_\star^{~k}\sigma_{\bu i}{1\over\beta}\psi(x)\rangle_B
+\mu\leftrightarrow\nu
+x\leftrightarrow 0
\nonumber
\end{eqnarray}
Using Eq. (\ref{sigmasigmas}) it is easy to see that the third term in the r.h.s. is $\sim {q_\perp^2\over s}$ while
the first two terms in the r.h.s. can be rewritten as
\begin{eqnarray}
&&\hspace{-1mm}
{g^\parallel_{\mu\nu}\over s^3}\langle\bsi A^i(x)\sigma_{\star j}{1\over \alpha}\psi(0)\rangle_A
\langle\bsi B^j(0)\sigma_{\bu i}{1\over\beta}\psi(x)\rangle_B+x\leftrightarrow 0
\nonumber\\
&&\hspace{-1mm}
=~{g^\parallel_{\mu\nu}\over 2s^3}\langle\bsi A^i(x)\sigma_{\star i}{1\over \alpha}\psi(0)\rangle_A
\langle\bsi B^j(0)\sigma_{\bu j}{1\over\beta}\psi(x)\rangle_B
\nonumber\\
&&\hspace{-1mm}
+~{g^\parallel_{\mu\nu}\over s^3}\langle\bsi\big(A^i(x)\sigma_{\star j}-{g_{ij}\over 2}A^k(x)\sigma_{\star k}\big) {1\over \alpha}\psi(0)\rangle_A
\langle\bsi B^j(0)\sigma_{\bu i}{1\over\beta}\psi(x)\rangle_B+x\leftrightarrow 0
\label{kalv3longi}
\end{eqnarray}
The corresponding contribution to $V^{\rm H}_{\mu\nu}(q)$ yields
\begin{eqnarray}
&&\hspace{-1mm}
V_{1\mu_\parallel\nu_\parallel}^{\rm (2a)H}(q)~=~-{g^\parallel_{\mu\nu}\over 2Q_\parallel^2}\!\int d^2k_\perp {k_\perp^2(q-k)_\perp^2\over m^2}
H^f(q,k_\perp)
\nonumber\\
&&\hspace{-1mm}
-~{g^\parallel_{\mu\nu}\over Q_\parallel^2}\!\int {d^2k_\perp\over m^2} \big[(k,q-k)_\perp^2-\half k_\perp^2(q-k)_\perp^2\big]
H_A^f(q,k_\perp)
\nonumber\\
&&\hspace{11mm}
\label{kalvklad3long}
\end{eqnarray}
where again we used QCD equations of motion (\ref{9.56}), (\ref{9.58}) and parametrization (\ref{maelsa}).

The result for $V_{1\mu\nu}^{\rm (2a)H}(q)$ is the sum of Eqs. (\ref{kalvklad3perp}) and (\ref{kalvklad3long}):
\begin{eqnarray}
&&\hspace{-11mm}
V_{1\mu\nu}^{\rm (2a)H}(q)~=~{g^\perp_{\mu\nu}-g^\parallel_{\mu\nu}\over Q_\parallel^2}
\!\int d^2k_\perp {k_\perp^2(q-k)_\perp^2\over 2m^2}
H^f(q,k_\perp)
\label{v21otvet}\\
&&\hspace{-1mm}
+~{1\over Q_\parallel^2}\!\int d^2k_\perp {1\over m^2}
\big\{[k^\perp_\mu(q-k)^\perp_\nu+\mu\leftrightarrow\nu](k,q-k)_\perp
-k_\perp^2(q-k)^\perp_\mu(q-k)^\perp_\nu
\nonumber\\
&&\hspace{-1mm}
-~(q-k_\perp)^2k^\perp_\mu k^\perp_\nu
-{g_{\mu\nu}^\perp\over 2}k_\perp^2(q-k_\perp)^2-g^\parallel_{\mu\nu}\big[(k,q-k)_\perp^2-\half k_\perp^2(q-k)_\perp^2\big]\big\}
H_A^f(q,k_\perp)
\nonumber
\end{eqnarray}
It is possible to demonstrate  that the contribution of
\begin{equation}
\hspace{-0mm}
\cheV_{2\mu\nu}^{\rm (2a)H}(x)~=~{N_c\over 2s}
\langle A,B|[\psi_A^m(x)\sigma_{\mu\xi}\Xi_1^n(0)][\psi_B^n(0)\sigma_\nu^{~\xi}\Xi_2^m(x)|A,B\rangle
+\mu\leftrightarrow\nu+x\leftrightarrow 0
\label{vh21}
\end{equation}
to $V_{\mu\nu}(q)$ doubles the result (\ref{v21otvet}) (see Ref. \cite{Balitsky:2020jzt}  for proof) so subtracting trace we obtain
\begin{eqnarray}
&&\hspace{-1mm}
W^{\rm (2a)T}_{\mu\nu}(q)~=~2V_{1\mu\nu}^{\rm (2a)H}(q)-{1\over 2}g_{\mu\nu}2V_{1\xi}^{\rm (2)H,\xi}(q)
\label{v2otvet}\\
&&\hspace{-1mm}
=~{g^\perp_{\mu\nu}-g^\parallel_{\mu\nu}\over Q_\parallel^2}\!\int d^2k_\perp {1\over m^2}k_\perp^2(q-k)_\perp^2
H^f(q,k_\perp)
\nonumber\\
&&\hspace{-1mm}
+~{2\over Q_\parallel^2N_c}\!\int d^2k_\perp {1\over m^2}\big\{[k^\perp_\mu(q-k)^\perp_\nu+\mu\leftrightarrow\nu](k,q-k)_\perp
-k_\perp^2(q-k)^\perp_\mu(q-k)^\perp_\nu
\nonumber\\
&&\hspace{-1mm}
-~(q-k_\perp)^2k^\perp_\mu k^\perp_\nu+g_{\mu\nu}^\perp(k,q-k)_\perp^2
-g_{\mu\nu}^\perp k_\perp^2(q-k_\perp)^2\big]\big\}
H_A(q,k_\perp)
\nonumber
\end{eqnarray}
As we will see below, cancellation of terms $\sim g^\parallel_{\mu\nu}$ proportional to $H_A$ in the r.h.s of this equation is actually a consequence of (EM) gauge invariance.

Let us now consider 
\begin{eqnarray}
&&\hspace{-7mm}
\cheV_{1\mu\nu}^{\rm (2b)H}(x)~=~{N_c\over 2s}
\langle A,B|[\psi_A^m(x)\sigma_{\mu\xi}\psi_A^n(0)][\Bxi_2^n(0)\sigma_\nu^{~\xi}\Xi_2^m(x)|A,B\rangle
+\mu\leftrightarrow\nu+x\leftrightarrow 0
\nonumber\\
&&\hspace{-7mm}
=~
{1\over s^3}
\Big\{-p_{1\mu} \langle \bar\psi(x)A_j(x)\sigma_{\nu k}A_i(0)\psi(0)\rangle_A
\langle\big(\bar\psi{1\over \beta}\big)(0)\gamma^i\sigma_\bu^{~k}
\gamma^j{1\over\beta}\psi(x)\rangle_B  
\nonumber\\
&&\hspace{-7mm} 
+~\langle \bar\psi(x)A_j(x)\sigma_{\mu \bu}A_i(0)\psi(0)\rangle_A
\langle\big(\bar\psi{1\over \beta}\big)(0)\gamma^i\sigma_{\bu\nu_\perp}
\gamma^j{1\over\beta}\psi(x)\rangle_B \Big\} +\mu\leftrightarrow\nu  ~+~x\leftrightarrow 0
 \nonumber\\
&&\hspace{-1mm} 
=~
-{4p_{1\mu}p_{1\nu}\over s^4}
\langle \bar\psi(x)A_j(x)\sigma_{\star k}A_i(0)\psi(0)\rangle_A
\langle\big(\bar\psi{1\over \beta}\big)(0)\gamma^i\sigma_\bu^{~k}
\gamma^j{1\over\beta}\psi(x)\rangle_B  
\label{4.55}
\end{eqnarray}
where we neglected terms shown in Ref. \cite{Balitsky:2020jzt}  to be small. Using Eq. (\ref{sisigaga}) the r.h.s. of this equation
can be rewritten as
\begin{equation}
\hspace{-0mm}
\cheV_{1\mu\nu}^{\rm (2b)H}(x)
~=~
-{4p_{1\mu}p_{1\nu}\over s^4}
\langle \bar\psi(x)\notA(x)\sigma_{\star k}\notA(0)\psi(0)\rangle_A
\langle\big(\bar\psi{1\over \beta}\big)(0)\sigma_\bu^{~k}
{1\over\beta}\psi(x)\rangle_B  
~+~x\leftrightarrow 0
\label{v5razotvet}
\end{equation}
so the corresponding contribution to $W_{\mu\nu}$ takes the form
\begin{equation}
\hspace{-0mm}
V_{1\mu\nu}^{\rm (2b)H}(q)
~=~
-{4p_{1\mu}p_{1\nu}\over \beta_q^2s^2}
\!\int\! d^2k_\perp
{1\over m^2} k_\perp^2(k,q-k)_\perp
H^f(q,k_\perp)
\label{v5otvet}
\end{equation}
where we used Eqs. (\ref{maelqg2}) and (\ref{AA2}).

Next, similarly to Eq. (\ref{w1t}), the contribution 
\begin{eqnarray}
&&\hspace{-7mm}
\cheV_{2\mu\nu}^{\rm (2b)H}(x)~=~{N_c\over 2s}
\langle A,B|[\Bxi_1^m(x)\sigma_{\mu\xi}\Xi_1^n(0)][\bsi_B^n(0)\sigma_\nu^{~\xi}\psi_B^m(x)|A,B\rangle
+\mu\leftrightarrow\nu+x\leftrightarrow 0
\end{eqnarray}
is obtained by the projectile $\leftrightarrow$ target replacement, namely $p_1\leftrightarrow p_2$, 
$\alpha_q\leftrightarrow\beta_q$ and $k_\perp\leftrightarrow (q-k)_\perp$, and we get contribution to
$W^{\rm T}_{\mu\nu}$ in the form
\begin{eqnarray}
&&\hspace{-7mm}
W_{\mu\nu}^{\rm (2b)T}(q)
~=~V_{1\mu\nu}^{\rm (2b)H}(q)+V_{2\mu\nu}^{\rm (2b)H}(q)
\label{v34otvet}\\
&&\hspace{10mm}
=~
-\!\int\! d^2k_\perp \Big[{4p_{1\mu}p_{1\nu}\over \beta_q^2s^2}{k_\perp^2\over m^2}
+{4p_{2\mu}p_{2\nu}\over \alpha_q^2s^2}{(q-k)_\perp^2\over m^2}\Big]
 (k,q-k)_\perp H(q,k_\perp)
\nonumber
\end{eqnarray}

Finally, it is easy to see that the two-gluon terms
\begin{eqnarray}
&&\hspace{-1mm}
 \cheW_{\mu\nu}^{\rm (2c)T}(x)~
=~
{N_c\over 2s}
\big(g_{\mu\nu}g^{\alpha\beta}-\delta_\mu^\alpha\delta_\nu^\beta-\delta_\nu^\alpha\delta_\mu^\beta\big)
\langle A,B|[\Bxi_1^m(x)\sigma_{\alpha\xi}\psi_A^n(0)][\psi_B^n(0)\sigma_\beta^{~\xi}\Xi_2^m(x)]
\nonumber\\
&&\hspace{16mm}
+~[\psi_A^m(x)\sigma_{\alpha\xi}\Xi_1^n(0)][\Xi_2^n(0)\sigma_\beta^{~\xi}\psi_B^m(x)]|A,B\rangle
+x\leftrightarrow 0
\label{chewt25}
\end{eqnarray}
are $O\big({1\over N_c^2}\big)$ so we neglect them. 

Summarizing, we get  
\begin{equation}
W_{\mu\nu}^{\rm T}(q)~=~{\rm  Eq. (\ref{w1t})+Eq. (\ref{v2otvet})+Eq. (\ref{v34otvet})}
\end{equation}
and adding $W_{\mu\nu}^{\rm G}(q)$ from Eq. (\ref{gotvet}) we finally get
\begin{equation}
W^{\rm H}_{\mu\nu}(q)~=~W^{\rm H}_{\mu\nu}(q)+W^{\rm H2}_{\mu\nu}(q)
\label{resulth}
\end{equation}
The first, gauge-invariant,  part is given by
\begin{eqnarray}
&&\hspace{-1mm}
W^{\rm H}_{\mu\nu}(q)~
=~\sum_f e_f^2W^{\rmH f}_{\mu\nu}(q),~~~~W^{\rmH f}_{\mu\nu}(q)~=~ 
\!\int\!d^2k_\perp H^f(q,k_\perp)\calW^H_{\mu\nu}(q,k_\perp)
\label{resulthginv}
\end{eqnarray}
where $H^f$ is given by Eq. (\ref{H}) and
\begin{eqnarray}
&&\hspace{-1mm}
m^2\calW_{\mu\nu}^H(q,k_\perp)~
\label{WH}\\
&&\hspace{-1mm}
=~-k^\perp_\mu(q-k)^\perp_\nu-k^\perp_\nu(q-k)^\perp_\mu-g_{\mu\nu}^\perp(k,q-k)_\perp
+2{\tilq_\mu\tilq_\nu-q^\parallel_\mu q^\parallel_\nu \over Q_\parallel^4}k_\perp^2(q-k)_\perp^2
\nonumber\\
&&\hspace{-1mm}
-~\Big({q^\parallel_\mu\over Q_\parallel^2}\big[k_\perp^2(q-k)^\perp_\nu+k^\perp_\nu(q-k)_\perp^2\big]
+~{\tilq_\mu\over Q_\parallel^2}\big[k_\perp^2(q-k)^\perp_\nu-k^\perp_\nu(q-k)_\perp^2\big]+\mu\leftrightarrow\nu\Big)
\nonumber\\
&&\hspace{-1mm}
-~{\tilq_\mu\tilq_\nu+q^\parallel_\mu q^\parallel_\nu \over Q_\parallel^4}\big[q_\perp^2-2(k,q-k)_\perp\big](k,q-k)_\perp
-~{q^\parallel_\mu\tilq_\nu+ \tilq_\mu q^\parallel_\nu \over Q_\parallel^4}(2k-q,q)_\perp(k,q-k)_\perp
\nonumber
\end{eqnarray}
where $q^\parallel_\mu\equiv \alpha_qp_1+\beta_qp_2$ and $\tilq_\mu\equiv \alpha_qp_1-\beta_qp_2$. 
It is easy to see that $q^\mu W^\rmH_{\mu\nu}~=~0$. 
  
The second part is
\begin{eqnarray}
&&\hspace{-1mm}
W^{\rm 2H}_{\mu\nu}(q)~=~\sum_f e_f^2W^{\rmH 2f}_{\mu\nu}(q),~~~~~~
\label{whneinv}\\
&&\hspace{-1mm}
W^{\rmH 2f}_{\mu\nu}(q)~=~{1\over Q^2}\!\int d^2k_\perp
\bigg[ {1\over m^2}\big\{[k^\perp_\mu(q-k)^\perp_\nu+\mu\leftrightarrow\nu](k,q-k)_\perp
-k_\perp^2(q-k)^\perp_\mu(q-k)^\perp_\nu
\nonumber\\
&&\hspace{-1mm}
-~(q-k_\perp)^2k^\perp_\mu k^\perp_\nu+g_{\mu\nu}^\perp(k,q-k)_\perp^2
-g_{\mu\nu}^\perp k_\perp^2(q-k_\perp)^2\big]\big\}H_A^f(q,k_\perp)
+O\big({1\over N_c}\big)\bigg]~+~O\big({Q_\perp^4\over Q^4}\big)
\nonumber
\end{eqnarray}
where $H_A$ is given by Eq. (\ref{HA}). 
These terms are not gauge invariant: $q^\mu W^2_{\mu\nu}(q)\neq 0$. The reason is that gauge invariance is restored after adding terms like ${m_\perp^2\over Q^2}\times{\rm Eq. (\ref{pc})}$  which we do not calculate in this paper. Indeed, for example,
\begin{eqnarray}
&&\hspace{-1mm}
q^\mu W^2_{\mu\nu}(q)~\sim~{q^\perp_\nu q_\perp^2\over\alpha_q\beta_qs}~~~~
{\rm and}~~~q^\mu \times~{p_2^\mu q_\perp^\nu q_\perp^2\over \alpha_q^2\beta_qs^2}
~=~{q^\perp_\nu q_\perp^2\over\alpha_q\beta_qs}
\end{eqnarray}
They are of the same order so one should expect that gauge invariance is restored after calculation of the terms $\sim{p_2^\mu q_\perp^\nu q_\perp^2\over \alpha_q^2\beta_qs^2}$ which are beyond the scope of this paper.
For the same reason we see that all structures in Eq. (\ref{pc}) except 
${g_{\mu\nu}^\perp q_\perp^2\over\alpha_q\beta_qs}$
and ${q^\perp_\mu q^\perp_\nu \over\alpha_q\beta_qs}$ are determined by leading-twist TMDs $f_1$ and $h_1^\perp$.

\subsection{Exchange-type power corrections  from $J_{A}^\mu(x)J_{B}^\nu(0)$ terms \label{sec:extype}}
Power corrections of the ``exchange''  type come from the terms
\begin{equation}
\hspace{-0mm} 
\chew^{{\rm ex}ff'}_{\mu\nu}(x)~=~{N_c\over s}\langle p_A,p_B|[\Bsi_1^f(x)\gamma_\mu\Psi^f_1(x)][\Bsi^{f'}_2(0)\gamma_\nu\Psi^{f'}_2(0)]
+\mu\leftrightarrow\nu|p_A,p_B\rangle
~+~x\leftrightarrow 0
\label{wexff}
\end{equation}
where $\Psi_1$ and $\Psi_2$ are given by Eq. (\ref{fildz0}). 
As demonstrated in Ref. \cite{Balitsky:2020jzt} , the nonzero contributions are
\begin{eqnarray}
&&\hspace{-1mm}
\chew^{{\rm ex}ff'}_{\mu\nu}~=~{N_c\over s}\langle p_A,p_B|
[\Bxi_{1}(x)\gamma_\mu\psi_A(x)\big]^f\big[\bar\psi_B(0)\gamma_\nu\Xi_{2}(0)\big]^{f'}
\label{wexsym}\\
&&\hspace{13mm}
+~[\bar\psi_A(x)\gamma_\mu\Xi_{1}(x)\big]^f\big[\Bxi_{2}(0)\gamma_\nu\psi_B(0)\big]^{f'}
+~[\Bxi_{1}(x)\gamma_\mu\psi_A(x)\big]^f\big[\Bxi_{2}(0)\gamma_\nu\psi_B(0)\big]^{f'}
\nonumber\\
&&\hspace{27mm}
+~[\bar\psi_A(x)\gamma_\mu\Xi_{1}(x)\big]^f\big[\bar\psi_B(0)\gamma_\nu\Xi_{2}(0)\big]^{f'}+\mu\leftrightarrow\nu|p_A,p_B\rangle
~+~x\leftrightarrow 0
\nonumber
\end{eqnarray}
with transverse $\mu$ and $\nu$.  

It is convenient to calculate traceless part and trace separately.  Let us start from traceless part.
Separating color-singlet contributions with the help of the formula 
\begin{equation}
\langle\bsi_m A^a_i\psi_n\rangle~=~{2t^a_{nm}\over N_c^2-1}\langle\bsi A_i\psi\rangle
\label{colors2}
\end{equation}
and using Eq. (\ref{formulas67}), we get the corresponding term in $\cheW_{\mu\nu}$ in the form
\begin{eqnarray}
&&\hspace{-1mm}
\cheW^{{\rm ex}ff'}_{\mu\nu}-{g_{\mu\nu}^\perp\over 2}(\cheW^{{\rm ex},ff'})^m_{m}
\label{wexbez}\\
&&\hspace{-1mm}
=~{N_c\over (N_c^2-1)s^3}\Big(
\langle\big(\bsi{1\over\alpha}\big)(x)\slashed{p}_2\brA_\mu(0)\psi(x)
+\bar\psi(x)\brA_\mu(0)\slashed{p}_2{1\over \alpha}\psi(x)\rangle_A^f
\nonumber\\
&&\hspace{-1mm}
\times~\langle\big(\bar\psi{1\over\beta}\big)(0)\slashed{p}_1\breB_\nu(x)\psi(0)
+\bar\psi(0)\breB_\nu(x)\slashed{p}_1{1\over\beta}\psi(0)\rangle_B^{f'}
+\mu\leftrightarrow \nu-{\rm trace}\Big)
~+~x\leftrightarrow 0
\nonumber
\end{eqnarray}
where we used notations
\begin{equation}
\brA_i\equiv A_i-i\tilA_i\gamma_5.~~~\breB_i\equiv B_i-i\tilB_i\gamma_5
\label{defbrevs}
\end{equation}
Using parametrization of matrix elements (\ref{paramjse}) and  (\ref{paramjset})
we get
\begin{eqnarray}
&&\hspace{-1mm}
W_{\mu\nu}^{{\rm ex}ff'}(q)-{\rm trace}
~=~{s/2\over(2\pi)^4}\int\!d^4x ~e^{-iqx} \big(\cheW_{\mu\nu}^{{\rm ex}ff'}(x)
-{\rm trace}\big),
\nonumber\\
&&\hspace{-1mm}
=~{N_c\over (N_c^2-1)Q_\parallel^2}\!\int\! d^2k_\perp
[k^\perp_\mu(q-k)^\perp_\nu+\mu\leftrightarrow\nu+g^\perp_{\mu\nu}(k,q-k)_\perp]
 J^{1ff'}_{--}(q,k_\perp)
\label{wextraceless}
\end{eqnarray}
where $J^{iff'}_{--}(q,k_\perp)$ are defined in Eq. (\ref{Js}).

The trace part can be obtained in a similar way. Using Eq. (\ref{gammas11}) one gets
\begin{eqnarray}
&&\hspace{-1mm}
g^{mn}\cheW^{{\rm ex}ff'}_{mn}~
=~{2N_c\over (N_c^2-1)s^3}\Big(
\langle\big(\bsi{1\over\alpha}\big)(x)\brA_m(0)\slashed{p}_2\psi(x)
+\bar\psi(x)\slashed{p}_2\brA_m(0){1\over \alpha}\psi(x)\rangle_A^f
\nonumber\\
&&\hspace{-1mm}
\times~\langle\big(\bar\psi{1\over\beta}\big)(0)\breB^m(x)\slashed{p}_1\psi(0)\rangle_B^{f'}
+\bar\psi(0)\slashed{p}_1\breB^m(x){1\over\beta}\psi(0)\rangle_B^{f'}\Big)
~+~x\leftrightarrow 0
\label{wextrac}
\end{eqnarray}
The corresponding contribution to trace part of 
$W(\alpha_q,\beta_q,x_\perp)$ takes the form
\begin{eqnarray}
&&\hspace{-1mm}
\half g^{mn}W_{mn}^{{\rm ex}ff'}(q)~=~{s/4\over(2\pi)^4}\int\!d^4x ~e^{-iqx} g^{mn}\cheW_{mn}^{{\rm ex}ff'}(x_\perp)
\label{wextrace}\\
&&\hspace{-1mm}
=~-{N_c\over (N_c^2-1)Q_\parallel^2}\!\int\! d^2k_\perp (k,q-k)_\perp J^{2ff'}_{--}(q,k_\perp)\nonumber
\end{eqnarray}
which agrees with Eq. (6.2) from Ref. \cite{Balitsky:2017gis} after 
replacements $j_2=j_2^{\rm tw 3}-i\tilj_2^{\rm tw 3}$ and $\barj_2=j_1^{\rm tw 3}+i\tilj_1^{\rm tw 3}$.
It should be noted that the difference between $j_1$ and $j_2$ in traceless {\it vs} trace part is due to difference in formulas 
(\ref{formulas67}) and (\ref{gammas11}).

The total  contribution of ``exchange'' power corrections is the sum of Eqs. (\ref{wextraceless}) and  (\ref{wextrace}), see Eq. (\ref{wexotvet}) below.

\subsection{Resulting hadronic tensor for photon-mediated DY process \label{sect:fotonresult}}
It is convenient to represent hadronic tensor as a sum of three parts
\begin{eqnarray}
&&\hspace{-1mm}
W^{\gamma}_{\mu\nu}(q)~=~W^{1}_{\mu\nu}(q)+W^{2\rmH}_{\mu\nu}(q)+W^{3}_{\mu\nu}(q)
\label{foton3parts}
\end{eqnarray}
The first, gauge-invariant, part has the form
\begin{eqnarray}
&&\hspace{-1mm}
W^{1}_{\mu\nu}(q)~=~\sum_f e_f^2\big[W^{\rmF f}_{\mu\nu}(q)+W^{\rmH f}_{\mu\nu}(q)\big]
\nonumber\\
&&\hspace{-1mm}
W^{\rmF f}_{\mu\nu}(q)~=~\!\int\!d^2k_\perp F^f(q,k_\perp)\calW^F_{\mu\nu}(q,k_\perp),
\nonumber\\
&&\hspace{-1mm}
W^{\rmH f}_{\mu\nu}(q)~=~ \!\int\!d^2k_\perp H^f(q,k_\perp)\calW^H_{\mu\nu}(q,k_\perp)
\label{resultginv}
\end{eqnarray}
where functions $F^f(q,k_\perp)$ and  $H^f(q,k_\perp)$ are given by Eqs. (\ref{F}) and (\ref{H}) 
while\\
 $\calW^{\rmF}_{\mu\nu}(q,k_\perp)$ and $\calW^{\rmH}_{\mu\nu}(q,k_\perp)$ are presented in 
Eqs. (\ref{WF}) and (\ref{WH}), respectively.
\footnote{It should be mentioned that $W^F$ part coincides with the result obtained in Refs. \cite{Nefedov:2018vyt,Nefedov:2020ugj} using 
parton Reggeization approach to DY process \cite{Nefedov:2012cq}.}
 Note that 
$q^\mu W^\rmF_{\mu\nu}$ and  $q^\mu W^\rmH_{\mu\nu}$ are exactly zero without any ${q_\perp^2\over Q^2}$ corrections. 
This is similar to usual ``forward'' DIS, but different from off-forward DVCS where the cancellations of right-hand sides of Ward identities 
involve infinite towers of twists \cite{Braun:2011dg,Braun:2011zr,Braun:2020zjm}
  
The second part is given by Eq. (\ref{whneinv}).  It has the same order in $N_c$ as the first part
but unfortunately is determined by quark-quark-gluon TMDs which are virtually unknown. 

The third, ``exchange'', part is given by the sum of Eqs. (\ref{wextraceless}) and  (\ref{wextrace}). 
\begin{eqnarray}
&&\hspace{-1mm}
W^{\rm ex}_{\mu\nu}(q)~=~\sum_{f,f'}e_fe_{f'}W_{\mu\nu}^{{\rm ex}ff'}(q)~
\nonumber\\
&&\hspace{-1mm}
W_{\mu\nu}^{{\rm ex}ff'}(q)~=~{N_c\over (N_c^2-1)Q_\parallel^2}\!\int\! d^2k_\perp\big\{
[k^\perp_\mu(q-k)^\perp_\nu+\mu\leftrightarrow\nu+g^\perp_{\mu\nu}(k,q-k)_\perp]
 J^{1ff'}_{--}(q,k_\perp)
 \nonumber\\
&&\hspace{18mm}
-~g^\perp_{\mu\nu}(k,q-k)_\perp J^{2ff'}_{--}(q,k_\perp)\big\}
\label{wexotvet}
\end{eqnarray}

Both second and third part come only from transverse indices. 

These terms are not gauge invariant: $q^\mu W^2_{\mu\nu}(q)\neq 0$. The reason is that gauge invariance is restored after adding terms like ${m_\perp^2\over Q^2}\times{\rm Eq. (\ref{pc})}$  which we do not calculate in this paper. Indeed, for example,
\begin{eqnarray}
&&\hspace{-1mm}
q^\mu W^{2,3}_{\mu\nu}(q)~\sim~{q^\perp_\nu q_\perp^2\over\alpha_q\beta_qs}~~~~
{\rm and}~~~q^\mu \times~{p_2^\mu q_\perp^\nu q_\perp^2\over \alpha_q^2\beta_qs^2}
~=~{q^\perp_\nu q_\perp^2\over\alpha_q\beta_qs}
\end{eqnarray}
They are of the same order so one should expect that gauge invariance is restored after calculation of the terms $\sim{p_2^\mu q_\perp^\nu q_\perp^2\over \alpha_q^2\beta_qs^2}$ which are beyond the scope of this paper.
For the same reason we see that all structures in Eq. (\ref{pc}) except 
${g_{\mu\nu}^\perp q_\perp^2\over\alpha_q\beta_qs}$
and ${q^\perp_\mu q^\perp_\nu \over\alpha_q\beta_qs}$ are determined by leading-twist TMDs $f_1$ and $h_1^\perp$.

In the remaining Sections we will use formulas from this Section as guidelines for calculation of other hadronic tensors in Eq. (\ref{ws}).

\section{$Z$-mediated hadronic tensor \label{sec:Z}}
In this Section we will consider the hadronic tensor corresponding to the part of DY cross section mediated by $Z$-boson. Let us start from the symmetric tensor $W^{\rm ZS}_{\mu\nu}$ defined in Eq. (\ref{warray})

\subsection{Symmetric part of $Z$-mediated hadronic tensor \label{sec:zsym}}

As in the photon case, 
after integration over central fields $\calj^\mu$ in Eq. (\ref{2.3}) is replaced by
\begin{eqnarray}
&&\hspace{-1mm}
-\calj^\mu~=~\calj^\mu_1+\calj^\mu_2+\calj^\mu_{12}+\calj^\mu_{21},
\nonumber\\
&&\hspace{-1mm}
\calj^\mu_1~=~\sum_f c_f\Bsi_1^f \gamma_\mu(a_f-\gamma_5) \Psi_1^f,~~~~~
\calj^\mu_{12}~=~\sum_f c_f\Bsi_1^f \gamma_\mu(a_f-\gamma_5) \Psi_2^f
\label{jeiz}
\end{eqnarray}
and similarly for $\calj^\mu_2$ and $\calj^\mu_{21}$. 

We start from ``annihilation-type'' contributions and consider 
\begin{eqnarray}
&&\hspace{-1mm}
\cheW^{\rm ZSa}_{\mu\nu}~=~{N_c\over s}
\langle A,B|\calj^\mu_{12}(x)\calj^\mu_{21}(0)+\mu\leftrightarrow\nu|A,B\rangle+x\leftrightarrow 0
\label{wzsa}
\end{eqnarray}
Similarly to the photon case, we will perform calculations for one flavor and sum over flavors later. 
After  Fierz transformations (\ref{fierz}), (\ref{fierz5sym})
one gets the hadronic tensor  (\ref{wzsa}) in the form
\begin{eqnarray}
&&\hspace{-5mm}
c_f^{-2}\cheW^{\rm ZSa}_{\mu\nu}~=~{N_c\over s}\langle A,B|a_f^2\Psi_1\gamma_\mu\Psi_2(x)\Psi_2\gamma_\nu\Psi_1(0)+
\Psi_1\gamma_\mu\gamma_5\Psi_2(x)\Psi_2\gamma_\nu\gamma_5\Psi_1(0)
\nonumber\\
&&\hspace{4mm}
-~a_f\Psi_1\gamma_\mu\Psi_2(x)\Psi_2\gamma_\nu\gamma_5\Psi_1(0)
-a_f\Psi_1\gamma_\mu\gamma_5\Psi_2(x)\Psi_2\gamma_\nu\Psi_1(0)|A,B\rangle
+\mu\leftrightarrow\nu+x\leftrightarrow 0
\nonumber\\
&&\hspace{18mm}
=~(a_f^2+1)\cheW_{\mu\nu}^{\rmF f}(x)+(a_f^2-1)\cheW_{\mu\nu}^{\rmH f}(x)
-2a_f\cheW_{\mu\nu}^{5f}(x)
\label{chekalwsym}
\end{eqnarray}
where $W_{\mu\nu}^{\rmF f}(x)$ and $W_{\mu\nu}^{\rmH f}(x)$ are defined by Eqs. (\ref{fotonf}) and (\ref{hoton}) whereas
\begin{eqnarray}
&&\hspace{-1mm}
 \cheW_{\mu\nu}^{\rm 5}(x)~=~
{N_c\over 2s}\big(g_{\mu\nu}g^{\alpha\beta}-\delta_\mu^\alpha\delta_\nu^\beta-\delta_\nu^\alpha\delta_\mu^\beta\big)
\Big\{ [\Bsi_1^m(x)\gamma_\alpha\gamma_5\Psi_1^n(0)][\Bsi_2^n(0)\gamma^\beta\Psi_2^m(x)]
\nonumber\\
&&\hspace{14mm}
+~\gamma_\alpha\gamma_5\otimes\gamma_\beta\leftrightarrow\gamma_\alpha\otimes\gamma_\beta\gamma_5\Big\}
+x\leftrightarrow 0.
\end{eqnarray}
for flavor $f$ which we are considering.  The results for $W_{\mu\nu}^{\rmF f}(q)$ and $W_{\mu\nu}^{\rmH f}(q)$ are given by Eqs. (\ref{resultf}) and (\ref{resulth})
from previous Section while $W_{\mu\nu}^{5f}(q)$ must be evaluated anew.

We will prove now that 
\begin{equation}
\cheW_{\mu\nu}^{\rm 5}(x)~=~0
\label{wf5}
\end{equation}
with our accuracy. It includes the same terms as we assembled to $\cheW_{\mu\nu}^{\rmF}(x)$
but with additional $\gamma_5$ attached to one of the fermion fields. Since extra $\gamma_5$ cannot change the power 
of $s$ we need to look how the terms which gave leading contribution to $\cheW^{\rmF}_{\mu\nu}(x)$ are affected by extra $\gamma_5$.  
First, note that the leading-twist first term in the r.h.s. of Eq. (\ref{chewlt})  with extra $\gamma_5$ can be neglected. 
Indeed, if one replaces $\gamma_\alpha$ by $\gamma_\alpha\gamma_5$
(or $\gamma_\beta$ by  $\gamma_\beta\gamma_5$ ) the term $\sim g_{\mu\nu}$ vanishes and two other 
terms are $\sim {1\over s}p_{2\beta}\epsilon_{\alpha j}q^j$ as seen from the parametrization (\ref{mael5}). 
Similarly, replacement $\gamma_\beta\rightarrow\gamma_\beta\gamma_5$ gives terms of order of $\sim {1\over s}p_{1\alpha}\epsilon_{\beta j}q^j$ 
which we neglect, see the discussion after Eq. (\ref{pc}). 

Next, let us consider sum of  terms in $\cheW^{\rm (1)F}_{1\mu\nu}(x)$ and $\cheW^{\rm (1)F}_{2\mu\nu}(x)$ which has the form
\begin{eqnarray}
&&\hspace{-1mm}
{p_{2\mu}\over s^2}\Big[
\langle 
\bar\psi(x)\notp_2{1\over\alpha}\psi(0)\rangle_A
\langle\bar\psi\notB(0)\notp_1\gamma^\perp_\nu\psi(x)\rangle_B
+~\langle\bsi{1\over \alpha}(x)\notp_2\psi(0)\rangle_A\langle\bsi(0)\gamma_\nu^\perp\notp_1\bar\notB(x)\psi(x)\rangle_B
\nonumber\\
&&\hspace{11mm}
+~\psi(0)\otimes\psi(x)\leftrightarrow\psi(0)\gamma_5\otimes\gamma_5\psi(x) \Big]
~+~\mu\leftrightarrow\nu~+~x\leftrightarrow 0,
\label{odd1}
\end{eqnarray}
see Eqs. (\ref{onexi19}) and (\ref{chewf12}). 
\footnote{In these equations we dropped vanishing terms $+~\psi(0)\otimes\psi(x)\leftrightarrow\psi(0)\gamma_5\otimes\gamma_5\psi(x)$
but now we need them.}
It is easy to see that the replacement $\psi(0)\rightarrow\gamma_5\psi(0)$ gives either
vanishing projectile matrix element or vanishing target matrix element
after using Eqs. (\ref{maelqg1}), (\ref{maelqg2}), (\ref{9.25}), and (\ref{9.26}). Similarly, the sum of  contributions 
to $\cheW^{\rm (1)F}_{3\mu\nu}(x)+\cheW^{\rm(1)F}_{4\mu\nu}(x)$
\begin{eqnarray}
&&\hspace{-1mm}
{p_{1\mu}\over s^2}
\big[
\langle\bar\psi\notA(x)\notp_2\gamma_{\nu_\perp}\psi(0)\rangle_A\langle\bar\psi(0)\notp_1{1\over \beta}\psi(x)\rangle_B
+\langle\bar\psi(x)\gamma_{\nu_\perp}\notp_2\notA(0)\psi(0)\rangle_A\langle\big(\bar\psi{1\over \beta}\big)(0)\notp_1\psi(x)\rangle_B
\nonumber\\
&&\hspace{5mm}
+~\psi(0)\otimes\psi(x)\leftrightarrow\psi(0)\gamma_5\otimes\gamma_5\psi(x) \Big]
~+~\mu\leftrightarrow\nu~+~x\leftrightarrow 0
\label{w1oddvanishes}
\end{eqnarray}
vanishes after replacement $\psi(0)\rightarrow\gamma_5\psi(0)$ due to QCD equations of motion mentioned above. 
Thus,  $\cheW^{(1)5}_{\mu\nu}(x)=0$

Let us now consider leading $W^{\rm (2)F}_{\mu\nu}$ terms with two gluon operators 
discussed in Sect. \ref{sec:f2g}, see Eq. (\ref{chew2odin}). As we saw, 
the leading contribution comes from transverse $\mu$ and $\nu$. It is given by sum of Eqs.  (\ref{kalvedvaperp}),  (\ref{kalv2glav}), 
and the corresponding terms coming from\\ $[\Bxi_1^m(x)\gamma_\alpha\gamma_5\psi_A^n(0)][\Bxi_2^n(0)\gamma_\beta\gamma_5\psi_B^m(x)]$
\begin{eqnarray}
&&\hspace{-11mm}
\cheW_{1\mu\nu}^{\rm (2)F}(x)~=~
{g_{\mu\nu}^\parallel\over s^3}\Big(\langle\bsi(x)\notA(x)\notp_2\gamma_i{1\over \alpha}\psi(0)\rangle_A
\langle\bsi(0)\notB(0)\notp_1\gamma^i{1\over\beta}\psi(x)\rangle_B
\nonumber\\
&&\hspace{11mm}
+~\langle\big(\bsi{1\over \alpha}\big)(x)\gamma_i\notp_2\notA(0)\psi(0)\rangle_A
\langle\big(\bsi{1\over\beta}\big)(0)\gamma^i\notp_1\notB(x)\psi(x)\rangle_B\Big)~+~x\leftrightarrow 0
\label{kalvedvaperpmoo}
\end{eqnarray}
As was mentioned in the end of Sect. \ref{sec:f2g}, from equations (\ref{gammas11}) it is clear that the second term 
in the r.h.s. gives the same contribution as the first term. If, however, one replaces  $\psi(0)\leftrightarrow\gamma_5\psi(0)$
(or, equivalently, $\psi(x)\leftrightarrow\gamma_5\psi(x)$), it is easy to see that the two contributions 
cancel so $\cheW_{\mu\nu}^{\rm (2)F5}(x)=0$. 

Next, the leading terms in $\cheW_{2\mu\nu}^{\rm (2)F}(x)$  are given by Eqs. (\ref{v4longa}) and (\ref{chew2dva})
\begin{eqnarray}
&&\hspace{-1mm}
\cheW_{2\mu\nu}^{\rm (2)F}(x)~=~
-{4p_{1\mu} p_{1\nu}\over s^4}\Big(
\langle \bar\psi(x)\notA(x)\notp_2\notA(0)\psi(0)\rangle_A\langle\big(\bar\psi{1\over \beta}\big)(0)\notp_1{1\over\beta}\psi(x)\rangle_B
\nonumber\\
&&\hspace{-1mm}
+~\psi(0)\otimes\psi(x)\leftrightarrow \gamma_5\psi(0)\otimes\gamma_5\psi(x)\Big)
-{4p_{2\mu} p_{2\nu}\over s^4}
\Big(\langle \bar\psi(x) \notB(x)\notp_1 \notB(0)\psi(0)\rangle_B
\nonumber\\
&&\hspace{-1mm}
\times~\langle\big(\bar\psi{1\over \alpha}\big)(0)\notp_2
{1\over\alpha}\psi(x)\rangle_A
+\psi(0)\otimes\psi(x)\leftrightarrow \gamma_5\psi(0)\otimes\gamma_5\psi(x)\Big)~+~x\leftrightarrow 0
\end{eqnarray}
If we now replace $\psi(0)\leftrightarrow\gamma_5\psi(0)$ it is easy to see from Eqs. (\ref{maelqg2}) and (\ref{AA2}) that 
$\cheW_{\mu\nu}^{(2b)5}(x)=0$. As to $\cheW_{3\mu\nu}^{\rm (2)F}(x)$, from Eq. (\ref{chew2tri}) we see that it is $O\big({1\over N_c^2}\big)$ so we neglect it. 

Thus, we obtain the ``annihilation part'' of symmetric hadronic tensor due to Z-boson currents in the form
\begin{equation}
\hspace{-0mm}
W^{\rm ZSan}_{\mu\nu}~=~e^2\sum_fc_f^2\big[(a_f^2+1)W_{\mu\nu}^{f\rm F}(q)+(a_f^2-1)W_{\mu\nu}^{f\rm H}(q)\big]
\label{wzsyma}
\end{equation}
where $W_{\mu\nu}^{f\rmF}(q)$ and  $W_{\mu\nu}^{f\rm H}(q)$ are given by Eqs. (\ref{resultf}) and  (\ref{resulth}). 
Note that it is gauge invariant up to $W_{\mu\nu}^{2\rmH}(q)$ term discussed in the end of Sect. \ref{sec:foton}.  

  \subsubsection{Exchange-type power corrections to $W^{ZS}_{\mu\nu}$ \label{zexsym}}
  Power corrections of the ``exchange''  type come from the terms
\begin{equation}
\hspace{-0mm}
(\cheW^{\rm ZS}_{ff'})^{\rm ex}_{\mu\nu}(x)~
=~c_fc_{f'}\big[a_fa_{f'}\chew^{ff'{\rm ex}}_{\mu\nu}(x)+(\chew^{{\rm S}ff'}_{55})^{\rm ex}_{\mu\nu}(x)
-a_f(\chew^{{\rm S}ff'}_{5\rma})^{\rm ex}_{\mu\nu}(x)-a_{f'}(\chew^{{\rm S}ff'}_{5\rmb})^{\rm ex}_{\mu\nu}(x)\big]
\label{chewzsex}
\end{equation}
where $\chekalw^{\rm ex}_{\mu\nu}(x)$ is given by  Eq. (\ref{wexsym})
while $(\chekalw^{\rm S}_{55})^{\rm ex}_{\mu\nu}(x)$
$(\chekalw^{\rm S}_{5A,B})^{\rm ex}_{\mu\nu}(x)$ are defined as
\begin{eqnarray}
&&\hspace{-3mm}
(\chew^{{\rm S}ff'}_{55})^{\rm ex}_{\mu\nu}(x)~\equiv~{N_c\over s}\langle A,B|
[\Bsi_1(x)\gamma_\mu\gamma_5\Psi_1(x)][\Bsi_2(0)\gamma_\nu\gamma_5\Psi_2(0)]
+\mu\leftrightarrow\nu|A,B\rangle
+x\leftrightarrow 0,
\nonumber\\
&&\hspace{-3mm}
(\chew^{{\rm S}ff'}_{5\rma})^{\rm ex}_{\mu\nu}(x)~\equiv~{N_c\over s}\langle A,B|[\Bsi_1(x)\gamma_\mu\gamma_5\Psi_1(x)][\Bsi_2(0)\gamma_\nu\Psi_2(0)]
+\mu\leftrightarrow\nu|A,B\rangle
~+~x\leftrightarrow 0,
\nonumber\\
&&\hspace{-3mm}
(\chew^{{\rm S}ff'}_{5\rmb})^{\rm ex}_{\mu\nu}(x)
~\equiv~{N_c\over s}\langle A,B|[\Bsi_1(x)\gamma_\mu\Psi_1(x)][\Bsi_2(0)\gamma_\nu\gamma_5\Psi_2(0)]
+\mu\leftrightarrow\nu|A,B\rangle
~+~x\leftrightarrow 0
\nonumber\\
\label{wzexs}
\end{eqnarray}

As seen from the of comparison parametrizations (\ref{paramjse}) and 
(\ref{paramjset5}),  the replacement $\psi\rightarrow \gamma_5\psi$ in the projectile matrix elements 
leads to $k_\mu j_1\rightarrow \pm i\epsilon_{\mu\nu}k^\nu j_1$, 
$k_\mu j_2\rightarrow \pm i\epsilon_{\mu\nu}k^\nu j_2$.  Similarly, the replacement  $\psi\rightarrow \gamma_5\psi$ in target 
matrix elements yields $(q-k)_\mu j_{1,2}\rightarrow \pm i\epsilon_{\mu}(q-k)^\nu j_{1,2}$.  
Looking at the result (\ref{wexotvet})
and taking care of signs of replacements $\psi\rightarrow \gamma_5\psi$ in Eqs. (\ref{paramjse}) and (\ref{paramjset}), we obtain
\begin{eqnarray}
&&\hspace{-1mm}
(\chew^{{\rm S}ff'}_{55})^{\rm ex}_{\mu\nu}(x)=~{N_c\over (N_c^2-1)s^3}\Big(
\langle\big(\bsi{1\over\alpha}\big)(x)\slashed{p}_2\brA_\mu(0)\gamma_5\psi(x)
+\bar\psi(x)\brA_\mu(0)\slashed{p}_2{1\over \alpha}\gamma_5\psi(x)\rangle_A^f
\nonumber\\
&&\hspace{22mm}
\times~\langle\big(\bar\psi{1\over\beta}\big)(0)\slashed{p}_1\breB_\nu(x)\gamma_5\psi(0)\rangle_B
+\bar\psi(0)\breB_\nu(x)\slashed{p}_1{1\over\beta}\gamma_5\psi(0)\rangle_B^{f'}
+\mu\leftrightarrow \nu-{\rm trace}
\nonumber\\
&&\hspace{11mm}
+~g_{\mu\nu}^\perp
\langle\big(\bsi{1\over\alpha}\big)(x)\brA_m(0)\slashed{p}_2\gamma_5\psi(x)
+\bar\psi(x)\slashed{p}_2\brA_m(0){1\over \alpha}\gamma_5\psi(x)\rangle_A^f
\nonumber\\
&&\hspace{22mm}
\times~\langle\big(\bar\psi{1\over\beta}\big)(0)\breB^m(x)\slashed{p}_1\gamma_5\psi(0)\rangle_B
+\bar\psi(0)\slashed{p}_1\breB^m(x){1\over\beta}\gamma_5\psi(0)\rangle_B^{f'}\Big)
~+~x\leftrightarrow 0
\nonumber\\
&&\hspace{-1mm}
\Rightarrow~(W_{\mu\nu}^{{\rm S}ff'})_{55}^{\rm ex}(q)~
=~-{N_c\over (N_c^2-1)Q_\parallel^2}\!\int\! d^2k_\perp\big\{
[k^\perp_\mu(q-k)^\perp_\nu+\mu\leftrightarrow\nu+g^\perp_{\mu\nu}(k,q-k)_\perp]
\nonumber\\
&&\hspace{22mm}
\times~J^{1ff'}_{++}(q,k_\perp)-~g^\perp_{\mu\nu} (k,q-k)_\perp J^{2ff'}_{++}(q,k_\perp)\big\}
\label{wex55}
\end{eqnarray}
and
\begin{eqnarray}
&&\hspace{-5mm}
(\chekalw^{{\rm S}ff'}_{5\rma})^{\rm ex}_{\mu\nu}(x)=~{N_c\over (N_c^2-1)s^3}\Big(
\langle\big(\bsi{1\over\alpha}\big)(x)\slashed{p}_2\brA_\mu(0)\gamma_5\psi(x)
+\bar\psi(x)\brA_\mu(0)\slashed{p}_2{1\over \alpha}\gamma_5\psi(x)\rangle_A^f
\nonumber\\
&&\hspace{-5mm}
\times~\langle\big(\bar\psi{1\over\beta}\big)(0)\slashed{p}_1\breB_\nu(x)\psi(0)\rangle_B
+\bar\psi(0)\breB_\nu(x)\slashed{p}_1{1\over\beta}\psi(0)\rangle_B^{f'}
+\mu\leftrightarrow \nu-{\rm trace}
\nonumber\\
&&\hspace{-5mm}
+~g_{\mu\nu}^\perp
\langle\big(\bsi{1\over\alpha}\big)(x)\brA_m(0)\slashed{p}_2\gamma_5\psi(x)
+\bar\psi(x)\slashed{p}_2\brA_m(0){1\over \alpha}\gamma_5\psi(x)\rangle_A^f
\nonumber\\
&&\hspace{-5mm}
\times~\langle\big(\bar\psi{1\over\beta}\big)(0)\breB^m(x)\slashed{p}_1\gamma_5\psi(0)\rangle_B
+\bar\psi(0)\slashed{p}_1\breB^m(x){1\over\beta}\gamma_5\psi(0)\rangle_B^{f'}\Big)
~+~x\leftrightarrow 0
\nonumber\\
&&\hspace{-5mm}
\Rightarrow~(W_{\mu\nu}^{{\rm S}ff'})_{5\rma}^{\rm ex}(q)~
=~-{N_c\over (N_c^2-1)Q_\parallel^2}\!\int\! d^2k_\perp[\epsilon_{\mu m}k^m(q-k)_\nu+\mu\leftrightarrow\nu]
I^{1ff'}_{+-}(q,k_\perp)
\label{wexa5}
\end{eqnarray}
\begin{eqnarray}
&&\hspace{-1mm}
(\chekalw^{{\rm S}ff'}_{5\rmb})^{\rm ex}_{\mu\nu}(x)~=~{N_c\over (N_c^2-1)s^3}\Big(
\langle\big(\bsi{1\over\alpha}\big)(x)\slashed{p}_2\brA_\mu(0)\psi(x)
+\bar\psi(x)\brA_\mu(0)\slashed{p}_2{1\over \alpha}\psi(x)\rangle_A^f
\nonumber\\
&&\hspace{-1mm}
\times~\langle\big(\bar\psi{1\over\beta}\big)(0)\slashed{p}_1\breB_\nu(x)\gamma_5\psi(0)\rangle_B
+\bar\psi(0)\breB_\nu(x)\slashed{p}_1\gamma_5{1\over\beta}\psi(0)\rangle_B^{f'}
+\mu\leftrightarrow \nu-{\rm trace}
\nonumber\\
&&\hspace{-1mm}
+~g_{\mu\nu}^\perp
\langle\big(\bsi{1\over\alpha}\big)(x)\brA_m(0)\slashed{p}_2\gamma_5\psi(x)
+\bar\psi(x)\slashed{p}_2\brA_m(0){1\over \alpha}\gamma_5\psi(x)\rangle_A^f
\nonumber\\
&&\hspace{-1mm}
\times~\langle\big(\bar\psi{1\over\beta}\big)(0)\breB^m(x)\slashed{p}_1\gamma_5\psi(0)\rangle_B
+\bar\psi(0)\slashed{p}_1\breB^m(x){1\over\beta}\gamma_5\psi(0)\rangle_B^{f'}\Big)
~+~x\leftrightarrow 0
\nonumber\\
&&\hspace{-1mm}
\Rightarrow~(W_{\mu\nu}^{{\rm S}ff'})_{5\rmb}^{\rm ex}(q)~
=~{N_c\over (N_c^2-1)Q_\parallel^2}\!\int\! d^2k_\perp[k_\mu\epsilon_{\nu n}(q-k)^n+\mu\leftrightarrow\nu]  
I^{1ff'}_{-+}(q,k_\perp)
\label{wexb5}
\end{eqnarray}
where 

Note that the two last contributions are traceless since
\begin{equation}
\epsilon_{ij}\!\int\! d^2k_\perp ~k^i(q-k)^j \phi_1(k_\perp^2)\phi_2((q-k)_\perp^2)~=~0
\label{eijqikj}
\end{equation}
for any functions $\phi_1$ and $\phi_2$.

\subsubsection{Results for symmetric hadronic tensor for Z-mediated DY process}
It is convenient to represent the hadronic tensor $W^{\rm ZS}_{\mu\nu}$ as a sum of three parts
\begin{eqnarray}
&&\hspace{-1mm}
W^{\rm ZS}_{\mu\nu}(q)~=~W^{\rm ZS1}_{\mu\nu}(q)+W^{\rm ZS2}_{\mu\nu}(q)+W^{\rm ZS3}_{\mu\nu}(q)
\label{resultwzs}
\end{eqnarray}
The first, gauge-invariant, part is given by Eq. (\ref{chekalwsym})
\begin{eqnarray}
&&\hspace{-1mm}
W^{\rm ZS1}_{\mu\nu}(q)~=~e^2\sum_f c_f^2
\big[(a_f^2+1)W_{\mu\nu}^{\rmF f}(q)+(a_f^2-1)W_{\mu\nu}^{\rmH f}(q)\big]
\nonumber\\
&&\hspace{-1mm}
W^{\rm fF}_{\mu\nu}(q)~=~{1\over N_c}\!\int\!d^2k_\perp F^f(q,k_\perp)\calW^F_{\mu\nu}(q,k_\perp),
\nonumber\\
&&\hspace{-1mm}
=~W^{\rm fH}_{\mu\nu}(q)~=~ {1\over N_c}\!\int\!d^2k_\perp H^f(q,k_\perp)\calW^H_{\mu\nu}(q,k_\perp)
\label{resultzsinv}
\end{eqnarray}
where $W_{\mu\nu}^{\rmF f}(q)$ is given by Eq. (\ref{resultf}) and $W_{\mu\nu}^{\rmH f}(q)$ 
by Eq. 
\begin{equation}
\hspace{-0mm}
W^{\rm ZSan}_{\mu\nu}~=~e^2\sum_fc_f^2\big[(a_f^2+1)W_{\mu\nu}^{f\rm F}(q)+(a_f^2-1)W_{\mu\nu}^{f\rm H}(q)\big]
\label{wzsyma}
\end{equation}
where $W_{\mu\nu}^{\rmF f}(q)$ and  $W_{\mu\nu}^{\rm H f}(q)$ are given by Eqs. (\ref{resultf}) and  (\ref{resulthginv}).

The second part is
\begin{eqnarray}
&&\hspace{-1mm}
W^{\rm ZS2}_{\mu\nu}(q)~
=~\sum_f c_f^2(a_f^2-1)W^{\rm 2Hf}_{\mu\nu}(q)
\label{resultzsnoninv}
\end{eqnarray}
where $W^{\rm 2Hf}_{\mu\nu}(q)$ is given by Eq. (\ref{whneinv}).  

The third, ``exchange'', part is given by the sum
\begin{eqnarray}
&&\hspace{-1mm}
W_{\mu\nu}^{\rm ZSex}(q)~=~{N_c\over (N_c^2-1)Q_\parallel^2}\sum_{f,f'}c_fc_{f'}
\!\int\! d^2k_\perp\Big\{
[k^\perp_\mu(q-k)^\perp_\nu+\mu\leftrightarrow\nu+g^\perp_{\mu\nu}(k,q-k)_\perp]
\nonumber\\
&&\hspace{15mm}
\times~\big(a_fa_{f'}J^{1ff'}_{--}-~J^{1ff'}_{++}\big)
-~g^\perp_{\mu\nu} (k,q-k)_\perp\big(a_fa_{f'}J^{2ff'}_{++}
-~J^{2ff'}_{++}\big)
\label{wexffsym}\\
&&\hspace{-1mm}
+~a_f[\epsilon_{\mu m}k^m(q-k)_\nu+\mu\leftrightarrow\nu]
I^{1ff'}_{+-}(q,k_\perp)-~a_{f'}
[k_\mu\epsilon_{\nu n}(q-k)^n+\mu\leftrightarrow\nu]  
I^{1ff'}_{-+}(q,k_\perp)\Big\}
\nonumber
\end{eqnarray}
where $J^i_{\pm\pm}$ are listed in Eq. (\ref{Js}) and $I^i_{\pm\pm}$ in Eq. (\ref{Is}).

As in the photon case, the exchange power corrections are non-zero only for transverse $\mu$ and $\nu$ 
in our approximation.

\subsection{Antisymmetric part of $Z$-boson hadronic tensor \label{sec:zasym}}
The antisymmetric part of  hadronic tensor for cross section mediated by Z-boson is defined in Eq. (\ref{warray}) where we should make substitution $\psi\rightarrow\Psi_1+\Psi_2$. Let us start from annihilation-type contribution
\begin{eqnarray}
&&\hspace{-1mm}
\cheW^{\rm ZAa}_{\mu\nu}(x)~=~{N_c\over s}
\langle A,B|\calj_{12\mu}(x)\calj_{21\nu}(0)-\mu\leftrightarrow\nu|A,B\rangle-x\leftrightarrow 0
\label{wzaa}
\end{eqnarray}
 Using Fierz transformations (\ref{fierzasy}) and (\ref{fierz5}) it can be rewritten as
\begin{eqnarray}
&&\hspace{-1mm}
\cheW^{\rm ZAa}_{\mu\nu}~=~\sum_f c_f^2\Big[{i\over 2}\epsilon_{\mu\nu}^{~~\alpha\beta}
\big[-2a_f\chepizW_{\alpha\beta}^{f\cal F}+(a_f^2+1)\chepizW_{\alpha\beta}^{ f5}\big]
+(a_f^2-1)\cheW^{\rm as}_{\mu\nu}\Big]
\label{waza}
\end{eqnarray}
where
\begin{eqnarray}
&&\hspace{-1mm}
\chepizw_{\mu\nu}^{\cal F}(x)~
=~{N_c\over 2s}\langle A,B|[\Bsi_1^m(x)\gamma_\mu\Psi_1^n(0)][\Bsi_2^n(0)\gamma_\nu\Psi_2^m(x)]
\label{WZantisym}\\
&&\hspace{14mm}
+~\gamma_\mu\otimes\gamma_\nu\leftrightarrow\gamma_\mu\gamma_5\otimes\gamma_\nu\gamma_5-\mu\leftrightarrow\nu|A,B\rangle
-x\leftrightarrow 0
\nonumber\\
&&\hspace{-1mm}
\chepizw_{\mu\nu}^{\rm 5}(x)~
=~{N_c\over 2s}\langle A,B|[\Bsi_1^m(x)\gamma_\mu\gamma_5\Psi_1^n(0)][\Bsi_2^n(0)\gamma_\nu\Psi_2^m(x)]
\nonumber\\
&&\hspace{14mm}
+~\gamma_\mu\gamma_5\otimes\gamma_\nu\leftrightarrow\gamma_\mu\otimes\gamma_\nu\gamma_5-\mu\leftrightarrow\nu|A,B\rangle-x\leftrightarrow 0,
\nonumber\\
&&\hspace{-1mm}
\cheW^{\rm as}_{\mu\nu}(x)~=~{iN_c\over 2s}
\langle A,B|-[\Bsi_1^m(x)\Psi_1^n(0)][\Bsi_2^n(0)\sigma_{\mu\nu}\Psi_2^m(x)]
\nonumber\\
&&\hspace{14mm}
+~[\Bsi_1^m(x)\sigma_{\mu\nu}\Psi_1^n(0)][\Bsi_2^n(0)\Psi_2^m(x)]
+[\Bsi_1^m(x)\gamma_5\Psi_1^n(0)][\Bsi_2^n(0)\sigma_{\mu\nu}\gamma_5\Psi_2^m(x)]
\nonumber\\
&&\hspace{14mm}
-[\Bsi_1^m(x)\sigma_{\mu\nu}\gamma_5\Psi_1^n(0)][\Bsi_2^n(0)\gamma_5\Psi_2^m(x)]\big\}
|A,B\rangle-x\leftrightarrow 0
\nonumber
\end{eqnarray}
for the flavor under consideration.

Let us start from the $\pizw_{\mu\nu}^{\cal F}(x)$ given by the first line in Eq. (\ref{WZantisym} ) and compare it to
$W_{\mu\nu}(x)$ for the photon case. It is easy to see that if we take $W^{\rm F}_{\mu\nu}(x)$ and and antisymmetrize with respect to $\mu$ and $\nu$ instead of symmetrization, we will get $\pizw_{\mu\nu}^{\cal F}(x)$.
Since  antisymmetrization {\it vs} symmetrization
does not affect power counting in ${q_\perp^2\over s}$ parameter (or $\alpha_q,\beta_q\ll 1$ parameter), we can consider only terms
that gave leading contribution to $W_{\mu\nu}(x)$.

First, consider the leading-twist contribution
\footnote{Recall that ``check'' means $W$'s in coordinate space multiplied by ${2N_c\over s}$, cf. Eq. (\ref{defW})}
\begin{eqnarray}
&&\hspace{-1mm}
\chepizw_{\mu\nu}^{\calf, {\rm lt}}(x)~=~{1\over 2s}\big(\langle\bar\psi(x_\bu,x_\perp)\gamma_\mu\psi(0)\rangle_A\langle\bsi_B(0)\gamma_\nu\psi(x_\star,x_\perp)\rangle_B
\nonumber\\
&&\hspace{22mm}
+~\gamma_\mu\otimes\gamma_\nu\leftrightarrow \gamma_\mu\gamma_5\otimes\gamma_\nu\gamma_5-\mu\leftrightarrow\nu\big)
~-~x\leftrightarrow 0.
\label{wflt}
\end{eqnarray}
 Using parametrizations (\ref{Amael}) and (\ref{barbmael}) we get
\begin{eqnarray}
&&\hspace{-1mm}
\pizw_{\mu\nu}^{\calf,{\rm lt}}(q)~=~{1\over 16\pi^4}\!\int\! dx_\bu dx_\star d^2x_\perp~
e^{-i\alpha_qx_\bu-i\beta_qx_\star+i(q,x)_\perp}\chepizw_{\mu\nu}^{\calf,lt}(x)
\nonumber\\
&&\hspace{16mm}
=~
-{2\over s}\big(p_{1\mu}p_{2\nu}-p_{1\nu}p_{2\mu}\big)\!\int\! d^2k_\perp~\pizf(q,k_\perp)
\label{ltasym2}
\end{eqnarray}
where 
\begin{equation}
\pizf^f(q,k_\perp)~=~f_1^f(\alpha_q,k_\perp)\barf_1^f(\beta_q,(q-k)_\perp)~-~f_1^f\leftrightarrow\barf_1^f
\label{pizf}
\end{equation}
As usually, term with $f_1\leftrightarrow \barf_1$ comes from $x\leftrightarrow 0$ contribution.

Next, we consider terms with gluon operators and separate them as  in Sect. \ref{sec:wf} according to number of gluon fields (contained in  $\Xi$'s ):
\begin{equation}
\chepizw_{\mu\nu}^{\cal F}(x)~{=}~\chekalw_{\mu\nu}^{\calf,lt}+\chepizw_{\mu\nu}^{1\calf}(x)
+\chepizw_{\mu\nu}^{(2a)\calf}
+\chepizw_{\mu\nu}^{(2b)\calf}
+\chepizw_{\mu\nu}^{(2c)\calf}
\end{equation}
where leading-twist term $\chekalw_{\mu\nu}^{\calf,lt}$ was considered above, and
\begin{eqnarray}
&&\hspace{-1mm}
\chepizw_{\mu\nu}^{1\calf}(x)~=~{N_c\over 2s}\langle A,B|\big[\bar\psi_A^m(x)\gamma_\mu\Xi_1^n(0)\big]\big[\bar\psi_B^n(0)\gamma_\nu\psi_B^m(x)\big]
\label{chepizw1}\\
&&\hspace{-1mm}
+~\big[\Bxi_1^m(x)\gamma_\mu\psi_A^n(0)\big]\big[\bar\psi_B^n(0)\gamma_\nu\psi_A^m(x)\big]
+\big[\bar\psi_A^m(x)\gamma_\mu\psi_A^n(0)\big]\big[\bar\psi_B^n(0)\gamma_\nu\Xi_2^m(x)\big]
\nonumber\\
&&\hspace{-1mm}
+~\big[\bar\psi_A^m(x)\gamma_\mu\psi_A^n(0)\big]\big[\Bxi_2^n(0)\gamma_\nu\psi_B^m(x)\big]
+~\gamma_\mu\otimes\gamma_\nu\leftrightarrow \gamma_\mu\gamma_5\otimes\gamma_\nu\gamma_5-\mu\leftrightarrow\nu|A,B\rangle~-~x\leftrightarrow 0
\nonumber
\end{eqnarray}
\begin{eqnarray}
&&\hspace{-1mm}
\chepizw_{\mu\nu}^{(2a)\calf}(x)~=~
{N_c\over 2s}\langle A,B|\big[\bar\psi_A^m(x)\gamma_\mu\Xi_1^n(0)\big]\big[\bar\psi_B^n(0)\gamma_\nu\Xi_2^m(x)\big]
\label{chepizw2a}\\
&&\hspace{-1mm}
+~[\Bxi_1^m(x)\gamma_\mu\psi_A^n(0)\big]\big[\Bxi_2^n(0)\gamma_\nu\psi_B^m(x)\big]
+~\gamma_\mu\otimes\gamma_\nu\leftrightarrow \gamma_\mu\gamma_5\otimes\gamma_\nu\gamma_5-\mu\leftrightarrow\nu|A,B\rangle~-~x\leftrightarrow 0
\nonumber
\end{eqnarray}
\begin{eqnarray}
&&\hspace{-1mm}
\chepizW_{\mu\nu}^{(2b)\calf}(x)~=~
{N_c\over 2s}\langle A,B|
\big[\bar\psi_A^m(x)\gamma_\mu\psi_A^n(0)\big]\big[\Bxi_2^n(0)\gamma_\nu\Xi_2^m(x)\big]
\label{chepizw2b}\\
&&\hspace{-1mm}
+~
\big[\Bxi_1^m(x)\gamma_\mu\Xi_1^n(0)\big]\big[\bar\psi_B^n(0)\gamma_\nu\psi_B^m(x)\big]
+~\gamma_\mu\otimes\gamma_\nu\leftrightarrow \gamma_\mu\gamma_5\otimes\gamma_\nu\gamma_5-\mu\leftrightarrow\nu|A,B\rangle~-~x\leftrightarrow 0
\nonumber
\end{eqnarray}
and
\begin{eqnarray}
&&\hspace{-1mm}
\chepizW_{\mu\nu}^{(2c)\calf}(x)~=~
{N_c\over 2s}\langle A,B|\big[\Bxi_1^m(x)\gamma_\mu\psi_A^n(0)\big]\big[\bar\psi_B^n(0)\gamma_\nu \Xi_2^m(x)\big]
\label{chepizw2c}\\
&&\hspace{-1mm}
+~\big[\bar\psi_A^m(x)\gamma_\mu\Xi_1^n(0)\big]\big[\Bxi_2^n(0)\gamma_\nu\psi_B^m(x)\big]
+~\gamma_\mu\otimes\gamma_\nu\leftrightarrow \gamma_\mu\gamma_5\otimes\gamma_\nu\gamma_5-\mu\leftrightarrow\nu|A,B\rangle~-~x\leftrightarrow 0
\nonumber
\end{eqnarray}

The corresponding contributions to $\pizW_{\mu\nu}(q)$ will be denoted $\pizW_{\mu\nu}^{(1)\calf}$,  $\pizW_{\mu\nu}^{(2a)\calf}$,  $\pizW_{\mu\nu}^{(2b)\calf}$, and 
 $\pizW_{\mu\nu}^{(2c)\calf}$, respectively.
We will consider these contributions in turn following the analysis in Sect. \ref{sec:wf}.

\subsubsection{One-gluon terms in $\pizw_{\mu\nu}^{\cal F}$}
First, we consider terms with one gluon operator and start with \\
$\big[\bar\psi_A^m(x)\gamma_\mu\Xi_1^n(0)\big]\big[\bar\psi_B^n(0)\gamma_\nu\psi_B^m(x)\big]$.
Using  $\Xi_1~=~-{\slashed{p}_2\over s}\gamma^iB_i{1\over \alpha}\psi_A$  and separating color-singlet terms, we get
\begin{eqnarray}
&&\hspace{-1mm}
\chepizw_{1\mu\nu}^{(1)\calf}(x)~=~
-{1\over 2s^2}
\big\{\langle\bsi(x)\gamma_\mu\notp_2\gamma_i{1\over \alpha}\psi(0)\rangle_A
\langle\bsi B^i(0)\gamma_\nu\psi(x)\rangle_B
\nonumber\\
&&\hspace{5mm}
~+~(\psi(0)\otimes\psi(x)\leftrightarrow\gamma_5\psi(0)\otimes\gamma_5\psi(x)~-~\mu\leftrightarrow\nu\big\}
~-~x\leftrightarrow 0
\label{asonexi2}
\end{eqnarray}
As we discussed above, we need to consider only terms  which gave leading contribution for symmetric case, 
i.e. with one index longitudinal and the other transverse. Similarly to Eq. (\ref{onexi19}), for longitudinal $\mu$
and transverse $\nu$ we get
\begin{eqnarray}
&&\hspace{-11mm}
\Big({2p_1^\mu p_2^{\mu'}\over s}+\mu\leftrightarrow\mu'\Big)\chepizw_{\mu\nu}^{(1)\calf}(x)~
\nonumber\\
&&\hspace{-1mm}
=~
-\Big({p_{2\mu} p_1^{\mu'}\over s^3}+\mu\leftrightarrow\mu'\Big)
\big\{\langle\bar\psi(x)\gamma_{\mu'}\notp_2\gamma_i{1\over \alpha}\psi(0)\rangle_A
\langle\bar\psi B^i(0)\gamma_{\nu_\perp}\psi(x)\rangle_B
\nonumber\\
&&\hspace{11mm}
+~(\psi(0)\otimes\psi(x)\leftrightarrow\gamma_5\psi(0)\otimes\gamma_5\psi(x)
~-~\mu'\leftrightarrow\nu\big\}~-~x\leftrightarrow 0
\label{asonexi12}
\end{eqnarray}
The term proportional to $p_{2\mu}$ in the r.h.s. can be expressed using Eq. (\ref{gammas1fild}) as follows:
\begin{eqnarray}
&&\hspace{-1mm}
{p_{2\mu}\over s^3}
\Big[
\langle \bar\psi(x)\gamma_{\nu_\perp}\notp_2\gamma_i{1\over \alpha}\psi(0)\rangle_A
\langle \bar\psi B^i(0)\notp_1\psi(x)\rangle_B
-\langle\bar\psi(x)\notp_1\notp_2\gamma_i{1\over \alpha}\psi(0)\rangle_A
\langle\bar\psi B^i(0)\gamma_{\nu_\perp}\psi(x)\rangle_B
\nonumber\\
&&\hspace{33mm}
+~(\psi(0)\otimes\psi(x)\leftrightarrow\gamma_5\psi(0)\otimes\gamma_5\psi(x)\Big]
~-~x\leftrightarrow 0
\nonumber\\
&&\hspace{-1mm}
=~{p_{2\mu}\over s^3}\Big[\langle\bar\psi(x)\notp_2{1\over \alpha}\psi(0) \rangle_A
\langle \bar\psi(0)\notB(0)\notp_1\gamma_{\nu_\perp}\psi(x)\rangle_B
\nonumber\\
&&\hspace{33mm}+~(\psi(0)\otimes\psi(x)\leftrightarrow\gamma_5\psi(0)\otimes\gamma_5\psi(x))\Big]
~-~x\leftrightarrow 0
\label{asonexi13}
\end{eqnarray}
since the second term in the first line is $O({q_\perp^2\over s}\big)$ with respect to the first one. 
Next, if $\nu$ is longitudinal and $\mu$ transverse, we consider $\chepizw_{1\nu\mu}^{(1)F}=-\chepizw_{1\mu\nu}^{(1)F}$, repeat the above calculation and get result (\ref{asonexi13}) with $\mu\leftrightarrow\nu$. Thus, the case with  longitudinal $\nu$ and transverse $\mu$ is obtained from (\ref{asonexi13}) by
$-(\mu\leftrightarrow\nu)$ replacement, so
using Eqs. (\ref{maelqg1}), (\ref{9.22}) and parametrizations from Sect. \ref{sec:paramlt} we obtain
\begin{eqnarray}
&&\hspace{-1mm}
\pizW_{1\mu\nu}^{(1)\calf}(q)~=~
{p_{2\mu}\over \alpha_qs}\!\int\! d^2k_\perp~(q-k)^\perp_\nu \pizf(q,k_\perp)~-~\mu\leftrightarrow\nu
\label{XiAs}
\end{eqnarray}
which is the same as  Eq. (\ref{onexi20}), only with antisymmetrization in $\mu,\nu$ and $f\leftrightarrow\barf$  instead of symmetrizaton.

The second term in the r.h.s. of Eq. (\ref{chepizw1})  with longitudinal $\mu$ and transverse $\nu$ can be obtained in a similar way.
Repeating steps from Eq. (\ref{asonexi2}) to Eq. (\ref{asonexi13}), we get
\begin{eqnarray}
&&\hspace{-1mm}
\chepizw_{2\mu\nu}^{(1)\calf}(x)~=~{N_c\over 2s}\langle A,B|\big[\Bxi_1^m(x)\gamma_\mu\psi_A^n(0)\big]\big[\bar\psi_B^n(0)\gamma_\nu\psi_A^m(x)\big]
|A,B\rangle
\nonumber\\
&&\hspace{22mm}
+~\gamma_\mu\otimes\gamma_\nu\leftrightarrow \gamma_\mu\gamma_5\otimes\gamma_\nu\gamma_5-\mu\leftrightarrow\nu|A,B\rangle~-~x\leftrightarrow 0
\nonumber\\
&&\hspace{-1mm}
=~
-{1\over 2s^2}
\big\{\langle\big(\bar\psi{1\over\alpha}\big)(x)\gamma_i\notp_2\gamma_\mu{1\over \alpha}\psi(0)\rangle_A
\langle\bsi(0)\gamma_\nu B^i(x)\psi(x)\rangle_B
\nonumber\\
&&\hspace{22mm}
+~(\psi(0)\otimes\psi(x)\leftrightarrow\gamma_5\psi(0)\otimes\gamma_5\psi(x)~-~\mu\leftrightarrow\nu\big\}
~-~x\leftrightarrow 0
\nonumber\\
&&\hspace{-1mm}
=~
-\Big({p_{2\mu} p_1^{\mu'}\over s^3}+\mu\leftrightarrow\mu'\Big)
\big\{\langle\bar\psi(x)\gamma_i\notp_2\gamma_{\mu'}{1\over \alpha}\psi(0)\rangle_A
\langle\bar\psi(0)\gamma_{\nu_\perp} B^i(x)\psi(x)\rangle_B
\nonumber\\
&&\hspace{22mm}
+~(\psi(0)\otimes\psi(x)\leftrightarrow\gamma_5\psi(0)\otimes\gamma_5\psi(x)
~-~\mu'\leftrightarrow\nu\big\}~-~x\leftrightarrow 0
\nonumber\\
&&\hspace{-1mm}
\simeq~{p_{2\mu}\over s^3}\Big[\langle\bar\psi(x)\notp_2{1\over \alpha}\psi(0) \rangle_A
\langle \bar\psi(0)\gamma_{\nu_\perp}\notp_1\notB(x)\psi(x)\rangle_B
\nonumber\\
&&\hspace{22mm}
+~(\psi(0)\otimes\psi(x)\leftrightarrow\gamma_5\psi(0)\otimes\gamma_5\psi(x))\Big]
~-~x\leftrightarrow 0
\label{asonexixz}
\end{eqnarray}
The opposite case with transverse $\mu$ and longitudinal $\nu$ is obtained by $-(\mu\leftrightarrow\nu)$
and therefore from Eq. (\ref{9.26}) we get
\begin{eqnarray}
&&\hspace{-1mm}
\pizW_{2\mu\nu}^{(1)\calf}(q)~=~
{p_{2\mu}\over \alpha_qs}\!\int\! d^2k_\perp~(q-k)^\perp_\nu \pizf(q,k_\perp)~-~\mu\leftrightarrow\nu
\label{BxiAs}
\end{eqnarray}
which doubles the result (\ref{XiAs}) similarly to the symmetric case.

Rewriting now the third  term in the r.h.s. of Eq. (\ref{chepizw1})  with longitudinal $\mu$ and transverse $\nu$ and repeating the above steps, 
we get (recall $\Xi_2~=~-{\slashed{p}_1\over s}\gamma^iA_i{1\over\beta}\psi_B$):
\begin{eqnarray}
&&\hspace{-1mm}
\chepizw_{1\mu\nu}^{(2)\calf}(x)~=~{N_c\over 2s}\langle A,B|
\big[\bar\psi_A^m(x)\gamma_\mu\psi_A^n(0)\big]\big[\bar\psi_B^n(0)\gamma_\nu\Xi_2^m(x)\big]
\label{asonexixz1}\\
&&\hspace{-1mm}
+~\gamma_\mu\otimes\gamma_\nu\leftrightarrow \gamma_\mu\gamma_5\otimes\gamma_\nu\gamma_5-\mu\leftrightarrow\nu|A,B\rangle~+~x\leftrightarrow 0
\nonumber\\
&&\hspace{-1mm}
=~
-{1\over 2s^2}
\big\{\langle\bar\psi(x)A^i(x)\gamma_\mu\psi(0)\rangle_A
\langle\bsi(0)\gamma_\nu\notp_1\gamma_i{1\over\beta}\psi(x)\rangle_B
\nonumber\\
&&\hspace{22mm}
+~(\psi(0)\otimes\psi(x)\leftrightarrow\gamma_5\psi(0)\otimes\gamma_5\psi(x)~-~\mu\leftrightarrow\nu\big\}
~-~x\leftrightarrow 0
\nonumber\\
&&\hspace{-1mm}
=~
-\Big({p_{1\mu} p_2^{\mu'}\over s^3}+\mu\leftrightarrow\mu'\Big)
\big\{\langle\bar\psi(x)A^i(x)\gamma_{\mu'}\psi(0)\rangle_A
\langle\bsi(0)\gamma_\nu\notp_1\gamma_i{1\over\beta}\psi(x)\rangle_B
\nonumber\\
&&\hspace{22mm}
+~(\psi(0)\otimes\psi(x)\leftrightarrow\gamma_5\psi(0)\otimes\gamma_5\psi(x)~-~\mu'\leftrightarrow\nu\big\}
~-~x\leftrightarrow 0
\nonumber\\
&&\hspace{-1mm}
=~{p_{1\mu}\over s^3}\big\{\langle\bar\psi(x)\notp_2\brA_{\nu_\perp}(x)\psi(0)\rangle_A\langle\bsi(0)\notp_1{1\over\beta}\psi(x)\rangle_B
\nonumber\\
&&\hspace{22mm}
+~(\psi(0)\otimes\psi(x)\leftrightarrow\gamma_5\psi(0)\otimes\gamma_5\psi(x)\big\}
~-~x\leftrightarrow 0
\nonumber
\end{eqnarray}
As above,  the case with  longitudinal $\nu$ and transverse $\mu$ is obtained from (\ref{asonexixz1}) by
$-(\mu\leftrightarrow\nu)$ replacement, so
using Eqs. (\ref{maelqg1}), (\ref{9.22}) and parametrizations from Sect. \ref{sec:paramlt} we obtain
\begin{eqnarray}
&&\hspace{-1mm}
\pizW_{1\mu\nu}^{(2)\calf}(q)~=~
-{p_{1\mu}\over \beta_qs}\!\int\! d^2k_\perp~k^\perp_\nu \pizf(q,k_\perp)~-~\mu\leftrightarrow\nu
\label{XiAs2}
\end{eqnarray}
Finally, similarly to Eq. (\ref{asonexixz}), it can be demonstrated that the forth term in the r.h.s. of Eq. (\ref{chepizw1}) $\sim\big[\bar\psi_A^m(x)\gamma_\mu\psi_A^n(0)\big]\big[\Bxi_2^n(0)\gamma_\nu\psi_B^m(x)\big]$ doubles the contribution (\ref{XiAs2}) of the third term,
so we get 
\begin{eqnarray}
&&\hspace{-1mm}
\pizW_{\mu\nu}^{(1)\calf}(q)~=~
2\!\int\! d^2k_\perp~\Big[{p_{2\mu}(q-k)^\perp_\nu\over \alpha_qs}-{p_{1\mu}k^\perp_\nu\over \beta_qs}\Big] \pizf(q,k_\perp)~-~\mu\leftrightarrow\nu
\label{XiBxiAs}
\end{eqnarray}
Note that it can be obtained from Eq. (\ref{w1fotvet}) by replacement of symmetrization in $\mu\leftrightarrow\nu$ and 
$f\leftrightarrow\barf$ with antisymmetrization.

\subsubsection{Two-gluon terms in $\pizw_{\mu\nu}^{\cal F}$ \label{sec:2gluonsF}}
We start from the first term in the r.h.s. of Eq. (\ref{chepizw2a})
\begin{eqnarray}
&&\hspace{-1mm}
\chepizw_{1\mu\nu}^{(2a)\calf}(x)~=~
{N_c\over 2s}\langle A,B|\big[\bar\psi_A^m(x)\gamma_\mu\Xi_1^n(0)\big]\big[\bar\psi_B^n(0)\gamma_\nu\Xi_2^m(x)\big]
\label{chepizw2a1}\\
&&\hspace{-1mm}
+~\gamma_\mu\otimes\gamma_\nu\leftrightarrow \gamma_\mu\gamma_5\otimes\gamma_\nu\gamma_5-\mu\leftrightarrow\nu|A,B\rangle~-~x\leftrightarrow 0
\nonumber
\end{eqnarray}
Separating color-singlet contributions one can rewrite Eq. (\ref{chepizw2a1}) as
\begin{eqnarray}
&&\hspace{-1mm}
\chepizw_{1\mu\nu}^{(2a)\calf}(x)~=~{1\over 2s^3}\big\{\langle\bsi A_i(x)\gamma_\mu\notp_2\gamma^j{1\over \alpha}\psi(0)\rangle_A
\langle\bsi B_j(0)\gamma_\nu\notp_1\gamma^i{1\over\beta}\psi(x)\rangle_B
\nonumber\\
&&\hspace{10mm}
+~\psi(0)\otimes\psi(x)\leftrightarrow\gamma_5\psi(0)\otimes\gamma_5\psi(x)-\mu\leftrightarrow\nu\}
~-~x\leftrightarrow 0
\label{kalve2}
\end{eqnarray}
Similar to the symmetric case discussed in Sect. \ref{sec:f2g}, the leading term comes from  $\mu$ and $\nu$ that are both transverse.
In this case we can use formula (\ref{formula9a}) and get
\begin{eqnarray}
&&\hspace{-1mm}
\chepizw_{1\mu_\perp\nu_\perp}^{(2a)\calf}(x)~
=~{1\over 2s^3}\big\{\langle\bsi A_i(x)\gamma_{\mu_\perp}\notp_2\gamma^j{1\over \alpha}\psi(0)\rangle_A
\langle\bsi B_j(0)\gamma_{\nu_\perp}\notp_1\gamma^i{1\over\beta}\psi(x)\rangle_B
\label{pizwedvaperp}\\
&&\hspace{10mm}
+~\psi(0)\otimes\psi(x)\leftrightarrow\gamma_5\psi(0)\otimes\gamma_5\psi(x)-\mu\leftrightarrow\nu\}
~-~x\leftrightarrow 0
\nonumber\\
&&\hspace{10mm}
=~-{1\over s^3}\Big(\langle\bsi(x)\notp_2\brA_\mu(x){1\over \alpha}\psi(0)\rangle_A
\langle\bsi(0)\notp_1\breB_\nu(0){1\over\beta}\psi(x)\rangle_B-\mu\leftrightarrow\nu\Big)~-~x\leftrightarrow 0
\nonumber
\end{eqnarray}
Using now Eqs. (\ref{maelqg1}), (\ref{9.22}) and (\ref{9.25}) we obtain contribution to $\pizW_{\mu\nu}^{F}(q)$
 in the form
\begin{eqnarray}
&&\hspace{-1mm}
\pizw_{1\mu_\perp\nu_\perp}^{(2a)\calf}(x)
=~{1\over 16\pi^4}\!\int\! dx_\bu dx_\star d^2x_\perp~e^{-i\alpha_qx_\bu-i\beta_qx_\star+i(q,x)_\perp}\chepizw_{1\mu_\perp\nu_\perp}^{(2a)F}(x)
\nonumber\\
&&\hspace{-1mm}
=~{1\over Q_\parallel^2}\int\! d^2k_\perp [k^\perp_\mu(q-k)^\perp_\nu-\mu\leftrightarrow\nu] \pizf(q,k_\perp)
~=~0.
\label{pizwe2glav}
\end{eqnarray}
This term vanishes after integration over $k_\perp$ but we will keep it for a while since we want to have
gauge invariance at the integrand level, see Eq. (\ref{pizwe}) below.

Next, second term in the r.h.s. of Eq. (\ref{chepizw2a}) 
\begin{eqnarray}
&&\hspace{-1mm}
\chepizw_{2\mu\nu}^{(2a)\calf}(x)~=~
{N_c\over 2s}\langle A,B|
[\Bxi_1^m(x)\gamma_\mu\psi_A^n(0)\big]\big[\Bxi_2^n(0)\gamma_\nu\psi_B^m(x)\big]
\nonumber\\
&&\hspace{-1mm}
+~\gamma_\mu\otimes\gamma_\nu\leftrightarrow \gamma_\mu\gamma_5\otimes\gamma_\nu\gamma_5-\mu\leftrightarrow\nu|A,B\rangle~-~x\leftrightarrow 0
\label{pizwedva2perp}
\end{eqnarray}
at transverse $\mu$ and $\nu$ can be transformed to
\begin{eqnarray}
&&\hspace{-2mm}
\chepizw_{2\mu_\perp\nu_\perp}^{(2a)\calf}(x)~=~{1\over 2s^3}\big\{\langle\big(\bsi{1\over \alpha}\big)(x)\gamma^j\notp_2\gamma_{\mu_\perp} A_i(0)\psi(0)\rangle_A
\langle\big(\bsi{1\over\beta}\big)(0)\gamma^i\notp_1\gamma_{\nu_\perp} B_j(x)\psi(x)\rangle_B
\label{pizwe2perpa}\\
&&\hspace{9mm}
+~\psi(0)\otimes\psi(x)\leftrightarrow\gamma_5\psi(0)\otimes\gamma_5\psi(x)-\mu\leftrightarrow\nu\}
~-~x\leftrightarrow 0
\nonumber\\
&&\hspace{9mm}
=~-{1\over s^3}\big\{\langle\big(\bsi{1\over \alpha}\big)(x)\brA_{\mu_\perp}(0)\notp_2\psi(0)\rangle_A
\langle\big(\bsi{1\over\beta}\big)(0)\breB_{\nu_\perp}(x)\notp_1\psi(x)\rangle_B
-\mu\leftrightarrow\nu\}
~-~x\leftrightarrow 0
\nonumber
\end{eqnarray}
where we used formula (\ref{formula9a}). It is easy to see that the corresponding contribution to 
$\pizw_{2\mu_\perp\nu_\perp}^{(2a)\calf}(q)$ doubles the result (\ref{pizwe2glav}), same as for Eq. (\ref{w2afotvet}) in the symmetric case, 
and we obtain
\begin{eqnarray}
&&\hspace{-1mm}
\pizw_{\mu_\perp\nu_\perp}^{(2a)\calf}(x)
~=~{2\over Q_\parallel^2}\int\! d^2k_\perp [k^\perp_\mu(q-k)^\perp_\nu-\mu\leftrightarrow\nu] \pizf(q,k_\perp).
\label{pizwe2}
\end{eqnarray}
Again, this integral vanishes, but we keep the integrand as a part of Eq. (\ref{pizwe}) in order to
have gauge invariance (\ref{ginvas}) visible at the integrand level.

As one can anticipate from  Eq. (\ref{w2botvet}) for the symmetric case, the contribution from Eq. (\ref{chepizw2b})
vanishes. Indeed, e.g. for the first term we get

\begin{eqnarray}
&&\hspace{-1mm}
{N_c\over 2s}\langle A,B|
\big[\bar\psi_A^m(x)\gamma_\mu\psi_A^n(0)\big]\big[\Bxi_2^n(0)\gamma_\nu\Xi_2^m(x)\big]
+~\gamma_\mu\otimes\gamma_\nu\leftrightarrow \gamma_\mu\gamma_5\otimes\gamma_\nu\gamma_5-\mu\leftrightarrow\nu|A,B\rangle
\nonumber\\
&&\hspace{-1mm}
=~{p_{1\nu}\over s^3}\Big(
\langle \bar\psi(x)A_j(x)\gamma_\mu A_i(0)\psi(0)\rangle_A\langle\big(\bar\psi{1\over \beta}\big)(0)\gamma^i\notp_1
\gamma^j{1\over\beta}\psi(x)\rangle_B
\nonumber\\
&&\hspace{-1mm}
+~\psi(0)\otimes\psi(x)\leftrightarrow \gamma_5\psi(0)\otimes\gamma_5\psi(x)\Big)
~-~\mu\leftrightarrow \nu
\label{v4razasy}
\end{eqnarray}
Next, 
if the index $\mu$ is transverse, the contribution of this  equation to $\pizw^{\calf}_{\mu\nu}$ is of order of $p_{1\nu}q^\perp_\mu{m_\perp^2\over \beta_q^2s^2}$
(cf. Eq. (\ref{v4})) which is $O\big({q_\perp^2\over\beta_qs}\big)$ in comparison to Eq. (\ref{XiBxiAs}). Also, if index $\nu$ is longitudinal, the contribution is
$\sim  p_{1\nu}p_{2\mu}{q_\perp^4\over\beta_q^2s^3}$ which is $O\big({\alpha_q^2m^4\over Q_\parallel^4}\big)$ in comparison
to leading-twist result (\ref{ltasym2}). 
Thus,
$\chepizW_{\mu\nu}^{(2b)\calf}(x)~=~0$ with our accuracy.
Finally, as discussed in Sect. \ref{sec:f2g}, the term $\chepizW_{\mu\nu}^{(2c)\calf}(x)$ is of order of ${1\over N_c^2}$ and can be neglected.

\subsubsection{Sum of $\pizw_{\mu\nu}^{\calf}$ terms}
Adding the contributions (\ref{ltasym2}), (\ref{XiBxiAs}), and (\ref{pizwe2}) we obtain 
\begin{eqnarray}
&&\hspace{-1mm}
\pizw_{\mu\nu}^{\calf}(q)~=~\int\! d^2k_\perp~\pizf(q,k_\perp)\pizw^{F}_{\mu\nu}(q,k_\perp)
\label{pizw}
\end{eqnarray}
where $\pizf^f(q,k_\perp)$ is given by Eq. (\ref{pizf}) and 
\begin{eqnarray}
&&\hspace{-1mm}
\pizw^{F}_{\mu\nu}(q,k_\perp)~=~{2p_{1\nu}p_{2\mu}\over s}+{2p_{2\mu}(q-k)^\perp_\nu\over \alpha_qs}+{2p_{1\nu}k^\perp_\mu\over \beta_qs}
+ {2k^\perp_\mu(q-k)^\perp_\nu\over \alpha_q\beta_qs}~-~\mu\leftrightarrow\nu
\nonumber\\
&&\hspace{-1mm}
=~{q_\mu\over Q_\parallel^2}(\tilq_\nu+q^\perp_\nu-2k^\perp_\nu)~-~\mu\leftrightarrow\nu
\label{pizwe}
\end{eqnarray}

The corresponding contribution to antisymmetric part of $Z$-boson hadronic tensor $\chekalw^{\rm asy}_{\mu\nu}(q)$ is proportional to 
$\epsilon_{\mu\nu\lambda\rho}\pizw^{\lambda\rho}(q)$ so we immediately see gauge invariance:
\begin{equation}
q_\mu\epsilon^{\mu\nu\alpha\beta}\pizw^{F}_{\alpha\beta}(q,k_\perp)~=~0
\label{ginvas}
\end{equation}

\subsubsection{$\pizw_{\mu\nu}^{\rm f5}$ contribution}
From Eq. (\ref{WZantisym}) we see that $\pizw_{\mu\nu}^{\rm \calf}$ and $\pizw_{\mu\nu}^{\rm 5}$ differ by replacement  $\psi(0)\rightarrow\gamma_5\psi(0)$.
Let us consider  terms assembled in $\pizw_{\mu\nu}^{\calf}$ and prove that they vanish after  such replacement.
Fist, for the leading-twist contribution it is evident from parametrizations (\ref{mael5}).
Second, let us write down
\begin{eqnarray}
&&\hspace{-11mm}
\chepizw_{1\mu\nu}^{(1)\calf}(x)+\chepizw_{2\mu\nu}^{(1)\calf}(x)~=~
\nonumber\\
&&\hspace{-5mm}
=~{p_{1\mu}\over s^3}\big\{
\langle\bar\psi(x)\notp_2{1\over \alpha}\psi(0) \rangle_A
\langle \bar\psi(0)\notB(0)\notp_1\gamma_{\nu_\perp}\psi(x)\rangle_B
+\langle\bar\psi(x)\notp_2{1\over \alpha}\psi(0) \rangle_A
\nonumber\\
&&\hspace{-1mm}
\times~\langle \bar\psi(0)\gamma_{\nu_\perp}\notp_1\notB(x)\psi(x)\rangle_B
+~(\psi(0)\otimes\psi(x)\leftrightarrow\gamma_5\psi(0)\otimes\gamma_5\psi(x)\big\}
~-~x\leftrightarrow 0
\end{eqnarray}
The contributions of these two terms are equal, but using equations from Sect. \ref{sec:qqgparam} 
it is easy to see that after replacement  $\psi(0)\rightarrow\gamma_5\psi(0)$ 
they cancel each other as in the Eq. (\ref{w1oddvanishes}) case so  $\chepizw_{1\mu\nu}^{(1)5}(x)=0$. Similarly,
one can demonstrate that
\begin{eqnarray}
&&\hspace{-1mm}
\cheW^{\rm(1)F5}_{3\mu\nu}(x)+\cheW^{\rm(1)F5}_{4\mu\nu}(x)~
=~
{p_{1\mu}\over s^2}
\Big[
\langle\bar\psi\notA(x)\notp_2\gamma_{\nu_\perp}\gamma_5\psi(0)\rangle_A\langle\bar\psi(0)\notp_1{1\over \beta}\psi(x)\rangle_B
\nonumber\\
&&\hspace{10mm}
+~\langle\bar\psi(x)\gamma_{\nu_\perp}\notp_2\notA(0)\gamma_5\psi(0)\rangle_A\langle\big(\bar\psi{1\over \beta}\big)(0)\notp_1\psi(x)\rangle_B
\nonumber\\
&&\hspace{22mm}
+~\psi(0)\otimes\psi(x)\leftrightarrow\psi(0)\gamma_5\otimes\gamma_5\psi(x) \Big]
~+~\mu\leftrightarrow\nu~+~x\leftrightarrow 0
\label{w1oddvanishes}
\end{eqnarray}
vanishes since the two terms in the r.h.s. cancel each other.

Let us turn now to terms with two gluon operators and start from $\chepizW_{\mu\nu}^{\rm (2a)F5}(x)$. The 
$\chepizW_{\mu\nu}^{(2a){\cal F}}(x)$ contribution is given by sum of Eqs. (\ref{pizwedvaperp}) and (\ref{pizwe2perpa})
\begin{eqnarray}
&&\hspace{-11mm}
\chepizW_{\mu\nu}^{(2a)\calf}(x)~=~
-{1\over s^3}\Big(\langle\bsi(x)\notp_2\brA_\mu(x){1\over \alpha}\psi(0)\rangle_A
\langle\bsi(0)\notp_1\breB_\nu(0){1\over\beta}\psi(x)\rangle_B
\nonumber\\
&&\hspace{-5mm}
+~\big(\bsi{1\over \alpha}\big)(x)\brA_{\mu_\perp}(0)\notp_2\psi(0)\rangle_A
\langle\big(\bsi{1\over\beta}\big)(0)\breB_{\nu_\perp}(x)\notp_1\psi(x)\rangle_B\Big)~-~\mu\leftrightarrow\nu~-~x\leftrightarrow 0
\label{pizwas}
\end{eqnarray}
As we saw in Sect. \ref{sec:2gluonsF}, the contributions of the two terms in the r.h.s. are equal. Now, when we replace  $\psi(0)\rightarrow\gamma_5\psi(0)$,
target matrix elements terms remain the same and projectile ones become of different sign as seen from Eq. (\ref{gamma5scancel}) so 
$\chepizW_{\mu\nu}^{\rm (2a)F5}(x)=0$

Finally, the contribution $\chepizW_{\mu\nu}^{\rm(2b)F5}(x)$ is small by power counting (see Eq. (\ref{v4razasy}) and subsequent discussion)
while   $\chepizW_{\mu\nu}^{\rm (2c)5}(x)$ has extra ${1\over N_c^2}$ so we neglect it. 
Thus, $\pizw_{\mu\nu}^{5}=0$ with our accuracy.

As to $W^{\rm as}_{\mu\nu}$ defined in Eq. (\ref{WZantisym}), it also vanishes with our accuracy but the calculation is more tedious.

\subsection{$W^{\rm as}_{\mu\nu}$ \label{wasvanish}}
In this Section we will prove that $W^{\rm as}_{\mu\nu}$ defined in Eq. (\ref{WZantisym}) is small in our approximation. As usual, for power counting we consider
$W^{\rm as}_{\mu\nu}(x)$ multiplied by $2N_c/s$:
\begin{eqnarray}
&&\hspace{-1mm}
\chew^{\rm as}_{\mu\nu}(x)~=~{iN_c\over 2s}\big\{
[\Bsi_1^m(x)\sigma_{\mu\nu}\Psi_1^n(0)][\Bsi_2^n(0)\Psi_2^m(x)]
-[\Bsi_1^m(x)\Psi_1^n(0)][\Bsi_2^n(0)\sigma_{\mu\nu}\Psi_2^m(x)]
\nonumber\\
&&\hspace{13mm}
-~\psi(0)\otimes\psi(x)\leftrightarrow\gamma_5\psi(0)\otimes\gamma_5\psi(x)\big\}-x\leftrightarrow 0
\label{chew0}
\end{eqnarray}
First, from parametrizations in Sect \ref{sec:paramlt} it is clear that the leading-twist contribution to $\chew^{\rm as}_{\mu\nu}$ is of order of
${p_{1\mu}q^\perp_\nu\over s}$ (or ${p_{2\mu}q^\perp_\nu\over s}$) which is $O(\beta_q)$ (or $O(\alpha_q)$) with respect to contribution (\ref{pizwe}).

\subsubsection{One-gluon terms}
Next, consider terms with one gluon field and start from term with $\Xi_1~=~-{\slashed{p}_2\over s}\gamma^iB_i{1\over \alpha}\psi_A$. We get the contribution
to the r.h.s.  of Eq. (\ref{chew0}) in the form
\begin{eqnarray}
&&\hspace{-1mm}
{i\over s^2}\big\{\langle\bsi(x)\notp_2\gamma_i{1\over\alpha}\psi(0)\rangle_A\langle\bsi(0)B^i(0)\sigma_{\mu\nu}\psi(x)\rangle_B
\label{chew1}\\
&&\hspace{-1mm}
-~\langle\bsi(x)\sigma_{\mu\nu}\notp_2\gamma^i{1\over\alpha}\psi(0)\rangle_A\langle\bsi(0)B_i(0)\psi(x)\rangle_B
-\psi(0)\otimes \psi(x)\leftrightarrow \gamma_5\psi(0)\otimes \gamma_5\psi(x)\big\}-x\leftrightarrow 0
\nonumber
\end{eqnarray}
Let us first take transverse $\mu$ and $\nu$ and consider the first term in the r.h.s.
Since the projectile matrix element $\langle\psi(x)\notp_2\gamma_i{1\over\alpha}\psi(0)\rangle_A$ is proportional to $x_i$ and the target one\\
$\langle\psi(0)B^i(0)\sigma^\perp_{\mu\nu}\psi(x)\rangle_B$ to $\delta_\mu^i x^\perp_\nu- \delta_\nu^i x^\perp_\mu$, this term vanishes.  
Since $\sigma^\perp_{\mu\nu}\gamma^i=i(\delta_\nu^i \gamma^\perp_\mu- \delta_\mu^i \gamma^\perp_\nu)$, the second term in the r.h.s. of Eq. (\ref{chew1}) 
vanishes for the same reason: target matrix element is proportional to $x_i$ and the projectile one to $\delta_\mu^i x^\perp_\nu- \delta_\nu^i x^\perp_\mu$. 
Next, consider terms with extra $\gamma_5$'s. Since $\langle\psi(x)\notp_2\gamma_i\gamma_5{1\over\alpha}\psi(0)\rangle_A\sim \epsilon_{ij}x^j$ and 
$\langle\psi(0)B^i(0)\sigma^\perp_{\mu\nu}\gamma_5\psi(x)\rangle_B={2i\over s}\epsilon^\perp_{\mu\nu}\langle\psi(0)B^i(0)\sigma^\perp_{\bu\star}\gamma_5\psi(x)\rangle_B
\sim x^i$, this term also gives no contribution. For the last term, since $\sigma_{\mu\nu}^\perp\gamma_5=i\epsilon_{\mu\nu}^\perp{2\over s}\sigma_{\star\bu}$, the projectile matrix elements proportional to $x^i$ and the target one to $\epsilon_{ij}x^j$ so the last term in the r.h.s. of Eq. (\ref{chew1}) vanishes. 
Now, terms with $\Bxi_1~=~-\big(\bar\psi_A{1\over\alpha}\big)\gamma^iB_i{\slashed{p}_2\over s}$ are 
similar to that of Eq. (\ref{chew1}) so they vanish for the same reason. Finally, the results for $\Bxi_2$ and $\Xi_2$ differ by usual projectile$\leftrightarrow$target 
replacements so we get the result that one-gluon contributions to $\chew^{\rm as}_{\mu\nu}$ vanish at transverse $\mu$ and $\nu$.

If now both $\mu$ and $\nu$ are longitudinal, $\sigma_{\mu\nu}={4\over s^2}(p_{1\mu}p_{2\nu}-\mu\leftrightarrow\nu)\sigma_{\star\bu}$. It is easy to see that
$\sigma_{\star\bu}$ in the target matrix element brings no factor of $s$ while in the projectile one  $\sigma_{\star\bu}\notp_2\gamma_i=s\sigma_{\bu i}$ 
can bring $s^2$. However, even in this case the corresponding contribution to r.h.s. of Eq. (\ref{chew1}) is proportional to 
${2\over\alpha_q s^2}(p_{1\mu}p_{2\nu}-\mu\leftrightarrow\nu)={\beta_q\over q_\parallel^2}\times{2\over s}(p_{1\mu}p_{2\nu}-\mu\leftrightarrow\nu)$ which is 
$O(\beta_q)$ smaller than the last term in Eq. (\ref{pizwe}) 
$\sim\epsilon_{\mu\nu ij}{k^i(q-k)^j\over q_\parallel^2}\sim {2\over s}(p_{1\mu}p_{2\nu}-\mu\leftrightarrow\nu){q_\perp^2\over q^2}$.

Finally, let us consider case when one of the indices is longitudinal and the other  transverse. The corresponding  contribution
to $\chew^{\rm as}_{\mu\nu}(x)$ is (cf. Eq. (\ref{onexi16}))
\begin{eqnarray}
&&\hspace{-1mm}
{2p_{2\mu}\over s^3}\big\{\langle\bsi(x)\sigma_{\star i}{1\over\alpha}\psi(0)\rangle_A\langle\bsi(0)B^i(0)\sigma_{\bu\nu_\perp}\psi(x)\rangle_B
-~\langle\bsi(x)\sigma_{\bu\nu_\perp}\sigma_{\star i}{1\over\alpha}\psi(0)\rangle_A
\nonumber\\
&&\hspace{-1mm}
\times~\langle\bsi(0)B^i(0)\psi(x)\rangle_B
-\psi(0)\otimes \psi(x)\leftrightarrow \gamma_5\psi(0)\otimes \gamma_5\psi(x)\big\}-\mu\leftrightarrow \nu-x\leftrightarrow 0
\label{chew2}
\end{eqnarray}
Using formulas (\ref{sigmasigmas}) it is easy to see that the second (and the fourth) term can be neglected since the  corresponding  contribution
to $\chew^{\rm as}_{\mu\nu}(q)$ is $\sim (p_{2\mu}q^\perp_\nu-\mu\leftrightarrow\nu){q_\perp^2\over \alpha_qs^2}$ which is $O(\alpha_q)$ in comparison
to the second term in the r.h.s. of Eq. (\ref{pizwe}) $\sim \epsilon_{\nu j}{p_{2\mu}q^j\over\alpha_q s}-\mu\leftrightarrow \nu$ so we are left with
\begin{eqnarray}
&&\hspace{-1mm}
{2p_{2\mu}\over s^3}\big\{\langle\bsi(x)\sigma_{\star i}{1\over\alpha}\psi(0)\rangle_A\langle\bsi(0)B^i(0)\sigma_{\bu\nu_\perp}\psi(x)\rangle_B
\nonumber\\
&&\hspace{11mm}
-~\langle\bsi(x)\sigma_{\star i}\gamma_5{1\over\alpha}\psi(0)\rangle_A\langle\bsi(0)B^i(0)\sigma_{\bu\nu_\perp}\gamma_5\psi(x)\rangle_B\big\}
-\mu\leftrightarrow \nu-x\leftrightarrow 0
\label{chew3}
\end{eqnarray}

Next, from Eq. (\ref{gammas13}) we get
\begin{eqnarray}
&&\hspace{-1mm}
\sigma_{\star i}\otimes\sigma_{\bu \nu_\perp}-\sigma_{\star i}\gamma_5\otimes\sigma_{\bu \nu_\perp}\gamma_5
\label{gammas13a}\\
&&\hspace{-1mm}
=~-g_{i\nu}\sigma_{\star l}\otimes\sigma_\bu^{~l}+\sigma_{\star \nu_\perp}\otimes\sigma_{\bu i}+\sigma_{\star i}\otimes\sigma_{\bu \nu_\perp}
-{s\over 4}g_{i\nu}\sigma_{mn}\otimes\sigma^{mn}+{s\over 2}\sigma_{\nu_\perp l}\otimes\sigma_i^{~l}
\nonumber
\end{eqnarray}
The two last terms can bring only factor $s$ to Eq. (\ref{chew3}) while the first three bring $\sim s^2$ so we get
\begin{eqnarray}
&&\hspace{-1mm}
{2p_{2\mu}\over s^3}\big\{\langle\bsi(x)\sigma_{\star i}{1\over\alpha}\psi(0)\rangle_A\langle\bsi(0)\big[B^i(0)\sigma_{\bu\nu_\perp}
-B_\nu(0)\sigma_\bu^{~ i}\big]\psi(x)\rangle_B
\nonumber\\
&&\hspace{11mm}
+~\langle\bsi(x)\sigma_{\star \nu_\perp}{1\over\alpha}\psi(0)\rangle_A\langle\bsi(0)B^i(0)\sigma_{\bu i}\psi(x)\rangle_B\big\}
-\mu\leftrightarrow \nu-x\leftrightarrow 0
\nonumber\\
&&\hspace{11mm}
=-~i\langle\bsi(x)\sigma_{\star \nu_\perp}{1\over\alpha}\psi(0)\rangle_A\langle\bsi(0)\notB(0)\notp_1\psi(x)\rangle_B
-\mu\leftrightarrow \nu-x\leftrightarrow 0
\label{chew5}
\end{eqnarray}
where we used Eq. (\ref{formula2}).

Let us calculate now the corresponding contribution to r.h.s. of Eq. (\ref{chew0}) coming from $\Bxi_1(x)$ (see Eq. (\ref{fildz0})):
\begin{eqnarray}
&&\hspace{-2mm}
{i\over s^2}\big\{\langle\big(\bsi(x){1\over\alpha}\big)\gamma_i\notp_2\psi(0)\rangle_A\langle\bsi(0)B^i(0)\sigma_{\mu\nu}\psi(x)\rangle_B
\label{chew6}\\
&&\hspace{-2mm}
-~\langle\big(\bsi{1\over\alpha}\big)(x)\sigma_{\mu\nu}\gamma_i\notp_2\psi(0)\rangle_A\langle\bsi(0)B_i(0)\psi(x)\rangle_B
-\psi(0)\otimes \psi(x)\leftrightarrow \gamma_5\psi(0)\otimes \gamma_5\psi(x)\big\}-x\leftrightarrow 0
\nonumber
\end{eqnarray}
The only difference from Eq. (\ref{chew0}) is the sign $\gamma_i\notp_2=-\notp_2\gamma^i$, replacement $B^i(0)\rightarrow B^i(x)$ and ${1\over\alpha}$ acting on 
$\bsi$ instead of $\psi$ so we get
\begin{eqnarray}
&&\hspace{-1mm}
-{2p_{2\mu}\over s^3}\big\{\langle\big(\bsi{1\over\alpha}\big)(x)\psi(0)\rangle_A\langle\bsi(0)\big[B^i(0)\sigma_{\bu\nu_\perp}
-B_\nu(0)\sigma_\bu^{~ i}\big]\psi(x)\rangle_B
\nonumber\\
&&\hspace{11mm}
+~\langle\big(\bsi{1\over\alpha}\big)(x)\sigma_{\star \nu_\perp}\psi(0)\rangle_A\langle\bsi(0)B^i(x)\sigma_{\bu i}\psi(x)\rangle_B\big\}
-\mu\leftrightarrow \nu-x\leftrightarrow 0
\nonumber\\
&&\hspace{11mm}
=~-i\langle\big(\bsi{1\over\alpha}\big)(x)\sigma_{\star \nu_\perp}\psi(0)\rangle_A\langle\bsi(0)\notp_1\notB(x)\psi(x)\rangle_B
-\mu\leftrightarrow \nu-x\leftrightarrow 0
\label{chew7}
\end{eqnarray}
and the sum of these contributions to $\chew^{\rm as}_{\mu\nu}(x)$ takes the form
\begin{eqnarray}
&&\hspace{-1mm}
-i\big(\langle\bsi(x)\sigma_{\star \nu_\perp}{1\over\alpha}\psi(0)\rangle_A\langle\bsi(0)\notB(0)\notp_1\psi(x)\rangle_B
\nonumber\\
&&\hspace{11mm}
+~\langle\big(\bsi{1\over\alpha}\big)(x)\sigma_{\star \nu_\perp}\psi(0)\rangle_A\langle\bsi(0)\notp_1\notB(x)\psi(x)\rangle_B\big)
-\mu\leftrightarrow \nu-x\leftrightarrow 0
\label{chew8}
\end{eqnarray}
Now, from Eqs. (\ref{maelqg1}),   (\ref{maelqg2}), and (\ref{9.58}) it is easy to see that the corresponding contribution to 
$\chew^{\rm as}_{\mu\nu}(q)$ vanishes due to cancellation between the two terms in the above equation.
Similarly, one can demonstrate that one-gluon contributions to $\chew^{\rm as}_{\mu\nu}(q)$ from 
$\Xi_2~=~-{\slashed{p}_1\over s}\gamma^iA_i{1\over\beta+i\epsilon}\psi_B$ and
$\Bxi_2~=~-\big(\bar\psi_B{1\over \beta-i\epsilon}\big)\gamma^iA_i{\slashed{p}_1\over s}$
cancel.

\subsubsection{Two-gluon terms}
Following analysis in Sect. \ref{sec:f2g}, let us start with the contribution to the  r.h.s. of Eq. (\ref{chew0}) 
coming from $\Xi_1$ and $\Xi_2$ (see Eq. (\ref{fildz0})).
\begin{eqnarray}
&&\hspace{-1mm}
{i\over 2s^3}\big\{\langle\bsi A^j(x)\sigma_{\star i}{1\over \alpha}\psi(0)\rangle_A\langle\bsi B^i(0)\sigma_{\mu\nu}\sigma_{\bu j}
{1\over\beta}\psi(x)\rangle_B~-~ \langle\bsi A^j(x)\sigma_{\mu\nu}\sigma_{\star i}{1\over \alpha}\psi(0)\rangle_A
\nonumber\\
&&\hspace{11mm}
\times~\langle\bsi B^i(0)\sigma_{\bu j}{1\over\beta}\psi(x)\rangle_B
-\psi(0)\otimes \psi(x)\leftrightarrow \gamma_5\psi(0)\otimes \gamma_5\psi(x)\big\}-x\leftrightarrow 0
\label{chew9}
\end{eqnarray}
The r.h.s. of Eq. (\ref{chew9}) at transverse $\mu$ and $\nu$ due to Eq. (\ref{sigmasigmas}) can be rewritten as
\begin{eqnarray}
&&\hspace{-3mm}
{\rm Eq. ~(\ref{chew9})}~=~
{1\over 2s^3}\big\{\langle\bsi A_\mu(x)\sigma_{\star i}{1\over \alpha}\psi(0)\rangle_A\langle\bsi B^i(0)\sigma_{\bu \nu_\perp}
{1\over\beta}\psi(x)\rangle_B-~ \langle\bsi A^i(x)\sigma_{\star \nu_\perp}{1\over \alpha}\psi(0)\rangle_A
\nonumber\\
&&\hspace{-3mm}
\times~\langle\bsi B_\mu(0)\sigma_{\bu i}
{1\over\beta}\psi(x)\rangle_B
-\psi(0)\otimes \psi(x)\leftrightarrow \gamma_5\psi(0)\otimes \gamma_5\psi(x)\big\}-\mu\leftrightarrow\nu-x\leftrightarrow 0.
\label{chew11}
\end{eqnarray}
Using now Eq. (\ref{gammas14a})  we get
\begin{eqnarray}
&&\hspace{-1mm}
{\rm Eq. ~(\ref{chew9})}~=~{1\over 2s^3}\big\{
\langle\bsi A_\mu(x)\sigma_{\star i}{1\over \alpha}\psi(0)\rangle_A\langle\bsi \big[B^i(0)\sigma_{\bu\nu_\perp}-B_\nu(0)\sigma_\bu^{~i}\big]
{1\over\beta}\psi(x)\rangle_B
\nonumber\\
&&\hspace{33mm}
+~\langle\bsi A_\mu(x)\sigma_{\star \nu_\perp}{1\over \alpha}\psi(0)\rangle_A\langle\bsi B^i(0)\sigma_{\bu i}{1\over\beta}\psi(x)\rangle_B
\nonumber\\
&&\hspace{-1mm}
+~\langle\bsi \big[A_\nu(x)\sigma_{\star i}-A^i(x)\sigma_{\star \nu_\perp}\big]{1\over \alpha}\psi(0)\rangle_A\langle\bsi B_\mu(0)\sigma_\bu^{~i}{1\over\beta}\psi(x)\rangle_B
\nonumber\\
&&\hspace{33mm}
-~\langle\bsi A^i(x)\sigma_{\star i}{1\over \alpha}\psi(0)\rangle_A\langle\bsi B_\mu(0)\sigma_{\bu \nu_\perp}{1\over\beta}\psi(x)\rangle_B\big\}
-\mu\leftrightarrow\nu-x\leftrightarrow 0
\nonumber\\
&&\hspace{-1mm}
=~{1\over 2s^3}\big\{\langle\bsi [A_\mu(x)\sigma_{\star \nu_\perp}-\mu\leftrightarrow\nu]{1\over \alpha}\psi(0)\rangle_A\langle\bsi B^i(0)\sigma_{\bu i}{1\over\beta}\psi(x)\rangle_B
\nonumber\\
&&\hspace{-1mm}
-~\langle\bsi  A^i(x)\sigma_{\star i}{1\over \alpha}\psi(0)\rangle_A\langle\bsi [B_\mu(0)\sigma_{\bu \nu_\perp}-\mu\leftrightarrow \nu]{1\over\beta}\psi(x)\rangle_B\big\}
-x\leftrightarrow 0~=~0
\label{chew11}
\end{eqnarray}
where we used Eq. (\ref{formula2}). Next,  the 
 contribution to  the r.h.s. of Eq. (\ref{chew0}) coming from $\Bxi_1$ and $\Bxi_2$  has the form
\begin{eqnarray}
&&\hspace{-1mm}
{i\over 2s^3}\big\{\langle\big(\bsi{1\over \alpha}\big)(x)\sigma_{\star i}A^j\psi(0)\rangle_A
\langle\big(\bsi{1\over\beta}\big)(0)\sigma_{\bu j} \sigma_{\mu\nu}B^i\psi(x)\rangle_B
~-~ \langle\big(\bsi{1\over \alpha}\big)(x)\sigma_{\star i}\sigma_{\mu\nu}A^j(0)\psi(0)\rangle_A
\nonumber\\
&&\hspace{11mm}
\times~
\langle\big(\bsi{1\over\beta}\big)(0)\sigma_{\bu j}B^i\psi(x)\rangle_B
-\psi(0)\otimes \psi(x)\leftrightarrow \gamma_5\psi(0)\otimes \gamma_5\psi(x)\big\}-x\leftrightarrow 0
\label{chew10a}
\end{eqnarray}
and vanishes for transverse $\mu$ and $\nu$ for the same reason as the r.h.s. of Eq. (\ref{chew11}).

If both $\mu$ and $\nu$ are longitudinal, we get contribution to $\chew^{\rm as}_{\mu\nu}(x)$ in the form
\begin{eqnarray}
&&\hspace{-1mm}
{\rm Eq. ~(\ref{chew9})}~=~{2i\over s^5}(p_{1\mu}p_{2\nu}-\mu\leftrightarrow\nu)
\big\{\langle\bsi A^j(x)\sigma_{\star i}{1\over \alpha}\psi(0)\rangle_A\langle\bsi B^i(0)\sigma_{\star\bu}\sigma_{\bu j}
{1\over\beta}\psi(x)\rangle_B
\nonumber\\
&&\hspace{-1mm}
-~ \langle\bsi A^j(x)\sigma_{\star\bu}\sigma_{\star i}{1\over \alpha}\psi(0)\rangle_A\langle\bsi B^i(0)\sigma_{\bu j}
{1\over\beta}\psi(x)\rangle_B
-\psi(0)\otimes \psi(x)\leftrightarrow \gamma_5\psi(0)\otimes \gamma_5\psi(x)\big\}
\nonumber\\
&&\hspace{-1mm}
-~x\leftrightarrow 0~=~{2\over s^4}(p_{1\mu}p_{2\nu}-\mu\leftrightarrow\nu)
\big\{\langle\bsi A^j(x)\sigma_{\star i}{1\over \alpha}\psi(0)\rangle_A\langle\bsi B^i(0)\sigma_{\bu j}
{1\over\beta}\psi(x)\rangle_B
\nonumber\\
&&\hspace{33mm}
-~\psi(0)\otimes \psi(x)\leftrightarrow \gamma_5\psi(0)\otimes \gamma_5\psi(x)\big\}-x\leftrightarrow 0
\nonumber\\
&&\hspace{-1mm}
=~{2\over s^4}(p_{1\mu}p_{2\nu}-\mu\leftrightarrow\nu)
\big\{\langle\bsi A^i(x)\sigma_{\star i}{1\over \alpha}\psi(0)\rangle_A\langle\bsi B^j(0)\sigma_{\bu j}
{1\over\beta}\psi(x)\rangle_B
\nonumber\\
&&\hspace{-1mm}
+~\langle\bsi A^j(x)\sigma_{\star i}{1\over \alpha}\psi(0)\rangle_A\langle\bsi [B^i(0)\sigma_{\bu j}-B^j(0)\sigma_{\bu i}
{1\over\beta}\psi(x)\rangle_B\big\}-x\leftrightarrow 0
\nonumber\\
&&\hspace{-1mm}
=~-{2\over s^4}(p_{1\mu}p_{2\nu}-\mu\leftrightarrow\nu)
\langle\bsi \notA(x)\notp_2{1\over \alpha}\psi(0)\rangle_A\langle\bsi \notB(0)\notp_1
{1\over\beta}\psi(x)\rangle_B-x\leftrightarrow 0
\label{chew12}
\end{eqnarray}
where again we used Eq. (\ref{gammas14a})  and Eq.  (\ref{formula2}).
Similarly, the corresponding term in Eq. (\ref{chew10a}) is 
\begin{eqnarray}
&&\hspace{-1mm}
{\rm Eq. ~(\ref{chew10a})}~=~{2i\over s^5}(p_{1\mu}p_{2\nu}-\mu\leftrightarrow\nu)
\big\{\langle\big(\bsi {1\over \alpha}\big)(x)\sigma_{\star i}A^j\psi(0)\rangle_A
\langle\big(\bsi{1\over\beta}\big)(0) \sigma_{\bu j}\sigma_{\star\bu}B^i\psi(x)\rangle_B
\nonumber\\
&&\hspace{-1mm}
-~\langle\big(\bsi {1\over \alpha}\big)(x)\sigma_{\star i}\sigma_{\star\bu}A^j\psi(0)\rangle_A
\langle\big(\bsi{1\over\beta}\big)(0) \sigma_{\bu j}B^i\psi(x)\rangle_B
-\psi(0)\otimes \psi(x)\leftrightarrow \gamma_5\psi(0)\otimes \gamma_5\psi(x)\big\}
\nonumber\\
&&\hspace{-1mm}
-~x\leftrightarrow 0~=~-{2\over s^4}(p_{1\mu}p_{2\nu}-\mu\leftrightarrow\nu)
\big\{\langle\big(\bsi{1\over \alpha}\big)(0) \sigma_{\star i}A^j\psi(0)\rangle_A\langle\big(\bsi{1\over\beta}\big)(0) \sigma_{\bu j}B^i\psi(x)\rangle_B
\nonumber\\
&&\hspace{33mm}
-~\psi(0)\otimes \psi(x)\leftrightarrow \gamma_5\psi(0)\otimes \gamma_5\psi(x)\big\}-x\leftrightarrow 0
\nonumber\\
&&\hspace{-1mm}
=~-{2\over s^4}(p_{1\mu}p_{2\nu}-\mu\leftrightarrow\nu)
\big\{\langle\big(\bsi{1\over \alpha}\big)(0) \sigma_{\star i}{1\over \alpha}A^i\psi(0)\rangle_A
\langle\big(\bsi{1\over\beta}\big)(0) \sigma_{\bu j}B^j\psi(x)\rangle_B
\nonumber\\
&&\hspace{-1mm}
+~\langle\big(\bsi{1\over \alpha}\big)(0) \sigma_{\star i}A^j\psi(0)\rangle_A\langle\big(\bsi{1\over\beta}\big)(0) [B^i(x)\sigma_{\bu j}-B^j(x)\sigma_{\bu i}]
\psi(x)\rangle_B\big\}-x\leftrightarrow 0
\nonumber\\
&&\hspace{-1mm}
=~-{2\over s^4}(p_{1\mu}p_{2\nu}-\mu\leftrightarrow\nu)
\langle\bsi A^i(x)\sigma_{\star i}{1\over \alpha}\psi(0)\rangle_A\langle\bsi B^j(0)\sigma_{\bu j}
{1\over\beta}\psi(x)\rangle_B-x\leftrightarrow 0
\nonumber\\
&&\hspace{-1mm}
=~
{2\over s^4}(p_{1\mu}p_{2\nu}-\mu\leftrightarrow\nu)
\big\{\langle\big(\bsi{1\over \alpha}\big)(0) \notp_2\notA\psi(0)\rangle_A
\langle\big(\bsi{1\over\beta}\big)(0) \notp_1\notB\psi(x)\rangle_B
\label{chew13}
\end{eqnarray}
Using QCD equations of motion from Sect. \ref{sec:qqgparam} we see that the contributions  (\ref{chew12}) and  (\ref{chew13})  cancel. 

Now suppose $\mu$ is longitudinal and $\nu$ transverse
\begin{eqnarray}
&&\hspace{-1mm}
{i\over s^4}\big\{p_{1\mu}\langle\bsi A^j(x)\sigma_{\star i}{1\over \alpha}\psi(0)\rangle_A\langle\bsi B^i(0)\sigma_{\star\nu_\perp}\sigma_{\bu j}
{1\over\beta}\psi(x)\rangle_B~-~ p_{2\mu}\langle\bsi A^j(x)\sigma_{\bu\nu_\perp}\sigma_{\star i}{1\over \alpha}\psi(0)\rangle_A
\nonumber\\
&&\hspace{11mm}
\times~\langle\bsi B^i(0)\sigma_{\bu j}{1\over\beta}\psi(x)\rangle_B
-\psi(0)\otimes \psi(x)\leftrightarrow \gamma_5\psi(0)\otimes \gamma_5\psi(x)\big\}-x\leftrightarrow 0
\label{chewxz}
\end{eqnarray}
Due to the first line in Eq. (\ref{sigmasigmas}), both projectile and target matrix elements can bring factor $s$ each, 
so the result is either $\sim p_{1\mu}q^\perp_\nu{q_\perp^2\over \alpha_q\beta_q s^2}$ or  $\sim p_{2\mu}q^\perp_\nu{q_\perp^2\over \alpha_q\beta_q s^2}$, both of which are small in comparison to  Eq. (\ref{XiBxiAs}).

Let us now consider term coming from $\Bxi_1$ and $\Xi_1$:
\begin{eqnarray}
&&\hspace{-1mm}
\cheW^{\rm as}_{\mu\nu}(x)~=~{i\over 2s^3}\big\{
\langle\big(\bsi{1\over\alpha}\big)(x)\gamma_i\notp_2\sigma_{\mu\nu}\notp_2\gamma_j{1\over\alpha}\psi(0)\rangle_A
\langle\bsi B^j(0)B^i\psi(x)\rangle_B
\nonumber\\
&&\hspace{22mm}
-~\psi(0)\otimes \psi(x)\leftrightarrow \gamma_5\psi(0)\otimes \gamma_5\psi(x)\big\}-x\leftrightarrow 0
\nonumber\\
&&\hspace{-1mm}
=~{i\over s^3}p_{2\nu}\big\{\langle\big(\bsi{1\over\alpha}\big)(x)\gamma_i\sigma_{\star\mu}\gamma_j{1\over\alpha}\psi(0)\rangle_A\langle\bsi B^j(0)B^i\psi(x)\rangle_B
\nonumber\\
&&\hspace{22mm}-~\psi(0)\otimes \psi(x)\leftrightarrow \gamma_5\psi(0)\otimes \gamma_5\psi(x)\big\}~-~\mu\leftrightarrow\nu~-~x\leftrightarrow 0
\label{chewxz1}
\end{eqnarray}
Since nonzero contribution comes from  transverse $\mu$, power counting for the r.h.s. of this equation gives ${p_{2\nu}q^\perp_\mu\over\alpha_q^2s^2}$
(see Eq. (\ref{maelqg2})) which is $O\big({q_\perp^2\over\alpha_q s}\big)$ in comparison   Eq. (\ref{XiBxiAs}). Similarly, the contribution 
to $\chew^{\rm as}_{\mu\nu}(x)$ coming from $\Bxi_2$ and $\Xi_2$ will be  $\sim {p_{1\nu}q^\perp_\mu\over\beta_q^2s^2}$ 
and hence negligible. Thus, we proved that  $W^{\rm as}_{\mu\nu}=0$ with our accuracy.

\subsubsection{Exchange-type power corrections to $W^{\rm ZA}_{\mu\nu}$ \label{sec:wzaex}}
Power corrections of the ``exchange''  type come from the terms
\begin{eqnarray}
&&\hspace{-1mm}
(\cheW^{\rm ZA}_{ff'})^{\rm ex}_{\mu\nu}(x)~=~{N_c\over s}
\langle A,B|\calj_{1\mu}(x)\calj_{2\nu}(0)
-\mu\leftrightarrow\nu|A,B\rangle-x\leftrightarrow 0
\label{chewzsex}\\
&&\hspace{-1mm}
=\sum_{f,f'}c_fc_{f'}\Big(a_fa_{f'}(\chew^{\rmA ff'})^{\rm ex}_{\mu\nu}(x)
+(\chew^{{\rm A}ff'}_{55})^{\rm ex}_{\mu\nu}(x)
-a_f(\chew^{{\rm A}ff'}_{5\rma})^{\rm ex}_{\mu\nu}(x)-a_{f'}(\chew^{{\rm A}ff'}_{5\rmb})^{\rm ex}_{\mu\nu}(x)
\Big)
\nonumber
\end{eqnarray}
where 
\begin{eqnarray}
&&\hspace{-3mm}
(\chew^{\rm A})^{\rm ex}_{\mu\nu}(x)~=~{N_c\over s}\langle A,B|
[\Bsi_1(x)\gamma_\mu\Psi^f_1(x)][\Bsi_2(0)\gamma_\nu\Psi^{f'}_2(0)]-\mu\leftrightarrow\nu|A,B\rangle
-x\leftrightarrow 0,
\nonumber\\
&&\hspace{-3mm}
(\chew^{\rm A}_{55})^{\rm ex}_{\mu\nu}(x)~\equiv~{N_c\over s}\langle A,B|
[\Bsi_1(x)\gamma_\mu\gamma_5\Psi_1(x)][\Bsi_2(0)\gamma_\nu\gamma_5\Psi_2(0)]
-\mu\leftrightarrow\nu|A,B\rangle
-x\leftrightarrow 0,
\nonumber\\
&&\hspace{-3mm}
(\chew^{\rm A}_{5\rma})^{\rm ex}_{\mu\nu}(x)~\equiv~{N_c\over s}\langle A,B|[\Bsi_1(x)\gamma_\mu\gamma_5\Psi_1(x)][\Bsi_2(0)\gamma_\nu\Psi_2(0)]
-\mu\leftrightarrow\nu|A,B\rangle
-x\leftrightarrow 0,
\nonumber\\
&&\hspace{-3mm}
(\chew^{\rm A}_{5\rmb})^{\rm ex}_{\mu\nu}(x)~\equiv~{N_c\over s}\langle A,B|[\Bsi_1(x)\gamma_\mu\Psi_1(x)][\Bsi_2(0)\gamma_\nu\gamma_5\Psi_2(0)]
-\mu\leftrightarrow\nu|A,B\rangle
-x\leftrightarrow 0
\nonumber\\
\label{defexasy}
\end{eqnarray}
To avoid cluttering formulas, we will omit trivial flavor indices until final result.

Let us start from 
$(W^{\rm A})^{\rm ex}_{\mu\nu}(x)$. Since replacement of symmetrization in $\mu,\nu$ by antisymmetrizaton
does not change power counting, we can start from analog of formula (\ref{wexsym})
\begin{eqnarray}
&&\hspace{-1mm}
(\chew^\rmA)^{{\rm ex}}_{\mu\nu}(x)~
=~{2N_c\over s}\langle p_A,p_B|
[\Bxi_{1}(x)\gamma_\mu\psi_A(x)+\bsi_A(x)\gamma_\mu\Xi_1(x)\big]
\label{wzexasym}\\
&&\hspace{-1mm}
\times~
\big[\Bxi_{2}(0)\gamma_\nu\psi_B(0)+\bar\psi_B(0)\gamma_\nu\Xi_{2}(0)\big]
|p_A,p_B\rangle
-\mu\leftrightarrow\nu-x\leftrightarrow 0
\nonumber\\
&&\hspace{-1mm}
=~{N_c\over (N_c^2-1)s^3}\Big(\langle
\psi(x)\gamma_\mu\slashed{p}_2\gamma^j A_k(0){1\over\alpha}\psi(x)
+\big(\bsi{1\over\alpha}\big)(x)\gamma^j\slashed{p}_2\gamma_\mu A_k(0)\psi(x)
\rangle_A
\nonumber\\
&&\hspace{-1mm}
\times~\langle\bsi(0)\gamma_\nu\slashed{p}_1\gamma^k B_j(x){1\over\beta}\psi(0)\rangle_B
+\big(\bsi{1\over\beta}\big)(0)\gamma^k\slashed{p}_1\gamma_\nu B_j(x)\psi_B(0)\rangle_B
-\mu\leftrightarrow\nu-x\leftrightarrow 0
\nonumber
\end{eqnarray}
with transverse $\mu$ and $\nu$. Using Eq. (\ref{A1.33}) we get
\begin{eqnarray}
&&\hspace{-1mm}
(\chew^\rmA)^{{\rm ex}}_{\mu\nu}(x)~=~-{N_c\over (N_c^2-1)s^3}
\langle\big(\bsi{1\over\alpha}\big)(x)\brA_\mu(0)\slashed{p}_2\psi(x)
+\bar\psi(x)\slashed{p}_2\brA_\mu(0){1\over \alpha}\psi(x)\rangle_A
\nonumber\\
&&\hspace{-1mm}
\times~\langle\big(\bar\psi{1\over\beta}\big)(0)\breB_\nu(x)\slashed{p}_1\psi(0)
+\bar\psi(0)\slashed{p}_1\breB_\nu(x){1\over\beta}\psi(0)\rangle_B
-\mu\leftrightarrow \nu
-x\leftrightarrow 0
\label{wzexasym}
\end{eqnarray}
and therefore
\begin{eqnarray}
&&\hspace{-1mm}
(W^A)_{\mu\nu}^{\rm ex}(q)~
=~-{N_c\over (N_c^2-1)Q_\parallel^2}\!\int\! d^2k_\perp
[k^\perp_\mu(q-k)^\perp_\nu-\mu\leftrightarrow\nu]
\nonumber\\
&&\hspace{33mm}
\times~[(j_2-\barj_2)(\alpha_q,k_\perp)(j^\ast_2-\barj^\ast_2)
(\beta_q,(q-k)_\perp)-{\rm c.c.}]
\label{wzexasy1}
\end{eqnarray}
where we used Eqs. (\ref{paramjse}) and (\ref{paramjset}). Since the functions $j_i(x,k_\perp)$ are actually
functions of $x$ and $k_\perp^2$, the
r.h.s. of the above equationcan be proportional only to $g^\perp_{\mu\nu}$ or $q^\perp_\mu q^\perp_\nu$ 
and hence it vanishes
\begin{equation}
(W^A)_{\mu\nu}^{\rm ex}(q)~=~0
\label{wavanish}
\end{equation}

Similarly to the symmetric case studied in Sect. \ref{zexsym}, 
the replacement $\psi\rightarrow \gamma_5\psi$ in the projectile matrix elements 
leads to $k_\mu j_{1,}\rightarrow \pm i\epsilon_{\mu\nu}k^\nu j_{1,2}$
 and the replacement  $\psi\rightarrow \gamma_5\psi$ in target 
matrix elements yields $(q-k)_\mu j_{1,2}\rightarrow \pm i\epsilon_{\mu}(q-k)^\nu j_{1,2}$.  
Looking at the result (\ref{wzexasy1})
and taking care of signs of replacements $\psi\rightarrow \gamma_5\psi$ 
in parametrizations (\ref{paramjse}) and 
(\ref{paramjset5}), we obtain
\begin{eqnarray}
&&\hspace{-1mm}
(\chew^{{\rm A}ff'}_{55})^{\rm ex}_{\mu\nu}(x)=~-{N_c\over (N_c^2-1)s^3}\Big(
\langle\big(\bsi{1\over\alpha}\big)(x)\brA_\mu(0)\slashed{p}_2\gamma_5\psi(x)
+\bar\psi(x)\slashed{p}_2\brA_\mu(0){1\over \alpha}\gamma_5\psi(x)\rangle_A^f
\nonumber\\
&&\hspace{22mm}
\times~\langle\big(\bar\psi{1\over\beta}\big)(0)\breB_\nu(x)\slashed{p}_1\gamma_5\psi(0)\rangle_B
+\bar\psi(0)\slashed{p}_1\breB_\nu(x){1\over\beta}\gamma_5\psi(0)\rangle_B^{f'}
-\mu\leftrightarrow \nu-x\leftrightarrow 0
\nonumber\\
&&\hspace{11mm}
\nonumber\\
&&\hspace{-1mm}
\Rightarrow~(W_{\mu\nu}^{{\rm A}ff'})_{55}^{\rm ex}(q)~
=~-{N_c\over (N_c^2-1)Q_\parallel^2}\!\int\! d^2k_\perp
[k^\perp_\mu(q-k)^\perp_\nu-\mu\leftrightarrow\nu]
\nonumber\\
&&\hspace{22mm}
\times~[(j_2+\barj_2)^f(\alpha_q,k_\perp)(j^\ast_2+\barj^\ast_2)^{f'}
(\beta_q,(q-k)_\perp)-{\rm c.c.}]~=~0
\label{wzexas55}
\end{eqnarray}
for the same reason as Eq. (\ref{wavanish}) above. The non-vanishing contributions come from
\begin{eqnarray}
&&\hspace{-1mm}
(\chew^{{\rm A}ff'}_{5\rma})^{\rm ex}_{\mu\nu}(x)~=~-{N_c\over (N_c^2-1)s^3}\Big(
\langle\big(\bsi{1\over\alpha}\big)(x)\brA_\mu(0)\slashed{p}_2\gamma_5\psi(x)
+\bar\psi(x)\slashed{p}_2\brA_\mu(0){1\over \alpha}\gamma_5\psi(x)\rangle_A^f
\nonumber\\
&&\hspace{-1mm}
\times~\langle\big(\bar\psi{1\over\beta}\big)(0)\breB_\nu(x)\slashed{p}_1\psi(0)\rangle_B
+\bar\psi(0)\slashed{p}_1\breB_\nu(x){1\over\beta}\psi(0)\rangle_B^{f'}
-\mu\leftrightarrow \nu-x\leftrightarrow 0
\end{eqnarray}
and 
\begin{eqnarray}
&&\hspace{-1mm}
(\chekalw^{{\rm A}ff'}_{5\rmb})^{\rm ex}_{\mu\nu}(x)~=~-{N_c\over (N_c^2-1)s^3}\Big(
\langle\big(\bsi{1\over\alpha}\big)(x)\brA_\mu(0)\slashed{p}_2\psi(x)
+\bar\psi(x)\slashed{p}_2\brA_\mu(0){1\over \alpha}\psi(x)\rangle_A^f
\nonumber\\
&&\hspace{-1mm}
\times~\langle\big(\bar\psi{1\over\beta}\big)(0)\breB_\nu(x)\slashed{p}_1\gamma_5\psi(0)\rangle_B
+\bar\psi(0)\slashed{p}_1\breB_\nu(x)\gamma_5{1\over\beta}\psi(0)\rangle_B^{f'}
-\mu\leftrightarrow \nu-x\leftrightarrow 0
\end{eqnarray}
Similarly to Eq. (\ref{wzexas55}), from parametrizations  (\ref{paramjse}) and 
(\ref{paramjset5}) one obtains
\begin{eqnarray}
&&\hspace{-1mm}
(W_{\mu\nu}^{{\rm A}ff'})_{5\rma}^{\rm ex}(q)~
=~-{iN_c\over (N_c^2-1)Q_\parallel^2}\!\int\! d^2k_\perp [\epsilon_{\mu m}k^m(q-k)_\nu-\mu\leftrightarrow\nu]
 J^{2ff'}_{+-}(q,k_\perp),
\nonumber\\
&&\hspace{-1mm}
(W_{\mu\nu}^{{\rm A}ff'})_{5\rmb}^{\rm ex}(q)~
=~{iN_c\over (N_c^2-1)Q_\parallel^2}\!\int\! d^2k_\perp [k_\mu\epsilon_{\nu n}(q-k)^n
-\mu\leftrightarrow\nu]  J^{2ff'}_{-+}(q,k_\perp)
\label{wzexasab5}
\end{eqnarray}
where $J^{2ff'}_{+-}$ and $J^{2ff'}_{-+}$ are defined in Eq. (\ref{Js}).

\subsection{Results for antisymmetric hadronic tensor for Z-mediated DY process}

The ``annihilation'' part is given by Eqs.  (\ref{waza}) and (\ref{pizwe})
\begin{eqnarray}
&&\hspace{-1mm}
W^{\rm ZAan}_{\mu\nu}(q)~=~-2i\epsilon_{\mu\nu\alpha\beta}{q^\alpha\over Q_\parallel^2}
\!\int\! d^2k_\perp (\tilq^\beta+q_\perp^\beta-2k_\perp^\beta)
\sum_fa_fc_f^2\calf^f(q,k_\perp)
\label{wzan}
\end{eqnarray}
It trivially satisfies $q^\mu W^{\rm ZAan}_{\mu\nu}(q)~=~0$. 

The ``exchange'' part is given by the sum of contributions (\ref{chewzsex}) calculated in previous Section
\begin{eqnarray}
\hspace{-1mm}
W^{\rm ZAex}_{\mu\nu}(q)&~=~&\sum_{f,f'}c_fc_{f'}(W^{\rm ZA}_{ff'})^{\rm ex}_{\mu\nu}(q)
\label{chewzaex}\\
\hspace{-1mm}
(W^{\rm ZA}_{ff'})^{\rm ex}_{\mu\nu}(q)&~=~&
{iN_c\over (N_c^2-1)Q_\parallel^2}\!\int\! d^2k_\perp\Big[
a_f[\epsilon_{\mu m}k^m(q-k)_\nu-\mu\leftrightarrow\nu]J^{2ff'}_{+-}(q,k_\perp)
\nonumber\\
\hspace{11mm}
&~-~&
a_{f'}[k_\mu\epsilon_{\nu n}(q-k)^n-\mu\leftrightarrow\nu]J^{2ff'}_{-+}(q,k_\perp)\Big]
\label{chewzaex}
\end{eqnarray}
where $J^{2ff'}_{+-}$ and $J^{2ff'}_{-+}$ are defined in Eq. (\ref{Js}). 

{\color{blue}
It can be parametrized as
\begin{eqnarray}
&&\hspace{-1mm}
W^{\rm ZAex}_{\mu\nu}(q)~=~{iN_c\over (N_c^2-1)Q_\parallel^2}\sum_{f,f'}c_fc_{f'}
\label{chewzaexe}\\
&&\hspace{-1mm}
\times~\Big[\Big(\epsilon_{\mu\nu}+{\epsilon_{\mu m}q^mq^\perp_\nu-\mu\leftrightarrow\nu \over q_\perp^2}\Big)
\big(a_f E^{2ff'}_{+-}+a_{f'}E^{2ff'}_{-+}\big)
-~\epsilon_{\mu\nu}\big(a_f \cale^{2ff'}_{+-}+a_{f'}\cale^{2ff'}_{-+}\big)\Big]
\nonumber
\end{eqnarray}
}

As usually, the exchange part is non-zero in our approximation only for transverse indices.

\section{Hadronic tensors for interference terms \label{sec:inter}}

\subsection{Symmetric part of interference hadronic tensor $W^{\rm I1}$}
From definitions (\ref{ws}) and (\ref{warray}) we get
\begin{eqnarray}
&&\hspace{-1mm}
\cheW^{\rm I1S}_{\mu\nu}(x)
~=~{N_c\over 2s}\sum_{f,f'}
\langle A,B|
(e_{f'}c_fa_f+e_fc_{f'}a_{f'})[\bsi(x)\gamma_\mu\psi(x)]^f[\bsi(0)\gamma_\nu\psi(0)]^{f'}
\nonumber\\
&&\hspace{15mm}
-~e_fc_{f'}[\bsi(x)\gamma_\mu\psi(x)]^f[\bsi(0)\gamma_\nu\gamma_5\psi(0)]^{f'}
\nonumber\\
&&\hspace{30mm}
-~e_{f'}c_f[\bsi(x)\gamma_\mu\gamma_5\psi(x)]^f[\bsi(0)\gamma_\nu\psi(0)]^{f'}|A,B\rangle
+\mu\leftrightarrow\nu
\label{wi2s}
\end{eqnarray}
where $\psi=\Psi_1+\Psi_2$ in our approximation.

\subsubsection{Annihilation-type terms}
Let us start from the ``annihilation'' part
\begin{eqnarray}
&&\hspace{-1mm}
\cheW^{\rm I1San}_{\mu\nu}(x)
~=~{N_c\over 2s}\sum_f e_fc_f
\langle A,B|2a_f[\Bsi_1^f\gamma_\mu\Psi_2^f(x)][\Bsi_2^f\gamma_\nu\Psi_1^f(0)]
\label{wi1sa}\\
&&\hspace{-1mm}
-~[\Bsi_1^f\gamma_\mu\Psi_2^f(x)][\Bsi_2^f\gamma_\nu\gamma_5\Psi_1^f(0)]
-[\Bsi_1^f\gamma_\mu\gamma_5\Psi_2^f(x)][\Bsi_2^f\gamma_\nu\Psi_1^f(0)]|A,B\rangle
+\mu\leftrightarrow\nu+x\leftrightarrow0
\nonumber
\end{eqnarray}

The first term can be copied from photon case (\ref{wefdef}) so we need to consider the last two terms. Using Fierz transformation (\ref{fierz5sym}) we get
\begin{eqnarray}
&&\hspace{-1mm}
{N_c\over 2s}\langle A,B|[\Bsi_1\gamma_\mu\Psi_2(x)][\Bsi_2\gamma_\nu\gamma_5\Psi(0)]
+[\Bsi_1\gamma_\mu\gamma_5\Psi_2(x)][\Bsi_2\gamma_\nu\Psi_1(0)]|A,B\rangle
+\mu\leftrightarrow\nu+x\leftrightarrow0
\nonumber\\
&&\hspace{-1mm}
=~
\big(g_{\mu\nu}g^{\alpha\beta}-\delta_\mu^\alpha\delta_\nu^\beta-\delta_\nu^\alpha\delta_\mu^\beta\big)
{N_c\over 2s}\langle A,B|[\Bsi_1(x)\gamma_\alpha\Psi_1(0)][\Bsi_2(0)\gamma_\nu\gamma_5\Psi_2(x)]
\nonumber\\
&&\hspace{33mm}
+~[\Bsi_1(x)\gamma_\alpha\gamma_5\Psi_1(0)][\Bsi_2(0)\gamma_\nu\Psi_2(x)]
|A,B\rangle+x\leftrightarrow0~=~ \cheW_{\mu\nu}^{\rm 5}(x)
\nonumber
\end{eqnarray}
Since we proved in Sect. \ref{sec:zsym} that $\cheW_{\mu\nu}^{\rm 5}$ is negligible (see Eq. (\ref{wf5})) we get
\begin{eqnarray}
&&\hspace{-1mm}
\cheW^{\rm I1San}_{\mu\nu}(x)
~=~\sum_f e_fc_f a_f\chew^f(x)
\nonumber\\
&&\hspace{-1mm}
\Rightarrow~W^{\rm I1San}_{\mu\nu}(q)~=~\sum_f e_fc_f a_f
\big[W^{\rmF f}_{\mu\nu}(q)+W^{\rmH f}_{\mu\nu}(q)+W^{\rmH 2f}_{\mu\nu}(q)\big]
\end{eqnarray}
where $W^{\rmF f}_{\mu\nu}$, $W^{\rmH f}_{\mu\nu}$, and $W^{\rmH 2f}_{\mu\nu}$ are given
by Eqs.  (\ref{resultf}),  (\ref{resulthginv}) , and  (\ref{whneinv}), respectively. 

\subsubsection{Exchange-type power corrections }
Let us consider now the exchange-type power corrections. From Eq. (\ref{wi2s}) we get 
\begin{eqnarray}
&&\hspace{-1mm}
\cheW^{\rm I1Sex}_{\mu\nu}(x)
~=~{N_c\over 2s}\sum_f 
\langle A,B|(e_{f'}c_fa_f+e_fc_{f'}a_{f'})[\Bsi_1^f\gamma_\mu\Psi_1^f(x)][\Bsi_2^{f'}\gamma_\nu\Psi_2^{f'}(0)]
\label{wi1sa}\\
&&\hspace{-1mm}
-~e_fc_{f'}[\Bsi_1^f\gamma_\mu\Psi_1^f(x)][\Bsi_2^{f'}\gamma_\nu\gamma_5\Psi_2^{f'}(0)]
\nonumber\\
&&\hspace{33mm}
-~c_fe_{f'}[\Bsi_1^f\gamma_\mu\gamma_5\Psi_1^f(x)][\Bsi_2^{f'}\gamma_\nu\Psi_2^{f'}(0)]|A,B\rangle
+\mu\leftrightarrow\nu+x\leftrightarrow0
\nonumber\\
&&\hspace{-3mm}
=~\sum_{f,f'} \Big(\half(e_{f'}c_fa_f+e_fc_{f'}a_{f'})\chew^{ff'{\rm ex}}_{\mu\nu}(x)
-{e_{f'}c_f\over 2}(\chew^{{\rm S}ff'}_{5\rma})^{\rm ex}_{\mu\nu}(x)
-{e_fc_{f'}\over 2}(\chew^{{\rm S}ff'}_{5\rmb})^{\rm ex}_{\mu\nu}(x)\Big)
\nonumber
\end{eqnarray}
where $\chew^{ff'{\rm ex}}_{\mu\nu}(x)$ is defined by  Eq. (\ref{wexsym}) 
while $(\chew^{{\rm S}ff'}_{5\rm a,b})^{\rm ex}_{\mu\nu}(x)$ are defined in Eq. (\ref{wzexs}). 
Thus,
\begin{eqnarray}
&&\hspace{-2mm}
W^{\rm I1Sex}_{\mu\nu}(q)=\sum_{f,f'} \Big[{e_{f'}c_fa_f+e_fc_{f'}a_{f'}\over 2}W^{ff'{\rm ex}}_{\mu\nu}(q)
-{e_{f'}c_f\over 2}(W^{\rmS ff'}_{5\rma})^{\rm ex}_{\mu\nu}(q)
-{e_fc_{f'}\over 2}(W^{\rmS ff'}_{5\rmb})^{\rm ex}_{\mu\nu}(q)\Big]
\nonumber\\
&&\hspace{-2mm}
=~{N_c\over 2(N_c^2-1)Q_\parallel^2}\!\int\! d^2k_\perp\sum_{f,f'} \Big[(e_{f'}c_fa_f+e_fc_{f'}a_{f'})
\Big([k^\perp_\mu(q-k)^\perp_\nu+\mu\leftrightarrow\nu+g^\perp_{\mu\nu}(k,q-k)_\perp]
\nonumber\\
&&\hspace{22mm}
\times~J^{1ff'}_{--}(q,k_\perp)
-~g^\perp_{\mu\nu} (k,q-k)_\perp J^{2ff'}_{--}(q,k_\perp)\Big)
\label{w1sexresult}\\
&&\hspace{-2mm}
-~e_{f'}c_f[\epsilon_{\mu m}k^m(q-k)_\nu+\mu\leftrightarrow\nu]
I^{1ff'}_{+-}(q,k_\perp)
+~e_fc_{f'}
[k_\mu\epsilon_{\nu n}(q-k)^n+\mu\leftrightarrow\nu]  
I^{1ff'}_{-+}(q,k_\perp)
\nonumber
\end{eqnarray}
where $W^{ff'{\rm ex}}_{\mu\nu}(q)$, $(W^{\rmS ff'}_{5\rma})^{\rm ex}_{\mu\nu}(q)$, and 
$(W^{\rmS ff'}_{5\rmb})^{\rm ex}_{\mu\nu}(q)$
 are given in Eqs. (\ref{wexotvet}), (\ref{wexa5}), and (\ref{wexb5}), respectively.

\subsubsection{The result for the symmetric part of $W^{\rm I1}_{\mu\nu}(x)$}
The result for $W^{\rm I1S}_{\mu\nu}(q)$ can be represented as a sum of ``annihilation'' and ``exchange'' parts
\begin{eqnarray}
&&\hspace{-1mm}
W^{\rm I1S}_{\mu\nu}(q)~=~W^{\rm I1San}_{\mu\nu}(q)~+~W^{\rm I1Sex}_{\mu\nu}(q)
\nonumber\\
&&\hspace{-1mm}
W^{\rm I1San}_{\mu\nu}(q)~
=~\sum_f e_fc_f a_f
\big[W^{\rm F f}_{\mu\nu}(q)+W^{\rm H f}_{\mu\nu}(q)+W^{\rm H 2f}_{\mu\nu}(q)\big]
\nonumber\\
&&\hspace{1mm}
W^{\rm I1Sex}_{\mu\nu}(q)~=~{\rm r.h.s.~of~Eq.~(\ref{w1sexresult})}
\label{wi1s}
\end{eqnarray}
where $W^{\rm F f}_{\mu\nu}(q)$ is given by Eq. (\ref{resultf}), 
$W^{\rm H f}_{\mu\nu}(q)$ by Eq. (\ref{resulth}), and $W^{\rm H2 f}_{\mu\nu}(q)$ by Eq. (\ref{whneinv}).

\subsection{Antiymmetric part of interference hadronic tensor $W^{\rm I1}$}
From definitions (\ref{ws}) and (\ref{warray}) we get
\begin{eqnarray}
&&\hspace{-1mm}
\cheW^{\rm I1A}_{\mu\nu}(x)
~=~{N_c\over 2s}\sum_{f,f'}
\langle A,B|
(e_{f'}c_fa_f+e_fc_{f'}a_{f'})[\bsi(x)\gamma_\mu\psi(x)]^f[\bsi(0)\gamma_\nu\psi(0)]^{f'}
\nonumber\\
&&\hspace{15mm}
-~e_fc_{f'}[\bsi(x)\gamma_\mu\psi(x)]^f[\bsi(0)\gamma_\nu\gamma_5\psi(0)]^{f'}
\nonumber\\
&&\hspace{30mm}
-~e_{f'}c_f[\bsi(x)\gamma_\mu\gamma_5\psi(x)]^f[\bsi(0)\gamma_\nu\psi(0)]^{f'}|A,B\rangle
-\mu\leftrightarrow\nu
\label{wi2s}
\end{eqnarray}
where $\psi=\Psi_1+\Psi_2$ in our approximation.

Let us start from the annihilation part. After Fierz transformations (\ref{fierz}) and (\ref{fierz5}) we get
\begin{eqnarray}
&&\hspace{-1mm}
\cheW^{\rm I1Aan}_{\mu\nu}(x)
~=~{N_c\over 2s}\sum_f e_fc_f
\langle A,B|2a_f[\Bsi_1^f\gamma_\mu\Psi_2^f(x)][\Bsi_2^f\gamma_\nu\Psi_1^f(0)]
\label{wi1sa}\\
&&\hspace{-1mm}
-~[\Bsi_1^f\gamma_\mu\Psi_2^f(x)][\Bsi_2^f\gamma_\nu\gamma_5\Psi_1^f(0)]
-[\Bsi_1^f\gamma_\mu\gamma_5\Psi_2^f(x)][\Bsi_2^f\gamma_\nu\Psi_1^f(0)]|A,B\rangle
-\mu\leftrightarrow\nu-x\leftrightarrow0
\nonumber\\
&&\hspace{-1mm}
=~\sum_f e_fc_f\big[ia_f\epsilon_{\mu\nu}^{~~\alpha\beta}\chepizw_{\alpha\beta}^{5f}(x)
+a_f\cheW^{{\rm as}f}_{\mu\nu}(x)
-{i\over 2}\epsilon_{\mu\nu}^{~~\alpha\beta}\chepizw_{\alpha\beta}^{\calf f}(x)\big]
\nonumber
\end{eqnarray}
where we used definitions (\ref{WZantisym}). As demonstrated in Sect. \ref{sec:zasym}, 
$\chepizw_{\mu\nu}^{5f}=\cheW^{{\rm as}f}_{\mu\nu}=0$ with our accuracy 
and $\pizw_{\alpha\beta}^{\calf f}(q)$ is given by Eq. (\ref{pizw}) so we obtain
\begin{eqnarray}
&&\hspace{-1mm}
W^{\rm I1Aan}_{\mu\nu}(q)
~=~-i\epsilon_{\mu\nu\alpha\beta}q^\alpha\!\int\! d^2k_\perp~(\tilq+q_\perp-2k_\perp)^\beta
\sum_f e_fc_f\pizf^f(q,k_\perp)
\end{eqnarray}

The exchange-type power corrections are
\begin{eqnarray}
&&\hspace{-1mm}
\cheW^{\rm I1Aex}_{\mu\nu}(x)~=~{N_c\over 2s}\sum_{f,f'}
\langle A,B|
(e_{f'}c_fa_f+e_fc_{f'}a_{f'})
[\Bsi_1(x)\gamma_\mu\Psi_1(x)]^f[\Bsi_2(0)\gamma_\nu\Psi_2(0)]^{f'}
\nonumber\\
&&\hspace{19mm}
-~e_fc_{f'}[\Bsi_1(x)\gamma_\mu\Psi_1(x)]^f[\Bsi_2(0)\gamma_\nu\gamma_5\Psi_2(0)]^{f'}
\\
&&\hspace{19mm}
-~e_{f'}c_f[\Bsi_1(x)\gamma_\mu\gamma_5\Psi_1(x)]^f[\Bsi_2(0)\gamma_\nu\Bsi_2(0)]^{f'}|A,B\rangle
-\mu\leftrightarrow\nu-x\leftrightarrow 0
\nonumber\\
&&\hspace{-1mm}
=~\sum_{f,f'}\Big[{e_{f'}c_fa_f+e_fc_{f'}a_{f'}\over 2}(\chew^{\rmA ff'})^{\rm ex}_{\mu\nu}(x)
-~{e_{f'}c_f\over 2}(\chew^{{\rm A}ff'}_{5\rma})^{\rm ex}_{\mu\nu}(x)
-{e_fc_{f'}\over 2}(\chew^{{\rm A}ff'}_{5\rmb})^{\rm ex}_{\mu\nu}(x)\Big]
\nonumber
\end{eqnarray}
where we used Eq. (\ref{defexasy}). In the momentum space this gives
\begin{eqnarray}
&&\hspace{-1mm}
W^{\rm I1Aex}_{\mu\nu}(q)~=~
{iN_c\over 2(N_c^2-1)Q_\parallel^2}\sum_{f,f'}\!\int\! d^2k_\perp
\Big\{c_fe_{f'}
[\epsilon_{\mu m}k^m(q-k)_\nu-\mu\leftrightarrow\nu]
J^{2ff'}_{+-}
\nonumber\\
&&\hspace{-1mm}
-~c_{f'}e_f[k_\mu\epsilon_{\nu n}(q-k)^n
-\mu\leftrightarrow\nu]  J^{2ff'}_{-+}\Big\}
\label{wi1aex}
\end{eqnarray}
where we used Eqs.  (\ref{wavanish}) and (\ref{wzexasab5}).

\subsubsection{The result for antisymmetric part of interference hadronic tensor $W^{\rm I1}$}
As usual, the result consists of ``annihilation'' and ``exchange'' parts
\begin{eqnarray}
&&\hspace{-1mm}
W^{\rm I1A}_{\mu\nu}(q)~=~W^{\rm I1Aan}_{\mu\nu}(q)+W^{\rm I1Aex}_{\mu\nu}(q),
\nonumber\\
&&\hspace{-1mm}
W^{\rm I1Aan}_{\mu\nu}(q)~=~-i\epsilon_{\mu\nu\alpha\beta}q^\alpha\!\int\! d^2k_\perp~(\tilq+q_\perp-2k_\perp)^\beta\sum_f e_fc_f\pizf^f(q,k_\perp),
\nonumber\\
&&\hspace{-1mm}
W^{\rm I1Aex}_{\mu\nu}(q)~=~{\rm r.h.s. ~of~Eq.~(\ref{wi1aex})}
\label{wi1asy}
\end{eqnarray}
where $\pizf^f(q,k_\perp)$ is given by Eq. (\ref{pizf}). Obviously, 
$q^\mu W^{\rm I1Aan}_{\mu\nu}(q)=0$.

\subsection{Symmetric part of interference hadronic tensor $W^{\rm I2}$}
From definitions (\ref{ws}) and (\ref{warray}) we get
\begin{eqnarray}
&&\hspace{-1mm}
\cheW^{\rm I2S}_{\mu\nu}(x)
~=~{N_c\over 2s}\sum_{f,f'}\Big((e_{f'}c_fa_f-e_fc_{f'}a_{f'})
\langle A,B|[\bsi(x)\gamma_\mu\psi(x)]^f[\bsi(0)\gamma_\nu\psi(0)]^{f'}
\nonumber\\
&&\hspace{15mm}
+~e_fc_{f'}\langle A,B|[\bsi(x)\gamma_\mu\psi(x)]^f[\bsi(0)\gamma_\nu\gamma_5\psi(0)]^{f'}
\nonumber\\
&&\hspace{15mm}
-~e_{f'}c_f\langle A,B|[\bsi(x)\gamma_\mu\gamma_5\psi(x)]^f[\bsi(0)\gamma_\nu\psi(0)]^{f'}|A,B\rangle\Big)
+\mu\leftrightarrow\nu
\label{wi2s}
\end{eqnarray}
where $\psi=\Psi_1+\Psi_2$ in our approximation.

\subsubsection{Annihilation-type power corrections}
Let us start with annihilation-type power corrections. Since in this case $f=f'$, the first term in the r.h.s. 
of Eq. (\ref{wi2s}) vanishes and the second turns to
\begin{eqnarray}
&&\hspace{-1mm}
\chew^{\rm I2San}_{\mu\nu}(x)~=~\sum_{f}e_fc_f\chew^{\rm If}_{\mu\nu}(x)
\nonumber\\
&&\hspace{-1mm}
\chew^{\rm If}_{\mu\nu}(x)
~=~{N_c\over 2s}\langle A,B|[\Bsi_1(x)\gamma_\mu\Bsi_2(x)][\Bsi_2(0)\gamma_\nu\gamma_5\Bsi_1(0)]
\nonumber\\
&&\hspace{15mm}
-~[\Bsi_1(x)\gamma_\mu\gamma_5\Bsi_2(x)][\Bsi_2(0)\gamma_\nu\Bsi_1(0)]
+\mu\leftrightarrow\nu|A,B\rangle-x\leftrightarrow 0
\label{wisa}
\end{eqnarray}
After Fierz transformation (\ref{fierz4}) it turns to
\begin{eqnarray}
&&\hspace{-1mm}
\cheW^{\rm I}_{\mu\nu}(x)~=~
-{N_c\over 4s} \langle p_A,p_B|[\Bsi_1^m(x)\sigma_{\mu\xi}\Psi^n_1(0)][\Bsi_2^n(0)\sigma_\nu^{~\xi}\gamma_5\Psi_2^m(x)]
\label{WI1}\\
&&\hspace{33mm}
-~\Bsi_1^m(x)\sigma_{\mu\xi}\gamma_5\Psi^n_1(0)][\Bsi_2^n(0)\sigma_\nu^{~\xi}\Psi_2^m(x)]
+\mu\leftrightarrow\nu|p_A,p_B\rangle
\nonumber\\
&&\hspace{-1mm}
+~
{N_cg_{\mu\nu}\over 2s} \langle p_A,p_B|[\Bsi_1^m(x)\Psi^n_1(0)][\Bsi_2^n(0)\gamma_5\Psi_2^m(x)]
\nonumber\\
&&\hspace{33mm}
-~[\Bsi_1^m(x)\gamma_5\Psi^n_1(0)][\Bsi_2^n(0)\Psi_1^m(x)]|p_A,p_B\rangle
~-~x\leftrightarrow0
\nonumber
\end{eqnarray}
where we suppressed flavor label.

Let us start from the second term. Obviously, the leading-twist contribution vanishes with our accuracy. 
Next, consider one-gluon terms and start from
\begin{eqnarray}
&&\hspace{-1mm}
{N_cg_{\mu\nu}\over 2s} \langle p_A,p_B|\big([\bsi_A^m(x)\Xi^n_1(0)]+[\Bxi_1^m(x)\Psi^n_1(0)\big)[\bsi_B^n(0)\gamma_5\psi_B^m(x)]
\\
&&\hspace{-1mm}
-~\big([\bsi_A^m(x)\gamma_5\Xi^n_1(0)]+[\Bxi_1^m(x)\gamma_5\Psi^n_1(0)\big)[\bsi_B^n(0)\psi_B^m(x)]|p_A,p_B\rangle
~-~x\leftrightarrow0
\nonumber\\
&&\hspace{-1mm}
=~{ig_{\mu\nu}\over 2s^2} \big\{
\langle \bsi(x)\sigma_{\star i}{1\over\alpha}\psi(0)\rangle_A\langle\bsi B^i(0)\gamma_5\psi(x)\rangle_B
-\langle \big(\bsi{1\over\alpha}\big)(x)\sigma_{\star i}\psi(0)\rangle_A\langle\bsi(0) \gamma_5B^i\psi(x)\rangle_B
\nonumber\\
&&\hspace{-1mm}
-~\langle \bsi(x)\sigma_{\star i}\gamma_5{1\over\alpha}\psi(0)\rangle_A\langle\bsi B^i(0)\psi(x)\rangle_B
+\langle \big(\bsi{1\over\alpha}\big)(x)\sigma_{\star i}\gamma_5\psi(0)\rangle_A\langle\bsi(0) B^i\psi(x)\rangle_B\big\}
~-~x\leftrightarrow0
\nonumber
\end{eqnarray}
The projectile matrix element can bring one factor of $s$ while the target one cannot so the contribution to 
$\cheW^{\rm I}_{\mu\nu}(x)$ 	is $\sim{g_{\mu\nu}\over \alpha_qs}$ which is $O(\beta_q)\times{q_{\mu\nu}\over Q_\parallel^2}$.
Similarly, the contributions coming from $\Xi_2$ and $\Bxi_1$ are  $\sim{g_{\mu\nu}\over \beta_qs}$ and can be neglected.

Let us consider now two-gluon term coming from $\Xi_1$ and $\Xi_2$
\begin{eqnarray}
&&\hspace{-1mm}
{N_cg_{\mu\nu}\over 2s} \langle p_A,p_B|[\bsi_A^m(x)\Xi^n_1(0)][\bsi_B^n(0)\gamma_5\Xi_2^m(x)]
\nonumber\\
&&\hspace{33mm}
-[\bsi_A^m(x)\gamma_5\Xi^n_1(0)][\bsi_B^n(0)\Xi_2^m(x)]|p_A,p_B\rangle
~-~x\leftrightarrow0
\nonumber\\
&&\hspace{-1mm}
=~{g_{\mu\nu}\over 2s^3} \big\{
\langle \bsi A^j(x)\sigma_{\star i}{1\over\alpha}\psi(0)\rangle_A\langle\bsi B^i(0)\sigma_{\bu j}\gamma_5{1\over\beta}\psi(x)\rangle_B
\nonumber\\
&&\hspace{33mm}
-~\langle \bsi A^j(x)\sigma_{\star i}{1\over\alpha}\psi(0)\rangle_A\langle\bsi B^i(0)\sigma_{\bu j}\gamma_5{1\over\beta}\psi(x)\rangle_B\big\}
~-~x\leftrightarrow0
\nonumber\\
&&\hspace{-1mm}
=~{g_{\mu\nu}\over 2s^3} \big\{
\langle \bsi(x) (A^j(x)\sigma_{\star i}-A^i(x)\sigma_{\bu j}){1\over\alpha}\psi(0)\rangle_A
\langle\bsi B^i(0)\sigma_{\bu j}\gamma_5{1\over\beta}\psi(x)\rangle_B
\nonumber\\
&&\hspace{33mm}
+~\langle \bsi A^i(x)\sigma_{\star i}{1\over\alpha}\psi(0)\rangle_A\langle\bsi B^j(0)\sigma_{\bu j}\gamma_5{1\over\beta}\psi(x)\rangle_B\big\}
~-~x\leftrightarrow0
\nonumber\\
&&\hspace{-1mm}
=~-{g_{\mu\nu}\over 2s^3} \langle \bsi(x)\notA(x)\notp_2{1\over\alpha}\psi(0)\rangle_A
\langle\bsi(0) \notB(0)\notp_1\gamma_5{1\over\beta}\psi(x)\rangle_B
~-~x\leftrightarrow0
\label{WI4}
\end{eqnarray}
where again we used Eq. (\ref{gammas13a}) without two last terms and Eq. (\ref{formula2}). 
Now, from equations of motion (\ref{eqm1}) and (\ref{eqm2}) we see that the target matrix element vanishes:
\begin{eqnarray}
&&\hspace{-1mm}
\int\! dx~e^{-i\beta_q x_\star+i(k,x)_\perp}\langle\bsi(0) \notB(0)\notp_1\gamma_5{1\over\beta}\psi(x)\rangle_B
\nonumber\\
&&\hspace{11mm}
=~{k^\perp_i\over\beta_q}\!\int\! dx~e^{-i\beta x_\star+i(k,x)_\perp}\langle\bsi(0)\gamma^i\notp_1\gamma_5\psi(x)\rangle_B
~\sim~{k^\perp_i\over\beta_q}s\epsilon^{ij}k_j~=~0
\label{b5vanishes}
\end{eqnarray}
It is easy to see that the  term coming from $\Bxi_1$ and $\Bxi_2$ vanishes for the same reason.
\begin{eqnarray}
&&\hspace{-1mm}
{N_cg_{\mu\nu}\over 2s} \langle p_A,p_B|[\Bxi_1^m(x)\Psi^n_1(0)][\Bxi_2^n(0)\gamma_5\psi_B^m(x)]
\label{WI6}\\
&&\hspace{33mm}
-[\Bxi_1^m(x)\gamma_5\Psi^n_1(0)][\Bxi_2^n(0)\psi_B^m(x)]|p_A,p_B\rangle
~-~x\leftrightarrow0
\nonumber\\
&&\hspace{-1mm}
=~-{g_{\mu\nu}\over 2s^3} \langle (\bsi{1\over\alpha}\big)(x)\notp_2\notA(0)\psi(0)\rangle_A
\langle\big(\bsi{1\over\beta}\big)(0) 
\notp_1\notB(x)\gamma_5\psi(x)\rangle_B
~-~x\leftrightarrow0~=~0
\nonumber
\end{eqnarray}
Next, two-gluon contribution from $\Bxi_1$ and $\Xi_1$ vanishes since $\Bxi_1 \Xi_1=0$, and  similarly
$\Bxi_2\Xi_2=0$. Finally, the contribution coming from $\Bxi_1$ and $\Xi_2$ is $\sim {1\over N_c^2}$ as demonstrated in Sect. 
\ref{sec:twogluonshoton} (see Eq. (\ref{chewt25})), so the second term in Eq. (\ref{WI1}) vanishes and we get
\begin{eqnarray}
&&\hspace{-1mm}
\cheW^{\rm I}_{\mu\nu}(x)~=~
-{N_c\over 4s} \langle p_A,p_B|[\Bsi_1^m(x)\sigma_{\mu\xi}\Psi^n_1(0)][\Bsi_2^n(0)\sigma_\nu^{~\xi}\gamma_5\Psi_2^m(x)]
\label{WI7}\\
&&\hspace{-1mm}
-~\Bsi_1^m(x)\sigma_{\mu\xi}\gamma_5\Psi^n_1)(0)][\Bsi_2^n(0)\sigma_\nu^{~\xi}\Psi_2^m(x)]
+\mu\leftrightarrow\nu|p_A,p_B\rangle~-~x\leftrightarrow0
\nonumber
\end{eqnarray}
This contribution is similar to Eq. (\ref{vh1}) up to extra $\gamma_5$ and relative signs. 
As we discussed above, extra
$\gamma_5$ cannot change the power of our small parameters ${q_\perp^2\over Q^2}$ and $\alpha_q,\beta_q$ so we can consider only terms
which gave leading contributions to $V^{\rm H}_{\mu\nu}$.

The leading-twist contribution
\begin{eqnarray}
&&\hspace{-1mm}
\cheW^{\rm I,l.t.}_{\mu\nu}(x)~=~
-{1\over 4s} \langle p_A,p_B|[\bsi_A(x)\sigma_{\mu\xi}\psi_A(0)][\bsi_B(0)\sigma_\nu^{~\xi}\gamma_5\psi_B(x)]
\label{WI1l}\\
&&\hspace{33mm}
-~[\bsi_A(x)\sigma_{\mu\xi}\gamma_5\psi_A(0)][\bsi_B(0)\sigma_\nu^{~\xi}\psi_B(x)]
+\mu\leftrightarrow\nu|p_A,p_B\rangle
~-~x\leftrightarrow0
\nonumber
\end{eqnarray}
 is easily obtained from parametrizations (\ref{hmael})
\begin{eqnarray}
&&\hspace{-1mm}
W^{\rm I,lt}_{\mu\nu}(x)(q)
~=~{1\over 16\pi^4}\!\int\! dx_\bu dx_\star d^2x_\perp~e^{-i\alpha_qx_\bu-i\beta_qx_\star+i(q,x)_\perp}\cheW^{\rm I,lt}_{\mu\nu}(x)
\nonumber\\
&&\hspace{-1mm}
~=~-{i\over 2m^2}\!\int\! d^2k_\perp 
\big(\epsilon_{\mu j}[k^j(q-k)^\perp_\nu+(q-k)_j k^\perp_\nu]+\mu\leftrightarrow\nu\big)\pizh^f(q,k_\perp)
\label{WIlt}
\end{eqnarray}
where 
\begin{eqnarray}
&&\hspace{-1mm}
\pizh^f(q,k_\perp)~=~h^\perp_{1}(\alpha_q,k_\perp)\barh^\perp_{1}(\beta_q,(q-k)_\perp) ~-~h_{1f}^\perp\leftrightarrow\barh_{1f}^\perp
\label{pizh}
\end{eqnarray}
As usually, the term with $h_{1f}^\perp\leftrightarrow\barh_{1f}^\perp$ comes from $x\leftrightarrow0$ contribution.

Next, we need to consider terms in Eq. (\ref{WI7}) with one or two gluon operators and generalize the calculations from Sect.  \ref{sec:wt} 
to our case. 

\subsubsection{One-gluon terms }
Let us first consider term coming from  $\Xi_1(0)={i\over s}\sigma_{\star i}B^i{1\over \alpha}\psi_A$. Separating color-singlet 
matrix elements in Eq. (\ref{WI1}), we get
\begin{eqnarray}
&&\hspace{-1mm}  
\cheW^{\rm 1I(1)}_{1\mu\nu}(x)~=~
-{i\over 4s^2}\big\{\langle\bsi(x)\sigma_{\mu\xi}\sigma_{\star i}{1\over\alpha}\psi(0)\rangle_A
\langle\bsi B^i(0)\sigma_\nu^{~\xi}\gamma_5\psi(x)]\rangle_B  
\label{WI9}\\
&&\hspace{-1mm}
-~\psi(0)\otimes\gamma_5\psi(x)\leftrightarrow\gamma_5\psi(0)\otimes\psi(x)+\mu\leftrightarrow\nu\big\}  
~-~x\leftrightarrow0
\nonumber
\end{eqnarray}
As we mentioned, this contribution is similar to the one considered in Sec. \ref{sec:onegluonhoton} so we 
need to take only the case 
of longitudinal $\mu$ and transverse $\nu$, or {\it vice versa} - all other cases are power-suppressed. We get 
\begin{eqnarray}
&&\hspace{-1mm}  
\cheW^{\rm 1I(1)}_{1\mu\nu}(x)~=~
-{ip_{2\mu}\over 2s^3}\big\{
\langle\bsi(x)\sigma_{\bu\xi}\sigma_{\star i}{1\over\alpha}\psi(0)\rangle_A
\langle\bsi B^i(0)\sigma_{\nu_\perp}^{~\xi}\gamma_5\psi(x)]\rangle_B  
\nonumber\\
&&\hspace{16mm}
+~\langle\bsi(x)\sigma_{\nu_\perp\xi}\sigma_{\star i}{1\over\alpha}\psi(0)\rangle_A
\langle\bsi B^i(0)\sigma_\bu^{~\xi}\gamma_5\psi(x)]\rangle_B   
\nonumber\\
&&\hspace{16mm}
-~\psi(0)\otimes\gamma_5\psi(x)\leftrightarrow\gamma_5\psi(0)\otimes\psi(x)\big\}  
~-~x\leftrightarrow0
\nonumber\\
&&\hspace{-1mm}  
=~
-{ip_{2\mu}\over 2s^3}\big\{
\langle\bsi(x)\sigma_{\nu_\perp j}\sigma_{\star i}{1\over\alpha}\psi(0)\rangle_A
\langle\bsi B^i(0)\sigma_\bu^{~j}\gamma_5\psi(x)]\rangle_B   
\nonumber\\\
&&\hspace{16mm}
-~i
\langle\bsi(x)\sigma_{\star i}{1\over\alpha}\psi(0)\rangle_A
\langle\bsi B^i(0)\sigma_{\nu_\perp\bu}\gamma_5\psi(x)]\rangle_B  
\nonumber\\\
&&\hspace{5mm}
+~
\langle\bsi(x)\sigma_{\bu j}\sigma_{\star i}{1\over\alpha}\psi(0)\rangle_A
\langle\bsi B^i(0)\sigma_{\nu_\perp}^{~j}\gamma_5\psi(x)]\rangle_B  
\nonumber\\
&&\hspace{16mm}
+~{2\over s}\langle\bsi(x)\sigma_{\nu_\perp\bu}\sigma_{\star i}{1\over\alpha}\psi(0)\rangle_A
\langle\bsi B^i(0)\sigma_{\bu\star}\gamma_5\psi(x)]\rangle_B   
\nonumber\\
&&\hspace{16mm}
-~\psi(0)\otimes\gamma_5\psi(x)\leftrightarrow\gamma_5\psi(0)\otimes\psi(x)\big\}  
~-~x\leftrightarrow0
\label{WI10}
\end{eqnarray}
where we used power-counting results from 
Ref. \cite{Balitsky:2020jzt}  to eliminate terms proportional to $p_{1\mu}$, cf Eq. (\ref{4.31}). 
Moreover, similarly to Eq. (\ref{4.31}) case,  Eq. (\ref{sigmasigmas}) shows that two last terms in the r.h.s. of Eq. (\ref{WI10}) 
are small and therefore
\begin{eqnarray}
&&\hspace{-1mm}  
\cheW^{\rm 1I(1)}_{1\mu\nu}(x)~=~
-{p_{2\mu}\over 2s^3}\big\{
\langle\bsi(x)\big[g_{i\nu_\perp}\sigma_{\star j}-g_{ij}\sigma_{\star \nu_\perp}\big]{1\over\alpha}\psi(0)\rangle_A
\langle\bsi B^i(0)\sigma_\bu^{~j}\gamma_5\psi(x)\rangle_B   
\nonumber\\\
&&\hspace{16mm}
-~\big[
\langle\bsi(x)\sigma_{\star i}{1\over\alpha}\psi(0)\rangle_A
\langle\bsi B^i(0)\sigma_{\bu\nu_\perp}\gamma_5\psi(x)]\rangle_B  
\nonumber\\\
&&\hspace{16mm}
-~\psi(0)\otimes\gamma_5\psi(x)\leftrightarrow\gamma_5\psi(0)\otimes\psi(x)\big]\big\}  
~-~x\leftrightarrow0
\nonumber\\\
&&\hspace{-1mm}
=~-{p_{2\mu}\over 2s^3}\big\{\langle\bsi(x)\sigma_{\star j}{1\over\alpha}\psi(0)\rangle_A
\langle\bsi(0)[B_\nu(0)\sigma_\bu^{~j}-B^j(0)\sigma_{\bu\nu_\perp}]\gamma_5\psi(x)\rangle_B 
\nonumber\\\
&&\hspace{16mm}
+~
\langle\bsi(x)\sigma_{\star \nu_\perp}\gamma_5{1\over\alpha}\psi(0)\rangle_A
\langle\bsi B^i(0)\sigma_{\bu i}\psi(x)\rangle_B  
\big\}  
~-~x\leftrightarrow0
\label{WI11}
\end{eqnarray}
where we used 
Eqs. (\ref{sigmasigmas}), (\ref{formula2}), and (\ref{b5vanishes}).
Next, the target matrix element in the first term in the r.h.s. can be rewritten as
\begin{eqnarray}
&&\hspace{-1mm}
\langle\bsi(0)[B_\nu(0)\sigma_\bu^{~j}-B^j(0)\sigma_{\bu\nu_\perp}]\gamma_5\psi(x)]\rangle_B 
~=~\epsilon_{\nu j}\langle\bsi(0)\notB(0)\notp_1\psi(x)\rangle_B
\end{eqnarray}
so we get
\begin{eqnarray}
&&\hspace{-1mm}  
\cheW^{\rm 1I(1)}_{1\mu\nu}(x)~
=~-{p_{2\mu}\over s^3}\epsilon_{\nu j}\langle\bsi(x)\sigma_{\star j}{1\over\alpha}\psi(0)\rangle_A
\langle\bsi(0)\notB(0)\notp_1\psi(x)\rangle_B
~-~x\leftrightarrow0
\label{WI13}
\end{eqnarray}
where we used $\sigma_{\star \nu_\perp}\gamma_5=i\epsilon_{\nu j}\sigma_\star^{~j}$. 
Now, from equation of motion (\ref{9.56}) and parametrization (\ref{hmael}) we get
the corresponding contribution to 
$\cheW^{\rm 1I}_{\mu\nu}(x)$ in the form
\begin{eqnarray}
&&\hspace{-1mm}  
\cheW^{\rm 1I(1)}_{1\mu\nu}(x)~
=~-i\epsilon_{\nu j}{p_{2\mu}\over \alpha_qsm^2}\!\int\! d^2k_\perp~k^j_\perp(q-k)_\perp^2\pizh(q,k_\perp)
\label{WI14}
\end{eqnarray}

The term in Eq. (\ref{WI7}) coming from $\Bxi_1$ is similar:
\begin{eqnarray}
&&\hspace{-1mm}  
\cheW^{\rm 1I(1)}_{2\mu\nu}(x)~=~
{i\over 4s^2}\big\{\langle\big(\bsi{1\over\alpha}\big)(x)\sigma_{\mu\xi}\sigma_{\star i}\psi(0)\rangle_A
\langle\bsi(0)\sigma_\nu^{~\xi}\gamma_5B^i(x)\psi(x)]\rangle_B  
\nonumber\\
&&\hspace{-1mm}
-~\psi(0)\otimes\gamma_5\psi(x)\leftrightarrow\gamma_5\psi(0)\otimes\psi(x)+\mu\leftrightarrow\nu\big\}  
~-~x\leftrightarrow0
\nonumber\\
&&\hspace{-1mm}
=~
{ip_{2\mu}\over 2s^3}\big\{
\langle\big(\bsi{1\over\alpha}\big)(x)\sigma_{\star i}\sigma_{\bu\xi}\psi(0)\rangle_A
\langle\bsi B^i(x)\sigma_{\nu_\perp}^{~\xi}\gamma_5\psi(x)]\rangle_B  
\nonumber\\
&&\hspace{16mm}
+~\langle\big(\bsi{1\over\alpha}\big)(x)\sigma_{\star i}\sigma_{\nu_\perp\xi}{1\over\alpha}\psi(0)\rangle_A
\langle\bsi B^i(x)\sigma_\bu^{~\xi}\gamma_5\psi(x)]\rangle_B   
\nonumber\\
&&\hspace{16mm}
-~\psi(0)\otimes\gamma_5\psi(x)\leftrightarrow\gamma_5\psi(0)\otimes\psi(x)\big\}  
~-~x\leftrightarrow0
\nonumber\\
&&\hspace{-1mm}  
=~
{ip_{2\mu}\over 2s^3}\big\{
\langle\big(\bsi{1\over\alpha}\big)(x)\sigma_{\star i}\sigma_{\nu_\perp j}\psi(0)\rangle_A
\langle\bsi B^i(x)\sigma_\bu^{~j}\gamma_5\psi(x)]\rangle_B   
\nonumber\\\
&&\hspace{16mm}
-~i
\langle\big(\bsi{1\over\alpha}\big)(x)\sigma_{\star i}{1\over\alpha}\psi(0)\rangle_A
\langle\bsi B^i(x)\sigma_{\bu\nu_\perp}\gamma_5\psi(x)]\rangle_B  
\nonumber\\\
&&\hspace{16mm}
-~\psi(0)\otimes\gamma_5\psi(x)\leftrightarrow\gamma_5\psi(0)\otimes\psi(x)\big\}  
~-~x\leftrightarrow0
\nonumber\\
&&\hspace{-1mm}  
=~
-{p_{2\mu}\over 2s^3}\big\{
\langle\big(\bsi{1\over\alpha}\big)(x)\big[g_{i\nu_\perp}\sigma_{\star j}-g_{ij}\sigma_{\star \nu_\perp}\big]\psi(0)\rangle_A
\langle\bsi B^i(x)\sigma_\bu^{~j}\gamma_5\psi(x)\rangle_B   
\nonumber\\\
&&\hspace{16mm}
-~\big[
\langle\bsi{1\over\alpha}\big)(x)\sigma_{\star i}\psi(0)\rangle_A
\langle\bsi B^i(x)\sigma_{\bu\nu_\perp}\gamma_5\psi(x)]\rangle_B  
\nonumber\\\
&&\hspace{16mm}
-~\psi(0)\otimes\gamma_5\psi(x)\leftrightarrow\gamma_5\psi(0)\otimes\psi(x)\big]\big\}  
~-~x\leftrightarrow0
\nonumber\\\
&&\hspace{-1mm}
=~-{p_{2\mu}\over 2s^3}\big\{\langle\bsi{1\over\alpha}\big)(x)\sigma_{\star j}\psi(0)\rangle_A
\langle\bsi(0)[B_\nu(x)\sigma_\bu^{~j}-B^j(x)\sigma_{\bu\nu_\perp}]\gamma_5\psi(x)\rangle_B 
\nonumber\\\
&&\hspace{16mm}
+~
\langle\big(\bsi{1\over\alpha}\big)(x)\sigma_{\star \nu_\perp}\gamma_5\psi(0)\rangle_A
\langle\bsi B^i(x)\sigma_{\bu i}\psi(x)\rangle_B  
\big\}  
~-~x\leftrightarrow0
\nonumber\\\
&&\hspace{16mm}
=~{p_{2\mu}\over s^3}\epsilon_{\nu j}\langle\big(\bsi{1\over\alpha}\big)(x)\sigma_{\star j}\psi(0)\rangle_A
\langle\bsi(0)\notp_1\notB(x)\psi(x)\rangle_B
~-~x\leftrightarrow0
\label{WI15}
\end{eqnarray}
where we left only the terms similar to the leading terms in Eq. (\ref{WI10})
and made the same transformations.

Now, from equation of motion (\ref{9.56}) and parametrization (\ref{hmael}) we get
the corresponding contribution to 
$\cheW^{\rm 1I}_{\mu\nu}(x)$ in the form
\begin{eqnarray}
&&\hspace{-1mm}  
\cheW^{\rm 1I(1)}_{2\mu\nu}(x)~
=~-i\epsilon_{\nu j}{p_{2\mu}\over \alpha_qsm^2}\!\int\! d^2k_\perp~k^j_\perp(q-k)_\perp^2\pizh(q,k_\perp)
\label{WI15a}
\end{eqnarray}
so it doubles the contribution (\ref{WI14}).

Let us now consider term in Eq. (\ref{WI7}) coming form $\Xi_2$. 
For longitudinal $\mu$ and transverse $\nu$ we get
\begin{eqnarray}
&&\hspace{-1mm}  
\cheW^{\rm 1I(2)}_{1\mu\nu}(x)~=~
-{i\over 4s^2}\big\{\langle\bsi A^i(x)\sigma_{\mu\xi}\psi(0)\rangle_A
\langle\bsi(0)\sigma_\nu^{~\xi}\sigma_{\bu i}\gamma_5{1\over\beta}\psi(x)\rangle_B  
\label{WI16}\\
&&\hspace{-1mm}
-~\psi(0)\otimes\gamma_5\psi(x)\leftrightarrow\gamma_5\psi(0)\otimes\psi(x)+\mu\leftrightarrow\nu\big\}  
~-~x\leftrightarrow0
\nonumber\\
&&\hspace{-1mm}
=~-{ip_{1\mu}\over 2s^3}\big\{\langle\bsi A^i(x)\sigma_{\star\xi}\psi(0)\rangle_A
\langle\bsi(0)\sigma_{\nu_\perp}^{~\xi}\sigma_{\bu i}\gamma_5{1\over\beta}\psi(x)\rangle_B  
\nonumber\\
&&\hspace{-1mm}
+~\langle\bsi A^i(x)\sigma_{\nu_\perp\xi}\psi(0)\rangle_A
\langle\bsi(0)\sigma_{\star}^{~\xi}\sigma_{\bu i}\gamma_5{1\over\beta}\psi(x)\rangle_B  
\nonumber\\
&&\hspace{33mm}
-~\psi(0)\otimes\gamma_5\psi(x)\leftrightarrow\gamma_5\psi(0)\otimes\psi(x)\big\}  ~-~x\leftrightarrow 0
\nonumber\\
&&\hspace{-1mm}
=~-{ip_{1\mu}\over 2s^3}\big\{\langle\bsi A^i(x)\sigma_{\star j}\psi(0)\rangle_A
\langle\bsi(0)\sigma_{\nu_\perp}^{~j}\sigma_{\bu i}\gamma_5{1\over\beta}\psi(x)\rangle_B  
\nonumber\\
&&\hspace{44mm}
-~i\langle\bsi A^i(x)\sigma_{\nu_\perp\star}\psi(0)\rangle_A
\langle\bsi(0)\sigma_{\bu i}\gamma_5{1\over\beta}\psi(x)\rangle_B  
\nonumber\\
&&\hspace{-1mm}
+~\langle\bsi A^i(x)\sigma_{\nu_\perp j}\psi(0)\rangle_A
\langle\bsi(0)\sigma_{\star}^{~j}\sigma_{\bu i}\gamma_5{1\over\beta}\psi(x)\rangle_B  
+~{2\over s}\langle\bsi A^i(x)\sigma_{\star\bu}\psi(0)\rangle_A
\nonumber\\
&&\hspace{-1mm}
\times~
\langle\bsi(0)\sigma_{\nu_\perp\star}\sigma_{\bu i}\gamma_5{1\over\beta}\psi(x)\rangle_B  
-~\psi(0)\otimes\gamma_5\psi(x)\leftrightarrow\gamma_5\psi(0)\otimes\psi(x)\big\}   ~-~x\leftrightarrow 0
\nonumber
\end{eqnarray}
where we neglected contribution $\sim p_{2\mu}$ since it is $\sim~p_{2\mu}q^\perp_\nu{m_\perp^2\over\beta_qs^2}$, see power counting (without $\gamma_5$) in Ref. \cite{Balitsky:2020jzt} .
Next, similarly to Eq. (\ref{WI10}),  the last two terms in the r.h.s. of Eq. (\ref{WI16}) are 
$\sim O\big({m_\perp^2\over s}\big)$ in comparison to the first two terms so we get
\begin{eqnarray}
&&\hspace{-1mm}  
\cheW^{\rm 1I(2)}_{1\mu\nu}(x)~
=~-{p_{1\mu}\over 2s^3}\big\{\langle\bsi A^i(x)\sigma_{\star j}\psi(0)\rangle_A
\langle\bsi(0)[g_{i\nu_\perp}\sigma_{\bu j}-g_{ij}\sigma_{\bu \nu_\perp}]\gamma_5{1\over\beta}\psi(x)\rangle_B  
\nonumber\\
&&\hspace{-1mm}
-~\langle\bsi A^i(x)\sigma_{\star\nu_\perp}\psi(0)\rangle_A
\langle\bsi(0)\sigma_{\bu i}\gamma_5{1\over\beta}\psi(x)\rangle_B  
-~\psi(0)\otimes\gamma_5\psi(x)\leftrightarrow\gamma_5\psi(0)\otimes\psi(x)\big\}   ~-~x\leftrightarrow 0
\nonumber\\
&&\hspace{-1mm}
=~{p_{1\mu}\over s^3}\big\{\langle\bsi(x)\big(A_\nu(x)\sigma_{\star j}-A_j(x)\sigma_{\star \nu_\perp}\big)\gamma_5\psi(0)\rangle_A
\langle\bsi(0)\sigma_\bu^{~ j}{1\over\beta}\psi(x)\rangle_B  
\nonumber\\
&&\hspace{-1mm} 
+~\langle\bsi A^i(x)\sigma_{\star i}\psi(0)\rangle_A
\langle\bsi(0)\sigma_{\bu\nu_\perp}\gamma_5{1\over\beta}\psi(x)\rangle_B\big\}  ~-~x\leftrightarrow 0
\nonumber\\
&&\hspace{-1mm}
=~-{ip_{1\mu}\over s^3}\epsilon_{\nu_\perp j}\langle\bsi A^i(x)\sigma_{\star i}\psi(0)\rangle_A\langle\bsi(0)\sigma_\bu^{~ j}{1\over\beta}\psi(x)\rangle_B
\label{WI18}
\end{eqnarray}
From Eq. (\ref{9.56}) we get the corresponding contribution to $W^{\rm I}_{\mu\nu}(q)$ in the form
\begin{eqnarray}
&&\hspace{-1mm}  
\cheW^{\rm 1I(2)}_{1\mu\nu}(q)~=~-i\epsilon_{\nu j}{p_{1\mu}\over \beta_qsm^2}\!\int\! d^2k_\perp~(q-k)^j_\perp k_\perp^2\pizh(q,k_\perp)
\label{WI19}
\end{eqnarray}

Similarly to Eq. (\ref{WI14}) one can demonstrate that the term coming from  $\Bxi_2$ doubles the result (\ref{WI19}) of $\Xi_2$
so we finally get
\begin{equation}
\hspace{-0mm}  
W^{\rm 1I}_{\mu\nu}(q)
=~-i{2\epsilon_{\nu j}\over Q_\parallel^2m^2}\!\int\! d^2k_\perp~[\beta p_{2\mu}k^j_\perp(q-k)_\perp^2+\alpha p_{1\mu}(q-k)^jk_\perp^2]\pizh(q,k_\perp)
+\mu\leftrightarrow\nu
\label{WI20}
\end{equation}
where we have added the contribution of transverse $\mu$ and longitudinal $\nu$.

\subsubsection{Two-gluon terms}
Let us start from the contribution to $\cheW^{\rm I}_{\mu\nu}(x)$ of Eq. (\ref{WI7}) coming from $\Xi_{A}$ and $\Xi_{B}$.  
After separation of color-singlet matrix elements, it takes the form
\begin{eqnarray}
&&\hspace{-1mm}  
\cheW^{\rm 2I(1)}_{\mu\nu}(x)~=~
{1\over 4s^3}\big\{\langle\bsi A^i(x)\sigma_{\mu\xi}\sigma_{\star j}{1\over \alpha}\psi(0)\rangle_A
\langle\bsi B^j(0)\sigma_\nu^{~\xi}\sigma_{\bu i}\gamma_5{1\over\beta}\psi(x)\rangle_B  
\label{WI21}\\
&&\hspace{-1mm}
-~\psi(0)\otimes\gamma_5\psi(x)\leftrightarrow\gamma_5\psi(0)\otimes\psi(x)+\mu\leftrightarrow\nu\big\}  
~-~x\leftrightarrow0
\nonumber
\end{eqnarray}
This equation resembles $\cheV_{1\mu\nu}^{\rm (2a)H}(x)$ of 
Eq. (\ref{4.44}) calculated in Sect. \ref{sec:twogluonshoton} so we will 
use power counting from that Section and 
consider only two transverse or two longitudinal indices.

First, let us consider  transverse $\mu$ and $\nu$.
\begin{eqnarray}
&&\hspace{-1mm}  
\cheW^{\rm 2I(1)}_{\mu\nu}(x)~=~
{1\over 4s^3}\big\{\langle\bsi A^i(x)\sigma_{\mu_\perp k}\sigma_{\star j}{1\over \alpha}\psi(0)\rangle_A
\langle\bsi B^j(0)\sigma_{\nu_\perp}^{~k}\sigma_{\bu i}\gamma_5{1\over\beta}\psi(x)\rangle_B  
\label{WI22}\\
&&\hspace{-1mm}
+~{2\over s}\langle\bsi A^i(x)\sigma_{\mu_\perp\bu}\sigma_{\star j}\psi(0)\rangle_A
\langle\bsi B^j(0)\sigma_{\nu_\perp\star}\sigma_{\bu i}\gamma_5{1\over\beta}\psi(x)\rangle_B  
\nonumber\\
&&\hspace{-1mm}
-~\psi(0)\otimes\gamma_5\psi(x)\leftrightarrow\gamma_5\psi(0)\otimes\psi(x)+\mu\leftrightarrow\nu\big\}  
~-~x\leftrightarrow0
\nonumber
\end{eqnarray}
Similarly to Eq. (\ref{4.45}),  from Eqs.  (\ref{sigmasigmas}) and  (\ref{formula2}),
we see that the second term in the r.h.s. 
can be neglected in comparison to the first one and after some algebra  we get 
\begin{eqnarray}
&&\hspace{-2mm}  
\cheW^{\rm 2I(1)}_{\mu_\perp\nu_\perp}(x)~
\nonumber\\
&&\hspace{-2mm}
=~
{-1\over 4s^3}\big\{\langle\bsi A^i(x)[g_{\mu j}\sigma_{\star k}-g_{jk}\sigma_{\star\mu_\perp}]{1\over \alpha}\psi(0)\rangle_A
\langle\bsi B^j(0)[g_{\nu i}\sigma_\bu^{~k}-\delta_i^k\sigma_{\bu\nu_\perp}]\gamma_5{1\over\beta}\psi(x)\rangle_B  
\nonumber\\
&&\hspace{11mm}
-~\psi(0)\otimes\gamma_5\psi(x)\leftrightarrow\gamma_5\psi(0)\otimes\psi(x)+\mu\leftrightarrow\nu\big\}  
~-~x\leftrightarrow0
\nonumber\\
&&\hspace{-2mm}
=~{1\over 2s^3}\Big[\langle\bsi(x)[A_\nu(x)\sigma_{\star i}-A_i(x)\sigma_{\star \nu}]{1\over \alpha}\psi(0)\rangle_A
\langle\bsi (B_\mu(0)\sigma_\bu^{~i}-B^i(0)\sigma_{\bu \mu})\gamma_5{1\over\beta}\psi(x)\rangle_B
\nonumber\\
&&\hspace{-2mm}
+~\langle\bsi A_\nu(x)\sigma_{\star k}{1\over \alpha}\psi(0)\rangle_A
\langle\bsi B^i(0)\sigma_{\bu j}\gamma_5{1\over\beta}\psi(x)\rangle_B(\delta_i^k\delta_\mu^j-\delta_i^j\delta_\mu^k)
\nonumber\\
&&\hspace{-2mm}
+~\langle\bsi A^i(x)\sigma_{\star j}{1\over \alpha}\psi(0)\rangle_A
\langle\bsi B_\mu(0)\sigma_{\bu k}\gamma_5{1\over\beta}\psi(x)\rangle_B(\delta_i^k\delta_\nu^j-\delta_\nu^k\delta_i^j)
\nonumber\\
&&\hspace{11mm}-~\psi(0)\otimes\gamma_5\psi(x)\leftrightarrow\gamma_5\psi(0)\otimes\psi(x)\Big]
~+~\mu\leftrightarrow\nu~-~x\leftrightarrow 0
\label{WI23}
\end{eqnarray}
Moreover, it is easy to see that the first term in the r.h.s. can be omitted: 
either projectile or target matrix element vanishes due to Eq. (\ref{formula2}). For the next two terms in the r.h.s. of Eq. (\ref{WI23}) we use
\begin{eqnarray}
&&\hspace{-2mm}  
{1\over 2s^3}\big\{\langle\bsi A_\nu(x)\sigma_{\star k}{1\over \alpha}\psi(0)\rangle_A
\langle\bsi B^i(0)\sigma_{\bu j}\gamma_5{1\over\beta}\psi(x)\rangle_B(\delta_i^k\delta_\mu^j-\delta_i^j\delta_\mu^k)
\nonumber\\
&&\hspace{22mm}
-~\psi(0)\otimes\gamma_5\psi(x)\leftrightarrow\gamma_5\psi(0)\otimes\psi(x)\big\}
\nonumber\\
&&\hspace{-2mm}
=~{1\over 2s^3}\big\{\langle\bsi(x)A_\nu(x)\sigma_\star^{~ i}\gamma_5{1\over\alpha}\psi(0)\rangle_A
\langle\bsi [B_\mu(0)\sigma_{\bu i}-B_i(0)\sigma_{\bu\mu_\perp}]{1\over\beta}\psi(x)\rangle_B
\nonumber\\
&&\hspace{22mm}
-~\langle\bsi A_\nu(x)\sigma_{\star\mu_\perp}{1\over \alpha}\psi(0)\rangle_A
\langle\bsi B^j(0)\sigma_{\bu j}\gamma_5{1\over\beta}\psi(x)\rangle_B
\nonumber\\
&&\hspace{-2mm}
+~{s\over 4}\langle\bsi A_\nu(x)\sigma_{mn}\gamma_5{1\over \alpha}\psi(0)\rangle_A \langle\bsi B_\mu(x)\sigma^{mn}{1\over\beta}\psi(x)\rangle_B
\nonumber\\
&&\hspace{22mm}
-~{s\over 2}\langle\bsi(x)A_\nu(x)\sigma_{\mu_\perp}^{~ j}\gamma_5{1\over\alpha}\psi(0)\rangle_A\big\}
\langle\bsi B_i(x)\sigma^{ij}{1\over\beta}\psi(x)\rangle_B~=~0
\label{WI24}
\end{eqnarray}
since the target matrix element in the first term vanishes due to Ee. (\ref{formula2}), the one in the second term due to Eq. (\ref{b5vanishes}),
and the last two terms are small by power counting, $\sim {q_\perp^2\over s}\big({ q^\perp_\mu q^\perp_\nu\over Q_\parallel^2}\big)$. 
Similarly,
\begin{eqnarray}
&&\hspace{-2mm}
{2\over s^3}\big\{\langle\bsi A^i(x)\sigma_{\star j}{1\over \alpha}\psi(0)\rangle_A
\langle\bsi B_\mu(0)\sigma_{\bu k}\gamma_5{1\over\beta}\psi(x)\rangle_B(\delta_i^k\delta_\nu^j-\delta_\nu^k\delta_i^j)
\nonumber\\
&&\hspace{22mm}
-~\psi(0)\otimes\gamma_5\psi(x)\leftrightarrow\gamma_5\psi(0)\otimes\psi(x)\big\}
\nonumber\\
&&\hspace{-2mm}
=~\langle\bsi(x)[A_i(x)\sigma_{\star \nu}-A_\nu\sigma_{\star i}{1\over \alpha}\psi(0)\rangle_A
\langle\bsi B_\mu(0)\sigma_\bu^{~i}\gamma_5{1\over\beta}\psi(x)\rangle_B
\nonumber\\
&&\hspace{22mm}
+~\langle\bsi A^i(x)\sigma_{\star i}\gamma_5{1\over \alpha}\psi(0)\rangle_A
\langle\bsi B_\mu(0)\sigma_{\bu\nu_\perp}{1\over\beta}\psi(x)\rangle_B
\nonumber\\
&&\hspace{-2mm}
-~{s\over 4}\langle\bsi(x)A_\nu\sigma_{mn}{1\over \alpha}\psi(0)\rangle_A
\langle\bsi B_\mu(0)\sigma^{mn}\gamma_5{1\over\beta}\psi(x)\rangle_B
\nonumber\\
&&\hspace{22mm}
+~{s\over 2}\langle\bsi(x)A^i\sigma_{ij}{1\over \alpha}\psi(0)\rangle_A
\langle\bsi B_\mu(0)\sigma_{\nu_\perp}^{~j}\gamma_5{1\over\beta}\psi(x)\rangle_B~=~0
\end{eqnarray}
for the same reason: the projectile matrix element in the first term in the r.h.s. vanishes due to Eq. (\ref{formula2}), 
the one in the second term due to Eq. (\ref{b5vanishes}),
and the last two terms are small by power counting. Thus, we get the result that there is no contribution to 
$W^{\rm 2I(1)}_{\mu_\perp\nu_\perp}(q)$ with our accuracy.

If both $\mu$ and $\nu$ are longitudinal,
 using formula
\begin{eqnarray}
&&\hspace{-1mm}
\sigma_{\mu_\parallel\alpha}\sigma_{\star j}\otimes\sigma_{\nu_\parallel}^{~\alpha}\sigma_{\bu j}+\mu\leftrightarrow\nu
~=~-2s g^\parallel_{\mu\nu}\big(\sigma_{\star j}\otimes\sigma_{\bu j}-{1\over s}\sigma_{\bu k}\sigma_{\star j}\otimes \sigma_\star^{~k}\sigma_{\bu j}\big)
\label{WI26}
\end{eqnarray}
we get
\begin{eqnarray}
&&\hspace{-1mm}  
\cheW^{\rm 2I(1)}_{\mu\nu}(x)~=~
-{g^\parallel_{\mu\nu}\over 2s^2}\big\{\langle\bsi A^i(x)\sigma_{\star j}{1\over \alpha}\psi(0)\rangle_A
\langle\bsi B^j(0)\sigma_{\bu i}\gamma_5{1\over\beta}\psi(x)\rangle_B  
\label{WI27}\\
&&\hspace{-1mm}
-~\psi(0)\otimes\gamma_5\psi(x)\leftrightarrow\gamma_5\psi(0)\otimes\psi(x)\big\}  
~-~x\leftrightarrow0
\nonumber
\end{eqnarray}
where we omitted contribution from the second term in the r.h.s. of Eq. (\ref{WI26}) due 
to power counting coming from Eq,  (\ref{sigmasigmas}). Now, using Eq. (\ref{gammas14a}) one obtains
\begin{eqnarray}
&&\hspace{-1mm}  
\cheW^{\rm 2I(1)}_{\mu\nu}(x)~=~
-{g^\parallel_{\mu\nu}\over 2s^2}\big\{\langle\bsi(x) [A_i(x)\sigma_{\star j}-A_j(x)\sigma_{\star i}]{1\over \alpha}\psi(0)\rangle_A
\langle\bsi B^j(0)\sigma_{\bu i}\gamma_5{1\over\beta}\psi(x)\rangle_B  
\nonumber\\
&&\hspace{-1mm}
+~\langle\bsi A^i(x)\sigma_{\star i}{1\over \alpha}\psi(0)\rangle_A
\langle\bsi B^j(0)\sigma_{\bu j}\gamma_5{1\over\beta}\psi(x)\rangle_B  \big\}  
~-~x\leftrightarrow0~=~0
\label{WI28}
\end{eqnarray}
due to Eqs. (\ref{formula2}) and  (\ref{b5vanishes}).  Thus, we get
\begin{eqnarray}
&&\hspace{-1mm}  
\cheW^{\rm 2I(1)}_{\mu\nu}(x)~=~0
\label{WI29}
\end{eqnarray}
Similarly, one can demonstrate that  the contribution to $\cheW^{\rm I}_{\mu\nu}(x)$ of Eq. (\ref{WI7}) coming from $\Bxi_{A}$ and $\Bxi_{B}$
vanishes.

Let us now consider term coming from  $\Bxi_2$ and  $\Xi_2$. After separating color-singlet contributions, it takes the form
\begin{eqnarray}
&&\hspace{-1mm}
\cheW^{\rm 3I}_{\mu\nu}(x)~=~
-{1\over 4s^3} \big\{\langle \bsi A^j(x)\sigma_{\mu\xi}A^i\psi(0)\rangle_A
\langle\big(\bar\psi_B{1\over \beta}\big)\sigma_{\bu i}\sigma_\nu^{~\xi}\sigma_{\bu j}\gamma_5{1\over\beta}\psi(x)\rangle_B
\nonumber\\
&&\hspace{-1mm}
+~\langle \bsi A^j(x)\sigma_{\mu\xi}\gamma_5A^i\psi(0)\rangle_A
\langle\big(\bar\psi_B{1\over \beta}\big)\sigma_{\bu i}\sigma_\nu^{~\xi}\sigma_{\bu j}{1\over\beta}\psi(x)\rangle_B\big\}
~-~x\leftrightarrow0
\label{WI30}
\end{eqnarray}
The power counting for similar Eq. (\ref{4.55}) in Sect. \ref{sec:twogluonshoton} shows that we need 
to take $\mu$ and $\nu$ both longitudinal: 
\begin{eqnarray}
&&\hspace{-1mm}
\cheW^{\rm 3I}_{\mu\nu}(x)~=~
-{p_{1\mu}p_{1\nu}\over s^5} \big\{\langle \bsi A^j(x)\sigma_{\star k}A^i\psi(0)\rangle_A
\langle\big(\bar\psi_B{1\over \beta}\big)\sigma_{\bu i}\sigma_\star^{~k}\sigma_{\bu j}\gamma_5{1\over\beta}\psi(x)\rangle_B
\nonumber\\
&&\hspace{33mm}
+~\langle \bsi A^j(x)\sigma_{\star k}\gamma_5A^i\psi(0)\rangle_A
\langle\big(\bar\psi_B{1\over \beta}\big)\sigma_{\bu i}\sigma_\star^{~k}\sigma_{\bu j}{1\over\beta}\psi(x)\rangle_B\big\}
\nonumber\\
&&\hspace{-1mm}
=~{ip_{1\mu}p_{1\nu}\over s^4} \big\{\langle \bsi A^j(x)\sigma_{\star k}A^i\psi(0)\rangle_A
\langle\big(\bar\psi_B{1\over \beta}\big)\gamma_i\sigma_\bu^{~k}\gamma_j\gamma_5{1\over\beta}\psi(x)\rangle_B
\nonumber\\
&&\hspace{33mm}
-~\psi(0)\otimes\gamma_5\psi(x)\leftrightarrow\gamma_5\psi(0)\otimes\psi(x)\big\}  
~-~x\leftrightarrow0
\nonumber\\
&&\hspace{-1mm}
=~{ip_{1\mu}p_{1\nu}\over s^4} \big\{\langle \bsi(x)\notA(x)\sigma_{\star k}\notA(0)\psi(0)\rangle_A
\langle\big(\bar\psi_B{1\over \beta}\big)\sigma_\bu^{~k}\gamma_5{1\over\beta}\psi(x)\rangle_B~=~0
\label{WI31}
\end{eqnarray}
where we used formula $\sigma_{\star k}\otimes\sigma_\bu^{~k}\gamma_5-\sigma_{\star k}\gamma_5\otimes\sigma_\bu^{~k}=0$ following from Eq.  (\ref{gammas13a}). Also, one can demonstrate that the contribution to $\cheW^{\rm I}_{\mu\nu}$ coming from $\Bxi_{A}$ and  $\Xi_{A}$ vanishes.
Finally, similarly to Eq. (\ref{chewt25}), we can neglect terms $\sim{1\over N_c^2}$ 
coming from $\Bxi_1,\Xi_{B}$ and  $\Bxi_2,\Xi_{A}$, see the discussion in Sect. 
\ref{sec:twogluonshoton}. Thus, we get the result that the contribution to  $\cheW^{\rm I}_{\mu\nu}$  coming from two quark-quark-gluon TMDs vanishes with 
our accuracy:
\begin{equation}
\cheW^{\rm 2I}_{\mu\nu}~=~0
\label{WI32}
\end{equation}

\newpage
\subsubsection{Exchange-type power corrections}
The exchange-type power corrections to eq. (\ref{wi2s}) are
\begin{eqnarray}
&&\hspace{-1mm}
\cheW^{\rm I2Sex}_{\mu\nu}(x)  
~=~{N_c\over 2s}\sum_{f,f'}\langle A,B|\Big((e_{f'}c_fa_f-e_fc_{f'}a_{f'})
[\Bsi_1(x)\gamma_\mu\Psi_1(x)]^f[\Bsi_2(0)\gamma_\nu\Psi_2(0)]^{f'}
\nonumber\\
&&\hspace{7mm}
-~e_{f'}c_f[\Bsi_1(x)\gamma_\mu\gamma_5\Psi_1(x)]^f
[\Bsi_2(0)\gamma_\nu\Psi_2(0)]^{f'}
\nonumber\\
&&\hspace{7mm}
+~e_fc_{f'}[\Bsi_1(x)\gamma_\mu\Psi_1(x)]^f
[\Bsi_2(0)\gamma_\nu\gamma_5\Psi_2(0)]^{f'}
|A,B\rangle\Big)
+\mu\leftrightarrow\nu-x\leftrightarrow 0
\label{wi2sex}
\end{eqnarray}
The terms in the r.h.s.  differ from those in 
Eqs. (\ref{wexff}),  (\ref{wexa5}), and  (\ref{wexb5}) by 
replacement of ``$+x\leftrightarrow 0$'' by
``$-x\leftrightarrow 0$'' which leads to change sign of ``$\pm$ ~c.c.'' terms in
those equations so we get instead of Eq. (\ref{wexffsym})
\begin{eqnarray}
&&\hspace{-2mm}
W^{\rm I2Sex}_{\mu\nu}(q)
~=~{iN_c\over 2(N_c^2-1)Q_\parallel^2}\sum_{f,f'}\!\int\! d^2k_\perp\Big\{
(e_fc_{f'}a_{f'}-e_{f'}c_fa_f)
[k^\perp_\mu(q-k)^\perp_\nu+k^\perp_\nu(q-k)^\perp_\mu
\nonumber\\
&&\hspace{17mm}    
+~g^\perp_{\mu\nu}(k,q-k)_\perp]I^{1ff'}_{--}(q,k_\perp)-g^\perp_{\mu\nu} (k,q-k)_\perp I^{2ff'}_{--}(q,k_\perp)
\label{wexi2s}\\
&&\hspace{-2mm}
+~c_fe_{f'}[\epsilon_{\mu m}k^m(q-k)_\nu+\mu\leftrightarrow\nu]J^{1ff'}_{+-}(q,k_\perp)
-c_{f'}e_f
[k_\mu\epsilon_{\nu n}(q-k)^n+\mu\leftrightarrow\nu]  J^{1ff'}_{-+}(q,k_\perp)\Big\}
\nonumber
\end{eqnarray}
where $J^i_{\pm\pm}$ and $I^i_{--}$ are defined in Eqs. (\ref{Js}) and (\ref{Is}).

\subsubsection{Result for symmetric interference term $W^{\rm I2San}$}
As usual, we represent the result for hadronic tensor $W^{\rm I2S}_{\mu\nu}(q)$ as a sum
of the ``annihilation'' and ``exchange'' parts:
\begin{eqnarray}
&&\hspace{-1mm}
W^{\rm I2S}_{\mu\nu}(q)~=~W^{\rm I2San}_{\mu\nu}(q)~+~W^{\rm I2Sex}_{\mu\nu}(q)
\nonumber\\
&&\hspace{-1mm}
\end{eqnarray}
where the exchange-type corrections are presented in Eq. (\ref{wexi2s}) above while
$W^{\rm I2San}_{\mu\nu}(q)$ is given by 
the sum of  Eqs. (\ref{WIlt}) and (\ref{WI20})
\begin{eqnarray}
&&\hspace{-1mm}
W^{\rm I2San}_{\mu\nu}(q)~=~\sum_{f}e_fc_fW^{\rm If}_{\mu\nu}(q),
\nonumber\\
&&\hspace{-1mm}
W^{\rm If}_{\mu\nu}(q)
~=~-{i\epsilon_{\mu j}\over 2m^2}\!\int\! d^2k_\perp 
\Big([k^j(q-k)^\perp_\nu+(q-k)_\perp^j k^\perp_\nu]
\nonumber\\
&&\hspace{-1mm}
+~{4\over Q_\parallel^2}[\beta p_{2\nu}k^j(q-k)_\perp^2+\alpha p_{1\nu}(q-k)^jk_\perp^2]\Big)\pizh^f(q,k_\perp)
+\mu\leftrightarrow\nu
\label{wi2sresult}
\end{eqnarray}
Let us check gauge invariance of annihilation part of interference hadronic tensor \\
$q^\mu W^{\rm I2San}_{\mu\nu}(q)~=~0$. First, let us rewrite it in as follows:
\begin{eqnarray}
&&\hspace{-1mm}
W^{\rm If}_{\mu\nu}(q)
~=~-{i\epsilon_{\mu j}\over 2m^2}\Big(\delta_\nu^i-2{q^\parallel_\nu q^i\over Q_\parallel^2}\Big)
\!\int\! d^2k_\perp 
[k^j(q-k)_i+(q-k)^j k_i+\delta_i^j(k,q-k)_\perp]\pizh^f(q,k_\perp)
\nonumber\\
&&\hspace{-1mm}
+~{i\epsilon_{\mu j}\over m^2}{\tilq_\nu\over Q_\parallel^2}\!\int\! d^2k_\perp 
[k^j(q-k)_\perp^2-(q-k)^jk_\perp^2]\pizh^f(q,k_\perp)
+\mu\leftrightarrow\nu
\label{wi2sresult}
\end{eqnarray}
The integral in the first term is proportional to $(2q_i q^j+\delta_i^q q_\perp^2)$ and it is easy to see
that
\begin{eqnarray}
&&\hspace{-1mm}
q^\mu\Big[\epsilon_{\mu j}\Big(\delta_\nu^i-2{q^\parallel_\nu q^i\over Q_\parallel^2}\Big)
(2q_iq^j+\delta_i^j) +\mu\leftrightarrow\nu\Big]~=~0
\end{eqnarray}
The integral in the second term is proportional to $q^j$ so
\begin{eqnarray}
&&\hspace{-1mm}
q^\mu\Big[\epsilon_{\mu j}q^j
+\mu\leftrightarrow\nu\Big]~=~0
\end{eqnarray}
and therefore $q^\mu W^{\rm If}_{\mu\nu}(q)=0$.

\subsection{Antisymmetric interference term of tensor $W^{\rm I2}$ \label{aninterm}}
Similarly to the symmetric case (\ref{wi2s}), 
from definitions (\ref{ws}) and (\ref{warray}) we get
\begin{eqnarray}
&&\hspace{-1mm}
\cheW^{\rm I2A}_{\mu\nu}(x)
~=~{N_c\over 2s}\sum_{f,f'}\Big((e_{f'}c_fa_f-e_fc_{f'}a_{f'})
\langle A,B|[\bsi(x)\gamma_\mu\psi(x)]^f[\bsi(0)\gamma_\nu\psi(0)]^{f'}
\nonumber\\
&&\hspace{15mm}
+~e_fc_{f'}\langle A,B|[\bsi(x)\gamma_\mu\psi(x)]^f[\bsi(0)\gamma_\nu\gamma_5\psi(0)]^{f'}
\nonumber\\
&&\hspace{15mm}
-~e_{f'}c_f\langle A,B|[\bsi(x)\gamma_\mu\gamma_5\psi(x)]^f[\bsi(0)\gamma_\nu\psi(0)]^{f'}|A,B\rangle\Big)
-\mu\leftrightarrow\nu
\label{wi2a}
\end{eqnarray}
\subsubsection{Annihilation-type power corrections \label{anni2vanishes}}
Again, let us start with annihilation-type power corrections. Since in this case $f=f'$, the first term in the r.h.s. 
of Eq. (\ref{wi2a}) vanishes and the second can be written as
\begin{eqnarray}
&&\hspace{-1mm}
\chew^{\rm I2Aa}_{\mu\nu}(x)~=~\sum_{f}e_fc_f\chew^{\rm IAf}_{\mu\nu}(x)
\nonumber\\
&&\hspace{-1mm}
\chew^{\rm IAf}_{\mu\nu}(x)
~=~{N_c\over 2s}\langle A,B|[\Bsi_1(x)\gamma_\mu\Bsi_2(x)][\Bsi_2(0)\gamma_\nu\gamma_5\Bsi_1(0)]
\nonumber\\
&&\hspace{15mm}
-~[\Bsi_1(x)\gamma_\mu\gamma_5\Bsi_2(x)][\Bsi_2(0)\gamma_\nu\Bsi_1(0)]
-\mu\leftrightarrow\nu|A,B\rangle+x\leftrightarrow 0
\label{wiaa}
\end{eqnarray}
where we made the usual replacement $\psi\rightarrow\Psi_1+\Psi_2$.
After Fierz transformation (\ref{fierz6}) the r.h.s. of the above equation turns to
\begin{eqnarray}
&&\hspace{-1mm}
\cheW^{\rm IA}_{\mu\nu}(x)~
\label{wi2aa}\\
&&\hspace{-1mm}=~
i{N_c\over 2s} \langle p_A,p_B|
[\Bsi_1^m(x)\Psi^n_1(0)][\Bsi_2^n(0)\sigma_{\mu\nu}\gamma_5\Psi_2^m(x)]
-[\Bsi_1^m(x)\sigma_{\mu\nu}\Psi^n_1(0)][\Bsi_2^n(0)\gamma_5\Psi_2^m(x)]
\nonumber\\
&&\hspace{-1mm}
-~[\Bsi_1^m(x)\gamma_5\Psi^n_1(0)][\Bsi_2^n(0)\sigma_{\mu\nu}\Psi_1^m(x)]
+[\Bsi_1^m(x)\sigma_{\mu\nu}\gamma_5\Psi^n_1(0)][\Bsi_2^n(0)\Psi_1^m(x)]|p_A,p_B\rangle
+x\leftrightarrow0
\nonumber
\end{eqnarray}
where we suppressed flavor label.

Let's us demonstrate that $\cheW^{\rm IA}_{\mu\nu}(x)$  is small in our approximation. 
After  Fierz transformation (\ref{fierz6})
one obtains
\begin{eqnarray}
&&\hspace{-1mm}
\cheW^{\rm IA}_{\mu\nu}(x)~
\nonumber\\
&&\hspace{-1mm}=~{iN_c\over 2s}
\langle A,B|-[\Bsi_1^m(x)\gamma_5\Psi_1^n(0)][\Bsi_2^n(0)\sigma_{\mu\nu}\Psi_2^m(x)]
+[\Bsi_1^m(x)\sigma_{\mu\nu}\gamma_5\Psi_1^n(0)][\Bsi_2^n(0)\Psi_2^m(x)]
\nonumber\\
&&\hspace{-1mm}
+~[\Bsi_1^m(x)\Psi_1^n(0)][\Bsi_2^n(0)\sigma_{\mu\nu}\gamma_5\Psi_2^m(x)]
-[\Bsi_1^m(x)\sigma_{\mu\nu}\Psi_1^n(0)][\Bsi_2^n(0)\gamma_5\Psi_2^m(x)]
|A,B\rangle+x\leftrightarrow 0
\nonumber\\
\label{chew0e}
\end{eqnarray}
It is convenient to convolute $W^{\rm IA}_{\mu\nu}(x)$ with $\epsilon_{\mu\nu\alpha\beta}$ and consider 
\begin{equation}
\pizW^{\rm IA}_{\mu\nu}(x)~=~{i\over 2}\epsilon_{\mu\nu}^{~~\alpha\beta}W^{\rm IA}_{\alpha\beta}(x)
\label{pizwei}
\end{equation}
then
\begin{eqnarray}
&&\hspace{-1mm}
\chepizW^{\rm Ia}_{\mu\nu}(x)~
\label{pizwei1}\\
&&\hspace{-1mm}
=~{iN_c\over 2s}\langle A,B|-[\Bsi_1^m(x)\gamma_5\Psi_1^n(0)][\Bsi_2^n(0)\sigma_{\mu\nu}\gamma_5\Psi_2^m(x)]
+[\Bsi_1^m(x)\sigma_{\mu\nu}\Psi_1^n(0)][\Bsi_2^n(0)\Psi_2^m(x)]
\nonumber\\
&&\hspace{-1mm}
+~[\Bsi_1^m(x)\Psi_1^n(0)][\Bsi_2^n(0)\sigma_{\mu\nu}\Psi_2^m(x)]
-[\Bsi_1^m(x)\sigma_{\mu\nu}\gamma_5\Psi_1^n(0)][\Bsi_2^n(0)\gamma_5\Psi_2^m(x)]
|A,B\rangle+x\leftrightarrow 0
\nonumber
\end{eqnarray}
This is similar to $\chew^{\rm as}_{\mu\nu}(x)$ of Eq. (\ref{chew0}) studied in previous Section. The only difference is
the relative sign between the first and the second term in the r.h.s. of these equations (and replacement of 
``$-x\leftrightarrow 0$'' by
``$+x\leftrightarrow 0$'' which does not change power counting). 
 Let us qiuckly check that this relative sign does not change the result that the contribution is small. 
 First, let us consider the one-gluon contribution similar to Eq. (\ref{chew1})
\begin{eqnarray}
&&\hspace{-1mm}
{i\over s^2}\big\{\langle\bsi(x)\notp_2\gamma_i{1\over\alpha}\psi(0)\rangle_A\langle\bsi(0)B^i(0)\sigma_{\mu\nu}\psi(x)\rangle_B
\label{pizwei2}\\
&&\hspace{-1mm}
+~\langle\bsi(x)\sigma_{\mu\nu}\notp_2\gamma^i{1\over\alpha}\psi(0)\rangle_A\langle\bsi(0)B_i(0)\psi(x)\rangle_B
-\psi(0)\otimes \psi(x)\leftrightarrow \gamma_5\psi(0)\otimes \gamma_5\psi(x)\big\}-x\leftrightarrow 0
\nonumber
\end{eqnarray}
As discussed in previous Section after Eq. (\ref{chew1}), the two  terms in the r.h.s. of Eq. (\ref{chew1}) vanish separately for transverse $\mu$ and $\nu$,
are both small for one longitudinal and one transverse index, and the term which changed sign is neglected in Eq. (\ref{chew3}) so in all cases
the relative sign does not matter. Similarly, one can check this for other one-gluon terms.

Let us now consider two-gluon term coming from $\Xi_1$ and $\Xi_2$ 
\begin{eqnarray}
&&\hspace{-1mm}
{i\over 2s^3}\big\{\langle\bsi A^j(x)\sigma_{\star i}{1\over \alpha}\psi(0)\rangle_A\langle\bsi B^i(0)\sigma_{\mu\nu}\sigma_{\bu j}
{1\over\beta}\psi(x)\rangle_B~+~ \langle\bsi A^j(x)\sigma_{\mu\nu}\sigma_{\star i}{1\over \alpha}\psi(0)\rangle_A
\nonumber\\
&&\hspace{11mm}
\times~\langle\bsi B^i(0)\sigma_{\bu j}{1\over\beta}\psi(x)\rangle_B
-\psi(0)\otimes \psi(x)\leftrightarrow \gamma_5\psi(0)\otimes \gamma_5\psi(x)\big\}+x\leftrightarrow 0
\label{pizwei3}
\end{eqnarray}
Again, this differs from Eq. (\ref{chew9}) by relative sign between two terms. Looking at the derivation of Eq. (\ref{chew11}) we see that 
both terms in the r.h.s. vanish separately due to Eq. (\ref{formula2}) so
\begin{eqnarray}
&&\hspace{-1mm}
{\rm Eq. ~(\ref{pizwei3})}
~=~{1\over 2s^3}\big\{\langle\bsi [A_\mu(x)\sigma_{\star \nu_\perp}-\mu\leftrightarrow\nu]{1\over \alpha}\psi(0)\rangle_A\langle\bsi B^i(0)\sigma_{\bu i}{1\over\beta}\psi(x)\rangle_B
\nonumber\\
&&\hspace{-1mm}
+~\langle\bsi  A^i(x)\sigma_{\star i}{1\over \alpha}\psi(0)\rangle_A\langle\bsi [B_\mu(0)\sigma_{\bu \nu_\perp}-\mu\leftrightarrow \nu]{1\over\beta}\psi(x)\rangle_B\big\}
+x\leftrightarrow 0~=~0
\label{pizwei4}
\end{eqnarray}
If $\mu$ and $\nu$ are longitudinal, we get
\begin{eqnarray}
&&\hspace{-1mm}
{\rm Eq. ~(\ref{pizwei3})}~=~{2i\over s^5}(p_{1\mu}p_{2\nu}-\mu\leftrightarrow\nu)
\big\{\langle\bsi A^j(x)\sigma_{\star i}{1\over \alpha}\psi(0)\rangle_A\langle\bsi B^i(0)\sigma_{\star\bu}\sigma_{\bu j}
{1\over\beta}\psi(x)\rangle_B
\nonumber\\
&&\hspace{20mm}
+~ \langle\bsi A^j(x)\sigma_{\star\bu}\sigma_{\star i}{1\over \alpha}\psi(0)\rangle_A\langle\bsi B^i(0)\sigma_{\bu j}
{1\over\beta}\psi(x)\rangle_B
\nonumber\\
&&\hspace{33mm}
-~\psi(0)\otimes \psi(x)\leftrightarrow \gamma_5\psi(0)\otimes \gamma_5\psi(x)\big\}-x\leftrightarrow 0
~=~0
\label{pizwei5}
\end{eqnarray}
If now one of indices is longitudinal and the other transverse, the contribution is small due to power counting as discussed 
after Eq. (\ref{chewxz}) so the  two-gluon term coming from $\Xi_1$ and $\Xi_2$ vanishes with our accuracy. 
The corresponding contribution coming from $\Bxi_1$ and $\Bxi_2$
vanishes for the same reason. 

Finally, as shown in Eq. (\ref{chewxz1}),  terms coming from $\Bxi_1$, $\Xi_1$ and from $\Bxi_{2B}$, $\Xi_{2B}$ are small due to power counting. 
Also, as usually we neglect the terms coming from  $\Bxi_1$, $\Xi_{2B}$ and from $\Bxi_{2B}$, $\Xi_1$ are $\sim{1\over N_c^2}$ so 
\begin{equation}
\chepizW^{\rm IA}_{\mu\nu}(q)~=~\cheW^{\rm IA}_{\mu\nu}(q)~=~0
\label{wiazero}
\end{equation}
with our accuracy.
%
\subsubsection{Exchange-type power corrections and the result for $W^{\rm I2}$}
The exchange-type power corrections to eq. (\ref{wi2a}) are
\begin{eqnarray}
&&\hspace{-1mm}
(\cheW^{\rm ex})^{\rm I2A}_{\mu\nu}(x)
~=~{N_c\over 2s}\sum_{f,f'}\langle A,B|\Big((e_{f'}c_fa_f-e_fc_{f'}a_{f'})
[\Bsi_1(x)\gamma_\mu\Psi_1(x)]^f[\Bsi_2(0)\gamma_\nu\Psi_2(0)]^{f'}\nonumber\\
&&\hspace{7mm}
+~e_fc_{f'}[\Bsi_1(x)\gamma_\mu\Psi_1(x)]^f
[\Bsi_2(0)\gamma_\nu\gamma_5\Psi_2(0)]^{f'}
\nonumber\\
&&\hspace{7mm}
-~e_{f'}c_f[\Bsi_1(x)\gamma_\mu\gamma_5\Psi_1(x)]^f
[\Bsi_2(0)\gamma_\nu\Psi_2(0)]^{f'}|A,B\rangle\Big)
-\mu\leftrightarrow\nu+x\leftrightarrow 0
\label{wi2aex}
\end{eqnarray}
Similarly to the symmetric case, one can use formulas from Sect. \ref{sec:wzaex} 
with change of signs of complex conjugations.
Indeed, the terms in the r.h.s. of Eq. (\ref{wi2aex}) differ from those in 
Eqs. (\ref{wzexasym}) and  (\ref{wzexasab5}) by 
replacement of ``$+x\leftrightarrow 0$'' by
``$-x\leftrightarrow 0$'' which leads to change sign of ``$\pm$ ~c.c.'' terms in
those equations.  We get
\begin{eqnarray}
&&\hspace{-1mm}
(W^{\rm I2A})^{\rm ex}_{\mu\nu}(q)~=~-{N_c\over (N_c^2-1)Q_\parallel^2}\sum_{f,f'}c_fc_{f'}
\!\int\! d^2k_\perp
\Big\{a_f[\epsilon_{\mu m}k^m(q-k)_\nu-\mu\leftrightarrow\nu]
I^{2ff'}_{+-}(q,k_\perp)
\nonumber\\
&&\hspace{21mm}
-~
a_{f'}[k_\mu\epsilon_{\nu n}(q-k)^n-\mu\leftrightarrow\nu]  I^{2ff'}_{-+}(q,k_\perp)
\Big\}
\label{wi2aexotvet}
\end{eqnarray}
cf. Eq. (\ref{chewzaex}). As usually, exchange power corrections exist only for transverse $\mu$ and $\nu$.

Finally, since we proved in previous Section that annihilation-type power corrections $W^{\rm I2A}_{\mu\nu}$ vanish, the total result for $W^{\rm I2A}_{\mu\nu}$ is equal to the ``exchange'' part:
\begin{equation}
W^{\rm I2A}_{\mu\nu}(q)~=~(W^{\rm I2A})^{\rm ex}_{\mu\nu}(q)~=~{\rm r.h.s~of~Eq.~(\ref{wi2aexotvet}) }
\end{equation}
%

\section{Results \label{sec:results}}

In this section we will take into account only gauge-invariant terms coming from 
annihilation-type terms proportional to TMDs $f_1$ and $h_1^\perp$.  The reason to neglect 
annihilation-type $\sim W^{\rm 2H}(q)$ is that the twist-three matrix elements (\ref{maelsa}) are virtually unknown, and exchange-type power corrections can presumably be neglected due to extra ${1\over N_c}$. 
Anyway, taking into account leading-twist contributions and their ``gauge-invariance-restoring'' 
counterparts appears to a good start for estimations of DY hadronic tensors. 

\subsection{Hadronic tensors in Collins-Soper frame}
In Collins-Soper frame the hadronic tensors are parametrized in terms of $q$ and three unit
vectors $X,Y,Z$ orthogonal to $q$ and to each other. In terms of Sudakov variables they are
\begin{eqnarray}
&&\hspace{-1mm}
Z~=~{\tilq\over Q_\parallel}~=~{1\over Q_\parallel}(\alpha_qp_1-\beta_qp_2),~~~~~~~
X~=~\Big({Q_\perp\over Q_\parallel Q}q+{Q\over Q_\perp Q_\parallel}q_\perp\Big)
\label{xyz}\\
&&\hspace{-1mm}
Y_\mu~=~-{1\over Q}\epsilon_{\mu\nu\lambda\rho}X^\nu Z^\lambda q^\rho
~=~\epsilon_{\mu i}{q_\perp^i\over Q_\perp}
\end{eqnarray}
where $Q_\perp\equiv |q_\perp|$.

\subsubsection{Hadronic tensor for photon-mediated DY process}
We  parametrize photon-mediated hadronic tensor in a standard way (up to extra $N_c$ and flavor factors), separately for $W^{\rmF f}_{\mu\nu}$ and $W^{\rmH f}_{\mu\nu}$ defined in Eq. (\ref{resultginv})
\begin{eqnarray}
&&\hspace{-1mm}
W^{\rmF f}_{\mu\nu}(q)
~=~-\big(g_{\mu\nu}-{q_\mu q_\nu\over q^2}\big)(W^{\rmF f}_T+W^{\rmF f}_{\dd})
-2X_\mu X_\nu W^{\rmF f}_\dd
\nonumber\\
&&\hspace{14mm}
+~Z_\mu Z_\nu(W^{\rmF f}_L-W^{\rmF f}_T-W^{\rmF f}_\dd)
-(X_\mu Z_\nu+X_\nu Z_\mu)W^{\rmF f}_\Delta
\label{WFdec}
\end{eqnarray}
and similarly 
\begin{eqnarray}
&&\hspace{-1mm}
W^{\rmH f}_{\mu\nu}(q)
~=~-\big(g_{\mu\nu}-{q_\mu q_\nu\over q^2}\big)(W^{\rmH f}_T+W^{\rmH f}_{\dd})
-2X_\mu X_\nu W^{\rmH f}_\dd
\nonumber\\
&&\hspace{14mm}
+~Z_\mu Z_\nu(W^{\rmH f}_L-W^{\rmH f}_T-W^{\rmH f}_\dd)
-(X_\mu Z_\nu+X_\nu Z_\mu)W^{\rmH f}_\Delta
\label{WHdec}
\end{eqnarray}
The expressions for $W_i(q)$ can be easily obtained from Eqs. (\ref{resultf}) and (\ref{resulthginv}):
\begin{eqnarray}
&&\hspace{-1mm}
W^{\rmF f}_T(q)~=~\!\int\! d^2k_\perp
\big[1-{q_\perp^2\over 2Q_\parallel^2}\big]F^f(q,k_\perp),~~~~~
W^{\rmF f}_L(q)~=~\!\int\! d k_\perp {(q-2k)_\perp^2\over Q_\parallel^2}F^f(q,k_\perp)
\label{WFs}\\
&&\hspace{-1mm}
W^{\rmF f}_\dd(q)~=~\!\int\!d^2k_\perp {q_\perp^2\over 2Q_\parallel^2}F^f(q,k_\perp),~~~~~
W^{\rmF f}_\Delta(q)~=~{Q\over Q_\parallel^2Q_\perp}
\!\int\!d^2k_\perp (q,q-2k)_\perp F^f(q,k_\perp)
\nonumber
\end{eqnarray}
and
\begin{eqnarray}
&&\hspace{-1mm}
W^{\rmH f}_T(q)
~=~{1\over 2m^2Q^2}\!\int\!d^2k_\perp \big(2k_\perp^2(q-k)_\perp^2+[k_\perp^2+(q-k)_\perp^2](k,q-k)_\perp\big)H^f(q,k_\perp),
\nonumber\\
&&\hspace{-1mm}
W_L^{\rmH f}(q)~=~
{1\over m^2Q^2}\!\int\! d k_\perp
\big(2k_\perp^2(q-k)_\perp^2-[k_\perp^2+(q-k)_\perp^2](k,q-k)_\perp\big)
H^f(q,k_\perp)
\nonumber\\
&&\hspace{-1mm}
W_\dd^{\rmH f}(q)~=~{1\over m^2}
\!\int\!d^2k_\perp
\Big({2(q,k)_\perp(q,q-k)_\perp\over q_\perp^2}-(k,q-k)_\perp
+{1\over Q^2}\big\{k_\perp^2(q-k)_\perp^2
\nonumber\\
&&\hspace{15mm}
+~\half (k,q-k)_\perp[k_\perp^2+(q-k)_\perp^2]+q_\perp^2(k,q-k)_\perp-2(q,k)(q,q-k)_\perp\big\}
\Big)H^f(q,k_\perp)
\nonumber\\
&&\hspace{-1mm}
W_\Delta^{\rmH f}(q)~=~-{Q\over Q_\parallel^2Q_\perp m^2}
\!\int\!d^2k_\perp (q,q-2k)(k,q-k)_\perp H^f(q,k_\perp)
\end{eqnarray}
These expressions were obtained in Ref. \cite{Balitsky:2020jzt} .

\subsubsection{Hadronic tensor for $Z$-mediated DY process}

The symmetric part of $W^{\rm ZS}_{\mu\nu}$ is given by Eq. (\ref{wzsyma})
\begin{equation}
\hspace{-0mm}
W^{\rm ZS}_{\mu\nu}~=~e^2\sum_fc_f^2\big[(a_f^2+1)W_{\mu\nu}^{f\rm F}(q)+(a_f^2-1)W_{\mu\nu}^{f\rm H}(q)\big]
\label{wzsymae}
\end{equation}
where $W_{\mu\nu}^{\rmF f}(q)$ and  $W_{\mu\nu}^{\rm H f}(q)$ are
expressed in terms of $X$ and $Z$ vectors in Eqs. (\ref{WFdec}) and  (\ref{WHdec}) above.

The antisymmetric part (\ref{wzan}) can be parametrized as
\begin{eqnarray}
&&\hspace{-1mm}
W^{\rm ZA}_{\mu\nu}(q)~=~-2i\epsilon_{\mu\nu\lambda\rho}q^\lambda\sum_f a_fc_f^2
\bigg[{Z^\rho\over Q_\parallel}\pizW_4^{\rmF f}(q)
+X^\rho{Q_\perp\over Q_\parallel Q}\pizW_3^{\rmF f}(q)\bigg]
\label{wasym}
\end{eqnarray}
where
\begin{equation}
\hspace{0mm}
\pizW_4^{\rmF f}(q)~=\int\! d^2k_\perp\pizf
^f(q,k_\perp),~~~~~
\pizW_3^{\rmF f}(q)~=\int\! d^2k_\perp\Big(1-{2(q,k)_\perp\over q_\perp^2}\Big)\pizf^f(q,k_\perp)
\label{pizwan}
\end{equation}
with $\pizf^f(q,k_\perp)$ given by Eq. (\ref{pizf}).

\subsubsection{Interference tensors }

The symmetric part of the interference tensor $W^{\rm I1}_{\mu\nu}(q)$ is given by Eq. (\ref{wi1s})
\begin{eqnarray}
&&\hspace{-1mm}
W^{\rm I1S}_{\mu\nu}(q)~
=~\sum_f e_fc_f a_f
\big[W^{\rm F f}_{\mu\nu}(q)+W^{\rm H f}_{\mu\nu}(q)\big]
\label{wi1sym}
\end{eqnarray}
where $W_{\mu\nu}^{\rmF f}(q)$ and  $W_{\mu\nu}^{\rm H f}(q)$ are
given by Eqs. (\ref{WFdec}) and  (\ref{WHdec}) above.

The antisymmetric part of $W^{\rm I1}_{\mu\nu}(q)$ is given by Eq. (\ref{wi1asy}) which can be represented 
similarly to Eq. (\ref{wasym})
\begin{eqnarray}
&&\hspace{-1mm}
W^{\rm I1A}_{\mu\nu}(q)~
=~-i\epsilon_{\mu\nu\lambda\rho}q^\lambda\sum_f e_fc_f
\bigg[{Z^\rho\over Q_\parallel}\pizW_A^{\rmF f}(q)
+X^\rho{Q_\perp\over Q_\parallel Q}\pizW_B^{\rmF f}(q)\bigg]
\label{wi1asym}
\end{eqnarray}
where $\pizW_{A,B}^{\rmF f}(q)$ are given by Eq. (\ref{pizwan}).

Next, the symmetric part of  $W^{\rm I2}_{\mu\nu}(q)$, given by Eq. (\ref{wi2sresult}), can be parametrized as
\begin{eqnarray}
&&\hspace{-1mm}
W^{\rm If}_{\mu\nu}(q)
~=~i\sum_f e_fc_f\Big({Q\over Q_\parallel}Y_\mu X_\nu W_1^{\rm If}(q)
+Y_\mu Z_\nu{Q_\perp\over Q_\parallel}W_2^{\rm If}(q)\Big)
+\mu\leftrightarrow\nu
\label{wi2sresulte}
\end{eqnarray}
where
\begin{eqnarray}
&&\hspace{-1mm}
W_1^{\rm If}(q)~=~{1\over m^2}\!\int\! d^2k_\perp\Big[{2(q,k)_\perp(q,q-k)_\perp\over q_\perp^2}-(k,q-k)_\perp\Big]\pizh^f(q,k_\perp)
\nonumber\\
&&\hspace{-1mm}
W_2^{\rm If}(q)~=~{1\over m^2q_\perp^2}\!\int\! d^2k_\perp\big[(q,k)_\perp(q-k)_\perp^2-(q,q-k)_\perp k_\perp^2\big]\pizh^f(q,k_\perp)
\end{eqnarray}

Finally, the antisymmetric part of  $W^{\rm I2}_{\mu\nu}(q)$ vanishes, see Eq. (\ref{wiazero}).
\begin{eqnarray}
&&\hspace{-1mm}
W^{\rm ZA}_{\mu\nu}(q)~=~-2i\epsilon_{\mu\nu\lambda\rho}q^\lambda\sum_f a_fc_f^2
\bigg[{Z^\rho\over Q_\parallel}\pizW_1^{\rmF f}(q)
+X^\rho{Q_\perp\over Q_\parallel Q}\pizW_2^{\rmF f}(q)\bigg]
\label{wasym}
\end{eqnarray}
\subsubsection{Angular coefficients}
Rewriting the differential cross section (\ref{desigma}) in terms of hadronic tensors listed in previous Section,
we obtain 
\begin{eqnarray}
&&\hspace{-1mm}
d\sigma~=~{d^4q\over q^4}{e^4\over N_cs\pi^2}\!\int\!d^3ld^3l'\delta(q-l-l')~\matW(q,l,l')
~=~{e^4\over 16\pi^2sN_c}{dQ^2\over Q^2}dYd^2q_\perp d\Omega_l~\matW(q,l,l')
\nonumber\\
\label{desigmae}
\end{eqnarray}
where
\begin{eqnarray}
&&\hspace{-1mm}
\matW(q,l,l')~=~{2\over Q^2}L^{\mu\nu}\sum_f e_f^2[W^{\rmF f}_{\mu\nu}(q)+W^{\rmH f}_{\mu\nu}(q)]
\nonumber\\
&&\hspace{-1mm}
+~{2c_e^2\over Q^2}\phi_1(Q^2)\Big\{(a_e^2+1)L^{\mu\nu}\sum_f
c_f^2\big[(a_f^2+1)W_{\mu\nu}^{f\rm F}(q)+(a_f^2-1)W_{\mu\nu}^{f\rm H}(q)\big]
\nonumber\\
&&\hspace{-1mm}
-~4a_ea_fc_f^2\epsilon^{\mu\nu\alpha\beta}l_\alpha {l'}_\beta\epsilon_{\mu\nu\lambda\rho}q^\lambda
\Big[{Z^\rho\over Q_\parallel}\pizW_1^{\rmF f}(q)
+X^\rho{Q_\perp\over Q_\parallel Q}\pizW_2^{\rmF f}(q)\Big]\Big\}
\nonumber\\
&&\hspace{-1mm}
+~{4c_e\over Q^2} \phi_2(Q^2)
\sum_f c_fe_f\Big\{a_e\big[W^{\rm F f}_{\mu\nu}(q)+W^{\rm H f}_{\mu\nu}(q)\big]
L^{\mu\nu}
\nonumber\\
&&\hspace{-1mm}
-~\epsilon^{\mu\nu\alpha\beta}l_\alpha {l'}_\beta\epsilon_{\mu\nu\lambda\rho}q^\lambda
\Big[{Z^\rho\over Q_\parallel}\pizW_1^{\rmF f}(q)
+X^\rho{Q_\perp\over Q_\parallel Q}\pizW_2^{\rmF f}(q)\Big]\Big\}
\nonumber\\
&&\hspace{-1mm}
-~{4c_ea_e\over Q^2}\phi_3(Q^2)
\sum_f c_fe_f\Big[{Q\over Q_\parallel}Y_\mu X_\nu W_1^{\rm If}(q)
+Y_\mu Z_\nu{Q_\perp\over Q_\parallel}W_2^{\rm If}(q)\Big]L^{\mu\nu}
\label{matwe}
\end{eqnarray}
and
\begin{equation}
\phi_1(Q^2)\equiv{Q^4\over  |m_Z^2-Q^2|^2+\Gamma_Z^2m_Z^2} ,~~~~
\phi_2(Q^2)={(Q^2-m_z^2)\over Q^2} f_1(Q^2),~~~~\phi_3(Q^2)={\Gamma_z m_z\over Q^2} f_1(Q^2)
\end{equation}

The angular dependence in the CS frame can be displayed using
the convolutions of leptonic and hadronic tensors are presented in Sect. \ref{sec:lhtproducts}. 
One obtains
\begin{eqnarray}
&&\hspace{-2mm}
\matW(q,l,l')~=~\sum_f\Big\{\big[e_f^2+c_e^2(a_e^2+1)c_f^2(a_f^2+1)\phi_1
+2c_ea_ec_fa_f\phi_2\big]
\nonumber\\
&&\hspace{-2mm}
\times~\big[\big(W_T^{\rm Ff}+{W_L^{\rm Ff}\over 2}\big)(1+\cos^2\theta)
+{W_L^{\rm Ff}\over 2}(1-3\cos^2\theta)
+W_\Delta^{\rm Ff}\sin2\theta\cos\phi 
+W_\dd^{\rm Ff}\sin^2\theta\cos2\phi \big]
\nonumber\\
&&\hspace{-2mm}
+~\big[e_f^2+c_e^2(a_e^2+1)c_f^2(a_f^2-1)\phi_1
+2c_ea_ec_fa_f\phi_2\big]
\nonumber\\
&&\hspace{-2mm}
\times~\big[\big(W_T^{\rm Hf}+{W_L^{\rm Hf}\over 2}\big)(1+\cos^2\theta)
+{W_L^{\rm Hf}\over 2}(1-3\cos^2\theta)
+W_\Delta^{\rm Hf}\sin2\theta\sin\phi 
+W_\dd^{\rm Hf}\sin^2\theta\cos2\phi \big]
\nonumber\\
&&\hspace{-2mm}
+~4c_ec_f\big[2a_ea_fc_ec_f\phi_1+e_f\phi_2\big]
\Big[\pizW_1^{\rmF f}\cos\theta+{Q_\perp\over Q}\pizW_2^{\rmF f}\sin\theta\cos\phi\Big]
\nonumber\\
&&\hspace{-2mm}
+~a_ea_fc_ec_f\phi_3\Big[W_1^{\rm If}\sin^2\theta\sin 2\phi
+{Q_\perp\over Q}W_2^{\rm If}\sin 2\theta\sin\phi\Big]
\Big\}
\label{wotvet}
\end{eqnarray}
where $\phi_i=\phi_i(Q^2)$ and $W^i=W^i(\alpha_q,\beta_q,Q^2)$.

The above formula (\ref{wotvet}) is the main result of the paper. It should be compared to
standard representation of angular distribution of DY leptons (\ref{Ais})

\subsection{Comparison with LHC measurements}

The LHC measurements are integrated over the region of invariant mass of DY pair between 80 and 100 GeV.
In this kinematic region the most important contribution comes from terms in Eq.  (\ref{wotvet}) multiplied
by $\phi_1$. Indeed, for an estimate we can consider $W^i(\alpha_q,\beta_q,Q^2)\simeq W^i(\alpha_q,\beta_q,m_z^2)$ and get
\begin{equation}
\hspace{-1mm}
\int_{80}^{100}{dQ^2\over Q^2}\simeq 0.2, ~~
\int_{80}^{100}{dQ^2\over Q^2}\phi_1(Q^2)\simeq 95, ~~
\int_{80}^{100}{dQ^2\over Q^2}\phi_2(Q^2)\simeq 0,~~
\int_{80}^{100}{dQ^2\over Q^2}\phi_3(Q^2)\simeq 2.6
\end{equation}
Since the accuracy of our small-x approximation is $\alpha_q=x_A,~\beta_q=x_B\sim 0.1$ we can neglect
all contributions coming form terms not multiplied by $\phi_1$. 
Moreover,  both theoretical \cite{Scimemi_2018} and phenomenological \cite{Barone:2009hw,Barone:2010gk} analysis  indicate  that $h_1^\perp$ is is of order of few percent of $f_1$ 
 and hence in numerical estimates we will disregard the contribution of $h_1^\perp$.
Introducing notations
\begin{eqnarray}
&&\hspace{-1mm}
\pizW^{\rmF f}(q)~=~\!\int\! d^2k_\perp F^f(q,k_\perp),~~~~~
\pizW^{\rmF f}_L(q)~=~\!\int\! d k_\perp {(q-2k)_\perp^2\over q_\perp^2}F^f(q,k_\perp)
\label{WFs}\\
&&\hspace{-1mm}
\pizW^{\rmF f}_1(q)~=~
\!\int\!d^2k_\perp {(q,q-2k)_\perp\over q_\perp^2} F^f(q,k_\perp)
\nonumber
\end{eqnarray}
we get
\begin{eqnarray}
&&\hspace{-1mm}
\matW(q,l,l')~=~c_e^2c_f^2\phi_1(Q^2)\sum_f\Big\{(a_e^2+1)(a_f^2+1)\Big(
\big[\pizw^{\rm Ff}-{Q_\perp^2\over 2Q^2}(\pizw^{\rm Ff}-\pizw_L^{\rm Ff})\big]
(1+\cos^2\theta)
\nonumber\\
&&\hspace{-1mm}
+~{Q_\perp^2\over 2Q^2}\pizW_L^{\rm Ff}(1-3\cos^2\theta)
+{Q_\perp\over Q}\pizW_1^{\rm Ff}\sin2\theta\cos\phi 
+{Q_\perp^2\over 2Q^2}\pizW^{\rm Ff}\sin^2\theta\cos2\phi \big]\Big)
\nonumber\\
&&\hspace{-1mm}
+~8a_ea_f\Big[{Q_\perp\over Q}\pizW_3^{\rmF f}\sin\theta\cos\phi+\pizW_4^{\rmF f}\cos\theta\Big]
\Big\}
\label{wotvete}
\end{eqnarray}
where  $\pizW^{\rmF f}_3$ and $\pizW^{\rmF f}_4$ are defined in  Eq. (\ref{pizwan}). 
Since we neglected exchange-type power corrections we should expect the accuracy of order of ${1\over N_c}\sim 30$\%.
Let us discuss now some qualitative and semi-quantitative predictions of this equation.
First, let us evaluate $\pizw_i$ at $Q_\perp^2\gg m^2$ following  Ref. \cite{Balitsky:2017gis}.

\subsubsection{Logarithmical estimates of $\pizW_i$ at $Q_\perp^2\gg m^2$\label{sec:logestimates}}
At $q_\perp^2\gg m^2$ we probe the perturbative tail of TMD $f_1$ which is $\sim{1\over k_\perp^2}$
So, as long as  $Q^2\gg q_\perp^2\gg m^2$ we can approximate
\begin{equation}
\hspace{-0mm}
f_1(\alpha_z,k_\perp^2)~\simeq~{f(\alpha_q)\over k_\perp^2},
~~~~~~\barf_1\simeq{\barf(\alpha_q)\over k_\perp^2}
\label{fhestimate1}
\end{equation}
(up to logarithmic corrections). Similarly, for the target we can use the estimate
\begin{equation}
f_1(\beta_z,k_\perp^2)~\simeq~{f(\beta_z)\over k_\perp^2},
~~~~~~\barf_1\simeq{\barf(\beta_z)\over k_\perp^2}
\label{fhestimate2}
\end{equation}
as long as $k_\perp^2\ll Q^2$.  Thus, we get an estimate
\begin{eqnarray}
&&\hspace{-1mm}
F^f(q,k_\perp)~\simeq~{F^f(\alpha_q,\beta_q)\over k_\perp^2(q-k)_\perp^2},
~~~~~F^f(\alpha_q,\beta_q)~\equiv~f^f(\alpha_q)\barf^f(\beta_q)+f^f\leftrightarrow\barf^f,
\nonumber\\
&&\hspace{-1mm}
\pizf^f(q,k_\perp)~\simeq~{\pizf^f(\alpha_q,\beta_q)\over k_\perp^2(q-k)_\perp^2},
~~~~~\pizf^f(\alpha_q,\beta_q)~\equiv~f^f(\alpha_q)\barf^f(\beta_q)-f^f\leftrightarrow\barf^f.
\label{festimate}
\end{eqnarray}
Due to Eqs. (\ref{fhestimate1}) and (\ref{fhestimate2}), the integrals over $k_\perp$ are logarithmic and should be cut from below
by $m^2_N$ and from above by $Q^2$ so we get an estimate
\begin{eqnarray}
&&\hspace{-0mm}
\!\int\! d^2k_\perp{1\over k_\perp^2(q-k)_\perp^2}~\simeq~{2\pi\over q_\perp^2}\ln{q_\perp^2\over m^2},~~~~~~
\!\int\! d^2k_\perp{(k,q-k)_\perp\over k_\perp^2(q-k)_\perp^2}~\simeq~-\pi\ln{Q^2\over q_\perp^2}
\label{loginteg1}
\end{eqnarray}
where we assumed that the first integral is determined by the logarithmical region $q_\perp^2\gg k_\perp^2\gg m^2_N$ and the second by $Q^2\gg k_\perp^2\gg q_\perp^2$.
Taking these integrals to Eq. (\ref{WFs})  one obtains
\begin{eqnarray}
&&\hspace{-1mm}
\pizW^{\rmF f}(q)~\simeq~{2\pi\over Q_\perp^2}\ln{q_\perp^2\over m^2} F^f(\alpha_q,\beta_q),~~~~~
\pizW_L^{\rmF f}(q)~\simeq~{2\pi\over Q_\perp^2}\Big[\ln{Q_\perp^2\over m^2}+2\ln{Q^2\over Q_\perp^2}\Big] F^f(\alpha_q,\beta_q)
\nonumber\\
&&\hspace{-1mm}
\pizW_4^{\rmF f}(q)~\simeq~{2\pi\over Q_\perp^2}\ln{q_\perp^2\over m^2} \pizf^f(\alpha_q,\beta_q),~~~~~~\pizW^{\rmF f}_1(q)~\simeq~\pizw_3^{\rmF f}(q)~\simeq~0.
\label{WFests}
\end{eqnarray}
For numerical estimates of $A_0$ and $A_2$ we will need the ratio
\begin{equation}
r(Q_\perp^2)~\equiv~{\pizw_L^{\rm Ff}(q)\over\pizw^{\rm Ff}(q)}~\simeq~1+2{\ln Q^2/Q_\perp^2\over\ln Q_\perp^2/m^2}~~~~
\label{pizratio}
\end{equation}
which does not depend on quark flavors. Substituting these formulas to Eq. (\ref{wotvet}), we get
we get
\begin{eqnarray}
&&\hspace{-1mm}
\matW(q,l,l')~=~c_e^2(a_e^2+1)c_f^2\phi_1(Q^2)\sum_f(a_f^2+1)\pizw^{\rm Ff}\Big\{\Big(
\Big[1-{Q_\perp^2\over 2Q^2}(1-r)\Big]
(1+\cos^2\theta)
\nonumber\\
&&\hspace{-1mm}
+~{Q_\perp^2\over 2Q^2}r(1-3\cos^2\theta)
+{Q_\perp^2\over 2Q^2}\sin^2\theta\cos2\phi \big]\Big)
+~{8a_ea_f\over (a_e^2+1)(a_f^2+1)}{\pizf^f\over F^f}\cos\theta
\Big\}
\label{westimate}
\end{eqnarray}

\subsubsection{Estimates of $A_0$ and $A_2$}
\begin{figure}[htb]
\begin{center}
\includegraphics[width=88mm]{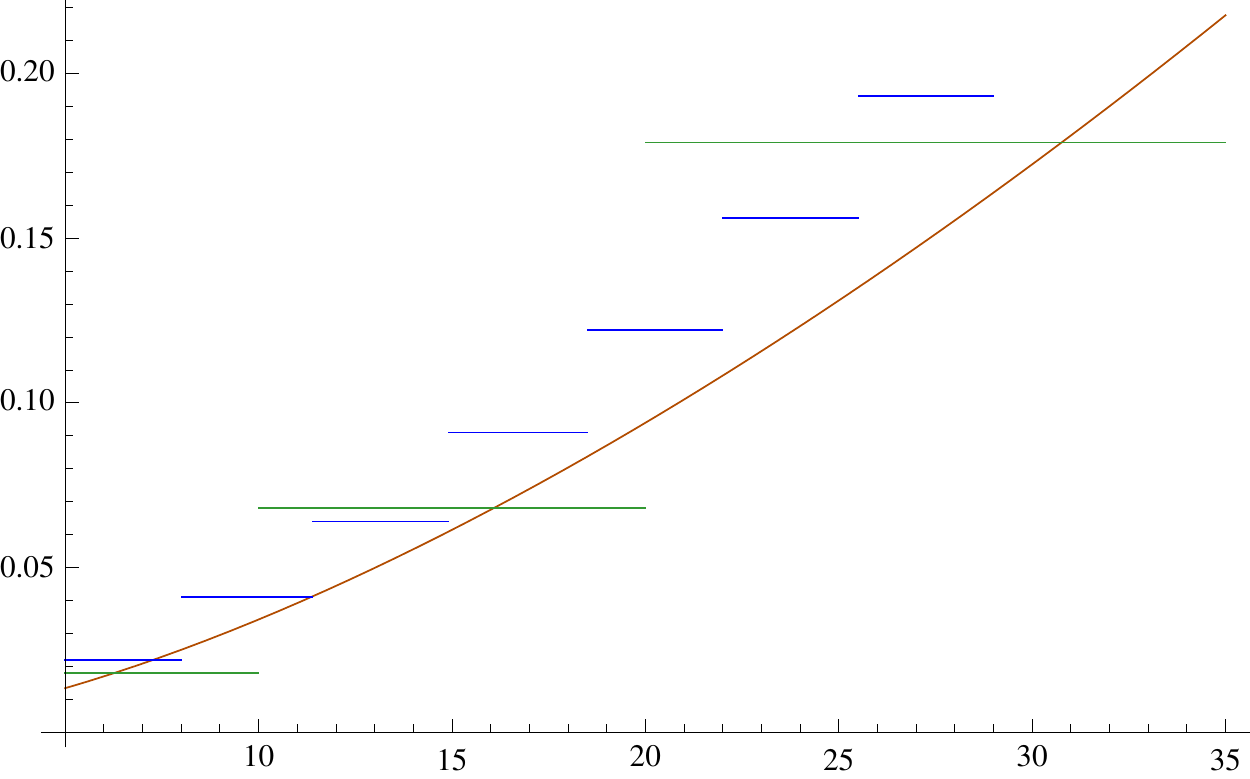}
\end{center}
\caption{Comparison of prediction (\ref{anot})  with  lines depicting angular coefficient $A_0$ in bins  of $Q_\perp$ and $Y<1$ from
Ref. \cite{Khachatryan:2015paa}. (long bins) and Ref. \cite{Aad:2016izn} (short bins). \label{fig:figaA0}}
\end{figure}

Since our estimate (\ref{pizratio}) does not depend on flavor, from Eq. (\ref{westimate}) we get
\begin{equation}
A_0~=~{Q_\perp^2\over m_z^2}{1+2{\ln m_z^2/Q_\perp^2\over\ln Q_\perp^2/m^2}\over 1+{Q_\perp^2\over m_z^2}{\ln m_z^2/Q_\perp^2\over\ln Q_\perp^2/m^2}}
\label{anot}
\end{equation}
where we replaced $Q^2$ by $m_Z^2$ since this function varies slowly between 80 and 100 GeV. This formula is compared to LHC measurements
in Fig. \ref{fig:figaA0}.
The accuracy of our estimate is about 20\% which is reasonable since we neglected corrections $\sim{1\over N_c}$.

\begin{figure}[htb]
\begin{center}
\includegraphics[width=88mm]{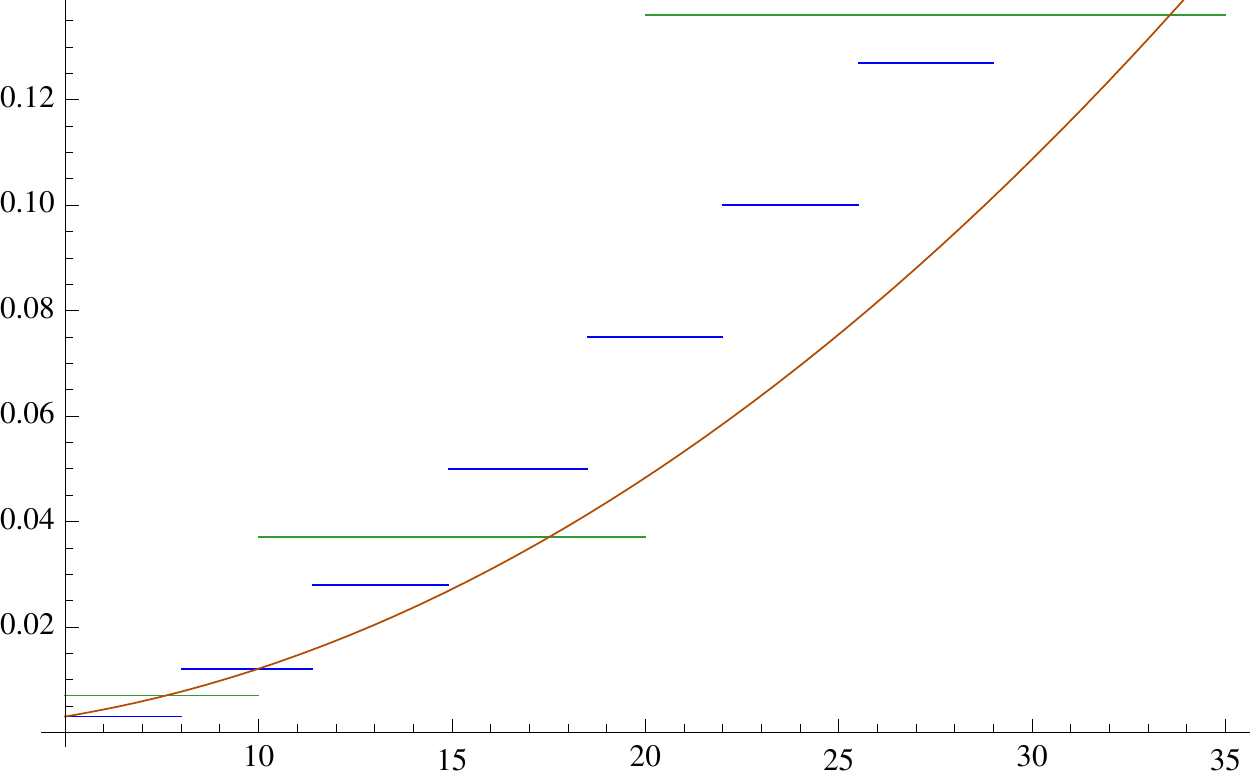}
\end{center}
\caption{Comparison of prediction (\ref{adva}) with  lines depicting angular coefficient $A_2$ in bins  of $Q_\perp$ and $Y<1$ from
Ref. \cite{Khachatryan:2015paa}. (long bins) and \cite{Aad:2016izn} (short bins). \label{fig:figaA2}}
\end{figure}

Next, in our logarithmic approximation
\begin{equation}
A_2~=~{Q_\perp^2/m_z^2\over 1+{Q_\perp^2\over m_z^2}{\ln m_z^2/Q_\perp^2\over\ln Q_\perp^2/m^2}}
\label{adva}
\nonumber
\end{equation}
The comparison with LHC data is shown in Fig. \ref{fig:figaA2}.
The accuracy here is about 30\%.

\subsubsection{Qualitative checks for other angular coefficients}
First, note that between 10 and 30 GeV the coefficient $A_1$ from Table 15 of Ref. \cite{Aad:2016izn} is an order of magnitude smaller than $A_0$ from table 14
(or $A_2$ from Table 16) in accordance with our estimate $A_1\simeq 0$. This is another argument in favor of factorization hypothesis  $f_1(x_B,k_\perp)\simeq f_1(x_B)g(k_\perp)$ 
which is frequently used in current TMD literature.

Second, from Eq. (\ref{westimate}) we see that our estimate for $A_4$ involves non-trivial flavor structure, but one thing immediately seen
is that $A_4$ does not depend on $Q_\perp$ in our region 10-30 GeV. As one can see from Table 18 of Ref. \cite{Aad:2016izn} , the experimental numbers for  $A_4$ 
almost do not change in this kinematical region.  Moreover, from Eq. (\ref{festimate})  $\pizf^f(\alpha_q,\beta_q)=0$ at $Y=0$ (i.e., when $\alpha_q=\beta_q$) so we should expect that 
$A_4$ in rapidity bin $2>Y>1$ is greater than in bin $Y<1$, and Table 18 confirms that. On the contrary,  the results for $A_0$ and $A_2$ do not change much between those 
rapidity beans in accordance with our prediction that $F^f(\alpha_q,\beta_q)$ does not change radically.  

Third, our estimate $A_3=0$ is in accordance to the fact that experimental numbers for $A_3$ from Table 17 of Ref. \cite{Aad:2016izn} are order 
of magnitude smaller than numbers $A_4$ from Table 18.

Finally, the coefficients $A_5$ to $A_7$ do not appear in our estimate (\ref{wotvet}), again in accordance with Tables 19-21 of Ref. \cite{Aad:2016izn}
where the numbers for $A_5,A_6$, and $A_7$ are actually two orders of magnitude smaller than the numbers for $A_0$ or $A_2$ in our kinematical region.

 Summarizing, it looks like our  $f_1$-based estimates  point in the right direction, but of course more phenomenological work 
 is required. Note also that in Ref. \cite{Balitsky:2020jzt}  it was demonstrated that TMD evolution  in the double-log approximation does not affect the predictions for angular coefficients,
 but the single-log corrections to TMD evolution may change the results for asymmetries. The study is in progress.

\section{Conclusions and outlook \label{sec:konklu}}

To my knowledge,
this is the first analysis of DY angular coefficients in the framework of TMD factorization, and, as we saw, the first signs are encouraging. 
 There are two questions about the TMD analysis of DY process: can we predict/explain angular dependence of DY cross 
 section from TMD factorization at LHC kinematical range and reciprocally, can we learn something about proton structure from this?

The answer to first question is probably yes. 
We see that when we have sufficient knowledge about TMDs responsible for a particular angular coefficient, that
coefficient comes out with reasonable accuracy even with back-of-the-envelope estimates.
For example, the very naive estimate of the coefficients $A_0$ and $A_2$ 
agrees with experiment at 30\% accuracy. The error  may be due to our approximation of TMDs by perturbative tails or 
maybe it is due to ${1\over N_c}$ corrections proportional to higher-twist operators. I hope that careful analysis 
involving established models of $f_1(x,q_\perp)$ will distinguish between these two possibilities.  
As to the rest of angular coefficients, more details about quark TMDs and at least some guesses about quark-quark-gluon
TMDs are necessary to make quantitative predictions. Also, one needs to take into account fiducial power corrections, see
recent paper \cite{Ebert:2020dfc} for a review.

This bring us to the second question, namely how much we can learn about proton structure from $Z$-boson experiments. 
First, as was demonstrated 
above, the coefficients $A_0$ and $A_2$ are determined by  TMD $f_1$ with ${1\over N_c}\sim 30$\% accuracy. 
We used the naive logarithmic estimates  and it is natural to assume  that  the realistic models for $f_1$ will move the curves in Figs.  
\ref{fig:figaA0} and \ref{fig:figaA2} closer to the experimental points. Also, it should be mentioned, tat the  ``factorization hypothesis'' 
for the LT TMDs like $f(x_B,k_\perp)~\simeq~f(x_B)g(k_\perp)$ seems to be confirmed by experimental data in  a sense that the 
angular coefficients which vanish in this approximation are smaller than the non-vanishing ones. 
The coefficients $A_1,A_3$, and $A_4$ are determined by non-factorized  part of $f_1$ or higher-twist exchange contributions.

Next, the coefficients $A_5$-$A_7$ seem to be determined by  exchange-type corrections proportional 
to higher-twist quark-quark-gluon TMDs. Unfortunately, comprehensive analysis of exchange-type power 
corrections requires calculation of higher-twist contributions restoring EM gauge invariance of these corrections,
see the discussion in Sect. \ref{sect:fotonresult}.

Let us also discuss the perturbative corrections to asymmetries. As demonstrated in Ref.  \cite{Balitsky:2020jzt}
our estimates of asymmetries  are not affected by summation of Sudakov double logs, but single logs
may bring some changes to tree-level results. It should be also emphasized that,
as discussed in Refs. \cite{Balitsky:2017flc,Balitsky:2017gis}, from the rapidity factorization (\ref{W5}) we get 
TMDs with rapidity-only cutoff $|\alpha|<\sigma_t$ or $|\beta|<\sigma_p$. Such cutoff, relevant for small-$x$ physics, is 
different from the combination of UV and rapidity cutoffs for TMDs used by moderate-$x$ community, see the analysis in two  \cite{Echevarria:2015byo,Li:2016axz,Luebbert:2016itl}   
and three \cite{Echevarria:2016scs} loops. 
This difference in cutoffs does not matter for the tree-level formulas of Sect. \ref{sec:results}, but if one goes beyond the tree level, 
one has to relate TMDs with rapidity-only cutoffs to the TMD models with conventional cutoffs. This requires calculations at the NLO level which are in progress.

The author is grateful to  A. Prokudin  and A. Vladimirov for valuable discussions. This  work is
 supported by Jefferson Science Associates, LLC under the U.S. DOE contract \#DE-AC05-06OR23177
 and by U.S. DOE grant \#DE-FG02-97ER41028.

\section{Appendix}
\subsection{Formulas with Dirac matrices \label{diracs}}
\subsubsection{Fierz transformations}
First, let us write down Fierz transformations for symmetric and antisymmetric combinations
\begin{eqnarray}
&&\hspace{-1mm}
\half[(\bsi\gamma_\mu\chi)(\bhi\gamma_\nu\psi)+\mu\leftrightarrow\nu]
\label{fierz}\\
&&\hspace{-1mm}
=~
-{1\over 4}\big(\delta_\mu^\alpha\delta_\nu^\beta+\delta_\nu^\alpha\delta_\mu^\beta-g_{\mu\nu}g^{\alpha\beta}\big)
\big[(\bsi\gamma_\alpha\psi)(\bhi\gamma_\beta\chi)
+(\bsi\gamma_\alpha\gamma_5\psi)(\bhi\gamma_\beta\gamma_5\chi)\big]
\nonumber\\
&&\hspace{-1mm}
+~{1\over 4}\big(\delta_\mu^\alpha\delta_\nu^\beta+\delta_\nu^\alpha\delta_\mu^\beta-\half g_{\mu\nu}g^{\alpha\beta}\big)
(\bsi\sigma_{\alpha\xi}\psi)(\bhi\sigma_\beta^{~\xi}\chi)-{g_{\mu\nu}\over 4}(\bsi\psi)(\bhi\chi)+{g_{\mu\nu}\over 4}(\bsi\gamma_5\psi)(\bhi\gamma_5\chi)
\nonumber
\end{eqnarray}
and 
\begin{eqnarray}
&&\hspace{-3mm}
\half[(\bsi\gamma_\mu\chi)(\bhi\gamma_\nu\psi)-\mu\leftrightarrow\nu]
~=~
{i\over 4}\epsilon_{\mu\nu\alpha\beta}[(\bsi\gamma_\alpha\gamma_5\psi)(\bhi\gamma_\beta\chi)
+(\bsi\gamma_\alpha\psi)(\bhi\gamma_\beta\gamma_5\chi)]
\nonumber\\
&&\hspace{-3mm}
-~{i\over 4}(\bsi\psi)(\bhi\sigma_{\mu\nu}\chi)+{i\over 4}(\bsi\gamma_5\psi)(\bhi\sigma_{\mu\nu}\gamma_5\chi)
+{i\over 4}(\bsi\sigma_{\mu\nu}\psi)(\bhi\chi)-{i\over 4}(\bsi\sigma_{\mu\nu}\gamma_5\psi)(\bhi\gamma_5\chi)
\label{fierzasy}
\end{eqnarray}
Using Eq. (\ref{formulaxz}) one can obtain the following formula
\begin{eqnarray}
&&\hspace{-1mm}
{1\over 4}[(\bsi\gamma_\mu\chi)(\bhi\gamma_\nu\gamma_5\psi)+(\bsi\gamma_\mu\gamma_5\chi)(\bhi\gamma_\nu\psi)+\mu\leftrightarrow\nu]
\nonumber\\
&&\hspace{-1mm}
=~
-{1\over 4}\big(\delta_\mu^\alpha\delta_\nu^\beta+\delta_\nu^\alpha\delta_\mu^\beta-g_{\mu\nu}g^{\alpha\beta}\big)
\big[(\bsi\gamma_\alpha\gamma_5\psi)(\bhi\gamma_\beta\chi)
+(\bsi\gamma_\alpha\psi)(\bhi\gamma_\beta\gamma_5\chi)\big]
\label{fierz5sym}
\end{eqnarray}
We need also
\begin{eqnarray}
&&\hspace{-1mm}
{1\over 4}\big[(\bsi\gamma_\mu\chi)(\bhi\gamma_\nu\gamma_5\psi)-(\bsi\gamma_\mu\gamma_5\chi)(\bhi\gamma_\nu\psi)+\mu\leftrightarrow\nu\big]
~=~{g_{\mu\nu}\over 4}(\bsi\psi)(\bhi\gamma_5\chi)
\label{fierz4}\\
&&\hspace{-1mm}
-~{g_{\mu\nu}\over 4}(\bsi\gamma_5\psi)(\bhi\chi)
-{1\over 8}
\big[(\bsi\sigma_{\mu\xi}\psi)(\bhi\sigma_\nu^{~\xi}\gamma_5\chi)-(\bsi\sigma_{\mu\xi}\gamma_5\psi)(\bhi\sigma_\nu^{~\xi}\chi)
+\mu\leftrightarrow\nu\big]
\nonumber
\end{eqnarray}
for symmetric tensors and
\begin{eqnarray}
&&\hspace{-3mm}
{1\over 4}\big[(\bsi\gamma_\mu\chi)(\bhi\gamma_\nu\gamma_5\psi)+(\bsi\gamma_\mu\gamma_5\chi)(\bhi\gamma_\nu\psi)-\mu\leftrightarrow\nu\big]
\label{fierz5}\\
&&\hspace{33mm}
=~
{i\over 4}\epsilon_{\mu\nu\alpha\beta}\big[(\bsi\gamma_\alpha\psi)(\bhi\gamma_\beta\chi)
+(\bsi\gamma_\alpha\gamma_5\psi)(\bhi\gamma_\beta\gamma_5\chi)\big]\nonumber
\end{eqnarray}
\begin{eqnarray}
&&\hspace{-3mm}
{1\over 4}\big[(\bsi\gamma_\mu\chi)(\bhi\gamma_\nu\gamma_5\psi)-(\bsi\gamma_\mu\gamma_5\chi)(\bhi\gamma_\nu\psi)-\mu\leftrightarrow\nu\big]
\label{fierz6}\\
&&\hspace{-3mm}
=~
-{i\over 4}(\bsi\gamma_5\psi)(\bhi\sigma_{\mu\nu}\chi)+{i\over 4}(\bsi\psi)(\bhi\sigma_{\mu\nu}\gamma_5\chi)
+{i\over 4}(\bsi\sigma_{\mu\nu}\gamma_5\psi)(\bhi\chi)-{i\over 4}(\bsi\sigma_{\mu\nu}\psi)(\bhi\gamma_5\chi)
\nonumber
\end{eqnarray}
for the antisymmetric ones.

\subsubsection{Formulas with $\sigma$-matrices \label{sigmaflas}}
It is convenient to define
\footnote{We use conventions from {\it Bjorken \& Drell} where $\epsilon^{0123}=-1$ and
$
\gamma^\mu\gamma^\nu\gamma^\lambda=g^{\mu\nu}\gamma^\lambda +g^{\nu\lambda}\gamma^\mu-g^{\mu\lambda}\gamma^\nu
-i\epsilon^{\mu\nu\lambda\rho}\gamma_\rho\gamma_5
$.
Also, with this convention $\tigma_{\mu\nu}\equiv\half \epsilon_{\mu\nu\lambda\rho}\sigma^{\lambda\rho}=i\sigma_{\mu\nu}\gamma_5$.
}
\begin{equation}
\epsilon_{ij}~\equiv~ {2\over s}\epsilon_{\star\bu ij}~=~ {2\over s}p_2^\mu p_1^\nu\epsilon_{\mu\nu ij}
\label{eps2}
\end{equation}
such that $\epsilon_{12}~=~1$ and $\epsilon_{ij}\epsilon_{kl}~=~g_{ik}g_{jl}-g_{il}g_{jk}$. 
The frequently used formula is 
\begin{equation}
\hspace{-1mm}
\sigma_{\mu\nu}\sigma_{\alpha\beta}~=~(g_{\mu\alpha}g_{\nu\beta}-g_{\mu\beta}g_{\nu\alpha})-i\epsilon_{\mu\nu\alpha\beta}\gamma_5
-i(g_{\mu\alpha}\sigma_{\nu\beta}-g_{\mu\beta}\sigma_{\nu\alpha}-g_{\nu\alpha}\sigma_{\mu\beta}+g_{\nu\beta}\sigma_{\mu\alpha})
\label{sigmasigma}
\end{equation}
with variations
\begin{eqnarray}
&&\hspace{-1mm}
{2\over s}\sigma_{\bu i}\sigma_{\star j}~=~g_{ij}-i\epsilon_{ij}\gamma_5 
-i\sigma_{ij}-{2i\over s}g_{ij}\sigma_{\bu\star},~~~
{2\over s}\sigma_{\star i}\sigma_{\bu j}~=~g_{ij}+i\epsilon_{ij}\gamma_5
-i\sigma_{ij}+{2\over s}g_{ij}\sigma_{\star\bu},
\nonumber\\
&&\hspace{-1mm}
\sigma_{i j}\sigma_{\bu k}=-\sigma_{\bu k}\sigma_{i j}~=-ig_{ik}\sigma_{\bu j}+ig_{jk}\sigma_{\bu i},~~~
\sigma_{i j}\sigma_{\star k}=-\sigma_{\star k}\sigma_{i j}~=-ig_{ik}\sigma_{\star j}+ig_{jk}\sigma_{\star i}  
\nonumber\\
\label{sigmasigmas}
\end{eqnarray}
We need also the following formulas with $\sigma$-matrices in different matrix elements
\begin{eqnarray}
&&\hspace{-1mm}
\tigma_{\mu\nu}\otimes\tigma_{\alpha\beta}~
=~-\half(g_{\mu\alpha}g_{\nu\beta}-g_{\nu\alpha}g_{\mu\beta})\sigma_{\xi\eta}\otimes\sigma^{\xi\eta}
\nonumber\\
&&\hspace{-1mm}
+~g_{\mu\alpha}\sigma_{\beta\xi}\otimes\sigma_\nu^{~\xi}-g_{\nu\alpha}\sigma_{\beta\xi}\otimes\sigma_\mu^{~\xi}-g_{\mu\beta}\sigma_{\alpha\xi}\otimes\sigma_\nu^{~\xi}+g_{\nu\beta}\sigma_{\alpha\xi}\otimes\sigma_\mu^{~\xi}
-\sigma_{\alpha\beta}\otimes\sigma_{\mu\nu}
\label{tigmi}
\end{eqnarray}
and 
\begin{eqnarray}
&&\hspace{-1mm}
\tigma_{\mu\xi}\otimes\tigma_\nu^{~\xi}~=~-{g_{\mu\nu}\over 2}\sigma_{\xi\eta}\otimes\sigma^{\xi\eta}+\sigma_{\nu\xi}\otimes\sigma_\mu^{~\xi}
,~~~~\sigma_{\xi\eta}\otimes\tigma^{\xi\eta}~=~\tigma_{\xi\eta}\otimes\sigma^{\xi\eta}
\label{formulaxz}\\
&&\hspace{-1mm}
\sigma_{\mu\xi}\gamma_5\otimes\sigma_\nu^{~\xi}\gamma_5+\mu\leftrightarrow\nu-{g_{\mu\nu}\over 2}\sigma_{\xi\eta}\gamma_5\otimes\sigma^{\xi\eta}\gamma_5
~=~-[\sigma_{\mu\xi}\otimes\sigma_\nu^{~\xi}+\mu\leftrightarrow\nu-{g_{\mu\nu}\over 2}\sigma_{\xi\eta}\otimes\sigma^{\xi\eta}]
\nonumber
\end{eqnarray}
\begin{eqnarray}
&&\hspace{-1mm}
\tigma_{\star j}\otimes\tigma_{\bu k}~=~-{s\over 4}g_{jk}\sigma_{\xi\eta}\otimes\sigma^{\xi\eta}
+{s\over 2}\sigma_{k\xi}\otimes\sigma_j^{~\xi}+g_{jk}\sigma_{\bu\xi}\otimes\sigma_\star^{~\xi}-\sigma_{\bu k}\otimes\sigma_{\star j}
\nonumber\\
&&\hspace{-1mm}
=~-{s\over 4}g_{jk}\sigma_{mn}\otimes\sigma^{mn}-g_{jk}\sigma_{\star l}\otimes\sigma_\bu^{~l}
+\sigma_{\star k}\otimes\sigma_{\bu j}
+{s\over 2}\sigma_{k l}\otimes\sigma_j^{~l}
\label{gammas13}
\end{eqnarray}
\begin{eqnarray}
&&\hspace{-1mm}
\sigma_{\star i}\otimes\sigma_{\bu j}-\sigma_{\star i}\gamma_5\otimes\sigma_{\bu j}\gamma_5
\label{gammas13a}\\
&&\hspace{-1mm}
=~-g_{ij}^\perp\sigma_{\star l}\otimes\sigma_\bu^{~l}+\sigma_{\star j}\otimes\sigma_{\bu i}+\sigma_{\star i}\otimes\sigma_{\bu j}
-{s\over 4}g_{ij}^\perp\sigma_{mn}\otimes\sigma^{mn}+{s\over 2}\sigma_{j l}\otimes\sigma_i^{~l}
\nonumber
\end{eqnarray}
\begin{eqnarray}
&&\hspace{-1mm}
\sigma_\star^{~k}\otimes\gamma_i\sigma_{\bu k}\gamma_j~=~\hatp_2\gamma^k\otimes\notp_1\gamma_i\gamma_k\gamma_j
~=~\hatp_2\gamma^k\otimes\notp_1(g_{ik}\gamma_j+g_{jk}\gamma_i-g_{ij}\gamma_k)
\nonumber\\
&&\hspace{-1mm}
=~\hatp_2(g_{ik}\gamma_j+g_{jk}\gamma_i-g_{ij}\gamma_k)\otimes \notp_1\gamma^k~=~(\gamma_j\sigma_\star^{~k}\gamma_i)\otimes\sigma_{\bu k}
\label{sisigaga}
\end{eqnarray}
We will need also
\begin{eqnarray}
&&\hspace{-1mm}
\notp_2\otimes\gamma_i\notp_1\gamma_j+\notp_2\gamma_5\otimes\gamma_i\notp_1\gamma_j\gamma_5
~=~\gamma_j\notp_2\gamma_i\otimes\notp_1+\gamma_j\notp_2\gamma_i\gamma_5\otimes\notp_1\gamma_5
\label{flagamma}
\end{eqnarray}
\subsubsection{Formulas with $\gamma$-matrices and one gluon field}
In the gauge $A_\bu=0$ the ``projectile'' field $A_i$ can be represented as 
\begin{equation}
A_i(x_\bu,x_\perp)~=~{2\over s}\!\int_{-\infty}^{x_\bu}\!dx'_\bu~F^{(A)}_{\star i}(x'_\bu,x_\perp)
\label{ai}
\end{equation}
and similarly for the ``target'' field
\begin{equation}
B_i(x_\bu,x_\perp)~=~{2\over s}\!\int_{-\infty}^{x_\star}\!dx'_\star~F^{(B)}_{\star i}(x'_\star,x_\perp)
\label{bi}
\end{equation}
in the $B_\bu=0$ gauge. It is convenient to define
\begin{equation}
\tilde{A}_i(x_\bu,x_\perp)~=~{2\over s}\!\int_{-\infty}^{x_\bu}\!dx'_\bu~\tilde{F}^{(A)}_{\star i}(x'_\bu,x_\perp),
~~~\tilde{B}_i(x_\star,x_\perp)~=~{2\over s}\!\int_{-\infty}^{x_\star}\!dx'_\star~\tilde{F}^{(B)}_{\bu i}(x'_\star,x_\perp),
\label{abefildz}
\end{equation}
 where $\tilF_{\mu\nu}=\half\epsilon_{\mu\nu\lambda\rho}F^{\lambda\rho}$ as usual. 
With this definition we have $\tilA_i=-\epsilon_{ij}A^j$  and $\tilB_i=\epsilon_{ij}B^j$ so 
\begin{equation}
\notp_2\brA_i~=~-\notA\notp_2\gamma_i,~~~~\brA_i\notp_2~=~-\gamma_i\notp_2\notA,~~~~
\notp_1\breB_i~=~-\notB\notp_1\gamma_i,~~~~\breB_i\notp_1~=~-\gamma_i\notp_1\notB
\label{glavla}
\end{equation}
We also used
\begin{eqnarray}
&&\hspace{-1mm}
A^i\notp_2\otimes\gamma_n\notp_1\gamma_i+~A^i\notp_2\gamma_5\otimes\gamma_n\notp_1\gamma_i\gamma_5
~=~-\notp_2\brA_n\otimes \notp_1-\notp_2\brA_n\gamma_5\otimes \notp_1\gamma_5
\nonumber\\
&&\hspace{-1mm}
A^i\notp_2\otimes\gamma_i\notp_1\gamma_n+~A^i\notp_2\gamma_5\otimes\gamma_i\notp_1\gamma_n\gamma_5
~=~-\brA_n\notp_2\otimes \notp_1-\brA_n\notp_2\gamma_5\otimes \notp_1\gamma_5
\nonumber\\
&&\hspace{-1mm}
\gamma_n\slashed{p}_2\gamma^i\otimes\slashed{p}_1 B_i
~+~ \gamma_n\slashed{p}_2\gamma^i\gamma_5 \otimes\slashed{p}_1\gamma_5 B_i
~=~-\slashed{p}_2\otimes \slashed{p}_1\breB_n -
\slashed{p}_2 \gamma_5\otimes \slashed{p}_1\breB_n\gamma_5
\nonumber\\
&&\hspace{-1mm}
\gamma^i\slashed{p}_2\gamma_n \otimes\slashed{p}_1B_i
~+~\gamma^i \slashed{p}_2\gamma_n\gamma_5 \otimes\slashed{p}_1\gamma_5 B_i
~=~-\slashed{p}_2\otimes \breB_n\slashed{p}_1 -
\slashed{p}_2 \gamma_5\otimes \breB_n\slashed{p}_1\gamma_5
\label{gammas1fild}
\end{eqnarray}
and
\begin{eqnarray}
&&\hspace{-3mm}
{2\over s}\big[\notp_1\notp_2\gamma_i\otimes B^i\gamma_n
+\notp_1\notp_2\gamma_i\gamma_5\otimes B^i\gamma_n\gamma_5\big]
~=~\gamma_i\otimes\gamma_n\breB_i+\gamma_i\gamma_5\otimes\gamma_n\breB_i\gamma_5
\nonumber\\
&&\hspace{-3mm}
{2\over s}\big[\gamma_i\notp_2\notp_1\otimes B^i\gamma_n
+\gamma_i\notp_2\notp_1\gamma_5\otimes B^i\gamma_n\gamma_5\big]
~=~\gamma_i\otimes\breB_i\gamma_n+\gamma_i\gamma_5\otimes\breB_i\gamma_n\gamma_5
\nonumber\\
&&\hspace{-3mm}
{2\over s}(\hatp_2\gamma^i\hatp_1\otimes \hatp_1 B_i+\hatp_2\gamma^i\hatp_1\gamma_5\otimes \hatp_1\gamma_5 B_i)
= -\gamma^i\otimes \breB_i\hatp_1-\gamma^i\gamma_5\otimes \breB_i\hatp_1\gamma_5
\nonumber\\
&&\hspace{-3mm}
\gamma_k\gamma^i\hatp_2\otimes B_i\gamma^k
+ \gamma_k\gamma^i\hatp_2\gamma_5\otimes B_i\gamma^k\gamma_5
~=~\hatp_2\otimes \gamma^i\breB_i
+\hatp_2\gamma_5\otimes\gamma_i\breB^i\gamma_5,
\nonumber\\
&&\hspace{-3mm}
 \slashed{p}_2 \gamma^j\gamma^i\otimes\slashed{p}_1 B_j
~+~ \slashed{p}_2\gamma_5 \gamma^i\gamma^j\otimes\slashed{p}_1\gamma_5 B_j
=\slashed{p}_2\otimes \slashed{p}_1\breB_i +
\slashed{p}_2\gamma_5 \otimes \slashed{p}_1\gamma_5\breB^i,
\nonumber\\
&&\hspace{-3mm}
 \slashed{p}_2 \gamma^i\gamma^j\otimes\slashed{p}_1 B_j
~+~ \slashed{p}_2\gamma_5 \gamma^i\gamma^j\otimes\slashed{p}_1\gamma_5 B_j
=\slashed{p}_2\otimes \slashed{p}_1\breB^i + \slashed{p}_2\gamma_5\otimes \slashed{p}_1\gamma_5\breB^i.
\label{gammas17}
\end{eqnarray}

\subsubsection{Formulas with $\gamma$-matrices and two gluon fields}

With definition (\ref{abefildz}), we have the following formulas
\begin{eqnarray}
&&\hspace{-1mm}
A_i\otimes \tilB_j=g_{ij}\tilA_k\otimes B^k-\tilA_j\otimes B_i,~~~\tilde{A}_i\otimes B_j=g_{ij}A_k\otimes \tilB^k-A_j\otimes \tilB_i
\label{abeznaki}\\
&&\hspace{-1mm}
\tilde{A}_i\otimes \tilB_j=-g_{ij}A_k\otimes B^k+A_j\otimes B_i, ~~~\Rightarrow~~~\tilde{A}_i\otimes\tilde{B}^i=-A_i\otimes B^i,~~\tilde{A}_i\otimes B^i= A_i\otimes\tilde{B}^i
\nonumber
\end{eqnarray}
In addition, it is convenient to define
\begin{equation}
\brA_i\equiv A_i-i\tilA_i\gamma_5,~~~~~~\breB_i\equiv B_i-i\tilB_i\gamma_5.
\end{equation}
Using these formulas, after some algebra one obtains
\begin{eqnarray}
&&\hspace{-2mm}
\gamma_m\notp_2\gamma_jA^i\otimes\gamma_n\notp_1\gamma_iB^j
+\gamma_m\notp_2\gamma_jA^i\gamma_5\otimes\gamma_n\notp_1\gamma_iB^j\gamma_5
=\notp_2\brA_n\otimes\notp_1\breB_m+\notp_2\brA_n\gamma_5\otimes\notp_1\breB_m\gamma_5
\nonumber\\
&&\hspace{-2mm}
\gamma_j\notp_2\gamma_mA^i\otimes\gamma_n\notp_1\gamma_iB^j
+\gamma_j\notp_2\gamma_mA^i\gamma_5\otimes\gamma_n\notp_1\gamma_iB^j\gamma_5
=\notp_2\brA_n\otimes\breB_m\notp_1+\notp_2\brA_n\gamma_5\otimes\breB_m\notp_1\gamma_5
\nonumber\\
&&\hspace{-2mm}
\gamma_m\notp_2\gamma_jA^i\otimes\gamma_i\notp_1\gamma_nB^j
+\gamma_m\notp_2\gamma_jA^i\gamma_5\otimes\gamma_i\notp_1\gamma_nB^j\gamma_5
=\brA_n\notp_2\otimes\notp_1\breB_m+\brA_n\notp_2\gamma_5\otimes\notp_1\breB_m\gamma_5
\nonumber\\
&&\hspace{-2mm}
\gamma_j\notp_2\gamma_mA^i\otimes\gamma_i\notp_1\gamma_nB^j
+\gamma_j\notp_2\gamma_mA^i\gamma_5\otimes\gamma_i\notp_1\gamma_nB^j\gamma_5
=\brA_n\notp_2\otimes\breB_m\notp_1+\brA_n\notp_2\gamma_5\otimes\breB_m\notp_1\gamma_5
\nonumber\\
\label{gammas7}
\end{eqnarray}
and
\begin{eqnarray}
&&\hspace{-1mm}
\notp_2\brA_m\otimes\notp_1\breB_n+\notp_2\brA_n\gamma_5\otimes\notp_1\breB_m\gamma_5~=~g_{mn}\notp_2\brA_k\otimes\notp_1\breB^k
\nonumber\\
&&\hspace{-1mm}
\notp_2\brA_m\otimes\breB_n\notp_1+\notp_2\brA_n\gamma_5\otimes\gamma_5\breB_m\notp_1~=~g_{mn}\notp_2\brA_k\otimes\breB^k\notp_1
\nonumber\\
&&\hspace{-1mm}
\brA_m\notp_2\otimes\notp_1\breB_n+\gamma_5\brA_n\notp_2\otimes\notp_1\breB_m\gamma_5~=~g_{mn}\brA_k\notp_2\otimes\notp_1\breB^k
\nonumber\\
&&\hspace{-1mm}
\brA_m\notp_2\otimes\breB_n\notp_1+\gamma_5\brA_n\notp_2\otimes\gamma_5\breB_m\notp_1~=~g_{mn}\brA_k\notp_2\otimes\breB^k\notp_1
\label{gammas6}
\end{eqnarray}
The corollary of Eq. (\ref{gammas6}) is
\begin{eqnarray}
&&\hspace{-11mm}
\notp_2\brA_k\gamma_5\otimes\notp_1\breB^k\gamma_5~=~\notp_2\brA_k\otimes\notp_1\breB^k,~~~~~~~~
\notp_2\brA_k\gamma_5\otimes\gamma_5\breB^k\notp_1~=~\notp_2\brA_k\otimes\breB^k\notp_1
\nonumber\\
&&\hspace{-11mm}
\gamma_5\brA_k\notp_2\otimes\notp_1\breB^k\gamma_5~=~\brA_k\notp_2\otimes\notp_1\breB^k,~~~~~~~~
\gamma_5\brA_k\notp_2\otimes\gamma_5\breB^k\notp_1~=~\brA_k\notp_2\otimes\breB^k\notp_1
\label{gammas6a}
\end{eqnarray}

From Eqs. (\ref{gammas7}) and (\ref{gammas6}) one easily obtains
\begin{equation}
\gamma_m\notp_2\gamma_jA^i\otimes\gamma_n\notp_1\gamma_iB^j
+\gamma_m\notp_2\gamma_jA^i\gamma_5\otimes\gamma_n\notp_1\gamma_iB^j\gamma_5
~+~m\leftrightarrow n~=~2g_{mn}\notp_2\brA_k\otimes\notp_1\breB^k
\label{formula9}
\end{equation}
and
\begin{eqnarray}
&&\hspace{-1mm}
\gamma_m\notp_2\gamma_jA^i\otimes\gamma_n\notp_1\gamma_iB^j
+\gamma_m\notp_2\gamma_jA^i\gamma_5\otimes\gamma_n\notp_1\gamma_iB^j\gamma_5
~-~m\leftrightarrow n~
\nonumber\\
&&\hspace{-1mm}
=~2\notp_2\brA_n\otimes\notp_1\breB_m~-~m\leftrightarrow n,
\nonumber\\
&&\hspace{-1mm}
\gamma_j\notp_2\gamma_mA^i\otimes\gamma_i\notp_1\gamma_nB^j
+\gamma_j\notp_2\gamma_mA^i\gamma_5\otimes\gamma_i\notp_1\gamma_nB^j\gamma_5
~-~m\leftrightarrow n~
\nonumber\\
&&\hspace{-1mm}
=~2\brA_n\notp_2\otimes\breB_m\notp_1~-~m\leftrightarrow n
\label{formula9a}
\end{eqnarray}
We need also the formula
\begin{eqnarray}
&&\hspace{-1mm}
{4\over s^2}A^i\notp_1\notp_2\gamma_j\otimes B^j\notp_1\notp_2\gamma_i
\nonumber\\
&&\hspace{-1mm}
=~
A^i\gamma_j\otimes B^j\gamma_i-iA^i\gamma_j\gamma_5\otimes \tilB^j\gamma_i
+i\tilA^i \gamma_j\otimes  B^j\gamma_i\gamma_5+\tilA^i\gamma_j\gamma_5\otimes \tilB^j\gamma_i\gamma_5
\label{fla1010}
\end{eqnarray}
and 
\begin{eqnarray}
&&\hspace{-1mm}
{4\over s^2}\big(A^i\notp_1\notp_2\gamma_j\otimes B^j\notp_1\notp_2\gamma_i
+A^i\notp_1\notp_2\gamma_j\gamma_5\otimes B^j\notp_1\notp_2\gamma_i\gamma_5\big)
\nonumber\\
&&\hspace{-1mm}
=~\gamma^j\brA_i\otimes\gamma^i\breB_j+\gamma^j\brA_i\gamma_5\otimes\gamma^i\breB_j\gamma_5
\label{gammas10}
\end{eqnarray}
$$
\gamma_i\brA_j\gamma_5\otimes\gamma_j\brA_i\gamma_5~=~\gamma_i\brA_j\otimes\gamma^i\breB^j-\gamma_i\brA^i\otimes\gamma_j\breB^j
$$
\begin{eqnarray}
&&\hspace{-1mm}
A_k\gamma_i\slashed{p}_2\gamma^j\otimes B_j\gamma^i\slashed{p}_1\gamma^k~=~
\slashed{p}_2\brA_i\otimes \slashed{p}_1\breB^i
~=~\notA\notp_2\gamma_i\otimes\notB\notp_1\gamma^i,
\nonumber\\
&&\hspace{-1mm}
A_k\gamma^j\slashed{p}_2\gamma_i\otimes B_j\gamma^k\slashed{p}_1\gamma^i~=~
\brA_i\slashed{p}_2\otimes \breB_i\slashed{p}_1
~=~\gamma_i\notp_2\notA\otimes\gamma^i\notp_1\notB,
\nonumber\\
&&\hspace{-1mm}
A_k\gamma_i\slashed{p}_2\gamma_j\otimes B^j\gamma^k\slashed{p}_1\gamma^i
~=~\slashed{p}_2\brA_i\otimes\breB^i\slashed{p}_1
~=~~\notA\notp_2\gamma_i\otimes\gamma^i\notp_1\notB,
\nonumber\\
&&\hspace{-1mm}
A_k\gamma_j\slashed{p}_2\gamma_i\otimes B^j\gamma^i\slashed{p}_1\gamma^k
~=~\brA_i\slashed{p}_2\otimes\slashed{p}_1\breB^i
~=~\gamma_i\notp_2\notA\otimes\notB\notp_1\gamma^i.
\label{gammas11}
\end{eqnarray}
We used also
\begin{eqnarray}
&&\hspace{-1mm}
A^k\gamma_m\slashed{p}_2\gamma_j\otimes B^j\gamma_n\slashed{p}_1\gamma_k +m\leftrightarrow n
-g_{mn}A^k\gamma_i\slashed{p}_2\gamma_j\otimes B^j\gamma^i\slashed{p}_1\gamma_k
\nonumber\\
&&\hspace{-1mm}
=~\brA_m\notp_2\otimes\breB_n\notp_1+m\leftrightarrow n-g_{mn}\brA_k\notp_2\otimes\breB^k\notp_1,
\nonumber\\
&&\hspace{-1mm}
A^k\gamma_j\slashed{p}_2\gamma_m\otimes B^j\gamma_n\slashed{p}_1\gamma_k+m\leftrightarrow n
-g_{mn}A^k\gamma_j\slashed{p}_2\gamma_i\otimes B^j\gamma_k\slashed{p}_1\gamma^i
~=~
\nonumber\\
&&\hspace{-1mm}
=~\notp_2\brA_m\otimes\breB_n\notp_1+m\leftrightarrow n-g_{mn}\notp_2\brA_k\otimes\breB^k\notp_1
\nonumber\\
&&\hspace{-1mm}
A^k\gamma_j\slashed{p}_2\gamma_m\otimes B^j\gamma_k\slashed{p}_1\gamma_n+m\leftrightarrow n
-g_{mn}A^k\gamma_j\slashed{p}_2\gamma_i\otimes B^j\gamma_k\slashed{p}_1\gamma^i
\nonumber\\
&&\hspace{-1mm}
=~\notp_2\brA_m\otimes\notp_1\breB_n+m\leftrightarrow n-g_{mn}\notp_2\brA_k\otimes\notp_1\breB^k,
\nonumber\\
&&\hspace{-1mm}
A^k\gamma_m\slashed{p}_2\gamma_j\otimes B^j\gamma_k\slashed{p}_1\gamma_n+m\leftrightarrow n
-g_{mn}A^k\gamma_j\slashed{p}_2\gamma_i\otimes B^j\gamma_k\slashed{p}_1\gamma^i
\nonumber\\
&&\hspace{-1mm}
=~\brA_m\notp_2\otimes\notp_1\breB_n+m\leftrightarrow n-g_{mn}\brA_k\notp_2\otimes\notp_1\breB^k,
\label{formulas67}
\end{eqnarray}
and
\begin{eqnarray}
&&\hspace{-1mm}
\gamma^j\slashed{p}_2\gamma_m A^k \otimes \gamma^k\slashed{p}_1\gamma_n B^j-m\leftrightarrow n
~=~-\brA_m\slashed{p}_2\otimes\breB_n\slashed{p}_1-m\leftrightarrow n
\nonumber\\
&&\hspace{-1mm}
\gamma_m\slashed{p}_2\gamma^j A^k \otimes \gamma^k\slashed{p}_1\gamma_n B^j-m\leftrightarrow n
~=~-\slashed{p}_2\brA_m\otimes\breB_n\slashed{p}_1-m\leftrightarrow n
\nonumber\\
&&\hspace{-1mm}
\gamma^j\slashed{p}_2\gamma_mA^k \otimes \gamma_n\slashed{p}_1\gamma^k B^j-m\leftrightarrow n
~=~-\brA_m\slashed{p}_2\otimes\slashed{p}_1\breB_n-m\leftrightarrow n
\nonumber\\
&&\hspace{-1mm}
\gamma_m\slashed{p}_2\gamma^j A^k \otimes \gamma_n\slashed{p}_1\gamma^k B^j-m\leftrightarrow n
~=~-\slashed{p}_2\brA_m\otimes\slashed{p}_1\breB_n-m\leftrightarrow n
\label{A1.33}
\end{eqnarray}

Next, using formula (\ref{gammas13a}) we get
\begin{eqnarray}
&&\hspace{-1mm}
A^i\sigma_{\star j}\otimes B^j\sigma_{\bu i}-A^i\sigma_{\star j}\gamma_5\otimes B^j\sigma_{\bu i}\gamma_5
\label{gammas14}\\
&&\hspace{-1mm}
=~(\barA^j\notp_2\gamma_k-\barA^k\notp_2\gamma_j)\otimes \barB^j\notp_1\gamma_k
-\bar\notA\notp_2\otimes \bar\notB\notp_1-{s\over 2}\bar\notA\gamma_i\otimes \bar\notB\gamma^i
-{s\over 4}\barA^i\sigma_{jk}\otimes \barB_i\sigma^{jk}
\nonumber
\end{eqnarray}
We frequently use this formula in matrix elements like
$\langle \bsi A^i(x)\sigma_{\star j}{1\over \alpha}\psi(0)\rangle_A
\langle\bsi B^j(0)\sigma_{\bu i}{1\over\beta}\psi(x)\rangle_B$. It is easy to see that in such matrix elements
the contribution of two last terms in Eq. (\ref{gammas14}) is $O\big({q_\perp^2\over s}\big)$ in comparison to the first two ones
so for our purposes
\begin{equation}
\hspace{-0mm}
A^i\sigma_{\star j}\otimes B^j\sigma_{\bu i}-A^i\sigma_{\star j}\gamma_5\otimes B^j\sigma_{\bu i}\gamma_5
\simeq~(\barA^j\notp_2\gamma_k-\barA^k\notp_2\gamma_j)\otimes \barB^j\notp_1\gamma_k
-\bar\notA\notp_2\otimes \bar\notB\notp_1
\label{gammas14a}
\end{equation}

We also need
\begin{eqnarray}
&&\hspace{-1mm}
{4\over s^2}\big[A_i\notp_1\notp_2\gamma_j\otimes B_j\notp_2\notp_1\gamma_i
+A_i\notp_1\notp_2\gamma_j\gamma_5\otimes B_j\notp_2\notp_1\gamma_i\gamma_5\big]~
\nonumber\\
&&\hspace{33mm}
=~\brA_i\gamma_j\otimes \breB_j\gamma_i+\brA_i\gamma_j\gamma_5\otimes \breB_j\gamma_i\gamma_5
\label{gammas15}
\end{eqnarray}
and
\begin{eqnarray}
&&\hspace{-1mm}
{2\over s}\big[A_i\notp_1\notp_2\gamma^j\otimes B_j\gamma_n\notp_1\gamma^i
+A_i\notp_1\notp_2\gamma^j\gamma_5\otimes B_j\gamma_{\nu_\perp}\notp_1\gamma^i\gamma_5\big]
\label{gammas16}\\
&&\hspace{-1mm}
=~-\gamma_i\brA_n\otimes \notp_1\breB^i - \gamma_i\brA_n\gamma_5\otimes \notp_1\breB^i\gamma_5
~=~\gamma_i\brA_n\otimes\notB\notp_1\gamma^i+\gamma_i\brA_n\gamma_5\otimes\notB\notp_1\gamma^i\gamma_5,
\nonumber\\
&&\hspace{-1mm} 
{2\over s}\big[A_i\gamma_n\notp_2\gamma^j\otimes B_j\notp_2\notp_1\gamma^i
+A_i\gamma_n\notp_2\gamma^j\gamma_5\otimes B_j\notp_2\notp_1\gamma^i\gamma_5\big]
\nonumber\\
&&\hspace{-1mm} 
=-\notp_2\brA_i\otimes\gamma^i\breB_n-\notp_2\brA_i\gamma_5\otimes\gamma^i\breB_n\gamma_5 
~=~\notA\notp_2\gamma_i\otimes\gamma^i\breB_n+\notA\notp_2\gamma_i\gamma_5\otimes\gamma^i\breB_n\gamma_5. 
\nonumber
\end{eqnarray}

The last formula which we need is
\begin{eqnarray}
&&\hspace{-2mm}
\gamma_m\notp_2\gamma_jA^i\otimes\gamma_n\notp_1\gamma_iB^j
+\gamma_m\notp_2\gamma_jA^i\gamma_5\otimes\gamma_n\notp_1\gamma_iB^j\gamma_5
=\notp_2\brA_n\otimes\notp_1\breB_m+\notp_2\brA_n\gamma_5\otimes\notp_1\breB_m\gamma_5
\nonumber\\
&&\hspace{-2mm}
\gamma_j\notp_2\gamma_mA^i\otimes\gamma_n\notp_1\gamma_iB^j
+\gamma_j\notp_2\gamma_mA^i\gamma_5\otimes\gamma_n\notp_1\gamma_iB^j\gamma_5
=\notp_2\brA_n\otimes\breB_m\notp_1+\notp_2\brA_n\gamma_5\otimes\breB_m\notp_1\gamma_5
\nonumber\\
&&\hspace{-2mm}
\gamma_m\notp_2\gamma_jA^i\otimes\gamma_i\notp_1\gamma_nB^j
+\gamma_m\notp_2\gamma_jA^i\gamma_5\otimes\gamma_i\notp_1\gamma_nB^j\gamma_5
=\brA_n\notp_2\otimes\notp_1\breB_m+\brA_n\notp_2\gamma_5\otimes\notp_1\breB_m\gamma_5
\nonumber\\
&&\hspace{-2mm}
\gamma_j\notp_2\gamma_mA^i\otimes\gamma_i\notp_1\gamma_nB^j
+\gamma_j\notp_2\gamma_mA^i\gamma_5\otimes\gamma_i\notp_1\gamma_nB^j\gamma_5
=\brA_n\notp_2\otimes\breB_m\notp_1+\brA_n\notp_2\gamma_5\otimes\breB_m\notp_1\gamma_5
\nonumber\\
\label{gammas17}
\end{eqnarray}

 \subsection{Parametrization of leading-twist matrix elements \label{sec:paramlt}}
 Let us  first consider matrix elements of operators without $\gamma_5$. The standard parametrization of quark TMDs reads
\begin{eqnarray}
&&\hspace{-1mm}
{1\over 16\pi^3}\!\int\!dx_\bu d^2x_\perp~e^{-i\alpha x_\bu+i(k,x)_\perp}
~\langle \hbsi_f(x_\bu,x_\perp)\gamma^\mu\hsi_f(0)\rangle_A
\label{Amael}\\
&&\hspace{27mm}
=~p_1^\mu f_1^f(\alpha,k_\perp)
+k_\perp^\mu f_\perp^f(\alpha,k_\perp)+p_2^\mu{2m^2_N\over s}f_3^f(\alpha,k_\perp),
\nonumber\\
&&\hspace{-1mm}
{1\over 16\pi^3}\!\int\!dx_\bu d^2x_\perp~e^{-i\alpha x_\bu+i(k,x)_\perp}
~\langle \hbsi_f(x_\bu,x_\perp)\hsi_f(0)\rangle_A
~=~m_Ne^f(\alpha,k_\perp)
\nonumber
\end{eqnarray}
for quark distributions in the projectile and 
\begin{eqnarray}
&&\hspace{-1mm}
{1\over 16\pi^3}\!\int\!dx_\bu d^2x_\perp~e^{-i\alpha x_\bu+i(k,x)_\perp}
~\langle \hbsi_f(0)\gamma^\mu\hsi_f(x_\bu,x_\perp)\rangle_A
\label{baramael}\\
&&\hspace{27mm}
=~-p_1^\mu \barf_1^f(\alpha,k_\perp)
-k_\perp^\mu\barf _\perp^f(\alpha,k_\perp)-p_2^\mu{2m^2_N\over s}\barf_3^f(\alpha,k_\perp),
\nonumber\\
&&\hspace{-1mm}
{1\over 16\pi^3}\!\int\!dx_\bu d^2x_\perp~e^{-i\alpha x_\bu+i(k,x)_\perp}
~\langle \hbsi_f(0)\hsi_f(x_\bu,x_\perp)\rangle_A
~=~
m_N\bare^f(\alpha,k_\perp)
\nonumber
\end{eqnarray}
for the antiquark distributions. 
\footnote{
In a general gauge for
projectile and target fields these matrix elements read 
\begin{eqnarray}
&&\hspace{-2mm}
\langle\hsi_{f}(x)\gamma_\mu \hsi_{f}(0)\rangle_A
~=~\langle\hsi_f(x_\bu,x_\perp)\gamma_\mu[x_\bu,-\infty_\bu]_x[x_\perp,0_\perp]_{-\infty_\bu}[-\infty_\bu,0_\bu]_0\hsi_f(0)\rangle_A,
\nonumber\\
&&\hspace{-2mm}
\langle\hsi_{f}(x)\gamma_\mu \hsi_{f}(0)\rangle_B
~=~\langle\hsi_f(x_\ast,x_\perp)\gamma_\mu[x_\ast,-\infty_\ast]_x[x_\perp,0_\perp]_{-\infty_\ast}[-\infty_\ast,0_\ast]_0\hsi_f(0)\rangle_B
\label{gaugelinks}
\end{eqnarray}
and similarly for other operators.
}

The corresponding matrix elements for the target are obtained by trivial replacements $p_1\leftrightarrow p_2$, $x_\bu\leftrightarrow x_\star$
and $\alpha\leftrightarrow\beta$:
\begin{eqnarray}
&&\hspace{-1mm}
{1\over 16\pi^3}\!\int\!dx_\star d^2x_\perp~e^{-i\beta x_\star+i(k,x)_\perp}
~\langle\hbsi_f(x_\star,x_\perp)\gamma^\mu\hsi_f(0)\rangle_B
\label{Bmael}\\
&&\hspace{27mm}
=~p_2^\mu f_1^f(\beta,k_\perp)+k_\perp^\mu f_\perp^f(\beta,k_\perp)
+p_1^\mu{2m^2_N\over s}f_3^f(\beta,k_\perp),
\nonumber\\
&&\hspace{-1mm}
{1\over 16\pi^3}\!\int\!dx_\star d^2x_\perp~e^{-i\beta x_\star+i(k,x)_\perp}
~\langle\hbsi_f(x_\star,x_\perp)\hsi_f(0)\rangle_B
~=~m_Ne^f(\beta,k_\perp),
\nonumber
\end{eqnarray}
and
\begin{eqnarray}
&&\hspace{-1mm}
{1\over 16\pi^3}\!\int\!dx_\star d^2x_\perp~e^{-i\beta x_\star+i(k,x)_\perp}
~\langle\hbsi_f(0)\gamma^\mu\hsi_f(x_\star,x_\perp)\rangle_B
\label{barbmael}\\
&&\hspace{27mm}
=~-p_2^\mu \barf_1^f(\beta,k_\perp)
-k_\perp^\mu\barf _\perp^f(\beta,k_\perp)-p_1^\mu{2m^2_N\over s}\barf_3^f(\beta,k_\perp),
\nonumber\\
&&\hspace{-1mm}
{1\over 16\pi^3}\!\int\!dx_\star d^2x_\perp~e^{-i\beta x_\star+i(k,x)_\perp}
~\langle\hbsi_f(0)\hsi_f(x_\star,x_\perp)\rangle_B
~=~
m_N\bare^f(\beta,k_\perp).
\nonumber
\end{eqnarray}

Matrix elements of operators with $\gamma_5$ are parametrized as follows: 
\begin{eqnarray}
&&\hspace{-1mm}
{1\over 16\pi^3}\!\int\!dx_\bu d^2x_\perp~e^{-i\alpha x_\bu+i(k,x)_\perp}
~\langle \hbsi_f(x_\bu,x_\perp)\gamma^\mu\gamma_5\hsi_f(0)\rangle_A
~=~-i\epsilon_{\mu_\perp i}k^ig^\perp_f(\alpha,k_\perp),
\nonumber\\
&&\hspace{-1mm}
{1\over 16\pi^3}\!\int\!dx_\bu d^2x_\perp~e^{-i\alpha x_\bu+i(k,x)_\perp}
~\langle \hbsi_f(0)\gamma^\mu\gamma_5\hsi_f(x_\bu,x_\perp)\rangle_A
~=~-i\epsilon_{\mu_\perp i}k^i\barg^\perp_f(\alpha,k_\perp)
\nonumber\\
\label{mael5}
\end{eqnarray}
The corresponding matrix elements for the target are obtained by trivial replacements $p_1\leftrightarrow p_2$, $x_\bu\leftrightarrow x_\star$
and $\alpha\leftrightarrow\beta$ similarly to eq. (\ref{barbmael}).

The parametrization of time-odd Boer-Mulders TMDs are
\begin{eqnarray}
&&\hspace{-1mm}
{1\over 16\pi^3}\!\int\!dx_\bu d^2x_\perp~e^{-i\alpha x_\bu+i(k,x)_\perp}
~\langle \hbsi_f(x_\bu,x_\perp)\sigma^{\mu \nu}\hsi_f(0)\rangle_A
\nonumber\\
&&\hspace{11mm}
=~{1\over m_N}(k_\perp^\mu p_1^\nu -\mu\leftrightarrow\nu)h_{1f}^\perp(\alpha,k_\perp)
+{2m_N\over s}(p_1^\mu p_2^\nu-\mu\leftrightarrow\nu)h_{f}(\alpha,k_\perp)
\nonumber\\
&&\hspace{33mm}
+~{2m_N\over s}(k_\perp^\mu p_2^\nu -\mu\leftrightarrow\nu)h_{3f}^\perp(\alpha,k_\perp),
\nonumber\\
&&\hspace{-1mm}
{1\over 16\pi^3}\!\int\!dx_\bu d^2x_\perp~e^{-i\alpha x_\bu+i(k,x)_\perp}
~\langle \hbsi_f(0)\sigma^{\mu \nu}\hsi_f(x_\bu,x_\perp)\rangle_A
\nonumber\\
&&\hspace{11mm}
=~-{1\over m_N}(k_\perp^\mu p_1^\nu -\mu\leftrightarrow\nu)\barh_{1f}^\perp(\alpha,k_\perp)
-{2m_N\over s}(p_1^\mu p_2^\nu-\mu\leftrightarrow\nu)\barh_{f}(\alpha,k_\perp)
\nonumber\\
&&\hspace{33mm}
-~{2m_N\over s}(k_\perp^\mu p_2^\nu -\mu\leftrightarrow\nu)\barh_{3f}^\perp(\alpha,k_\perp)
\label{hmael}
\end{eqnarray}
and similarly for the target with usual replacements   $p_1\leftrightarrow p_2$, $x_\bu\leftrightarrow x_\star$
and $\alpha\leftrightarrow\beta$.

Note that
the coefficients in front of $f_3$,  $g^\perp_f$, $h$ and $h_3^\perp$ in eqs. (\ref{Amael}),  (\ref{Bmael}), (\ref{mael5}),  and  (\ref{hmael}) 
 contain an extra ${1\over s}$ since $p_2^\mu$ enters only through the direction
of gauge link so the result should not depend on rescaling $p_2\rightarrow\lambda p_2$. For this reason,  these functions do not contribute to $W(q)$ in our approximation.

Last but not least, an important point in our analysis is that any $f(x,k_\perp)$ may have only logarithmic dependence on Bjorken $x$ but
not the power dependence $\sim{1\over x}$. Indeed, at small $x$ the cutoff of corresponding longitudinal integrals comes from 
the rapidity cutoff $\sigma_a$, see footnote \ref{foot:smallx} and corresponding discussion in Ref. \cite{Balitsky:2020jzt}. Thus, at small $x$ one can safely put $x=0$ and the corresponding 
logarithmic contributions would be proportional to powers of $\alpha_s\ln\sigma_a$ (or, in some cases, $\alpha_s\ln^2\sigma_a$, see e.g. 
ref. \cite{Kovchegov:2015zha}). 

\subsection{Matrix elements of quark-quark-gluon operators \label{sec:qqgparam}}

 In this section we will demonstrate that matrix elements of quark-antiquark-gluon operators 
 from section \ref{sec:foton} can be expressed in terms of leading-power
 matrix elements from section \ref{sec:paramlt}. First, let us note that operators ${1\over\alpha}$ and ${1\over\beta}$ in Eqs. (\ref{3.25}) are
 replaced by $\pm{1\over\alpha_q}$ and $\pm{1\over\beta_q}$ in forward matrix elements. Indeed
\begin{eqnarray}
&&\hspace{-1mm}
\!\int\!dx_\bu ~e^{-i\alpha_q x_\bu}\langle\bar\Phi(x_\bu,x_\perp)\Gamma{1\over \alpha+\ie}\psi(0)\rangle_A
\label{maelqg1}\\
&&\hspace{-1mm}
=~-i\!\int\!dx_\bu \!\int_{-\infty}^0 \!\!\!dx'_\bu~e^{-i\alpha_q x_\bu}\langle\bar\Phi(x_\bu,x_\perp)\Gamma\psi(x'_\bu,0_\perp)\rangle_A
={1\over\alpha_q}\!\int\!dx_\bu ~e^{-i\alpha x_\bu}\langle\bar\Phi(x_\bu,x_\perp)\Gamma\psi(0)\rangle_A
\nonumber
\end{eqnarray}
where $\bar\Phi(x_\bu,x_\perp)$ can be $\bsi(x_\bu,x_\perp)$ or  $\bsi(x_\bu,x_\perp)A_i(x_\bu,x_\perp)$ and $\Gamma$ can be any $\gamma$-matrix.
Similarly,
\begin{eqnarray}
&&\hspace{-1mm}
\!\int\!dx_\bu ~e^{-i\alpha_q x_\bu}\langle\big(\bar\psi(x_\bu,x_\perp){1\over \alpha-\ie}\big)\Gamma\Phi(0)\rangle_A
=~{1\over\alpha_q}\!\int\!dx_\bu ~e^{-i\alpha x_\bu}\langle\bar\psi(x_\bu,x_\perp)\Gamma\Phi(0)\rangle_A
\label{maelqg2}\\
&&\hspace{-1mm}
\!\int\!dx_\bu ~e^{-i\alpha_q x_\bu}\langle\big(\bar\psi{1\over \alpha-\ie}\big)(x_\bu,x_\perp)\Gamma{1\over \alpha+\ie}\psi(0)\rangle_A
=~{1\over\alpha_q^2}\!\int\!dx_\bu ~e^{-i\alpha_q x_\bu}\langle\bar\psi(x_\bu,x_\perp)\Gamma\psi(0)\rangle_A
\nonumber
\end{eqnarray}
where $\Phi(x_\bu,x_\perp)$ can be $\psi(x_\bu,x_\perp)$ or  $A_i(x_\bu,x_\perp)\psi(x_\bu,x_\perp)$. We need also
\begin{eqnarray}
&&\hspace{-7mm}
\!\int\!dx_\bu ~e^{-i\alpha_q x_\bu}\langle\big(\bar\psi(0){1\over \alpha-\ie}\big)\Gamma\Phi(x_\bu,x_\perp)\rangle_A
=~-{1\over\alpha_q}\!\int\!dx_\bu ~e^{-i\alpha x_\bu}\langle\bar\psi()\Gamma\Phi(x_\bu,x_\perp)\rangle_A
\nonumber\\
&&\hspace{-7mm}
\!\int\!dx_\bu ~e^{-i\alpha_q x_\bu}\langle\bar\Phi(0)\Gamma{1\over \alpha+\ie}\psi(x_\bu,x_\perp)\rangle_A
=~-{1\over\alpha_q}\!\int\!dx_\bu ~e^{-i\alpha x_\bu}\langle\bar\Phi(0)\Gamma\psi(x_\bu,x_\perp)\rangle_A
\label{maelqg3}
\end{eqnarray}
The corresponding formulas for target matrix elements are obtained by substitution $\alpha\leftrightarrow\beta$ (and $x_\bu\leftrightarrow x_\star$).

Next, we will use QCD equation of motion to reduce quark-quark-gluon TMDs to leading-twist TMDs (cf. Ref. \cite{Mulders:1995dh}).
Let us start with matrix element 
\begin{eqnarray}
&&\hspace{-1mm}
\!\int\! dx_\bu dx_\perp~e^{-i\alpha_qx_\bu+i(k,x)_\perp}\langle \bar\psi(x_\bu,x_\perp)\slashed{p}_2\brA_i(x_\bu,x_\perp)\psi(0)\rangle_A
\label{tw3mael1}\\
&&\hspace{-1mm}
=~-\!\int\! dx_\bu dx_\perp~e^{-i\alpha_qx_\bu+i(k,x)_\perp}\langle \bar\psi(x_\bu,x_\perp)\notA(x_\bu,x_\perp)\slashed{p}_2\gamma_i\psi(0)\rangle_A
\nonumber\\
&&\hspace{-1mm}
=~\int\! dx_\bu dx_\perp~e^{-i\alpha_qx_\bu+i(k,x)_\perp}
\nonumber\\
&&\hspace{31mm}
\times~\big[\langle \hbsi(x_\bu,x_\perp)\slashed{k}_\perp\slashed{p}_2\gamma_i\hsi(0)\rangle_A
+i
\langle \hbsi(x_\bu,x_\perp)\!\stackrel{\leftarrow}{\notD}_\perp\!\gamma_j\slashed{p}_2\gamma_i\hsi(0)\rangle_A\big].
\nonumber
\end{eqnarray}
Using QCD equations of motion we can rewrite the r.h.s. of eq. (\ref{tw3mael1}) as
\begin{eqnarray}
&&\hspace{-2mm}
\int\! dx_\bu dx_\perp~e^{-i\alpha_qx_\bu+i(k,x)_\perp}\big[\langle \hbsi(x_\bu,x_\perp)\slashed{k}_\perp\slashed{p}_2\gamma_i\hsi(0)\rangle_A
+\alpha_q\langle \hbsi(x_\bu,x_\perp)\slashed{p}_1\slashed{p}_2\gamma_i\hsi(0)\rangle_A\big]
\nonumber\\
&&\hspace{-1mm}
=~\int\! dx_\bu dx_\perp~e^{-i\alpha_qx_\bu+i(k,x)_\perp}
\Big[
-k_i\langle \hbsi(x_\bu,x_\perp)\slashed{p}_2\hsi(0)\rangle_A
+~\alpha_q{s\over 2} \langle \hbsi(x_\bu,x_\perp)\gamma_i\hsi(0)\rangle_A
\nonumber\\
&&\hspace{6mm}
-~i\epsilon_{ij}k^j\langle \hbsi(x_\bu,x_\perp)\slashed{p}_2\gamma_5\hsi(0)\rangle_A
+i{s\over 2}\alpha\epsilon_{ij}
\langle \hbsi(x_\bu,x_\perp)\gamma^j\gamma_5\hsi(0)\rangle_A\Big]
\nonumber\\
&&\hspace{14mm}
=~-k_i8\pi^3sf_1(\alpha_q,k_\perp)+8\pi^3s\alpha_q k_i\big[ f_\perp(\alpha_q,k_\perp)+g^\perp(\alpha_q,k_\perp)\big],
\label{tw3mael2}
\end{eqnarray}
where we used parametrizations (\ref{Amael}) and (\ref{mael5}) for the leading power matrix elements. 

Now, the second term in eq. (\ref{tw3mael2}) contains extra $\alpha_q$ with respect to the first term
\footnote{It can be demonstrated that $f_1(x,k_\perp^2)$, $f_\perp(x,k_\perp^2)$, and $f_\perp(x,k_\perp^2)$ have the same type of (logarithmic) behavior at low $x$. 
Indeed, the low-$x$ behavior is determined by interaction with pomeron.  This interaction is specified by so-called impact factor 
 and it is easy to check that the impact factor for all three TMDs is similar.
\label{foot:smallx}}
, so
 it should be neglected in our kinematical region $s\gg Q^2\gg q_\perp^2$  and we get 
\begin{eqnarray}
&&\hspace{-1mm}
{g\over 8\pi^3s}\!\int\! dx_\bu dx_\perp~e^{-i\alpha_qx_\bu+i(k,x)_\perp}\langle \bar\psi^f(x_\bu,x_\perp)\slashed{p}_2\brA_i(x_\bu,x_\perp)\psi^f(0)\rangle_A
\label{9.22}\\
&&\hspace{-1mm}
=~{-g\over 8\pi^3s}\!\int\! dx_\bu dx_\perp~e^{-i\alpha_qx_\bu+i(k,x)_\perp}\langle \bar\psi^f(x_\bu,x_\perp)\notA(x_\bu,x_\perp)\slashed{p}_2\gamma_i\psi^f(0)\rangle_A
~=-k_if_1^f(\alpha_q,k_\perp)
\nonumber
\end{eqnarray}
By complex conjugation
\begin{eqnarray}
&&\hspace{-1mm}
{g\over 8\pi^3s}\!\int\! dx_\perp dx_\bu~e^{-i\alpha_q x_\bu+i(k,x)_\perp}
\langle \hat{\bsi}_f(x_\bu,x_\perp)\brA_i(0)\slashed{p}_2\hsi_f(0)\rangle_A
\label{8.48}\\
&&\hspace{-1mm}
=~{-g\over 8\pi^3s}\!\int\! dx_\bu dx_\perp~e^{-i\alpha_qx_\bu+i(k,x)_\perp}\langle \bar\psi^f(x_\bu,x_\perp)\gamma_i\slashed{p}_2\notA(0)\psi^f(0)\rangle_A
=~-k_if_{1f}(\alpha_q,k_\perp).
\nonumber
\end{eqnarray}

For the  corresponding antiquark distributions we get 
\begin{eqnarray}
&&\hspace{0mm}
{g\over 8\pi^3s}\!\int\! dx_\perp dx_\bu~e^{-i\alpha x_\bu+i(k,x)_\perp}
\langle \bsi_f(0)\brA_i(x_\bu,x_\perp)\slashed{p}_2\hsi_f(x_\bu,x_\perp)\rangle_A
\nonumber\\
&&\hspace{5mm}
=~{1\over 8\pi^3s}\!\int\! dx_\bu dx_\perp e^{-i\alpha_qx_\bu+i(k,x)_\perp}\Big[-\langle \hat{\bar\psi}(0)\gamma_i\slashed{p}_2\slashed{k}_\perp\hat{\psi}(x_\bu,x_\perp)\rangle_A
\nonumber\\
&&\hspace{11mm}
-~i\langle \hbsi(0)\gamma_i\slashed{p}_2\notD_\perp\hsi(x_\bu,x_\perp)\rangle_A\Big]
~=~-k_i\barf_{1f}(\alpha_q,k_\perp)
\label{9.25}
\end{eqnarray}
and
\begin{equation}
\hspace{0mm}
{g\over 8\pi^3s}\!\int\! dx_\perp dx_\bu~e^{-i\alpha_q x_\bu+i(k,x)_\perp}
\langle \hat{\bsi}_f(0)\slashed{p}_2\brA_i(0)\hsi_f(x_\bu,x_\perp)\rangle_A
~=~
-k_i\barf_{1f}(\alpha_q,k_\perp) .
\label{9.26}
\end{equation}

The corresponding target matrix elements are obtained by trivial replacements 
$x_\star\leftrightarrow x_\bu$, $\alpha_q\leftrightarrow\beta_q$ and
$\slashed{p}_2\leftrightarrow\slashed{p}_1$.

Next, let us consider
\begin{eqnarray}
&&\hspace{-1mm}
{g\over 8\pi^3s}\!\int\! dx_\bu dx_\perp~e^{-i\alpha_qx_\bu+i(k,x)_\perp}
\langle \hsi(x_\bu,x_\perp)\slashed{p}_2\notA(x_\bu,x_\perp)\hsi(0)\rangle_A
\label{eqm1}\\
&&\hspace{11mm}
=~{1\over 8\pi^3s}\!\int\! dx_\bu dx_\perp~e^{-i\alpha_qx_\bu+i(k,x)_\perp}
\nonumber\\
&&\hspace{22mm}
\times~\Big[\langle \hsi(x_\bu,x_\perp)\slashed{k}_\perp\slashed{p}_2\hsi(0)\rangle_A
+i\langle \hsi(x_\bu,x_\perp)\stackrel{\leftarrow}{\notD}_\perp\slashed{p}_2\hsi(0)\rangle_A\Big].
\nonumber
\end{eqnarray}
Using QCD equation of motion and parametrization (\ref{hmael}), one can rewrite the r.h.s. of this equation as
\begin{eqnarray}
&&\hspace{-3mm}
{1\over 8\pi^3s}\!\int\! dx_\bu dx_\perp~e^{-i\alpha_qx_\bu+i(k,x)_\perp}
\Big[\langle \hbsi(x_\bu,x_\perp)\slashed{k}_\perp\slashed{p}_2\hsi(0)\rangle_A
+\alpha_q\langle \hbsi(x_\bu,x_\perp)\slashed{p}_1\slashed{p}_2\hsi(0)\rangle_A\Big]
\nonumber\\
&&\hspace{-1mm}
=~i{k_\perp^2\over m_N}h_{1}^\perp(\alpha_q,k_\perp)+\alpha_q m_N\big[e(\alpha,k_\perp)+i h(\alpha,k_\perp)\big].
\label{eqm2}
\end{eqnarray}
Again, only the first term contributes in our kinematical region so we finally get
\begin{eqnarray}
&&\hspace{-2mm}
{g\over 8\pi^3s}\!\int\! dx_\bu dx_\perp~e^{-i\alpha_qx_\bu+i(k,x)_\perp}
\langle \hbsi^f(x_\bu,x_\perp)\slashed{p}_2\notA(x_\bu,x_\perp)\hsi^f(0)\rangle_A
~=~i{k_\perp^2\over m_N}h_{1f}^\perp(\alpha_q,k_\perp).
\nonumber\\
\label{9.56}
\end{eqnarray}
By complex conjugation we obtain
\begin{eqnarray}
&&\hspace{-1mm}
{g\over 8\pi^3s}\!\int\! dx_\bu dx_\perp~e^{-i\alpha_qx_\bu+i(k,x)_\perp}
\langle \hbsi^f(x_\bu,x_\perp)\slashed{p}_2\notA(0)\hsi^f(0)\rangle_A
~=~i{k_\perp^2\over m_N}h_{1f}^\perp(\alpha_q,k_\perp).
\nonumber\\
\label{9.57}
\end{eqnarray}
For corresponding antiquark distributions one gets in a similar way
\begin{eqnarray}
&&\hspace{-2mm}
{g\over 8\pi^3s}\!\int\! dx_\bu dx_\perp~e^{-i\alpha_qx_\bu+i(k,x)_\perp}
\langle \hbsi^f(0)\slashed{p}_2\notA(x_\bu,x_\perp)\hsi^f(x_\bu,x_\perp)\rangle_A
=~
i{k_\perp^2\over m_N}\barh_{1f}^\perp(\alpha_q,k_\perp),
\nonumber\\
&&\hspace{-2mm}
{g\over 8\pi^3s}\!\int\! dx_\bu dx_\perp~e^{-i\alpha_qx_\bu+i(k,x)_\perp}
\langle \hbsi^f(0)\slashed{p}_2\notA(0)\hsi^f(x_\bu,x_\perp)\rangle_A
=~
i{k_\perp^2\over m_N}\barh_{1f}^\perp(\alpha_q,k_\perp).
\nonumber\\
\label{9.58}
\end{eqnarray}
The target matrix elements are obtained by usual replacements 
$x_\star\leftrightarrow x_\bu$, $\alpha_q\leftrightarrow\beta_q$ and
$\slashed{p}_2\leftrightarrow\slashed{p}_1$.

For the Z-boson hadronic tensor we need these operators with extra $\gamma_5$. From formula (\ref{tw3mael2}) we see that
\begin{eqnarray}
&&\hspace{-1mm}
{1\over 8\pi^3s}\!\int\! dx_\bu dx_\perp~e^{-i\alpha_qx_\bu+i(k,x)_\perp}
\langle \bar\psi^f(x_\bu,x_\perp)\slashed{p}_2\brA_i(x_\bu,x_\perp)\gamma_5\psi^f(0)\rangle_A=-i\epsilon_{ij}k^jf_1^f(\alpha_q,k_\perp)
\nonumber\\
&&\hspace{-1mm}
{1\over 8\pi^3s}\!\int\! dx_\bu dx_\perp~e^{-i\alpha_qx_\bu+i(k,x)_\perp}
\langle \bar\psi^f(x_\bu,x_\perp)\brA_i(0)\slashed{p}_2\gamma_5\psi^f(0)\rangle_A=~i\epsilon_{ij}k^jf_1^f(\alpha_q,k_\perp)
\label{gamma5scancel}
\end{eqnarray}

Finally, we need
\begin{eqnarray}
&&\hspace{-1mm}
{1\over 8\pi^3s}\!\int\! dx_\bu dx_\perp~e^{-i\alpha_qx_\bu+i(k,x)_\perp}
\langle \hsi(x_\bu,x_\perp)\notA(x_\bu,x_\perp)\slashed{p}_2\notA(0)\hsi(0)\rangle_A
\nonumber\\
&&\hspace{1mm}
=~{1\over 8\pi^3s}\!\int\! dx_\bu dx_\perp~e^{-i\alpha_qx_\bu+i(k,x)_\perp}
\langle \hsi(x_\bu,x_\perp)\Big(\slashed{k}_\perp
+i\stackrel{\leftarrow}{\notD}\big)\slashed{p}_2
\big(\slashed{k}_\perp
-i{\notD}\big)\hat{\psi}(0)\rangle_A
\nonumber\\
&&\hspace{22mm}
=~{k_\perp^2\over 16\pi^3}f_1(\alpha_q,k_\perp)~+~O(\alpha_q,\beta_q)
\label{AA1}
\end{eqnarray}
and similarly
\begin{eqnarray}
&&\hspace{-1mm}
{1\over 8\pi^3s}\!\int\! dx_\bu dx_\perp~e^{-i\alpha_qx_\bu+i(k,x)_\perp}
\langle \hsi(x_\bu,x_\perp)\notA(x_\bu,x_\perp)\sigma_{\star i}\notA(0)\hsi(0)\rangle_A
\nonumber\\
&&\hspace{1mm}
=~{1\over 8\pi^3s}\!\int\! dx_\bu dx_\perp~e^{-i\alpha_qx_\bu+i(k,x)_\perp}
\langle \hsi(x_\bu,x_\perp)\Big(\slashed{k}_\perp
+i\stackrel{\leftarrow}{\notD}\big)\sigma_{\star i}
\big(\slashed{k}_\perp
-i{\notD}\big)\hat{\psi}(0)\rangle_A
\nonumber\\
&&\hspace{1mm}
=~{1\over 16\pi^3}{k_ik_\perp^2\over m}h_1^\perp(\alpha_q,k_\perp)~+~O(\alpha_q,\beta_q)
\label{AA2}
\end{eqnarray}

For corresponding antiquark distributions we get
\begin{eqnarray}
&&\hspace{-1mm}
{g\over 8\pi^3s}\!\int\! dx_\bu dx_\perp~e^{-i\alpha_qx_\bu+i(k,x)_\perp}
\langle \hsi(0)\notA(0)\slashed{p}_2\notA(x_\bu,x_\perp)\hsi(x_\bu,x_\perp)\rangle_A
\nonumber\\
&&\hspace{22mm}
=~-{k_\perp^2\over 16\pi^3}\barf_1(\alpha_q,k_\perp)~+~O(\alpha_q,\beta_q)
\nonumber\\
&&\hspace{-1mm}
{g\over 8\pi^3s}\!\int\! dx_\bu dx_\perp~e^{-i\alpha_qx_\bu+i(k,x)_\perp}
\langle \hsi(x_\bu,x_\perp)\notA(x_\bu,x_\perp)\sigma_{\star i}\notA(0)\hsi(0)\rangle_A
\nonumber\\
&&\hspace{22mm}
=~-{1\over 16\pi^3}{k_ik_\perp^2\over m}h_1^\perp(\alpha_q,k_\perp)~+~O(\alpha_q,\beta_q)
\end{eqnarray}

Also, as we saw in Sect. \ref{sec:twogluonshoton}, at the leading order in $N_c$ there is one quark-antiquark-gluon operator that does not reduce to twist-2 distributions. It can be parametrized as follows
\begin{eqnarray}
&&\hspace{-1mm}
{1\over 16\pi^3}{2\over s}\!\int\!dx_\bu d^2x_\perp~e^{-i\alpha x_\bu+i(k,x)_\perp}
~\langle \hbsi_f(x_\bu,x_\perp)\big[A_i(x)\sigma_{\star j}-\half g_{ij}A^k\sigma_{\star k}(x)\big]\hsi_f(0)\rangle_A~
\nonumber\\
&&\hspace{33mm}
=~-(k_ik_j+\half g_{ij}k_\perp^2){1\over m}h_A^f(\alpha,k_\perp),
\nonumber\\
&&\hspace{-1mm}
{1\over 16\pi^3}{2\over s}\!\int\!dx_\bu d^2x_\perp~e^{-i\alpha x_\bu+i(k,x)_\perp}
~\langle \hbsi_f(0)[A_i(0)\sigma_{\bu j}-\half g_{ij}A^k\sigma_{\bu k}(0)]\hsi_f(x_\bu,x_\perp)\rangle_A~
\nonumber\\
&&\hspace{33mm}
=~-(k_ik_j+\half g_{ij}k_\perp^2){1\over m}\barh_A^f(\alpha,k_\perp)
\label{maelsa}
\end{eqnarray}
and similarly for the target matrix element.

\subsection{Matrix elements for exchange-type power corrections \label{sec:maexs}}

We  parametrize ``exchange'' TMDs  as follows
\begin{eqnarray}
&&\hspace{-1mm}
{1\over 4\pi^3s^2}\!\int\! d^2x_\perp dx_\bu~e^{-i\alpha x_\bu+i(k,x)_\perp}
\!\int_{-\infty}^{x_\bu}\! dx'_\bu\langle\bsi (x_\bu,x_\perp)\breF_{\star i}(0)\slashed{p}_2\psi(x'_\bu,x_\perp)\rangle_A
~=~k_ij_1(\alpha,k_\perp),
\nonumber\\
&&\hspace{-1mm}
{1\over 4\pi^3s^2}\!\int\! d^2x_\perp dx_\bu~e^{-i\alpha x_\bu+i(k,x)_\perp}
\!\int_{-\infty}^{x_\bu}\! dx'_\bu~\langle \bsi (x_\bu,x_\perp)\slashed{p}_2\breF_{\star i}(0)\psi(x'_\bu,x_\perp)\rangle_A
~=~k_ij_2(\alpha,k_\perp),
\nonumber\\
&&\hspace{-1mm}
{1\over 4\pi^3s^2}\!\int\! d^2x_\perp dx_\bu~e^{-i\alpha x_\bu+i(k,x)_\perp}
\!\int_{-\infty}^{x_\bu}\! dx'_\bu~\langle\bsi (x'_\bu,x_\perp)\slashed{p}_2\breF_{\star i}(0)\psi(x_\bu,x_\perp)\rangle_A
~=~k_i\barj_1(\alpha,k_\perp),
\nonumber\\
&&\hspace{-1mm}
{1\over 4\pi^3s^2}\!\int\! d^2x_\perp dx_\bu~e^{-i\alpha x_\bu+i(k,x)_\perp}
\!\int_{-\infty}^{x_\bu}\! dx'_\bu~\langle\bsi (x'_\bu,x_\perp)\breF_{\star i}(0)\slashed{p}_2\psi(x_\bu,x_\perp)\rangle_A
~=~k_i\barj_2(\alpha,k_\perp)
\nonumber\\
\label{paramj}
\end{eqnarray}
where $\breF_{\mu\nu}~\equiv~F_{\mu\nu}-i\gamma_5\tilde{F}_{\mu\nu}$. 
By complex conjugation we get
\begin{eqnarray}
&&\hspace{-1mm}
{1\over 4\pi^3s^2}\!\int\! d^2x_\perp dx_\bu~e^{-i\alpha x_\bu+i(k,x)_\perp}
\!\int_{-\infty}^0\! dx'_\bu~\langle\bar\psi(x'_\bu,0_\perp)\slashed{p}_2\breF_{\star i}(x_\bu,x_\perp)\psi(0)\rangle_A
~=~k_i j_1^\ast(\alpha,k_\perp),
\nonumber\\
&&\hspace{-1mm}
{1\over 4\pi^3s^2}\!\int\! d^2x_\perp dx_\bu~e^{-i\alpha x_\bu+i(k,x)_\perp}
\!\int_{-\infty}^0\! dx'_\bu~\langle\bar\psi(x'_\bu,0_\perp)\breF_{\star i}(x_\bu,x_\perp)\slashed{p}_2\psi(0)\rangle_A
~=~k_ij_2^\ast(\alpha,k_\perp),
\nonumber\\
&&\hspace{-1mm}
{1\over 4\pi^3s^2}\!\int\! d^2x_\perp dx_\bu~e^{-i\alpha x_\bu+i(k,x)_\perp}
\!\int_{-\infty}^0\! dx'_\bu\langle\bar\psi(0)\breF_{\star i}(x_\bu,x_\perp)\slashed{p}_2\psi(x'_\bu,0_\perp)\rangle_A
~=~k_i \barj_1^\ast(\alpha,k_\perp),
\nonumber\\
&&\hspace{-1mm}
{1\over 4\pi^3s^2}\!\int\! d^2x_\perp dx_\bu~e^{-i\alpha x_\bu+i(k,x)_\perp}
\!\int_{-\infty}^0\! dx'_\bu\langle\bar\psi(0)\slashed{p}_2\breF_{\star i}(x_\bu,x_\perp))\psi(x'_\bu,0_\perp)\rangle_A
~=~k_i\barj_2^\ast(\alpha,k_\perp)
\nonumber\\
\label{paramjstar}
\end{eqnarray}
Note that unlike two-quark matrix elements,
quark-quark-gluon ones may have  imaginary parts.
Target matrix elements are obtained by usual substitutions 
$\alpha\leftrightarrow\beta$, $\slashed{p}_2\leftrightarrow\slashed{p}_1$, $x_\bu\leftrightarrow x_\star$, and $\breF_{\star i}\leftrightarrow \breF_{\bu i}$.
\footnote{
For completeness let us present the explicit form of the gauge links in an arbitrary gauge:
\begin{eqnarray*}
&&\hspace{-3mm}
\bsi (x'_\bu,x_\perp)
F_{\star i}(0)\psi(x_\bu,x_\perp)
~\rightarrow~\bsi (x'_\bu,x_\perp)[x'_\bu,-\infty_\bu]_x[x_\perp,0_\perp]_{-\infty_\bu}
\\
&&\hspace{40mm}
\times~[-\infty_\bu,0]_{0_\perp}F_{\star i}(0)
[0,-\infty_\bu]_{0_\perp}[0_\perp,x_\perp]_{-\infty_\bu}[-\infty_\bu,x_\bu]_x\psi(x_\bu,x_\perp).
\nonumber
\end{eqnarray*}
}

Let us now consider the  corresponding matrix elements from Sect. \ref{sec:extype}. As shown in Ref. \cite{Balitsky:2020jzt} ,
Eqs. (\ref{paramj}) and (\ref{paramjstar}) lead to (here $x=(x_\bu,x_\perp)$)
\begin{eqnarray}
&&\hspace{-3mm}
{1\over 8\pi^3s}\!\int\! d^2x_\perp dx_\bu~e^{-i\alpha x_\bu+i(k,x)_\perp}
\langle\bsi(x)
\bigg[\begin{array}{c}
\brA_i(0)\slashed{p}_2\\
\slashed{p}_2\brA_i(0)
\end{array}\bigg]
{1\over\alpha}\psi(x)\rangle_A~
=~{k_i\over\alpha}
\Big[
\begin{array}{c}j_1(\alpha,k_\perp)\\
j_2(\alpha,k_\perp)
\end{array}\Big],
\nonumber\\
&&\hspace{-3mm}
{1\over 8\pi^3s}\!\int\! d^2x_\perp dx_\bu~e^{-i\alpha x_\bu+i(k,x)_\perp}
\langle\big(\bsi {1\over\alpha}\big)(x_\bu,x_\perp)
\bigg[
\begin{array}{c}
\slashed{p}_2\brA_i(0)\\
\brA_i(0)\slashed{p}_2
\end{array}\bigg]
\psi(x_\bu,x_\perp)\rangle_A~
=~-{k_i\over\alpha}\Big[
\begin{array}{c}
\barj_1(\alpha,k_\perp)\\
\barj_2(\alpha,k_\perp)
\end{array}\Big]
\nonumber\\
&&\hspace{-3mm}
{1\over 8\pi^3s}\!\int\! dx_\bu~e^{-i\alpha x_\bu+i(k,x)_\perp}
\langle\bsi(0)
\bigg[\begin{array}{c}
\brA_i(x)\slashed{p}_2\\
\slashed{p}_2\brA_i(x)
\end{array}\bigg]
{1\over\alpha}\psi(0)\rangle_A
~=~-{k_i\over\alpha}\Big[
\begin{array}{c}
\barj_1^\ast(\alpha,k_\perp)\\
\barj_2^\ast(\alpha,k_\perp)
\end{array}\Big],
\nonumber\\
&&\hspace{-3mm}
{1\over 8\pi^3s}\!\int\! d^2x_\perp dx_\bu~e^{-i\alpha x_\bu+i(k,x)_\perp}
\langle\big(\bsi {1\over\alpha}\big)(0)
\bigg[\begin{array}{c}
\slashed{p}_2\brA_i(x)\\
\brA_i(x)\slashed{p}_2
\end{array}
\bigg]
\psi(0)\rangle_A
~=~{k_i\over\alpha}
\Big[\begin{array}{c}j_1^\ast(\alpha,k_\perp)\\
j_2^\ast(\alpha,k_\perp)
\end{array}\Big]
\label{paramjse}
\end{eqnarray}
For the target matrix elements, we obtain ($x\equiv x_\star,x_\perp$)
\begin{eqnarray}
&&\hspace{-1mm}
{1\over 8\pi^3s}\!\int\! d^2x_\perp dx_\star~e^{-i\beta x_\star+i(k,x)_\perp}\langle\bsi(x)
\bigg[
\begin{array}{c}
\breB_i(0)\slashed{p}_1\\
\slashed{p}_1\breB_i(0)
\end{array}\bigg]
{1\over\beta}\psi(x\rangle_B
~=~{k_i\over\beta}\Big[\begin{array}{c}j_1(\beta,k_\perp)\\
j_2(\beta,k_\perp)
\end{array}\Big],
\nonumber\\
&&\hspace{-1mm}
{1\over 8\pi^3s}\!\int\! d^2x_\perp dx_\star~e^{-i\alpha x_\star+i(k,x)_\perp}
\langle\big(\bsi {1\over\beta}\big)(x)
\bigg[
\begin{array}{c}
\slashed{p}_1\breB_i(0)\\
\breB_i(0)\slashed{p}_1
\end{array}\bigg]
\psi(x)\rangle_B
~=~-{k_i\over\beta}\Big[
\begin{array}{c}
\barj_1(\beta,k_\perp)\\
\barj_2(\beta,k_\perp)
\end{array}\Big],
\nonumber\\
&&\hspace{-1mm}
{1\over 8\pi^3s}\!\int\! d^2x_\perp dx_\star~e^{-i\beta x_\star+i(k,x)_\perp}
\langle\bsi(0)
\bigg[
\begin{array}{c}
\breB_i(x)\slashed{p}_1\\
\slashed{p}_1\breB_i(x)
\end{array}\bigg]
{1\over\beta}\psi(0)\rangle_B~
=~-{k_i\over\beta}\Big[
\begin{array}{c}
\barj_1^\ast(\beta,k_\perp)\\
\barj_2^\ast(\beta,k_\perp)
\end{array}\Big],
\nonumber\\
&&\hspace{-1mm}
{1\over 8\pi^3s}\!\int\! d^2x_\perp dx_\star~e^{-i\beta x_\star+i(k,x)_\perp}
\langle\big(\bsi {1\over\beta}\big)(0)
\bigg[
\begin{array}{c}
\slashed{p}_1\breB_i(x)\\
\breB_i(x)\slashed{p}_1
\end{array}\Big]
\psi(0)\rangle_B
~=~{k_i\over\beta}\Big[\begin{array}{c}j_1^\ast(\beta,k_\perp)\\
j_2^\ast(\beta,k_\perp)
\end{array}\Big]
\label{paramjset}
\end{eqnarray}

We need also parametrization of matrix elements with extra $\gamma_5$. Since $\brA_i\gamma_5=i\epsilon_{ij}\brA^j$ and 
$\breB_i\gamma_5=i\epsilon_{ij}\breB^j$ we get from Eqs. (\ref{paramjse}), (\ref{paramjset})
\begin{eqnarray}
&&\hspace{-3mm}
{1\over 8\pi^3s}\!\int\! d^2x_\perp dx_\bu~e^{-i\alpha x_\bu+i(k,x)_\perp}
\langle\bsi(x)
\bigg[\begin{array}{c}
\brA_i(0)\slashed{p}_2\\
\slashed{p}_2\brA_i(0)
\end{array}\bigg]
{\gamma_5\over\alpha}\psi(x_\bu,x_\perp)\rangle_A~
=~i\epsilon_{ij}{k^j\over\alpha}
\Big[
\begin{array}{c}
-j_1(\alpha,k_\perp)\\
j_2(\alpha,k_\perp)
\end{array}\Big],
\nonumber\\
&&\hspace{-3mm}
{1\over 8\pi^3s}\!\int\! d^2x_\perp dx_\bu~e^{-i\alpha x_\bu+i(k,x)_\perp}
\langle\big(\bsi {1\over\alpha}\big)(x)
\bigg[
\begin{array}{c}
\slashed{p}_2\brA_i(0)\\
\brA_i(0)\slashed{p}_2
\end{array}\bigg]\gamma_5
\psi(x)\rangle_A~
=~i\epsilon_{ij}{k^j\over\alpha}\Big[
\begin{array}{c}
-\barj_1(\alpha,k_\perp)\\
\barj_2(\alpha,k_\perp)
\end{array}\Big]
\nonumber\\
&&\hspace{-3mm}
{1\over 8\pi^3s}\!\int\! dx_\bu~e^{-i\alpha x_\bu+i(k,x)_\perp}
\langle\bsi(0)
\bigg[\begin{array}{c}
\brA_i(x)\slashed{p}_2\\
\slashed{p}_2\brA_i(x)
\end{array}\bigg]
{\gamma_5\over\alpha}\psi(0)\rangle_A
~=~i\epsilon_{ij}{k^j\over\alpha}\Big[
\begin{array}{c}
\barj_1^\ast(\alpha,k_\perp)\\
-\barj_2^\ast(\alpha,k_\perp)
\end{array}\Big],
\\
&&\hspace{-3mm}
{1\over 8\pi^3s}\!\int\! d^2x_\perp dx_\bu~e^{-i\alpha x_\bu+i(k,x)_\perp}
\langle\big(\bsi {1\over\alpha}\big)(0)
\bigg[\begin{array}{c}
\slashed{p}_2\brA_i(x)\\
\brA_i(x)\slashed{p}_2
\end{array}
\bigg]\gamma_5
\psi(0)\rangle_A
~=~i\epsilon_{ij}{k^j\over\alpha}
\Big[\begin{array}{c}
j_1^\ast(\alpha,k_\perp)\\
-j_2^\ast(\alpha,k_\perp)
\end{array}\Big],
\nonumber
\end{eqnarray}
\begin{eqnarray}
&&\hspace{-3mm}
{1\over 8\pi^3s}\!\int\! d^2x_\perp dx_\bu~e^{-i\alpha x_\bu+i(k,x)_\perp}
\langle\bsi(x)
\bigg[\begin{array}{c}
\brA_i(0)\slashed{p}_2\\
\slashed{p}_2\brA_i(0)
\end{array}\bigg]
{\gamma_5\over\alpha}\psi(x)\rangle_A~
=~i\epsilon_{ij}{k^j\over\alpha}
\Big[
\begin{array}{c}
-j_1(\alpha,k_\perp)\\
j_2(\alpha,k_\perp)
\end{array}\Big],
\nonumber\\
&&\hspace{-3mm}
{1\over 8\pi^3s}\!\int\! d^2x_\perp dx_\bu~e^{-i\alpha x_\bu+i(k,x)_\perp}
\langle\big(\bsi {1\over\alpha}\big)(x)
\bigg[
\begin{array}{c}
\slashed{p}_2\brA_i(0)\\
\brA_i(0)\slashed{p}_2
\end{array}\bigg]\gamma_5
\psi(x)\rangle_A~
=~i\epsilon_{ij}{k^j\over\alpha}\Big[
\begin{array}{c}
-\barj_1(\alpha,k_\perp)\\
\barj_2(\alpha,k_\perp)
\end{array}\Big]
\nonumber\\
&&\hspace{-3mm}
{1\over 8\pi^3s}\!\int\! dx_\bu~e^{-i\alpha x_\bu+i(k,x)_\perp}
\langle\bsi(0)
\bigg[\begin{array}{c}
\brA_i(x)\slashed{p}_2\\
\slashed{p}_2\brA_i(x)
\end{array}\bigg]
{\gamma_5\over\alpha}\psi(0)\rangle_A
~=~i\epsilon_{ij}{k^j\over\alpha}\Big[
\begin{array}{c}
\barj_1^\ast(\alpha,k_\perp)\\
-\barj_2^\ast(\alpha,k_\perp)
\end{array}\Big],
\label{paramjse5}\\
&&\hspace{-3mm}
{1\over 8\pi^3s}\!\int\! d^2x_\perp dx_\bu~e^{-i\alpha x_\bu+i(k,x)_\perp}
\langle\big(\bsi {1\over\alpha}\big)(0)
\bigg[\begin{array}{c}
\slashed{p}_2\brA_i(x)\\
\brA_i(x)\slashed{p}_2
\end{array}
\bigg]\gamma_5
\psi(0)\rangle_A
~=~i\epsilon_{ij}{k^j\over\alpha}
\Big[\begin{array}{c}
j_1^\ast(\alpha,k_\perp)\\
-j_2^\ast(\alpha,k_\perp)
\end{array}\Big]
\nonumber
\end{eqnarray}
where $x\equiv x_\bu,x_\ast$, 
and
\begin{eqnarray}
&&\hspace{-1mm}
{1\over 8\pi^3s}\!\int\! d^2x_\perp dx_\star~e^{-i\beta x_\star+i(k,x)_\perp}\langle\bsi(x)
\bigg[
\begin{array}{c}
\breB_i(0)\slashed{p}_1\\
\slashed{p}_1\breB_i(0)
\end{array}\bigg]
{\gamma_5\over\beta}\psi(x)\rangle_B
~=~i\epsilon_{ij}{k^j\over\beta}\Big[\begin{array}{c}
-j_1(\beta,k_\perp)\\
j_2(\beta,k_\perp)
\end{array}\Big],
\label{paramjset5}\\
&&\hspace{-1mm}
{1\over 8\pi^3s}\!\int\! d^2x_\perp dx_\star~e^{-i\alpha x_\star+i(k,x)_\perp}
\langle\big(\bsi {1\over\beta}\big)(x)
\bigg[
\begin{array}{c}
\slashed{p}_1\breB_i(0)\\
\breB_i(0)\slashed{p}_1
\end{array}\bigg]\gamma_5
\psi(x)\rangle_B
~=~i\epsilon_{ij}{k^j\over\beta}\Big[
\begin{array}{c}
-\barj_1(\beta,k_\perp)\\
\barj_2(\beta,k_\perp)
\end{array}\Big],
\nonumber\\
&&\hspace{-1mm}
{1\over 8\pi^3s}\!\int\! d^2x_\perp dx_\star~e^{-i\beta x_\star+i(k,x)_\perp}
\langle\bsi(0)
\bigg[
\begin{array}{c}
\breB_i(x)\slashed{p}_1\\
\slashed{p}_1\breB_i(x)
\end{array}\bigg]
{\gamma_5\over\beta}\psi(0)\rangle_B~
=~i\epsilon_{ij}{k^j\over\beta}\Big[
\begin{array}{c}
\barj_1^\ast(\beta,k_\perp)\\
-\barj_2^\ast(\beta,k_\perp)
\end{array}\Big],
\nonumber\\
&&\hspace{-1mm}
{1\over 8\pi^3s}\!\int\! d^2x_\perp dx_\star~e^{-i\beta x_\star+i(k,x)_\perp}
\langle\big(\bsi {1\over\beta}\big)(0)
\bigg[
\begin{array}{c}
\slashed{p}_1\breB_i(x)\\
\breB_i(x)\slashed{p}_1
\end{array}\bigg]\gamma_5
\psi(0)\rangle_B
~=~i\epsilon_{ij}{k^j\over\beta}\Big[\begin{array}{c}j_1^\ast(\beta,k_\perp)\\
-j_2^\ast(\beta,k_\perp)
\end{array}\Big]
\nonumber
\end{eqnarray}
where $x\equiv x_\star,x_\ast$

It is convenient to introduce the following combinations for the parametrization of the exchange-type power corrections
\begin{eqnarray}
&&\hspace{-1mm}
J^{1ff'}_{++}(q,k_\perp)~=~(j_1+\barj_1)^f(\alpha_q,k_\perp)(j^\ast_1+\barj^\ast_1)^{f'}
(\beta_q,(q-k)_\perp)~+~{\rm c.c.},
\nonumber\\
&&\hspace{-1mm}
J^{2ff'}_{++}(q,k_\perp)~=~(j_2+\barj_2)^f(\alpha_q,k_\perp)(j^\ast_2+\barj^\ast_2)^{f'}
(\beta_q,(q-k)_\perp)~+~{\rm c.c.},
\nonumber\\
&&\hspace{-1mm}
J^{1ff'}_{--}(q,k_\perp)~=~(j_1-\barj_1)^f(\alpha_q,k_\perp)(j^\ast_1-\barj^\ast_1)^{f'}
(\beta_q,(q-k)_\perp)~+~{\rm c.c.},
\nonumber\\
&&\hspace{-1mm}
J^{2ff'}_{--}(q,k_\perp)~=~(j_2-\barj_2)^f(\alpha_q,k_\perp)(j^\ast_2-\barj^\ast_2)^{f'}
(\beta_q,(q-k)_\perp)~+~{\rm c.c.},
\nonumber\\
&&\hspace{-1mm}
J^{1ff'}_{+-}(q,k_\perp)~=~(j_1+\barj_1)^f(\alpha_q,k_\perp)(j^\ast_1-\barj^\ast_1)^{f'}(\beta_q,(q-k)_\perp)~+~{\rm c.c.},
\nonumber\\
&&\hspace{-1mm}
J^{1ff'}_{-+}(q,k_\perp)~=~(j_1-\barj_1)^f(\alpha_q,k_\perp)(j^\ast_1+\barj^\ast_1)^{f'}(\beta_q,(q-k)_\perp)~+~{\rm c.c.},
\nonumber\\
&&\hspace{-1mm}
J^{2ff'}_{+-}(q,k_\perp)~=~(j_2+\barj_2)^f(\alpha_q,k_\perp)(j^\ast_2-\barj^\ast_2)^{f'}(\beta_q,(q-k)_\perp)~+~{\rm c.c.},
\nonumber\\
&&\hspace{-1mm}
J^{2ff'}_{-+}(q,k_\perp)~=~(j_2-\barj_2)^f(\alpha_q,k_\perp)(j^\ast_2+\barj^\ast_2)^{f'}(\beta_q,(q-k)_\perp)~+~{\rm c.c.},
\label{Js}
\end{eqnarray}
and
\begin{eqnarray}
&&\hspace{-1mm}
I^{1ff'}_{--}(q,k_\perp)~=~i(j_1-\barj_1)^f(\alpha_q,k_\perp)(j^\ast_1-\barj^\ast_1)^{f'}
(\beta_q,(q-k)_\perp)~+~{\rm c.c.},
\nonumber\\
&&\hspace{-1mm}
I^{2ff'}_{--}(q,k_\perp)~=~i(j_2-\barj_2)^f(\alpha_q,k_\perp)(j^\ast_2-\barj^\ast_2)^{f'}
(\beta_q,(q-k)_\perp)~+~{\rm c.c.},
\nonumber\\
&&\hspace{-1mm}
I^{1ff'}_{+-}(q,k_\perp)~=~i(j_1+\barj_1)^f(\alpha_q,k_\perp)(j^\ast_1-\barj^\ast_1)^{f'}
(\beta_q,(q-k)_\perp)~+~{\rm c.c.},
\nonumber\\
&&\hspace{-1mm}
I^{1ff'}_{-+}(q,k_\perp)~=~i(j_1-\barj_1)^f(\alpha_q,k_\perp)(j^\ast_1+\barj^\ast_1)^{f'}
(\beta_q,(q-k)_\perp)~+~{\rm c.c.},
\nonumber\\
&&\hspace{-1mm}
I^{2ff'}_{+-}(q,k_\perp)~=~i(j_2+\barj_2)^f(\alpha_q,k_\perp)(j^\ast_2-\barj^\ast_2)^{f'}
(\beta_q,(q-k)_\perp)~+~{\rm c.c.}
\nonumber\\
&&\hspace{-1mm}
I^{2ff'}_{-+}(q,k_\perp)~=~i(j_2-\barj_2)^f(\alpha_q,k_\perp)(j^\ast_2+\barj^\ast_2)^{f'}
(\beta_q,(q-k)_\perp)~+~{\rm c.c.}
\label{Is}
\end{eqnarray}
%

\subsection{Products of leptonic tensor and hadronic tensors' structures \label{sec:lhtproducts}}
The lepton momenta in the Collins-Soper frame are parametrized as
\begin{equation}
\hspace{0mm}
l~=~{Q\over 2}(1,\sin\theta\cos\phi,\sin\theta\sin\phi,\cos\theta),~~~
l'~=~{Q\over 2}(1,-\sin\theta\cos\phi,-\sin\theta\sin\phi,-\cos\theta)
\end{equation}
The  products of the symmetric leptonic tensor with unit vectors $X$, $Y$, and $Z$ are
\begin{eqnarray}
&&\hspace{-1mm}
L^{\mu\nu}\big({q_\mu q_\nu\over q^2}-g_{\mu\nu}\big)~=~2{(q\cdot l)(q\cdot l')\over q^2}+l\cdot l'
~=~Q^2
\nonumber\\
&&\hspace{-1mm}
L^{\mu\nu}X_\mu X_\nu~=~2(l\cdot X)(l'\cdot X)+l\cdot l'~=~{Q^2\over 2}\big[-\sin^2\theta\cos^2\phi+1\big]
\nonumber\\
&&\hspace{-1mm}
L^{\mu\nu}Z_\mu Z_\nu~=~2(l\cdot Z)(l'\cdot Z)+l\cdot l'~=~{Q^2\over 2}\big[-\cos^2\theta+1\big]
~=~{Q^2\over 2}\sin^2\theta
\nonumber\\
&&\hspace{-1mm}
L^{\mu\nu}X_\mu Z_\nu~=~(l\cdot X)(l'\cdot Z)+(l'\cdot X)(l\cdot Z)
~=~-{Q^2\over 4}\sin2\theta\cos\phi
\nonumber\\
&&\hspace{-1mm}
L^{\mu\nu}Y_\mu Z_\nu~=~(l\cdot Y)(l'\cdot Z)+(l'\cdot Y)(l\cdot Z)~=~-{Q^2\over 4}\sin 2\theta\sin\phi
\nonumber\\
&&\hspace{-1mm}
L^{\mu\nu}Y_\mu X_\nu~=~(l\cdot Y)(l'\cdot X)+(l'\cdot Y)(l\cdot X)~=~-{Q^2\over 4}\sin^2\theta\sin2\phi
\label{WHdec}
\end{eqnarray}
and therefore
\begin{eqnarray}
&&\hspace{-1mm}
{1\over Q^2}L^{\mu\nu}\Big(-\big(g_{\mu\nu}-{q_\mu q_\nu\over q^2}\big)(W_T+W_{\dd})
-2X_\mu X_\nu W_\dd
\label{WFdec}\\
&&\hspace{14mm}
+~Z_\mu Z_\nu(W_L-W_T-W_\dd)
-(X_\mu Z_\nu+X_\nu Z_\mu)W_\Delta\Big)
\nonumber\\
&&\hspace{-1mm}
=~\half(W_T+W_L)\bigg[1+\cos^2\theta{W_T-W_L\over W_T+W_L}
+\sin2\theta\sin\phi{W_\Delta\over W_T+W_L}
+\sin^2\theta\cos2\phi {W_\dd\over W_T+W_L}\bigg]
\nonumber
\end{eqnarray}
We also need  products of antisymmetric leptonic tensor with hadron sctructures
\begin{eqnarray}
&&\hspace{-5mm}
\epsilon_{\mu\nu\alpha\beta}l^\alpha {l'}^\beta\epsilon^{\mu\nu\lambda\rho}q_\lambda Z_\rho~
\\
&&\hspace{-5mm}=~
-2[(l\cdot q)(l'\cdot Z)-(l'\cdot q)(l\cdot Z)]~=~Q[(l\cdot Z)-(l'\cdot Z)]~=~-Q^2\cos\theta,
\nonumber\\
&&\hspace{-5mm}
\epsilon_{\mu\nu\alpha\beta}l^\alpha {l'}^\beta\epsilon^{\mu\nu\lambda\rho}q_\lambda X_\rho~
\nonumber\\
&&\hspace{-5mm}=~
-2[(l\cdot q)(l'\cdot X)-(l'\cdot q)(l\cdot X)]~=~Q[(l\cdot X)-(l'\cdot X)]~=~-Q^2\sin\theta\cos\phi
\nonumber
\end{eqnarray}
and therefore ($Q_\parallel=\sqrt{Q^2+Q_\perp^2}$)
\begin{eqnarray}
&&\hspace{-1mm}
\epsilon^{\mu\nu\alpha\beta}l_\alpha {l'}_\beta\epsilon_{\mu\nu\lambda\rho}q^\lambda
\Big[{Z^\rho\over Q_\parallel}\pizW_1^{\rmF f}(q)
+X^\rho{Q_\perp\over Q_\parallel Q}\pizW_2^{\rmF f}(q)\Big]
\nonumber\\
&&\hspace{-1mm}
=~-Q^2\Big[\pizW_1^{\rmF f}(q)\cos\theta+{Q_\perp\over Q}\pizW_2^{\rmF f}(q)\sin\theta\cos\phi\Big]\Big[1+O\big({Q_\perp^2\over Q^2}\big)\Big].
\end{eqnarray}
\bibliography{fact1}
\bibliographystyle{JHEP}

\end{document}